%% file: PhDThesis.tex
\documentclass[a4paper]{book}
\usepackage{a4wide}
\usepackage{amsmath,amsthm,amssymb,amsfonts}
\usepackage[T1]{fontenc}
\usepackage{mathptmx}
\usepackage[pdftex,pdftitle={Dynamical Parton Distributions of the Nucleon up to NNLO of QCD}, pdfauthor={P. Jimenez--Delgado}, colorlinks=true, linkcolor=black, citecolor=black, urlcolor=black]{hyperref}
\usepackage[all]{hypcap}
\usepackage{graphics,graphicx}
\usepackage{rotating,subfigure}
\usepackage{float}
\usepackage[numbers,sort&compress]{natbib}
\usepackage[small]{caption}
\usepackage{color}
\definecolor{red}{rgb}{1,0,0}
\definecolor{blue}{rgb}{0,0,0.4}
\definecolor{darkred}{rgb}{0.6,0.2,0.2}
\usepackage{fancyhdr}

\renewcommand{\arraystretch}{1.2}

\begin{document}
\fontsize{11}{15}
\selectfont
\pagestyle{empty}
\title{\vspace{-2cm}\fontsize{28}{45} \bf Dynamical Parton Distributions\\ of the Nucleon up to NNLO of QCD}
\author{\\[4cm]\Large{ \bf Dissertation}\\[0.5cm]
zur Erlangung des Grades eines\\
Doktors der Naturwissenschaften\\
der Abteilung Physik\\
der Technische Universit\"at Dortmund\\[2.5cm]
vorgelegt von\\[0.3cm]
\Large{ \bf Pedro Jim\'enez Delgado}\\[2.7cm]}
\date{Dezember 2008}
\maketitle
\cleardoublepage
\tableofcontents
\cleardoublepage
\pagestyle{fancy}
\fancyhf{}
\fancyfoot[CE,CO]{\thepage}
\fontsize{11}{15.4}
\selectfont
\chapter*{Introduction}
\chaptermark{Introduction}
\addcontentsline{toc}{chapter}{Introduction}
The work described in this thesis deals with the structure of the nucleon within the context of perturbative \emph{Quantum Chromodynamics} (QCD). We determine a new generation of unpolarized dynamical parton distributions using the latest experimental and theoretical advances in the field and paying special attention to the implications of their dependence on the different prescriptions utilized in their derivation. Most of the results are to a large extent based on references \cite{Gluck:2005aq, Gluck:2005xh, Gluck:2007ck, Gluck:2008gs, JimenezDelgado:2008hf}.

While being confined within hadrons, quarks behave in hard scattering processes as almost free particles, i.e., they break the scaling expected in the \emph{naive parton model} only logarithmically. A fundamental property of QCD is the ability to accommodate these rather different aspects of the strong interaction in a consistent picture. Qualitatively, the strong force gets stronger as quarks move apart; the eventual creation of new quarks and their combination into color neutral hadrons (\emph{confinement}) gets favored before reaching the hadronic size. On the other hand, it gets weaker as their distance goes to zero (\emph{asymptotic freedom}), which permits a reliable perturbative description of short--distance (large momentum transfer) phenomena. Long--range (soft) interactions are related to the structure of hadrons and are not to be described perturbatively.

This separation of scales or \emph{factorization} means in practice that hard scattering cross--sections involving initial hadrons are described as products of factors describing the collisions sequentially: the interactions happening ``before" the collision, i.e., the structure of the incoming hadrons, the collision between partons and, eventually, the formation of the detected hadrons. In the case of hadronic final states we will only consider inclusive reactions; for this reason, it suffices for our purposes to consider the first two factors.

The interaction between partons is described by hard scattering partonic cross--sections (\emph{coefficient functions}) which contain only short--distance interactions and therefore are (in principle) calculable in perturbative QCD (pQCD). The structure of incoming hadrons is described by parton distribution functions (PDFs). Roughly speaking (at lowest order), they represent the probability of finding a particular parton (with certain kinematical properties) within the described hadron. It should be clear by now that they contain long--range non--perturbative interactions and therefore are not calculable in perturbation theory but rather to be extracted from experiment. They are, however, independent of the scattering process under consideration (\emph{universality}), which means that, once they are determined from a reduced set of data, they can be used to calculate observables; in this sense pQCD is a fully predictive theory.

Although in principle the partonic description of hadrons is completely general (and even extendable to other objects, e.g. photons), the most relevant PDFs are by far those of the nucleon. They constitute an ineludible ingredient in any calculation involving initial nucleons, for instance, at the \emph{Large Hadron Collider} (LHC), which will be the center of attention for particle physicists over the next years. Their optimal determination is therefore essential for future achievements and we concentrate on them in this work.

Even after three decades of intensive research and the fact that there are several groups working on the field, there are still outstanding issues on the determination of nucleon PDFs, which is to some extent model dependent and, in any case, different for different analysis. Which sets to use for a particular calculation is somehow a matter of taste but the reader should be aware that there is a definite  correspondence between partonic cross--sections and PDFs sets, so that one has to choose PDFs suitable for combination with the particular partonic cross--section to be used. This is sometimes ignored or neglected in the literature, leading to theoretically inconsistent (or simply wrong) results.

First of all, since the PDFs are determined from (global) fits to experimental data, the results depend on the kind of data (and the particular sets) which are used and the sensibility of the distributions to them. Furthermore, the experimental errors of the data induce an intrinsic ``error" of the resulting PDFs which is now possible to estimate. The distributions are usually extracted as parametrizations of contour conditions for a set of differential equations (\emph{evolution}), and the results are therefore biased by the initial parametrization. We use the so--called \emph{dynamical} (or \emph{radiative}) model, which has proved to be a reliable and highly predictive approach to the determination of PDFs.

There are also intrinsic theoretical uncertainties on the determination of the PDFs which are translated into the reliability of the predictions of observables. As mentioned, one works in perturbation theory, so that there is always an uncertainty of the size of the last terms neglected on the perturbative series (and even a certain freedom in whether or not to include some terms and effects). In general a leading order (LO) description is not satisfactory, the standard at the moment is the next--to--LO (NLO) approximation which works quite well for most observables. The (massless) evolution of the PDFs and some important processes can now be calculated up to NNLO and some (few) calculations have been carried out to even higher orders. Other sources of dependence/uncertainty are the factorization and renormalization prescriptions and the treatment of heavy quark masses. We address all these issues through the thesis.

After introducing basic elements of pQCD and establishing our theoretical framework, we discuss to some extent the \emph{dynamical} model and compare it extensively with the approach to PDFs used by most other groups (\emph{``standard''}) by examining the results of analyses carried out under identical conditions in both approaches. Furthermore, parton distributions sets (with uncertainties) at LO, NLO and NNLO of QCD, using different factorization schemes ($\overline{\text{MS}}$, DIS) and different treatments of heavy quark masses (FFNS, VFNS) are extracted and compared.

In the last part of the thesis we turn to the analysis of some of the applications of our results. We briefly outline the astrophysical implications of the dynamical predictions before focusing on collider phenomenology. There we study to some extent the relevance and perturbative stability of the longitudinal structure function of the nucleon and analyze extensively the role of heavy quark flavors in high--energy colliders by comparing the predictions obtained for several processes utilizing different prescriptions for the treatment of heavy quark masses. In addition, we show how isospin violations in the nucleon help to explain the so--called ``NuTeV anomaly''.

\fancyhead[CO]{\nouppercase\rightmark}
\fancyhead[CE]{\bfseries\nouppercase\leftmark}

\chapter*{1. Elements of QCD}
\setcounter{chapter}{1}
\setcounter{section}{0}
\chaptermark{Elements of QCD}
\addcontentsline{toc}{chapter}{1. Elements of pQCD}

\section{Renormalization and the Running Coupling}\label{Sec.Renormalization}
In renormalizable quantum field theories the logarithmic \emph{ultraviolet}\footnote{Related to very high ($\to \infty$) energy/momentum or equivalently short ($\to 0$) distance phenomena.} divergences stemming from virtual--particle loops can be summed to all orders using the renormalization group. This procedure replaces the (unphysical) \emph{bare} coupling constant appearing in the formulation of the theory with an effective or \emph{renormalized} coupling constant $\alpha_s\!\equiv\!\tfrac{g^2_{s,\text{REN}}}{4\pi}$ which depends on the substraction point, an arbitrary finite mass scale known as the renormalization scale $\mu_R$.

The independence of physical quantities on $\mu_R$ is reflected in the so--called renormalization group equations, which are solved by implicitly introducing a function $\alpha_s(\mu^2)$ known as the \emph{running} coupling and given by:
\begin{equation}
\alpha_s\equiv\alpha_s(\mu^2_R)
,\qquad
\beta(\alpha_s)\equiv\frac{\partial\alpha_s}{\partial\ln\mu^2_R}
,\qquad
\frac{d\alpha_s(\mu^2)}{d\ln\mu^2} = \beta(\alpha_s(\mu^2))
\end{equation}
The observables appearing in scattering processes are usually expressed as functions of ratios of invariants (\emph{scaling variables}) and a single scale ($\mu$) with dimensions of energy (typically related to the center--of--mass energy $s$, e.g. $\mu\!\simeq\!\sqrt{s}$). After renormalization all the dependence of observables on the energy scale appears through the running coupling in a well prescribed form (for more details see, for instance, \cite{Yndurain, BargerPhillips, Pink, Reya:1979zk, Altarelli:1981ax}).

In contrast to other theories, the $\beta$\emph{--function} is negative for \emph{QCD} and therefore the coupling decreases as the scale increases (\emph{asymptotic freedom}), which makes a perturbative treatment of the strong interactions at high scales plausible. We can then expand the \mbox{$\beta$\emph{--function}} itself in powers series; the running coupling is obtained from:
\begin{equation}
\label{RGas}
 a_s(\mu^2)\equiv\frac{\alpha_s(\mu^2)}{4\pi}, \qquad \frac{da_s(\mu^2)}{d\ln\mu^2} =-\sum_{m=0}^\infty\beta_{m} \ a_s^{m+2}(\mu^2)
\end{equation}
where the expansion coefficients $\beta_m$ depend in general (the first two are independent) on the renormalization scheme used. In this work we will use the $\overline{\text{MS}}$ scheme, in which in addition to the divergences only $\ln4\pi - \gamma_E$ is absorbed into the running coupling; the first coefficients are given \cite{Gross:1973id, Politzer:1973fx, Jones:1974mm, Caswell:1974gg, Tarasov:1980au} by:
\begin{equation}
\beta_0=11-\tfrac{2}{3}n_R, \qquad \beta_1=102-\tfrac{38}{3}n_R, \qquad \beta_2^{\overline{\text{MS}}}=\tfrac{2857}{2} - \tfrac{5033}{18}n_R + \tfrac{325}{54}n_R^2
\end{equation}
being $n_R$ the appropriate number of \emph{active} quark flavors at the scale $\mu$, which is determined by convention.

In principle, it is possible to fix $n_R$ to the number of light flavors and treat the heavy quark contributions as separated corrections of the form $\ln\tfrac{m^2}{\mu^2}$, being $m$ the mass of a generic heavy quark. However, the most common approach in the $\overline{\text{MS}}$ scheme is the use of a variable number of active flavors taken to be the number of flavors which mass satisfies $m^2\!\leq\!\mu^2$. The $\beta_m$ coefficients are then discontinuous at the thresholds $\mu\!=\!m$ and the solutions for $a_s$ at both sides of the thresholds are subject to adequate matching conditions \cite{Bernreuther:1981sg,Larin:1994va,Chetyrkin:1997sg}. Up to two loops ($m\!=\!1$) the running coupling is continuous; at three loops a marginal term $\big(-\frac{22}{9}a_s^3\big)$ appears, although its effects are numerically insignificant (some few per mil difference in the relevant regions) and we have neglected it in our NNLO analyses. Using this variable number of active flavors the contributions arising from heavy quarks are automatically included (resummed) and consequently the stability of the perturbative expansion is improved.

Turning now to the integration of Eq.\,\ref{RGas}, it is conventional to introduce a parameter $\Lambda_{\text{QCD}}$ (henceforth just $\Lambda$) to be extracted from data, i.e., determined by a contour condition at a reference value, often chosen to be $\alpha_s(M_Z^2)$. $\Lambda$ depends in general on the order considered in the expansion, the renormalization scheme used, the number of active flavors and the explicit form of the solution adopted. Up to next--to--leading--order (NLO) a straightforward solution is:
\begin{equation}
\label{asNLO}
\ln\frac{\mu_R^2}{\Lambda^2} = \frac{1}{\beta_0a_s} -\frac{\beta_1}{\beta_0^2}\ln\Big(\frac{1}{\beta_0a_s}+\frac{\beta_1}{\beta_0^2}\Big)
\end{equation}
where the first term alone is a solution of the leading order (LO) equation\footnote{of course $\Lambda^{\text{LO}}\!\neq\!\Lambda^{\text{NLO}}$, $a_s\!=\!a_s(\mu^2)$, etc.; we try to keep the notation as compact as possible}. This equation constitutes the definition of the $\Lambda$ parameter since it is determined from Eq.\,\ref{RGas} up to a multiplicative constant. In the same way, a solution of the NNLO ($m\!=\!2$) equation may be found by direct integration:
\begin{equation}
\label{asNNLO}
\ln\frac{\mu_R^2}{\Lambda^2} = \frac{1}{\beta_0a_s} -\frac{\beta_1}{2\beta_0^2}\ln\left(\frac{1}{\beta_0^2a_s^2}+\frac{\beta_1}{\beta_0^3a_s} + \frac{\beta_2}{\beta_0^3}\right) + \frac{2\beta_0\beta_2-\beta_1^2}{\beta_0^2\sqrt{4\beta_0\beta_2-\beta_1^2}}\arctan\left(\frac{2\beta_2a_s+\beta_1}{\sqrt{4\beta_0\beta_2-\beta_1^2}}\right)
\end{equation}
Note that for $n_R\!=\!6$ the argument of the square root is negative and this formula does not give $a_s$; in this case a solution is given by:
\begin{equation}\label{asNNLO6}
\begin{array}{l}
  \displaystyle \ln\frac{\mu_R^2}{\Lambda^2} = \frac1{\beta_0a_s} + \frac{\beta_1}{\beta_0^2}\ln a_s
- \frac{+\beta_1^2-\beta_0\beta_2-\beta_1\beta_2\Delta^{+}}{\beta_0^2\beta_2\Delta^{+}-\beta_0^2\beta_2\Delta^-}\ln\left|a_s+\Delta^{+}\right|
\hspace{12em}\\[1em]\displaystyle\hspace{9.9em}
- \frac{+\beta_1^2-\beta_0\beta_2-\beta_1\beta_2\Delta^{-}}{\beta_0^2\beta_2\Delta^--\beta_0^2\beta_2\Delta^+}\ln\left|a_s+\Delta^-\right|,\;
\Delta^\pm=\frac{\beta_1}{2\beta_2}\pm\sqrt{\frac{\beta_1^2}{4\beta_2^2}-\frac{\beta_0}{\beta_2}}
\end{array}
\end{equation}

Since Eqs.\,\ref{asNLO} to \ref{asNNLO6} constitute implicit solutions for $a_s$, a numerical iteration is required for the exact determination of the coupling beyond LO and one could equally integrate numerically Eq.\,\ref{RGas} directly, without introducing the $\Lambda$ parameter. In principle one could also expand $a_s(\mu^2)$ in a power series in $a_s(\mu^2_o)$ for some fixed $\mu^2_o$; up to $\mathcal{O}(a^3_s)$ we get:
\begin{equation}
\label{asMR}
 a_s(\mu^2) = a_s(\mu^2_o) - a^2_s(\mu^2_0)\,\beta_0\ln\tfrac{\mu^2}{\mu^2_o}
 + a^3_s(\mu^2_0) \Big(\beta^2_0 \ln^2 \tfrac{\mu^2}{\mu^2_o} - \beta_1 \ln \tfrac{\mu^2}{\mu^2_o} \Big),
\end{equation}
however, the uncertainty introduced by higher order terms makes these kind of expansions not accurate enough and they are not generally used for the determination of $a_s(\mu^2)$. On the other hand this relation is useful for the study of the uncertainties which are introduced by the choice of particular scales; we will discuss this in the next section.

Another alternative to the numerical evaluation of the coupling which is often used \cite{Amsler:2008zz} is the expansion of the solutions in inverse powers of $L\equiv\ln\tfrac{\mu^2}{\Lambda^2}$, so that the first term in Eqs.\,\ref{asNLO} to \ref{asNNLO6} is the LO solution and higher terms are found recursively. Again, there is a certain freedom in the definition of $\Lambda$; up to NNLO we choose:
\begin{equation}
\label{asAPP}
a_s(\mu^2) = \frac{1}{\beta_0L} - \frac{b_1\ln L}{(\beta_0L)^2} + \frac{b_1^2(\ln^2L-\ln L -1)+b_2}{(\beta_0L)^3},
\end{equation}
where $b_m\equiv\frac{\beta_m}{\beta_0}$. As mentioned, although theoretically consistent, these approximate \emph{asymptotic} solutions differ considerably from the exact ones at low scales (say, $\mu^2\!\leq$2 $\text{ GeV}^2$). Since, as we will see, this region is relevant for the dynamical approach to parton distribution functions we will use the exact iterated solutions of Eqs. Eq.\,\ref{asNLO} and Eq.\,\ref{asNNLO6} all throughout this work.

\section{Factorization and the RG Evolution Equations}\label{Sec.Factorization}
The application of pQCD relies in what are known as \emph{factorization theorems} \cite{Yndurain, BargerPhillips, Pink, Reya:1979zk, Altarelli:1981ax}, which prove that factorization is possible for particular processes and observables, i.e., that the singularities which appear in the calculations either cancel or can be absorbed in the definitions of physical quantities. As we have seen, the ultraviolet divergences stemming from virtual radiative corrections are contained in the running coupling constant and are automatically included in the renormalized theory by changing the bare coupling for the running coupling $\alpha_s(\mu^2)$. Depending on the gauge used in the calculations other ultraviolet divergences may emerge but they cancel.

In general, they appear also \emph{infrared} divergences with different (or mixed) origins: \emph{soft}, related to the emission of gluons with vanishing energy $\big(\frac{E_g}{\mu}\to 0\big)$, or \emph{collinear/mass} singularities, related to collinear radiation off massless partons. If masses are introduced, the soft singularities remain while the collinear ones are regularized given rise to a logarithmic dependence in the masses. After the combination of all (real and virtual) contributions, the soft divergences cancel while the remaining collinear singularities are included in the definition of the parton distributions at a certain (arbitrary) finite mass scale known as the factorization scale $\mu_F$. As part of the proof of factorization theorems precise definitions of the \emph{parton distribution functions} $f_i$ (where $f_i$ stands for the distribution of the different kinds of partons, i.e., massless quarks and gluons) and \emph{coefficient functions} (or partonic cross--sections) are formulated (for details see, for instance, \cite{Yndurain, BargerPhillips, Pink, Reya:1979zk, Altarelli:1981ax}).

A major point of factorization is that the infrared divergences appear at the partonic level in an \emph{universal} way, i.e. with independence of the scattering process considered. Therefore the factorization scale dependence of the parton distributions is also of \emph{universal} validity, it reflects the  independence of physical quantities on $\mu_F$ and it is given by the \emph{evolution} or \emph{renormalization group equations} (RGE):
\begin{equation}
\frac{\partial f_i(x,\mu_F^2)}{\partial\ln \mu_F^2}=\sum_j\int_x^1\frac{dy}{y} P_{ij} \Big(\tfrac{x}{y},\mu_F^2\Big) f_j(y,\mu_F^2)
=\sum_{j}P_{ij}\otimes f_j
\label{RGE}
\end{equation}
where the Bjorken--$x$ is (related to) the longitudinal momentum fraction of the partons. The evolution kernels $P_{ij}$ or \emph{splitting functions} stem from the collinear divergences absorbed in the distributions and represent the (collinear) resolution of a parton $i$ in a parton $j$. They are calculated perturbatively as a series expansion in $a_s(\mu^2_F)$:
\begin{equation}
\label{Pexp}
P_{ij}\big(\tfrac{x}{y},\mu_F^2\big) \equiv P_{ij}\big(\tfrac{x}{y},a_s(\mu_F^2)\big) \equiv \sum_{m=0}^{\infty} a_s^{m+1}\!(\mu_F^2)\; P_{ij}^{(m)}\!\big(\tfrac{x}{y}\big)
\end{equation}
At the moment they have been calculated up to 3 loops ($m\!=\!2$), which is the necessary accuracy for a NNLO analysis; we summarize the expressions that we use in Appendix B.

Since we work in fixed--order perturbation theory ($m\!\leq\!2$), the last equality holds only up to higher order terms and the splitting functions (and therefore the PDFs) depend effectively on the renormalization scale as well. We can make explicit this dependence by expanding $a_s(\mu^2_F)$ in terms of $a_s(\mu^2_R)$; using Eq.\,\ref{asMR} we get up to NNLO:
\begin{equation}
\label{PexpRF}
\begin{array}{l}
\displaystyle P_{ij}\Big(\tfrac{x}{y},\ln \tfrac{\mu_F^2}{\mu_R^2},a_s(\mu^2_R)\Big) = P_{ij}^{(0)} a_s(\mu_R^2)+   \Big(P_{ij}^{(1)} - \beta_0 P_{ij}^{(0)} \ln \tfrac{\mu_F^2}{\mu_R^2}\Big)a_s^2(\mu_R^2)\\
\displaystyle \hspace{9.5em}+ \left( P_{ij}^{(2)} - \Big( 2\beta_0 P_{ij}^{(1)} + \beta_1 P_{ij}^{(0)} \Big) \ln \tfrac{\mu_F^2}{\mu_R^2}
+ \beta_0^2 P_{ij}^{(0)} \ln^2 \tfrac{\mu_F^2}{\mu_R^2} \right)a_s^3(\mu_R^2),
\end{array}
\end{equation}
which is useful to investigate the uncertainties due to variations on the renormalization and factorization scales separately. However, all available sets of parton distribution functions have been generated using $\mu_F\!=\!\mu_R$ and we will also set them equal.

The coefficient functions (also known as Wilson coefficients or simply as partonic cross--sections) are in general different for different processes but depend only on short--distance interactions and therefore are calculable in pQCD. They are expanded in analogy to Eqs.\,\ref{Pexp} and \ref{PexpRF} (with different starting powers depending on the process) and thus depend, besides on the physical energy scale ($\mu^2$), on the factorization ($\mu^2_F$) and renormalization ($\mu^2_R$) scales as well. In an obviously symbolic notation, and writing only the dependence on the different scales, a general observable is calculated in pQCD as:
\begin{equation}
\label{Fac}
\begin{array}{l}
\displaystyle \sigma(\mu^2) = \sum_i \mathbf{C}_i\Big(\!\ln\tfrac{\mu^2}{\mu_F^2},a_s(\mu_F^2)\!\Big)\otimes\mathbf{PDF}_i(a_s(\mu_F^2))\\
\displaystyle \hspace{2.85em}  = \sum_i \mathbf{C}_i\Big(\ln\tfrac{\mu^2}{\mu_F^2},\ln\tfrac{\mu_F^2}{\mu_R^2},a_s(\mu^2_R)\Big)\otimes \mathbf{PDF}_i\Big(\!\ln\tfrac{\mu_F^2}{\mu_R^2},a_S(\mu_R^2)\Big)
\end{array}
\end{equation}
where $i$ denotes a particular contribution (subprocess) and \mbox{$\mathbf{PDF}_i$} represent the appropriate combination of PDFs to be combined with the partonic cross--section $\mathbf{C}_i$.

Beyond LO there is a certain freedom on where to include the finite contributions so that a factorization convention or \emph{factorization scheme} is needed for a complete definition of the factors appearing in the factorization formulae. The most extended prescription is the so--called $\overline{\text{MS}}$, where in addition to the divergent pieces only an ubiquitous $\ln4\pi - \gamma_E$ is absorbed into the PDFs. Another commonly used scheme is the so--called DIS scheme, in which all the (light quark flavor) finite contributions to the nucleon structure function $F_2$ (Sec.\,\ref{Sec.DISstructurefs}) are absorbed into the parton distributions. We discuss the factorization scheme invariance and the transformations between these two schemes in Sec.\,\ref{Sec.DISscheme}. Global fits in both factorization schemes are presented in Sec.~2.

By solving the RGE we determine the parton distributions at a certain scale from the distributions at another scale and therefore we can make predictions for all $\mu^2$ once the $x$--dependence of the functions has been extracted from experiment at one scale known as the input scale. This is usually done by parametrizing the $x$--dependence of the parton distributions at the input scale and determining the parameters by means of global QCD fits, i.e., fits to significative data sets which (ideally) constrain all the PDFs for all $x$.

An alternative to the direct numerical integration of the RGE Eq.\,\ref{RGE} is to work in the so-called \emph{Mellin $n$--space}. The symbol $\otimes$ in Eqs.\,\ref{RGE} and \ref{Fac} denotes the \emph{Mellin convolution}, which reduces to multiplication for the \emph{Mellin} (\emph{n}--)moments of the quantities involved, i.\,e.,
\begin{equation}
a^n\equiv\int_0^1dx\,x^{n-1}a(x) \Rightarrow (a\otimes b)^n=a^nb^n.
\end{equation}
Hence, in the \emph{Mellin space}, the evolution equations reduce to ordinary differential equations which are solved analytically; the $x$--space solution may also be obtained via a numerical \emph{Mellin inversion}. On the other hand, this technique requires analytic continuations of non--trivial functions to complex values of $n$. We use the Mellin space to solve the RGE and for the calculation of nucleon structure functions. Other quantities, for which the moments of the coefficient functions are not known, are calculated in ordinary $x$--space.

\section{The Flavor Decomposition}\label{Sec.Flavordecomposition}
The calculations appearing in pQCD and in particular the RGE can be greatly simplified by expressing them in terms of particular combinations of parton distributions. For instance, the splitting functions are constrained by charge conjugation and flavor symmetry invariance, so that they can be written as:
\begin{equation}
\begin{array}{l}
\displaystyle P_{gq_i}=P_{g\bar{q}_i}\equiv P_{gq}\\[0.5em]
\displaystyle P_{q_ig}=P_{\bar{q}_ig}\equiv \frac{P_{qg}}{2 n_F}
\end{array}
\hspace{2em};\hspace{1.5em}
\begin{array}{l}
\displaystyle P_{q_iq_j}=P_{\bar{q}_i\bar{q}_j}\equiv \delta_{ij}P^V_{qq}+P^S_{qq}\\[0.9em]
\displaystyle P_{q_i\bar{q}_j}=P_{\bar{q}_iq_j}\equiv \delta_{ij}P^V_{q\bar{q}}+P^S_{q\bar{q}}\\[0.4em]
\end{array}
\end{equation} 
where $n_F$ is the number of quark flavors entering in the evolution (massless), which is not necessarily equal to the number $n_R$ considered in the running coupling \cite{Gluck:2006ju}. 

In view of the general structure of the splitting functions, it is possible to introduce some flavor combinations in order to decouple the RGE as far as possible. Note for instance\footnote{We work in the Mellin space although we do not explicitly write the \emph{n}--dependence of the moments.}:
\begin{equation}
\label{RGEvalence}
 q^\pm\equiv q_i \pm \bar{q}_i \Rightarrow \left\lbrace
\begin{array}{l}
\displaystyle \frac{\partial q^-_i}{\partial \ln \mu_F^2} = P^-_{\text{NS}} \,q^-_i + \frac{P^S_{\text{NS}}}{n_F}  \sum_{k=1}^{n_F}q^-_k\\[2em]
\displaystyle \frac{\partial q^+_i}{\partial \ln \mu_F^2} = P^+_{\text{NS}} \,q^+_i + \frac{P_{\text{PS}}}{n_F} \sum_{k=1}^{n_F}q^+_k + \frac{P_{qg}}{n_F} g\\
\end{array} \right.
, \text{where}
\left\lbrace
\begin{array}{l}
P^\pm_\text{NS}\equiv P^V_{qq} \pm P^V_{q\bar{q}}\\[0.5em]
P^S_\text{NS}\equiv n_F(P^S_{qq}-P^S_{q\bar{q}})\\[0.5em]
P_\text{PS}\equiv n_F(P^S_{qq}+P^S_{q\bar{q}})\\[0.5em]
\end{array}
\right.
\end{equation}
The minus combinations decouple from the gluon, whereas the $q^+$ combination maximally coupled to it:
\begin{equation}
\begin{array}{l}
\displaystyle q_S\equiv q^+ \equiv \sum^{n_F}_{k=1} q^+_k \hspace{0.5em}\Rightarrow\hspace{0.5em}
\frac{\partial q_S}{\partial \ln \mu_F^2} = P_{qq}\,q_S + P_{qg}\,g \hspace{1em}, \text{where}\hspace{1em} P_{qq}\equiv P^+_\text{NS} + P_\text{PS}\\[0.2em]
\displaystyle \hspace{9.5em} \frac{\partial g}{\partial \ln \mu_F^2} = P_{gq}\,q_S + P_{gg}\,g
\end{array}
\end{equation}
this combination together with the gluon are known as the (flavor) singlet parton distributions. Defining the vector $\mathbf{q}$ and a matrix of splitting functions $\mathbf{P}$, they evolve as:
\begin{equation}
\mathbf{q}\equiv \left(\!\!\!\begin{array}{l}q_S\\g\end{array}\!\!\!\right),\hspace{1em}
\mathbf{P}\equiv\left(\!\!\!\begin{array}{l}P_{qq} \hspace{1em} P_{qg}\\P_{gq} \hspace{1em} P_{gg} \end{array}\!\!\!\right)
\hspace{0.5em}\Rightarrow\hspace{0.5em} \frac{\partial \mathbf{q}}{\partial \ln \mu_F^2} = \mathbf{P}\,\mathbf{q}
\end{equation}
Simple non--singlet combinations which evolve independently are:
\begin{equation}
\begin{array}{l}
\displaystyle q^\pm_{\text{NS},ij}\equiv q^\pm_i - q^\pm_j  \hspace{0.5em}\Rightarrow\hspace{0.5em} \frac{\partial q^\pm_{\text{NS},ij}}{\partial \ln \mu_F^2} = P^\pm_\text{NS} \,q^\pm_{\text{NS},ij},\hspace{1em}\text{with }\,i,k=1,...n_F\\[1em]
\displaystyle q^V_\text{NS}\equiv q^- \equiv \sum^{n_F}_{k=1} q^-_k \hspace{0.5em}\Rightarrow\hspace{0.5em} \frac{\partial q^V_\text{NS}}{\partial \ln \mu_F^2} = P^V_\text{NS} \, q^V_\text{NS},\hspace{1em}\text{where }\; P^V_\text{NS}\equiv P^-_\text{NS} + P^S_\text{NS}
\end{array}
\end{equation}
However, together with $q^V_\text{NS}$ we will use for the evolution in the non--singlet sector even more convenient combinations given by \cite{Blumlein:1997em}: 
\begin{equation}\label{Flavorcombinations}
\begin{array}{l}
\displaystyle v^\pm_{i^2-1}\equiv \sum_{k=1}^i q^\pm_k - iq^\pm_i  \hspace{0.5em}\Rightarrow\hspace{0.5em} \frac{\partial v^\pm_{i^2-1}}{\partial \ln \mu_F^2} = P^\pm_\text{NS} \,v^\pm_{i^2-1}, \\[1em] 
\hspace{-10em} \text{or inversely:} \hspace{4.5em} \displaystyle q^\pm_i = \frac{1}{n_F} q^\pm - \frac{1}{i} v^\pm_{i^2-1} + \sum_{k=i+1}^{n_F}\frac{1}{k(k-1)}v^\pm_{k^2-1}
\end{array}
\end{equation}

In the power expansion, $P^V_{qq}$ starts at leading order while $P^V_{q\bar{q}}$, $P^S_{qq}$ and $P^S_{q\bar{q}}$ start at NLO, however $P^S_{qq}=P^S_{q\bar{q}}$ at this order, being different for the first time at NNLO. This allow for some simplifications at lower orders. As mentioned, at leading order $P^{V(0)}_{q\bar{q}}$, $P^{S(0)}_{qq}$ and $P^{S(0)}_{q\bar{q}}$ vanish and the structure of the splitting functions reduces to:
\begin{equation}
\left.
\begin{array}{l}
P^{V(0)}_{q\bar{q}}=0\\[0.5em]
P^{S(0)}_{qq}=P^{S(0)}_{q\bar{q}}
\end{array}\right\rbrace \Rightarrow \left\lbrace
\begin{array}{l}
P_{\text{PS}}=P^{S}_{\text{NS}}=0\\[0.5em]
P_{qq}=P^V_{\text{NS}}=P^\pm_{\text{NS}}=P^{V(0)}_{qq}\equiv P^{(0)}_{qq}
\end{array}
\right.
\end{equation}
At NLO $P^{S(1)}_{q\bar{q}}\!=\!P^{S(1)}_{qq}$, i.e., $P^{S}_{q\bar{q}}\!=\!P^{S}_{qq}$ (hence $P^S_{\text{NS}}\!=\!0$ and $P^V_{\text{NS}}\!=\!P^-_{\text{NS}}$) and some simplifications still hold. For instance, in view of Eq.\,\ref{RGEvalence} it is clear that each valence distribution ($u^-$, $d^-$) evolves independently and any initial symmetric sea (e.g., $s^-\!=\!0$) is preserved by the evolution up to NLO. Beyond NLO we find the complete structure of the evolution equations where the $q^-_i$ combinations evolve together in what is called the \emph{non--singlet sector} and the $q^+_i$ combinations evolve in conjunction with the singlet distributions in what is called the \emph{singlet sector}.

\section{Solutions of the RGE}\label{Sec.RGEsolution}
Since it is clear how to recover the explicit dependence on the renormalization and factorization scales separately, we will set here $\mu_F^2=\mu_R^2$ (henceforth simply $\mu^2$) for simplicity in the discussion. With the definitions of last section all the evolution equations are of the same form, namely:
\begin{equation}
\frac{\partial q(\mu^2)}{\partial \ln \mu^2} = P_q(\mu^2)\;q(\mu^2)
\end{equation}
where for the singlet distributions this equation is a matrix equation and one has to take care of the commutativity of the subsequent factors.

Because all of the scale dependence of the splitting functions appears through $a_s(\mu^2)$, it is convenient to combine this equation with Eq.\,\ref{RGas} and use $a_s\!\equiv\! a_s(\mu^2)$ as independent variable. At leading order we can integrate immediately;formally:
\begin{equation}
\frac{\partial q}{\partial a_s}=-\tfrac{1}{\beta_0 a_s}P^{(0)}_{q}q\hspace{1em}\Rightarrow\hspace{1em} 
q\big(a_s(\mu^2)\big)=\big(\tfrac{a_s}{a_o}\big)^{-R_{0,q}}\!q_o\!\equiv L(a_s,a_o)\,q_o\;,\hspace{1em} R_{0,q}\equiv \tfrac{P^{(0)}_{q}}{\beta_0}
\end{equation}
where $a_o\!\equiv\! a_s(\mu^2_o)$ and $q_o\!\equiv\! q(\mu^2_o)$, being $\mu^2_o$ some reference scale at which the distributions are known. For the singlet distributions $\mathbf{R}_0$ is a matrix and the above solution is only symbolic. We can find an explicit solution by diagonalizing; the eigenvalues, projection matrices and some useful relations are:
\begin{equation}
\begin{split}
&\mbox{Eigenvalues:}\hspace{1em}r^{\pm}\equiv\frac{1}{2}\left(R_{0,qq}+R_{0,gg}\pm
\sqrt{(R_{0,qq}-R_{0,gg})^2 + 4 \,R_{0,qg}\, R_{0,gq}}\right)\\
&\mbox{Projectors:}\hspace{1em}\mathbf{R}^{\pm}\equiv\frac{\mathbf{R}_{0}-r_\mp}{r_\pm - r_\mp} \hspace{0.5em} \Rightarrow \hspace{0.5em} (\mathbf{R}^\pm)^2=\mathbf{R}^\pm,\hspace{0.5em} \mathbf{R}^+\!+\mathbf{R}^-\!=\!1, \hspace{0.5em} \mathbf{R}^+ \mathbf{R}^-\!=\!\mathbf{R}^-\mathbf{R}^+\!=\!0\\
&\mbox{For an analytic function $f$:}\hspace{1em}f(\mathbf{R}_{0})=f(r^+)\,\mathbf{R}^+ + f(r^-)\,\mathbf{R}^-
\end{split}
\label{R0}
\end{equation}
The last relation holds in particular for $L(a_s,a_o)$, which immediately gives an explicit solution for the singlet distributions at LO.

To solve the equations at an arbitrary fixed order $m$ \cite{Ellis:1993rb, Blumlein:1997em} we combine the evolution equations at this order with Eq.\,\ref{RGas} at the same order $m$:
\begin{equation}
 \frac{\partial q(a_s)}{\partial a_s} = -\frac{1}{a_s}\left( \sum_{m=0}^\infty a_s^m R_{m,q}\right)\,q(a_s),\hspace{1em} 
R_{m,q}\equiv \tfrac{P^{(m)}}{\beta_0}-\sum_{i=1}^m b_i\,R_{m-i,q}
\end{equation}
where again $b_m\equiv\frac{\beta_m}{\beta_0}$. Next, we make a series expansion around the LO solution as:
\begin{equation}
q(a_s) \equiv U_q(a_s)\,L(a_s,a_o)\,U^{-1}(a_o)\,q_o,\hspace{0.5em} \text{with}\hspace{0.5em} 
U_q(a_s)\equiv\sum_{m=0}^\infty a_s^m\,U_q^{(m)}, \hspace{0.5em} U_q^{(0)}= 1
\end{equation}
so that the problem is now reduced to the one of finding the operators $U_q^{(m)}$. By inserting this ansatz in the evolution equations we get:
\begin{equation}
 U_q^{(0)}R_{0,q}+\sum_{m=1}^\infty a_s^m \big( U_q^{(m)}R_{0,q} - m U_q^{(m)}\big) = 
 \left( R_{0,q} + \sum_{m=1}^\infty a_s^m R_{m,q}\right)\left(\sum_{m=0}^\infty a_s^m\,U_q^{(m)} \right),
\end{equation}
which order by order gives the commutation relations:
\begin{equation}
 \big[U_q^{(m)}, R_{0,q}\big] = \tilde{R}_{m,q} + m\,U_q^{(m)}, \hspace{1em} \tilde{R}_{m,q} \equiv \sum_{k=1}^m R_{k,q}\,U_q^{(m-k)},
\end{equation}
which determine the $U_q^{(m)}$ operators. For non--singlet combinations the commutators vanish and the operators are directly $U_q^{(m)}\!\!=\!\!-\tfrac{\tilde{R}_{m,q}}{m}$. For the singlet combinations we can use the properties of the projection operators (second line in Eq.\,\ref{R0}) to express them as:
\begin{equation}
 \mathbf{U}^{(m)}=-\frac{1}{m}(\mathbf{R}^+\tilde{\mathbf{R}}_m\mathbf{R}^+ + \mathbf{R}^-\tilde{\mathbf{R}}_m\mathbf{R}^-)
 + \frac{\mathbf{R}^+\tilde{\mathbf{R}}_m\mathbf{R}^-}{r^--r^+-m} + \frac{\mathbf{R}^-\tilde{\mathbf{R}}_m\mathbf{R}^+}{r^+-r^--m}
\end{equation}

In addition to higher--order in the expansions, \emph{redundant} terms (${\cal O}(a_s^n)$ with $n\!>\!m$) arise from the multiplications in the ansatz. One can remove these terms in advance by expanding in powers of $a_o$ the factor $U^{-1}(a_o)$ and carrying out the multiplications keeping only terms up to the desired order for consistency with the fixed--order perturbative approach. Up to NNLO we get:
\begin{equation}
q(a_s(\mu^2))=\Big(L + a_s U_q^{(1)}L -a_o L U_q^{(1)} + a_s^2 U_q^{(2)}L - a_s a_o U_q^{(1)}LU_q^{(1)} + a_o^2 L (U_q^{(1)}U_q^{(1)}-U_q^{(2)})\Big)\!q_o
\end{equation}
where, again, for the singlet distributions one has to take care of the order of the different factors. This defines what are called the \emph{truncated solutions}, which are the ones employed in our analyses since, as mentioned, they are more consistent with the fixed--order perturbative treatment.

We have tested our singlet and non--singlet evolution codes using the \texttt{PEGASUS} program \cite{Vogt:2004ns} for generating the truncated solutions together with the commonly used toy input of the Les Houches and HERA--LHC Workshops \cite{Giele:2002hx,Dittmar:2005ed}. For $10^{-7}\!<\!x\!<\!0.9$ we achieved an agreement of up to four decimal places in most cases, which is similar to the required high--accuracy benchmarks advocated in \cite{Giele:2002hx,Dittmar:2005ed}.

Note that the truncated solutions are not directly comparable with $x$--space evolutions or the so--called \emph{iterated} $n$--space solutions, where all the terms are kept. Since the differences are of higher orders, whether or not to include which redundant terms is somehow a matter of taste. Notice, however, that these differences are not negligible, specially at NNLO, where the precision aimed is of a few percent\footnote{For instance, using the toy input mentioned, we get for the NNLO\,(NLO) iterated/truncated ratios at $\mu^2\!=\!10 \textrm{ GeV}^2$ and \mbox{$x\!=\!10^{-5}$}, $10^{-3}$, $0.1$ respectively, the values 0.973\,(0.940), 0.992\,(0.975), 1.001\,(1.002) for the sea quark ($\bar{u}+\bar{d}$) distributions and of 1.043\,(1.049), 1.006\,(1.016), 0.998\,(0.996) for the gluon ones; the differences are smaller in the non--singlet sector.}.

\fontsize{11}{16}
\selectfont

\section{Isospin Violations in the Nucleon}\label{Sec.Isospinviolations}
Electroweak radiative corrections may be included together with the QCD description of the structure of the nucleon \cite{Kripfganz:1988bd,Gluck:1994vy,Spiesberger:1994dm,Roth:2004ti}. In complete analogy with the emission of gluons, the emission of photons by quarks introduce new mass singularities which are factorized in the parton content of the nucleon. The evolution equations of Eq.\,\ref{RGE} are consequently modified to include electromagnetic splitting terms; in general:
\begin{equation}
\label{QEDDGLAP}
\frac{\partial f_i(x,\mu^2)}{\partial\ln \mu^2} = \sum_{j}\left(P_{ij,\text{QCD}}\otimes f_j + P_{ij,\text{QED}}\otimes f_j \right)
\end{equation}
where the sum over partons includes now  additional distributions, e.g., the photon parton distribution function $\gamma(x,\mu^2)$.

The modified evolution equations contain additional mixing between distributions and flavor symmetry violations since different quark flavors have different charges and therefore evolve differently in what concerns the QED terms.  In particular the standard isospin symmetry relations for the parton distributions of the nucleon, i.e., vanishing ``majority'' and ``minority'' asymmetries defined via
\begin{equation}
\begin{array}{l}
\delta u(x,\mu^2) =  u^p(x,\mu^2)-d^n(x,\mu^2),\\
\delta d(x,\mu^2) =  d^p(x,\mu^2)-u^n(x,\mu^2),
\end{array}
\end{equation}
and similarly for the antiquarks $\delta\bar{q}$, are violated. Note that these asymmetries must observe the quark number and momentum conservation rules, e.g., separating, as usual, between valence and sea contributions, the first moments of the valence asymmetries must vanish.

Although in principle one should solve the modified evolution equations, it is possible to get a first estimate, sufficient for most purposes, by working in LO QED; from Eq.\,\ref{QEDDGLAP}, the leading corrections are given by:
\begin{equation}
\label{QEDcorr}
\begin{array}{l}
\displaystyle \frac{d}{d\ln \mu^2}\, \delta u(x,\mu^2) = \frac{\alpha}{4\pi}(e_u^2-e_d^2)\int_x^1
 \frac{dy}{y}\,P_{qq}^{\gamma (0)}\!\big(\tfrac{x}{y}\big) \,u(y,\mu^2)\\[1em]
\displaystyle\frac{d}{d\ln \mu^2}\, \delta d(x,\mu^2) = -\frac{\alpha}{4\pi}(e_u^2-e_d^2)\int_x^1
 \frac{dy}{y}\,P_{qq}^{\gamma (0)}\!\big(\tfrac{x}{y}\big) \,d(y,\mu^2)
\end{array}
\end{equation}
where $e_f$ is the electric charge and $P_{qq}^{\gamma (0)}\!=\!\frac{P_{qq}^{(0)}}{C_F}$; similar evolution equations hold for the isospin asymmetries of sea quarks $\delta\bar{u}$ and $\delta\bar{d}$. The evolution  equation for the gluon distribution remains unaltered at this order. Notice that the addition \cite{Roth:2004ti, Martin:2004dh} of further terms proportional to $\alpha e_q^2 P_{q\gamma}\otimes\gamma$ would actually amount to a subleading ${\cal{O}}(\alpha^2)$ contribution since the photon distribution $\gamma(x,\mu^2)$ of the nucleon is of  ${\cal{O}}(\alpha)$ \cite{Kniehl:1990iv, Blumlein:1993ef, DeRujula:1998yq, Gluck:2002fi}.

In principle one could expect the ${\cal{O}}(\alpha)$ corrections to be of the order of NNLO QCD; however the effects turn out \cite{Kripfganz:1988bd, Spiesberger:1994dm, Roth:2004ti} to be rather small (at most 1\% in the relevant regions) and irrelevant for usual QCD applications. Furthermore they often mix with non--perturbative effects like differences in quark masses. Nevertheless they may be relevant for high precision experiment in which they appear somehow isolated.

Since usually the data used in global QCD fits are rather insensitive to the isospin asymmetries, some plausible contour conditions for the above equations, i.e., some reasonable (non--perturbative) value for the asymmetries, must be assumed in order to integrate them. In this way it is possible to express the (LO QED) isospin asymmetries as functions of usual isospin symmetric parton distributions. We make such estimates in Sec.\,\ref{Sec.NuTeV} within the framework of the dynamical model and in connexion with the so--called ``NuTeV anomaly'', where they turn out to contribute significantly.

\section{Unpolarized DIS Structure Functions}\label{Sec.DISstructurefs}
\begin{figure}
\centering
\includegraphics[width=0.4\textwidth]{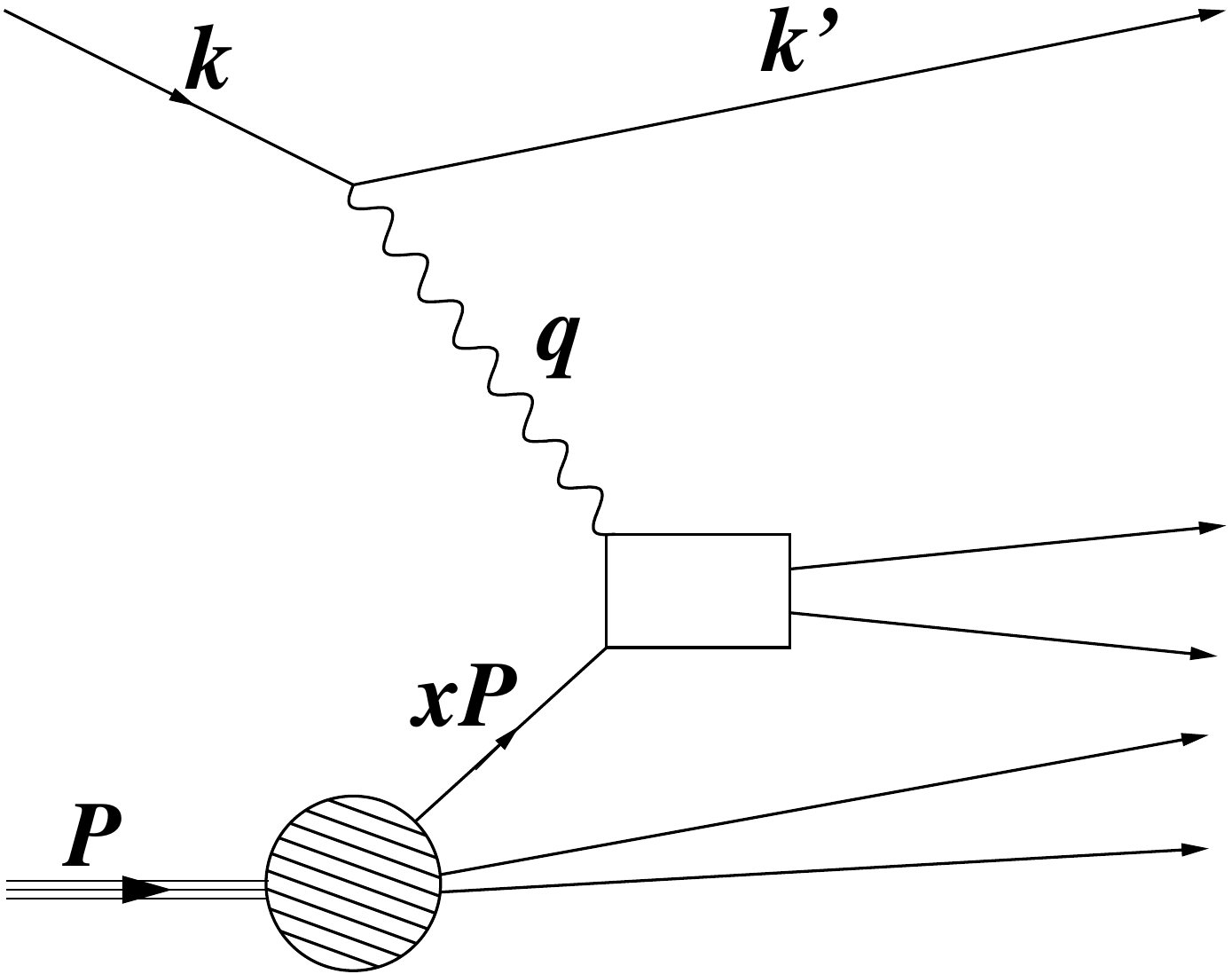}
\caption{Parton model picture of DIS. The filled circle represents the structure of the proton and the rectangle the hard partonic interaction. The exchanged gauge boson (drawn as a photon) can be $\gamma,\, Z^0$ or $W^\pm$.\label{Fig101}}
\end{figure}

Lepton--nucleon deep inelastic scattering (DIS) is the most powerful available tool for the determination of nucleon PDFs. DIS cross--sections are described in terms of standard kinematics \cite{Yndurain, BargerPhillips, Pink, Reya:1979zk, Altarelli:1981ax}; we recall here only the definitions of the variables that we use (cf. Fig.\,\ref{Fig101}):
\begin{equation}
s \equiv (P+k)^2,\quad W^2 \equiv (P+q)^2,\quad Q^2\equiv-q^2,\quad x\equiv\frac{Q^2}{2P\cdot q},\quad y\equiv\frac{P\cdot q}{P\cdot k};
\end{equation}
$W$ is the invariant mass of the recoiling system, $x$ is known as the \emph{Bjorken--$x$} and $y$ is often referred to as the \emph{inelasticity}. For fixed center--of--mass energy ($s$), there are only two independent variables and the following relations hold:
\begin{equation}\label{W2andQ2}
 W^2 = M^2 + \frac{Q^2}{x}(1-x),\qquad Q^2=xys,
\end{equation}
where $M$ is the mass of the nucleon and the DIS regime implies $Q^2\!\gg\! M^2$ (deep) and $W^2\!\gg\! M^2$ (inelastic).

The differential cross--section for DIS can be written as products of leptonic and hadronic tensors. The leptonic tensors describe the lepton--gauge boson vertex and are well known \cite{Yndurain, BargerPhillips, Pink, Reya:1979zk, Altarelli:1981ax}. The hadronic tensor for unpolarized DIS is usually parametrized as:
\begin{equation}
W_{\mu \nu}(P,q)\equiv (-g_{\mu  \nu}+\frac{q_\mu q_\nu}{q^2})\,F_1(x,Q^2) + \frac{\hat{P}_\mu\hat{P}_\nu}{P\cdot q}\,F_2(x,Q^2) 
+i\varepsilon_{\mu \nu \alpha \beta} \frac{P^\alpha q^\beta}{2P\cdot q}\,F_3(x,Q^2)
\end{equation}
with $\hat{P}_\mu\!\equiv\! P_\mu - \tfrac{P\cdot q}{q^2}q_\mu $. An alternative is to expand it using the polarization vectors of the (virtual) gauge boson; one gets then the so--called \emph{longitudinal structure function}:
\begin{equation}
 F_L(x,Q^2)\equiv (1+4\frac{M^2}{Q^2}x^2)\,F_2(x,Q^2)-2xF_1(x,Q^2)\simeq F_2(x,Q^2)-2xF_1(x,Q^2)
\end{equation}
which is frequently used instead of $F_1$. A (sometimes) more convenient alternative definition of the structure functions is:
\begin{equation}
 \mathcal{F}_{1,2,3}\equiv(2F_1,\frac{F_2}{x},F_3) \Rightarrow \mathcal{F}_L\equiv\frac{F_L}{x}=\mathcal{F}_2-\mathcal{F}_1
\end{equation}
The differential cross--section for DIS is given in terms of the structure functions as:
\begin{equation}
 \frac{d^2\!\sigma^i}{dx\,dy}=\frac{2\pi\alpha^2}{xyQ^2}\left((Y_++2x^2y^2\tfrac{M^2}{Q^2})F_2^i-y^2F_L^i\mp Y_-xF_3^i\right)
\end{equation}
where $Y_\pm \equiv 1\pm(1-y)^2$, $i=\text{\rm NC},\text{\rm CC}$ and $-$ is taken for an incoming \emph{antiparticle} ($l^+$ or $\bar{\nu}$) and + for an incoming \emph{particle} ($l^-$ or $\nu$). The factors stemming from the couplings and propagators of the gauge bosons have been included in the structure functions as ratios to those of the photon \cite{Klein:1983vs}. For charged currents:
\begin{equation}
F^{\text{\rm CC}}_{1,2,L,3}=\eta \kappa^2_W(Q^2)\,F^W_{1,2,L,3},\quad \text{with} \quad \kappa_W(Q^2)=\kappa_W\!(\infty)\,\tfrac{Q^2}{Q^2+M_W^2},\quad
\kappa_W(\infty)=\tfrac{1}{2} \tfrac{G_FM_W^2}{2\sqrt 2\pi\alpha}
\end{equation}
where $\eta\!=\!2$ for unpolarized charged lepton scattering and $\eta\!=\!4$ for neutrinos. For NC:
\begin{equation}
\label{neutralcurrent}
\begin{array}{l}
F^{\text{\rm NC}}_{1,2,L}=F^{\gamma}_{1,2,L} - v_e \kappa_Z(Q^2)\,F^{\gamma Z}_{1,2,L} + (v_e^2+a_e^2) \kappa^2_Z(Q^2)\,F^{Z}_{1,2,L}\\[0.7em]
  F^{\text{\rm NC}}_{3}=\hspace{3.5em} - a_e \kappa_Z(Q^2)\,F^{\gamma Z}_{3} + 2 v_e a_e \kappa^2_Z(Q^2)\,F^{Z}_{3}
\end{array};
\left\{
\begin{array}{l}
 \kappa_Z(Q^2)=\kappa_W\!(\infty)\,\frac{Q^2}{Q^2+M_Z^2},\\
 \kappa_Z(\infty)=\frac{G_FM_Z^2}{2\sqrt{2}\pi\alpha}\\
 v_e=-\frac{1}{2}+2\sin^2\theta_W, \;a_e=-\frac{1}{2}
\end{array}
\right.
\end{equation}
The $Z^0$ and $W^\pm$ contributions contain suppression factors which make the neutral--current cross--section being dominated by electromagnetic ($\gamma$) exchange except at high $Q^2$. Furthermore, the longitudinal structure function contributes sizably only for large inelasticity; a quantity which is often used is the \emph{``reduced'' cross--section}:
\begin{equation}
\label{sigmar}
\sigma_r^{\text{\rm NC}}\equiv\left(\tfrac{2\pi\alpha^2}{xyQ^2}Y_+\right)^{-1}\!\frac{d^2\!\sigma^{\text{\rm NC}}}{dx\,dy}=
F_2^{\text{\rm NC}}-\tfrac{y^2}{Y_+}F_L^{\text{\rm NC}}\mp \tfrac{Y_-}{Y_+}xF_3^{\text{\rm NC}}
\end{equation}
which, as mentioned, approximately equals $F_2^{\gamma}$ for most of the data used in our analysis (Sec~\ref{Sec.DataFormalism}).

In DIS there is a natural energy scale $Q^2$ which by definition (deep) is much larger than the masses of the quarks flavors $u$, $d$ and $s$ (therefore called \emph{light}), while it may be comparable to the masses of the $c$, $b$ and $t$ quarks (heavy). It is therefore convenient to consider separately the contributions to DIS structure functions stemming from light and heavy flavors.

For the light quark contributions a massless description (in particular the massless evolution of Sec.\,\ref{Sec.RGEsolution}) works fine and $Q^2$ provides an appropriate natural choice for the factorization and renormalization scales; we give more details on their calculation in Sec.\,\ref{Sec.Flight}. The situation is not that clear for the heavy quark contributions since, besides $Q^2$, the heavy quark masses appear as well as natural energy scales in the problem; we discuss these contributions in Sec.\,\ref{Sec.Fheavy}.

Furthermore, since, as already mentioned, the DIS regime implies $Q^2\!\gg\! M^2$ and $W^2\!\gg\! M^2$, the mass of the nucleon is often neglected. Its effects are only sizable at low $Q^2$, say $Q^2\!\!\leq\!100\text{ GeV}^2$ and medium to large $x$. In this region modifications to the usual ($M\!\!=\!0$) expressions or \emph{target mass corrections} (TMC) must be taken in to account. The well--known expressions for the dominant ``light'' $F^{\text{\rm NC}}_2$ structure function are given in $n$--space by \cite{Georgi:1976ve}:
\begin{equation}\label{TMC}
F^{\rm NC}_{2,\rm TMC}(n,Q^2) = \sum_{\ell =0}^2 \left(\tfrac{M^2}{Q^2}\right)^{\ell} \frac{(n+\ell)!}{\ell !(n-2)!}\,\, \frac{F^{\rm NC}_2(n+2\ell ,Q^2)}{(n+2\ell )(n+2\ell -1)}
     + {\cal{O}}\Big(\Big(\tfrac{M^2}{Q^2}\Big)^3\,\Big)
\end{equation}
where higher powers than $(\frac{M^2}{Q^2})^2$ are negligible for the relevant $x<0.8$ region, as can straightforwardly be shown by comparing the above expression with the exact one in $x$--space \cite{Georgi:1976ve}.

\fontsize{11}{14.8}
\selectfont

\section{Light Quark Contributions to NC Structure Functions}\label{Sec.Flight}
After factorization, the (leading--twist) light quark contributions to the neutral--current DIS structure functions are given as\footnote{We work in Mellin space although we do not explicitly write the $n$--dependence. Recall also that we have set $\mu^2\!=\!\mu_F^2\!=\!\mu_R^2$.}:
\begin{equation}
\mathcal{F}_{j=1,2,L,3}^{i=\gamma,\gamma Z, Z}(Q^2)=\sum_{q}q_q^{i,j}\bigg(C_{j,qq}\Big(\ln\tfrac{Q^2}{\mu^2},a_s(\mu^2)\Big)q(\mu^2)+
C_{j,qg}\Big(\ln\tfrac{Q^2}{\mu^2},a_s(\mu^2)\Big)g(\mu^2)\bigg),
\end{equation}
where the sum runs over all the light quark \emph{and} antiquark flavors, $q_q^{i,j}$ are the appropriate electroweak coupling factors and we have obviated the renormalization scale dependence. The coefficient functions are expanded in power series as usual; for $j=1,2,3$:
\begin{equation}
\begin{array}{l}
\displaystyle C_{j,qq}(a_s) = \sum_{m=0}^{\infty} a_s^m C_{j,qq}^{(m)}, \quad C_{j,qq}^{(0)} = 1;\qquad
\displaystyle C_{j,qg}(a_s) = \sum_{m=0}^{\infty} a_s^m C_{j,qg}^{(m)}, \quad C_{j,qg}^{(0)} = 0
\end{array}
\end{equation}
where $m$ coincides in this case with the number of loops entering in the calculations\footnote{For the longitudinal structure function $C_{L,qq}^{(0)}\!\!=\!C_{2,qq}^{(0)}\!-C_{1,qq}^{(0)}\!=0$; therefore $F_L$ appears at the one--loop level, i.e., at LO is already of order $a_s$, which is in agreement with the naive parton model expectation $F_2\!\!=\!2xF_1$ known as the Callan--Gross relation.}.

Combining the electroweak factors related with the exchanged boson and the quarks respectively, the charge factors appropriate for the complete NC structure functions are:
\begin{equation}\label{chargefactors}
\begin{array}{l}
a_q^+\equiv a_q^{1,2,L}\equiv e_q^2 - 2 e_q v_e v_q \kappa_Z(Q^2) + (v_e^2+a_e^2)\kappa_Z^2(Q^2),\quad a^{+}_{\bar{q}} =a^{+}_q\\
a_q^- \equiv a_q^3\equiv \hspace{2.8em} - 2 e_q a_q \kappa_Z(Q^2) + 4v_ea_ev_qa_q\kappa_Z^2(Q^2),\quad \hspace{0.5em} a^-_{\bar{q}} =-a^-_q, \qquad \text{with:}\\[1em]
\hspace{12em} e_U=\frac{2}{3}, \hspace{0.8em} \quad v_U=\frac{1}{2} - \frac{4}{3} \sin^2\theta_W, \quad a_U=\frac{1}{2}\\[2pt]
\hspace{12em} e_D=-\frac{1}{3}, \quad v_D=-\frac{1}{2} + \frac{2}{3} \sin^2\theta_W, \quad a_D=-\frac{1}{2}
 \end{array}
\end{equation}
where U(D) refers to u(d)--type quarks. Note that the couplings for quarks and antiquarks are either equal (1,2,L) or opposite (3), this allows for the rewriting of the structure functions as a sum over the quark flavors:
\begin{equation}
\mathcal{F}_{j=1,2,L}^{\text{\rm NC}}=\sum_{i=1}^{n_F}a_{q_i}^{+}\big(C_{j,qq}q^+_i + 2C_{j,qg}g\big), \qquad \mathcal{F}_{3}^{\text{\rm NC}}=\sum_{i=1}^{n_F}a_{q_i}^{-}C_{3,qq}q^-_i
\end{equation}
where $n_F$ is the number of flavors entering in the evolution, i.e., light. To express them in terms of the flavors combinations introduced in Sec.\,\ref{Sec.Flavordecomposition} we consider the following identity:
\begin{equation}
\sum_{i=1}^{n_F}a^\pm_{q_i}(q^\pm_i-\frac{q^\pm}{n_F})=\sum_{i=2}^{n_F}\tfrac{1}{i(i-1)}\big(\sum_{k=1}^i a^\pm_{q_k}-ia^\pm_{q_i}\big)v^{\pm}_{i^2-1}\equiv \sum_{i=2}^{n_F}a^\pm_{i^2-1}v_{i^2-1}^\pm
\end{equation}
Defining as well $a^\pm\!\!\equiv\!\frac{1}{n_F}\!\!\sum_{i=1}^{n_F}\!a^\pm_{q_i}$ the appropriate combinations of couplings reproduce (upon normalization) those of the distributions in Eq.\,\ref{Flavorcombinations}.

It is important to note that the coefficient functions for quarks contain a pure singlet part stemming from interactions which cancel for the non--singlet combinations ($C_{j,qq}\!\!\equiv\! C_{j,\text{NS}}$ for non--singlet), and a non--singlet part which contributes both for singlet and non--singlet combinations ($C_{j,qq}\!\!\equiv\! C_{j,\text{S}}\!\!\equiv\! C_{j,\text{NS}}\!+C_{j,\text{PS}}$ for singlet); redefining the gluon coefficient function as $C_{j,G}\!\!\equiv\! 2 n_F C_{j,qg}$ and using the above identity we get\footnote{Actually the standard notation is somewhat more complicated, as for the splitting functions, there appear ``+'' and ``-'' combinations of coefficient functions. Since for NC only the ``+'' (``-'') combinations are relevant for $F_{2,L}$ ($F_3$) we omit these signs.}:
\begin{equation}
\begin{array}{l}
\displaystyle \mathcal{F}_{j=1,2,L}^{\text{\rm NC}}=C_{j,\text{NS}}\sum_{i=2}^{n_F}a_{i^2-1}^{+}v^+_{i^2-1} +a^{+}\big(C_{j,\text{S}}\,q_S + C_{j,G}\,g\big)\\
\displaystyle \hspace{2.5em}=C_{j,\text{NS}}\big(a^{+}q_S + \sum_{i=2}^{n_F}a_{i^2-1}^{+}v^+_{i^2-1}\big) +a^{+}\big(C_{j,\text{PS}}\,q_S + C_{j,G}\,g\big)\\
\displaystyle\mathcal{F}_{3}^\text{\rm NC}=C_{3,\text{NS}}\big(a^{-}q^V_{NS} + \sum_{i=2}^{n_F}a_{i^2-1}^{-}v^-_{i^2-1}\big)
 \end{array}
\end{equation}
The combinations stemming from one flavor only ($q_i$) are given in this notation for the coefficient functions by:
\begin{equation}
\begin{array}{l}
\mathcal{F}_{j=1,2,L}^{\text{\rm NC},q_i}=a_{q_i}^{+}\big(C_{j,qq}q^+_i+2C_{j,qg}g\big)=a_{q_i}^{+}\big(C_{j,\text{NS}}\,q^+_i+
\tfrac{C_{j,\text{PS}}}{n_F}q_S+\tfrac{C_{j,\text{G}}}{n_F}g\big)\\[0.5em]
\mathcal{F}_{3}^{\text{\rm NC},q_i}=a_{q_i}^{-}C_{3,qq}q^-_i=a_{q_i}^{-}C_{3,\text{NS}}\,q^-_i
\end{array}
\end{equation}

At the moment all the necessary coefficient functions for a NNLO analysis are known: up to one--loop we use the well--known expressions, summarized for example in \cite{Floratos:1981hs}, and for the 2-- and 3--loop (for $F_L$) coefficient functions we use the parametrizations of \cite{Vermaseren:2005qc, Moch:2004xu}, except for $C_{3,\rm NS}^{(2)}$ for which we use the parametrizations given in \cite{vanNeerven:1999ca}.

\fontsize{11}{15.5}
\selectfont

\section{The DIS Factorization Scheme}\label{Sec.DISscheme}
The DIS factorization scheme is defined by the requirement that the light quark contributions to the function $F_2$ retain their leading order form at all orders in perturbation theory. We can derive transformation relations between factorization schemes by comparing expressions for physical quantities, which must be, order by order, factorization scheme independent \cite{Gluck:1983mm}.

In order to manipulate simultaneously the contributions of quarks and gluons, it is convenient to regard the contributions to the structure function $F_2$ stemming from a combination of quarks ($q$), and its related gluon contribution, as the quark sector of the product of a matrix of coefficient functions $\mathbf{C}_2$ times a general (quark and gluon sectors) distribution $\mathbf{q}$; in analogy with the notation introduced for the singlet sector in Sec.\,\ref{Sec.Flavordecomposition}. Since there is no corresponding gluon sector in the structure functions, only the upper part of the products is to be considered, i.e., beyond LO the gluon sector of the matrix of coefficients is undefined at this point. Keeping this in mind the mentioned contributions are given as:
\begin{equation}\label{SFmatrix}
\mathcal{F}^{q}_{2} \propto \mathbf{C}_2\mathbf{q}, \qquad \mathbf{C}_2 = \sum_{m=0}^{\infty} a_s^m \mathbf{C}_{2}^{(m)}, \quad \mathbf{C}_{2}^{(0)} \equiv \mathbf{1}
\end{equation}
Combining this with Eqs.\,\ref{RGas} and \ref{RGE} and setting $\mu_F^2\!\!=\!\mu_R^2\!\!=Q^2$ for simplicity, we can express its derivative as function of the structure functions itself:
\begin{equation}
\frac{d\,\mathcal{F}^q_2}{d\ln Q^2} = (\beta \frac{d\,\mathbf{C}_2}{d\,a_s}\mathbf{C}_2^{-1} + \mathbf{P} + [\mathbf{C}_2,\mathbf{P}]\mathbf{C}_2^{-1})\mathcal{F}^q_2,
\end{equation}
where $\mathbf{P}$ denotes the matrix of splitting functions appropriate for $\mathbf{q}$. This makes clear that the quantity in brackets must be invariant under factorization convention transformations. Hence, if the coefficient functions are changed so that $\mathbf{C}_2^{(m)}\!\!\rightarrow\!\mathbf{C}_2^{(m)}\!+\!\Delta\mathbf{C}_2^{(m)}$, the splitting functions must be conveniently changed as $\mathbf{P}^{(m)}\!\!\rightarrow\!\mathbf{P}^{(m)}\!+\!\Delta\mathbf{P}^{(m)}$. Expanding up to $\mathcal{O}(a_s^3)$ and requiring invariance:
\begin{equation}
\begin{array}{l}
\Delta \mathbf{P}^{(1)}=\beta_0\Delta\mathbf{C}_2^{(1)}-[\Delta\mathbf{C}_2^{(1)},\mathbf{P}^{(0)}]\\
\Delta \mathbf{P}^{(2)}=-2\beta_0\mathbf{C}_2^{(1)}\Delta\mathbf{C}_2^{(1)}
                           -\beta_0\Delta\mathbf{C}_2^{(1)}\Delta\mathbf{C}_2^{(1)}
                          +2\beta_0\Delta\mathbf{C}_2^{(2)}
                          +\beta_1\Delta\mathbf{C}_2^{(1)}
                          -\mathbf{P}^{(0)}\mathbf{C}_2^{(1)}\Delta\mathbf{C}_2^{(1)}\\
\hspace{4em}            +\Delta\mathbf{C}_2^{(1)}\Delta\mathbf{C}_2^{(1)}\mathbf{P}^{(0)}
                          -\Delta\mathbf{C}_2^{(1)}\mathbf{P}^{(0)}\Delta\mathbf{C}_2^{(1)}
                          +\mathbf{C}_2^{(1)}\Delta\mathbf{C}_2^{(1)}\mathbf{P}^{(0)}
                          -\Delta\mathbf{C}_2^{(1)}\mathbf{P}^{(1)}\\
\hspace{4em}            +\mathbf{P}^{(1)}\Delta\mathbf{C}_2^{(1)}
                          -[\Delta\mathbf{C}_2^{(2)},\mathbf{P}^{(0)}]
\end{array}
\end{equation}

On the other hand, the structure functions themselves must be (order by order) invariant under factorization convention transformations. Hence the parton distributions must change up to NNLO as $\mathbf{q} \rightarrow \mathbf{q} + a_s\Delta\mathbf{q}^{(1)} + a_s^2\Delta\mathbf{q}^{(2)}$; again expanding and requiring invariance:
\begin{equation}
\label{DIStransformation}
  \Delta \mathbf{q}^{(1)}=-\Delta\mathbf{C}_2^{(1)}\mathbf{q},\qquad
  \Delta \mathbf{q}^{(2)}=\big(\mathbf{C}_2^{(1)}\Delta\mathbf{C}_2^{(1)} +
                               \Delta\mathbf{C}_2^{(1)}\Delta\mathbf{C}_2^{(1)}-\Delta\mathbf{C}_2^{(2)}\big)\mathbf{q}
\end{equation}

As mentioned, the DIS factorization scheme is defined by $\mathbf{C}_{2}\!\!\equiv\! \mathbf{1}$ at all orders, hence the transformation of the $F_2$ coefficient functions from any other scheme (in practice from $\overline{\text{MS}}$) is given by $\Delta\mathbf{C}_{2}^{(m)}\!\!\equiv\!\mathbf{C}_{2}^{(m)\text{DIS}}\!-\mathbf{C}_{2}^{(m)}\!\!=\!-\mathbf{C}_{2}^{(m)},\;\forall \, m\geq1$. The general results for the splitting functions and the parton distributions reduces then to:
\begin{equation}
 \begin{array}{l}
  \Delta\mathbf{P}^{(1)}=-\beta_0 \mathbf{C}_2^{(1)}+[\mathbf{C}_2^{(1)},\mathbf{P}^{(0)}]\\
  \Delta\mathbf{P}^{(2)}= \beta_0\mathbf{C}_2^{(1)}\mathbf{C}_2^{(1)}
                         -2\beta_0\mathbf{C}_2^{(2)}
                         -\beta_1\mathbf{C}_2^{(1)}
                         -[\mathbf{C}_2^{(1)},\mathbf{P}^{(0)}]\mathbf{C}_2^{(1)}
                         +[\mathbf{C}_2^{(1)},\mathbf{P}^{(1)}]
                         +[\mathbf{C}_2^{(2)},\mathbf{P}^{(0)}]\\
  \Delta \mathbf{q}^{(1)}=\mathbf{C}_2^{(1)}\mathbf{q},\qquad
  \Delta \mathbf{q}^{(2)}=\mathbf{C}_2^{(2)}\mathbf{q}

 \end{array}
\end{equation}

These relations refer to a general distribution $\mathbf{q}$, it is convenient to make explicit the transformations for the flavor combinations introduced in Sec.\,\ref{Sec.Flavordecomposition}. For non--singlet combinations the gluonic contributions cancel and therefore the above relations may be applied directly to get the respective non--singlet splitting functions and distributions in the DIS factorization scheme; note that the commutators vanish. In the singlet sector, however, these relations do not completely fix the transformations because the gluon sector of the matrix of coefficient functions is still undetermined. The conventional choice is to extend the restriction on the coefficient functions stemming from momentum conservation to all moments; this results in:
\begin{equation}
 q_S^{\text{DIS}} + g^{\text{DIS}} = q_S + g \Rightarrow 
\mathbf{C}_2^{(m)}
\!=\!\left(\!\!\!\begin{array}{l}\hspace{0.5em}C_{2,qq}^{(m)} \hspace{2em} 2n_FC_{2,qg}^{(m)}\\[0.5em]
                                              -C_{2,qq}^{(m)} \hspace{0.8em} -2n_FC_{2,qg}^{(m)} \end{array}\!\!\!\right)
\!=\!\left(\!\!\!\begin{array}{l}\hspace{0.5em}C_{2,S}^{(m)} \hspace{2em} C_{2,G}^{(m)}\\[0.5em]
                                              -C_{2,S}^{(m)} \hspace{0.8em} -C_{2,G}^{(m)} \end{array}\!\!\!\right)
,\quad m=1,2
\end{equation}

Since in order to invert the transformations a change $\Delta\mathbf{C}_{2}^{(m)}\!\!=\!-\mathbf{C}_{2}^{(m)}$ is needed in the coefficient functions, the same expressions with the appropriate sign changes give the inverse transformation.

\section{Heavy Quark Contributions and their Resummation}\label{Sec.Fheavy}
The contributions to hard scattering processes involving heavy quarks $h\!=\!c, b, t$ become increasingly important above the threshold for the production of a quark--antiquark pair of heavy flavor (e.g. $W^2\!\geq 4m^2$ in DIS, where $m$ stands for  the mass of the heavy quark). Although in principle there could exist an \emph{intrinsic} (initial--state) heavy quark content of the nucleon, all data are well described through \emph{extrinsic} heavy quark production only (Sec.\,2), i.e., being \emph{all} the heavy quarks generated from the (initial--state) light quarks ($u$, $d$, $s$) and gluons, so that it appears that the intrinsic heavy quark content of the nucleon, if any, is marginal. The extrinsic production of heavy quarks is in principle calculable in \emph{fixed--order} perturbation theory and is known as the \emph{fixed flavor number scheme} (FFNS) since, in contrast to \emph{variable flavor number schemes} (VFNS) to be discussed below, the number of light quark flavors (partons) does not change with the factorization scale.

In the case of the neutral--current (we drop here the superscript NC) DIS structure functions of Sec.\,\ref{Sec.DISstructurefs}, the (subleading) heavy quark contributions enter as:
\begin{equation}
\label{Flight_heavy}
 F_{2, \rm L}(x,Q^2)=F_{2, \rm L}^{\rm light} + F_{2, \rm L}^{\rm heavy},\qquad F_{2, \rm L}^{\rm heavy} = F_{2, \rm L}^c +F_{2, \rm L}^b
\end{equation}
where ``light'' refers to the contributions of Sec.\,\ref{Sec.Flight} and top contributions are negligible at present energies. Their contributions to $F_3$ vanish at LO \cite{Leveille:1978px} and are negligibly small at higher orders\footnote{This can be guessed by observing that at the relevant large $Q^2$ values $F^h_3\propto h-\bar{h}$, where the meaning of the \emph{effective} heavy quark distributions $h$ will be clarified below.}.

\begin{figure}
\centering
\includegraphics[width=0.9\textwidth]{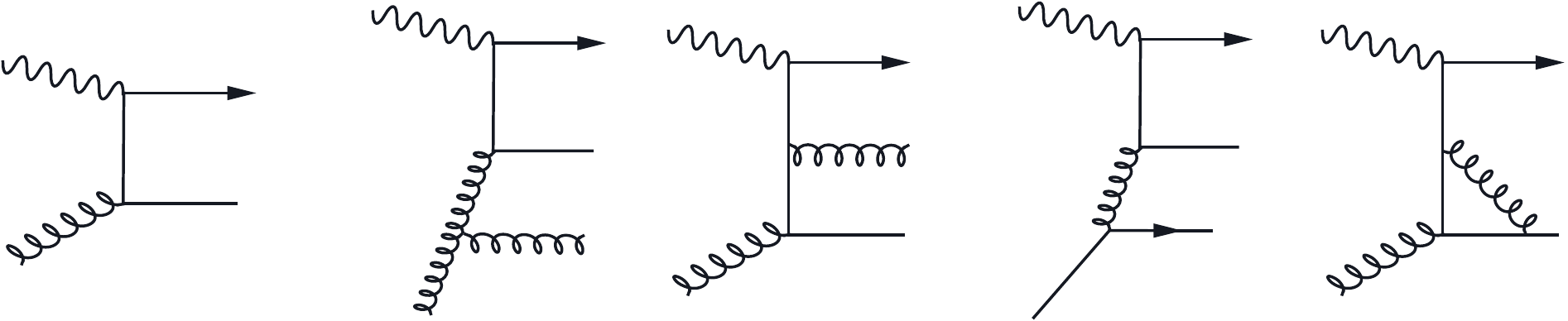}
\caption{$\mathcal{O}(\alpha_s)$ (leftmost) and some illustrative $\mathcal{O}(\alpha_s^2)$ diagrams contributing to the virtual photon--gluon fusion process $\gamma^*g\to h\bar{h}$. Note that beyond LO other production mechanisms, e.g. $\gamma^* q \to h \bar{h} q$, etc. appear. The figure has been taken from \cite{Martin:1994kk}.\label{Fig102}}
\end{figure}

The ${\cal{O}}(\alpha_s)$ FFNS contributions due to the virtual photon--gluon fusion subprocess \mbox{$\gamma^*g\to h\bar{h}$} are well known \cite{Witten:1975bh, Shifman:1977yb, Leveille:1978px, Gluck:1979aw} and their QCD corrections (cf. Fig.\,\ref{Fig102}) have been calculated so far up to ${\cal{O}}(\alpha_s^2)$ \cite{Laenen:1992zk, Riemersma:1994hv}. Although complete analytic expressions are not available for all the coefficient functions, the results are contained in computer programs. In particular we use the code offered in \cite{Riemersma:1994hv}, which combines some known analytic expressions together with grids for the more complicated coefficient functions; they are represented in the $\overline{\text{MS}}$ factorization scheme. Up to NLO the relevant heavy quark contributions to DIS structure functions are given by\footnote{We use the expressions in \cite{Riemersma:1994hv}, which were slightly modified respect to those in \cite{Laenen:1992zk} due to an additional mass divergence which appears when the virtual photon coupled to the light quark goes on--mass shell, leading to the appearance of the function $\bar{d}^{(1)}_{2,i} (\bar{d}^{(1)}_{L,i}\!=\!0)$.}:
\begin{equation}\label{Fheavyformula}
\begin{array}{l}
\displaystyle F^h_{k=2,L}(x,Q^2,m^2) = \frac{Q^2 \alpha_s(\mu^2)}{4\pi^2 m^2} \int_x^{\frac{Q^2}{Q^2+4m^2}} \frac{dz}{z}  \left\{ \,a_h^+ \, c^{(0)}_{k,g}(\eta,\xi)\,g\big(\tfrac{x}{z},\mu^2\big) \right.\\[1em]
\displaystyle \hspace{5em} +4\pi\alpha_s(\mu^2) \left[a_h^+\,\Big(c^{(1)}_{k,g}(\eta,\xi) + \bar c^{(1)}_{k,g}(\eta,\xi) \ln\tfrac{\mu^2}{m^2}\Big)\,g\big(\tfrac{x}{z},\mu^2\big) \right. \\[1em]
\displaystyle \hspace{10em} +\sum_{q} \left( a_h^+ \,\Big(c^{(1)}_{k,q}(\eta,\xi) + \bar c^{(1)}_{k,q}(\eta,\xi) \ln\tfrac{\mu^2}{m^2}\Big)\,q\big(\tfrac{x}{z},\mu^2\big) \right.\\
\displaystyle \hspace{12em} \left. \left. \left. +a_q^+\,\Big(d^{(1)}_{k,q}(\eta,\xi) + \bar d^{(1)}_{k,q}(\eta,\xi) \ln\tfrac{\mu^2}{m^2}\Big)\,q\big(\tfrac{x}{z},\mu^2\big) \right) \right] \right\},
\end{array}
\end{equation}
where the sum runs over all the light quark \emph{and} antiquark flavors and the charge factors $a_{q}^+$ appropriate for the complete NC structure functions are given in Eq.\,\ref{chargefactors}. Note that the coefficient functions indicated by $c$'s ($d$'s) originate from partonic subprocesses where the virtual photon is coupled to the heavy (light) quark and that those indicated by a bar appear through mass factorization \cite{Laenen:1992zk, Riemersma:1994hv}. The scaling variables ($\eta,\xi$) are defined as:
\begin{equation}
\eta \equiv \frac{\hat{s}}{4m^2} - 1, \qquad\xi \equiv \frac{Q^2}{m^2};
\end{equation}
being $\hat{s}=\frac{Q^2}{z}(1-z)$ the square of the c.m. energy of the virtual photon--parton subprocess (i.e., the partonic version of $W^2$, cf. Eq.\,\ref{W2andQ2}). With these definitions the \emph{threshold} region ($\hat{s} \sim 4m^2$) is characterized by $\eta \to 0$, while the \emph{asymptotic} regime ($Q^2\!\gg\! m^2$) is given as $\xi \to \infty$. Notice that these are two independent limits. In principle one can consider the asymptotic structure function defined by $F^{h,\text{\scriptsize ASYMP}}_k(x,Q^2,m^2) \equiv \lim_{Q^2\gg m^2} F^h_k(x,Q^2,m^2)$, however, in contrast to what is sometimes believed, even in this limit $F^h_k$ receives contributions from the threshold region as is clear from Eq.\,\ref{Fheavyformula}, where the collinear momentum fraction ($z$) is integrated out, i.e., $\hat{s}$ takes all its possible values $4m^2\leq\hat{s}\leq \frac{Q^2}{x}(1-x)\simeq W^2$, in particular those close to the threshold.

\fontsize{11}{15.8}
\selectfont

Further, $\mu$ in Eq.\,\ref{Fheavyformula} stands for the factorization scale, which has been set equal to the renormalization scale for simplicity (recall Sec.\,\ref{Sec.Factorization}). A dedicated study \cite{Gluck:1993dpa} of the perturbative stability of heavy quark production at high energy colliders concluded that the factorization scale in Eq.\,\ref{Fheavyformula} should preferably be chosen to be $\mu^2\!=\!4 m^2$ at LO, while the NLO results are rather insensitive to this choice (usually either $\mu^2\!=\!4m^2$ or $\mu^2\!=\!Q^2 + 4m^2$); even choosing a very large scale like $\mu^2\!=\!4(Q^2+4m^2)$ leaves the NLO results essentially unchanged, in particular at small--$x$. Furthermore, they concluded that the fixed--order calculation for heavy quark production is entirely reliable and perturbatively stable, provided that one employ consistently parton distributions and strong coupling constant values of the appropriate order \cite{Gluck:1993dpa}. As has already been mentioned, the extrinsic heavy quark production is, in addition, experimentally required, in particular near the threshold region.

The ${\cal{O}}(\alpha_s^3)$ 3--loop corrections to $F_L^h$ and first rudimentary contributions to $F_2^h$ have been calculated recently \cite{Blumlein:2006mh, Bierenbaum:2007rg, Bierenbaum:2008yu} in the limit $Q^2\!\gg\!m^2$. However these asymptotic results are neither applicable for our present investigations nor relevant for the majority of presently available data at lower values of $Q^2$. There exist also soft gluon resummations \cite{Laenen:1998kp, Alekhin:2008hc} which include the logarithmically enhanced terms near threshold up to NLL, which improve the convergence of the perturbative series in the very small--x region and gives a first approximation (NLO + NLL) towards the NNLO results. Our ignorance\footnote{It should be mentioned that we have attempted to mimic the NNLO contributions in the relevant kinematical regions by naively assuming them to be down by one power of $\alpha_s$ respect to the NLO ones and being given by a constant $K$--factor times different combinations of the lower order expressions evaluated using parton distributions of different orders. The results (Sec.~2) were however insensitive to such \emph{ad hoc} corrections and, furthermore, this approach (guess) appears to be not appropriate since playing the same game at NLO can not reproduce the correct results in the relevant kinematical regions.} of the full fixed--order ${\cal{O}}(\alpha_s^3)$ corrections to $F_{2,L}^h$ constitute the major drawback for any NNLO analysis of DIS data\footnote{Note however that for totally inclusive (light + heavy) structure functions the heavy quark contributions, although important, enter at subleading levels.}.

The coefficient functions appearing in Eq.\,\ref{Fheavyformula} contain terms with factors of the form $\ln\frac{Q^2}{m^2}$ which in principle could vitiate the stability of the calculation and might suggest that these contributions should be resummed \cite{Collins:1986mp, Buza:1996wv}. In its simplest version, what is known as the \emph{zero mass} (ZM) \emph{variable flavor number scheme} (VFNS), the \emph{asymptotic} FFNS heavy quark coefficient functions are separated into the \emph{massless} Wilson coefficients of the light partons and \emph{massive} operator matrix elements which are used as transition functions to define new parton distributions for ``heavy'' quarks, which are subsequently treated as \emph{massless} partons within the nucleon; i.e. whose contributions to DIS structure functions follows Sec.\,\ref{Sec.Flight} and whose distributions $h(x,\mu^2)$ are treated as those of the light quarks in Sec.\,\ref{Sec.RGEsolution}, being usually (up to NLO and choosing $m\!=\!\mu$) generated from the boundary conditions $h(x,m^2)\!=\!\bar{h}(x,m^2)\!=\!0$.

In general the matching conditions are fixed by continuity relations \cite{Collins:1986mp, Buza:1996wv} at the unphysical threshold $\mu_F^2\!=\!m^2$, where the ``heavy'' $n_F\!>\!3$ quark distributions are generated from the $n_F-1$ ones via the massless renormalization group equations of Secs.\,\ref{Sec.Factorization} to \ref{Sec.RGEsolution}. Hence, this factorization scheme is characterized by increasing the number of massless partons $n_F$ by one unit at the unphysical ``thresholds`` $\mu^2\!\equiv\!\mu_F^2\!=\!m^2$, in a similar way as is done in the $\overline{\text{MS}}$ renormalization scheme for the strong running coupling (Sec.\,\ref{Sec.Renormalization}); note however that, as already mentioned, in general $n_F\!\neq\!n_R$ \cite{Gluck:2006ju}.

Despite its theoretical problems (e.g. the transmutation of a final--state quark into an initial--state one), a major general advantage of the ZM--VFNS is that it simplifies considerably the calculations, which in many situations become unduly complicated in the FFNS. For example, the single top production process at hadron colliders via $W^\pm$--gluon fusion requires in the FFNS the calculation of the subprocess $u\,g\to d\,t\,\bar{b}$ at LO and of $u\,g\to d\,t\,\bar{b}\,g$, etc. at NLO while in the ZM--VFNS one needs merely $u\,b\to d\,t$ at LO and $u\,b\to d\,tg$, etc. at the NLO of perturbation theory \cite{Stelzer:1997ns}. Obviously this approach is convenient because of its simplicity, as is clear in particular for processes for which the FFNS coefficient functions containing explicitly the complete mass dependence are not known. It should nevertheless be regarded as an \emph{effective} treatment of heavy quarks, keeping always in mind that the existence of an intrinsic heavy quark content of the nucleon is experimentally disfavored and that, being generated from the \emph{asymptotic} heavy quark contributions, it cannot reproduce the fully massive result (FFNS), not even at very high $Q^2$ \cite{Gluck:1993dpa}.

There exist also more involved schemes with a variable number of active flavors (for a recent review see \cite{Thorne:2008xf}), sometimes referred as \emph{general mass} (GM) VFNS, where mass--dependent corrections are maintained in the hard cross--sections and a model--dependent interpolation between the ZM--VFNS (for the asymptotic regime) and the FFNS (near threshold as required experimentally) is achieved. Since the threshold behaviour of the FFNS is included in these models, they are generally able to reach a good agreement with data; in fact there are no observable signatures which allow to uniquely distinguish between these GM--VFNS schemes and the FFNS. On the other hand, the universality of the distributions is somehow put in danger due to the inclusion of non--collinear terms in the partonic picture. Furthermore, the partonic cross--sections for most processes are either calculated in the simplest ZM--VFNS or already known for the FFNS, so that ultimately the user \emph{must} choose one of these two schemes. Needless to say that the combination of GM-VFNS distributions with these expressions, despite being frequent, is inconsistent and should be avoided.

In our opinion the stability of the FFNS renders attempts to resum supposedly ``large logarithms'' in heavy quark production cross--sections superfluous. Since besides no intrinsic heavy quark content of the nucleon is needed experimentally, only the light $u,d,s$ quark flavors and gluons constitute the ``intrinsic'' genuine partons of the nucleon and the heavy $c,b,t$ quark flavors should not be included in its parton structure. For applications for which the FFNS expressions are not known, it is possible to generate ZM--VFNS distributions (based on FFNS fits) to be used consistently in these calculations at large enough scales. We compare FFNS and ZM-VFNS predictions for several illustrative processes in Secs.\,\ref{Sec.HeavyDIS} and \ref{Sec.WandHiggs} (cf. \cite{Gluck:2008gs}), and show that the ZM--VFNS can be employed for calculating processes where the invariant mass of the produced system is sizeably larger than the mass of the participating heavy quark flavor, for instance for LHC phenomenology.

\section{Hadroproduction of Vector Bosons and Jets}\label{Sec.Hadroproduction}
In spite of its title, the aim of this section is not to provide a detailed discussions of hadronic cross--sections in QCD, moreover since the major experimental input for the determination of our dynamical distributions in Sec.~2 comes from deep inelastic scattering experiments. For completeness and reference we include here a brief overview of the hadron--hadron scattering processes that we use in addition to the DIS structure functions, namely Drell-Yan dimuon production as well as $W^\pm$ and inclusive jet production.

As discussed in Sec.\,\ref{Sec.Factorization}, a fundamental property of the QCD partonic description of the hadronic structure is that the collinear divergences appearing in the calculation of any hard-scattering process (which originate the evolution equations) are \emph{universal}, i.e., the same ones in all processes. Therefore the parton distributions appearing, for example, in the expressions for DIS structure functions are the same that describe the structure of the incoming hadrons in hadronic production. In the latter case the general factorization formula of Eq.\,\ref{Fac} reduces to:
\begin{equation}
\label{hadronicFac}
\displaystyle \sigma(P_1,P_2) = \sum_{i,j}\int dx_1\,dx_2\,\hat{\sigma}_{ij}\Big(x_1,x_2,\ln\tfrac{M^2}{\mu^2},a_s(\mu^2)\Big)\,f_i(x_1,\mu^2)\,f_j(x_2,\mu^2)
\end{equation}
where $M$ is the relevant energy scale (e.g. the invariant mass of the muon pair for Drell-Yan, the $W^\pm$ mass, etc.) and $\mu$ is the factorization and renormalization scale, which have been set equal for simplicity (recall Eq.\,\ref{asMR}). Again $\hat{\sigma}_{ij}$ is calculable in fixed--order perturbation theory as a series expansion in $a_s(\mu^2)$ which starts which different powers depending on the process.

\begin{figure}
\centering
\includegraphics[width=0.4\textwidth]{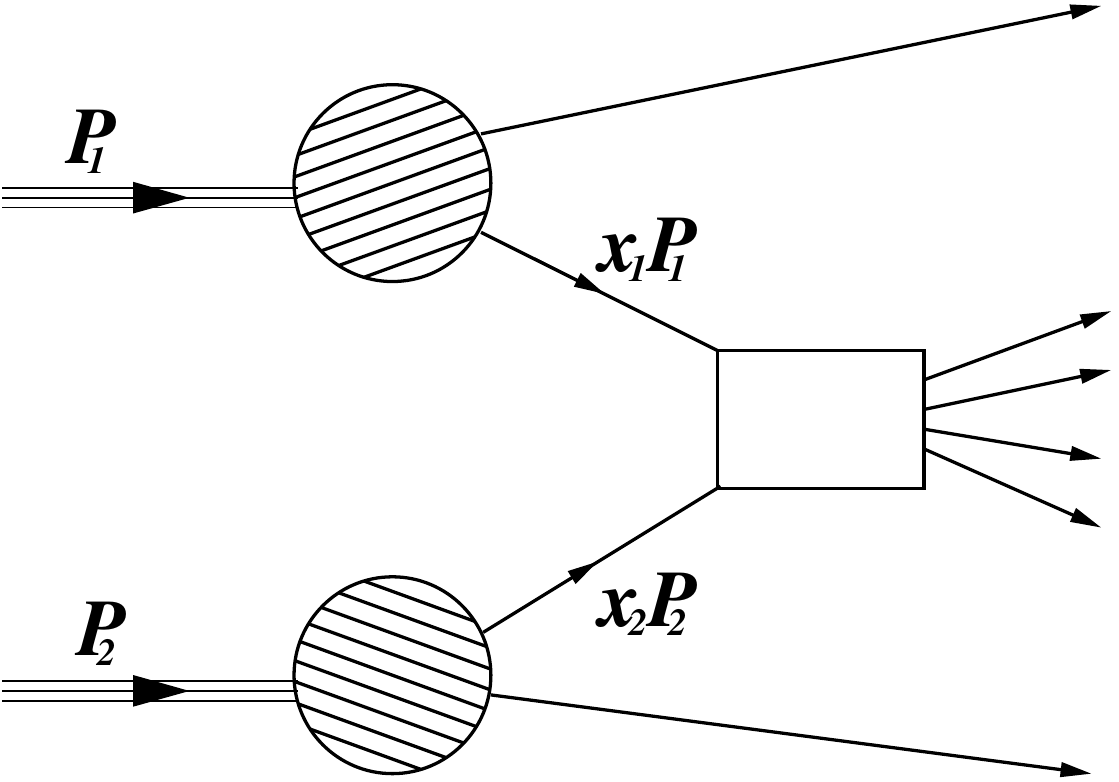}
\caption{Parton model picture of hadron--hadron scattering. The filled circles represent the structure of the proton and the rectangle the hard partonic interaction.\label{Fig103}}
\end{figure}

As in the case of DIS, hadron--hadron collisions are described in terms of standard kinematics \cite{Yndurain, BargerPhillips, Pink}; we recall here some definitions (cf. Fig.\,\ref{Fig103}):
\begin{equation}
s\equiv (P_1+P_2)^2,\quad M^2\equiv\hat{s}\equiv(x_1P_1+x_2P_2)^2\simeq x_1x_2s,\quad \tau\equiv\frac{M^2}{s}\simeq x_1x_2
\end{equation}
the four--momentum of the produced system is conveniently given as:
\begin{equation}
 (E,p_x,p_y,p_z)\equiv\big(m_T\cosh{y},p_T\sin{\phi},p_T\cos{\phi},m_T\sinh{y}\big)
\end{equation}
where $p_T$ is the transverse momentum, $m_T^2\!\equiv\!M^2+p_T^2$ and $\phi$ is the azimuthal angle. The rapidity $y$ and, equivalently, the Feynman--$x$ are defined by:
\begin{equation}
 y\equiv\frac{1}{2}\ln{\left(\tfrac{E+p_z}{E-p_z}\right)},\quad x_F\equiv\frac{p_Z}{p_Z^{\rm max}}\simeq\frac{2p_Z}{\sqrt{s}},
\end{equation}
which lead to the relations:
\begin{equation}
x_F=2\Big(\tau+\tfrac{p_T^2}{s}\Big)^{\frac{1}{2}}\sinh{y} \Longleftrightarrow y={\rm arcsinh} \bigg( \tfrac{1}{2} x_F \Big(\tau+\tfrac{p_T^2}{s}\Big)^{-\frac{1}{2}}\bigg)
\end{equation}

Depending on the detected final state it is possible to define different cross--sections for hadron--hadron collisions. In the Drell-Yan mechanism \cite{Drell:1970yt} a quark from one hadron and an antiquark from the other one annihilate into an intermediate vector boson ($\gamma$, $Z^0$ or $W^\pm$) which subsequently decays into a lepton pair. Considering lepton pairs of invariant mass $M\!\ll\!M_Z$ the process is dominated by virtual photon exchange. The experiments usually consist on proton\,(beam)--nucleon\,(proton or some other nuclei) collisions and the cross--sections are extracted from the detection of muon pairs (dimuon production) produced in the decay of the virtual photons\footnote{There are also prompt/direct photon experiments which detect a real photon with high $p_T$; we do not consider them any further.}.

The relevant LO/NLO double--differential distributions $\frac{d^2\sigma}{dM\,dx_F}$, $\frac{d^2\sigma}{dM\,dy}$ for the Drell-Yan process have been summarized in the Appendix of \cite{Sutton:1991ay}\footnote{There is an error in Eq.~(A.8) of \cite{Sutton:1991ay}, which has to be modified \cite{Furmanski:1981cw, Gluck:1992tq} in order to conform with the usual $\overline{\rm MS}$ convention for the number of gluon polarization states, $2(1-\epsilon)$ in $4-2\epsilon$ dimensions.}. More recently the NNLO corrections to the rapidity distribution  have been calculated as well \cite{Anastasiou:2003yy, Anastasiou:2003ds}. Combinations of parton distributions like\footnote{To keep the notation compact we use here $q_i\!\equiv\!f_{q}(x_i,\mu^2)$ with $i\!=\!1,2$ and below in this paragraph $\sigma^{pN}\!\equiv\!\frac{d^2 \sigma^{pN}}{dM\,dx_F}$.} $q_1\bar{q_2}+q_2\bar{q_1}$, etc. appear in these expressions and therefore the Drell--Yan process is sensitive separately to sea and valence distributions, which is in contrast to the NC DIS cross--sections, where they usually enter as $q+\bar{q}$ (cf. Sec.\,\ref{Sec.Flight}). This is the main reason for the inclusion of Drell--Yan data in global QCD analyses of PDFs; in particular they are instrumental in fixing $\bar{d}/\bar{u}$ (or $\bar{d}-\bar{u}$), for example $\frac{\sigma^{pd}}{2\sigma^{pp}}\simeq\frac{1}{2}\big(1+\frac{\bar{d}_2}{\bar{u}_2}\big)$ for $x_1\gg x_2$ at LO. Another alternative is the use of CC (neutrino) deep inelastic scattering structure functions.

A process closely related to Drell--Yan dilepton production is the hadronic production of electroweak bosons \cite{BargerPhillips, Pink, Reya:1979zk}. Here we briefly outline the calculation of total rates for $W^\pm$ production, in particular their dependence on the factorization scale, because we use them in Sec.\,\ref{Sec.WandHiggs} in connexion with the perturbative stability of different treatments of heavy quark masses at high--energy colliders. Since the decay widths of the electroweak bosons are small compared to their masses, the narrow width approximation may be used to integrate the differential mass distribution $\frac{d\sigma}{dM}$ \cite{Hamberg:1990np}. Using an arbitrary factorization scale $\mu^2\!=\!{\cal{O}}(M_W^2)$ we get at LO\footnote{In Sec.\,\ref{Sec.WandHiggs} we evaluate these rates up to NLO, however, since the expressions are somehow lengthy and cumbersome and, moreover, the corresponding FFNS heavy quark contributions are only known at LO, we limit the discussion here to LO; which is sufficient for the exhibition of the main features. The reader interested in the NLO expressions is referred to \cite{Hamberg:1990np}.}:
\begin{equation}
\sigma^{W^\pm}\!(s)=\frac{\pi}{3}\sqrt{2}G_F\tau_W\hspace{-1em}\sum_{e_{q_i}+e_{\bar{q}_j}=\pm1}\hspace{-1em} V_{q_i\,q_j}^2\int_{\tau_W}^1\frac{dx_1}{x_1}q_i(x_1,\mu^2)\bar{q}_j(\tfrac{x_1}{\tau_W},\mu^2)
\end{equation}
where $\tau_W\!\equiv\!\frac{M_W^2}{s}$, $V_{q_i\,q_j}$ are the CKM matrix elements and the sum runs over all light quark and antiquark flavors in both hadrons, i.e. the contributions considered originate from partonic subprocess involving light flavors ($u\,\bar{d}\to W^+,\, u\,\bar{s}\to W^+$, etc.).

The heavy quark flavor contributions to the total $W^\pm$ production rate in the ZM--VFNS are calculated in analogy to the light contributions \big($c\,\bar{s}(\bar{d})\! \to \!W^+$, $\bar{b}\,u\!\to\!W^+$, etc.\big) while in the FFNS they are, like in DIS, gluon induced (i.e. proceed via $g\,\bar{s}(\bar{d})\!\to\!\bar{c}\,W^+$, $g\,u\!\to\!b\,W^+$, etc.), for example, for the subprocess $g\,s\!\to\!c\,W^-$ they are given by:
\begin{equation}
\sigma^{p\bar{p}\!\to\!cW^-X}(s) = \int_{\tau}^1 dx_1 \int_{\tau/x_1}^1 dx_2\, [g(x_1,\mu^2)s(x_2,\mu^2)+(1\leftrightarrow2)]\,\hat{\sigma}^{gs\!\to\! cW^-}\!\!\!(x_1 x_2 s)
\end{equation}
where now $\tau\!=\!\frac{(m_c+M_W)^2}{s}$ and $\hat{\sigma}^{gs\!\to\!cW^-}\hspace{-0.5em}$ can be found in \cite{Halzen:1987qm, Gluck:2008gs}. Unfortunately, the NLO corrections to this (massive) FFNS cross--section are not available in the literature. Only quantitative LO and NLO results for the analogous process $g\,b\!\to\!t\,W^-$ have been presented in \cite{Zhu:2002uj}, but questioned in \cite{Campbell:2005bb}. The NLO/LO $K$-- factor is expected \cite{Campbell:2005bb} to be in the range of 1.2 -- 1.3\,.

Another important process in hadron--hadron collisions is the production of high--$p_T$ jets resulting from the hadronization of partons produced in hard scattering processes. Although non straightforwardly, it is possible to define infrared--safe cross--sections and to compare data and theoretical predictions \cite{Yndurain, BargerPhillips, Pink}. Of particular interest for us is the $p_T$ dependence of these cross--sections, which fall strongly as $p_T$ increases. These distributions help to constrain the strong coupling and are specially important for the determination of the gluon distribution at high $x$, where no other experimental information directly sensitive to the gluon distribution is available.

Since the detected objects are not directly the (colored) partons but rather the result of their fragmentation into bunches of color neutral particles (jets), any description of jet cross--sections is strongly dependent on the (Monte Carlo) algorithm used to define the jets; needless to say that it is crucial for an appropriate comparison that both data and theory use the same jet definitions. We use the so-called \emph{cone algorithm} as implemented up to NLO in the \texttt{fastNLO} package \cite{Kluge:2006xs}, in which the relevant LO/NLO quantities needed for the calculation of jet cross--sections are tabulated for each particular data set in a way that permits the evaluation of the cross--sections using arbitrary parton distributions. Jet cross--sections beyond NLO have not yet been calculated.

\chapter*{2. Analysis and Results}
\setcounter{chapter}{2}
\setcounter{section}{0}
\setcounter{figure}{0}
\setcounter{table}{0}
\setcounter{equation}{0}
\setcounter{footnote}{0}
\chaptermark{Analysis and Results}
\addcontentsline{toc}{chapter}{2. Analysis and Results}
\fontsize{11}{16.5}
\selectfont
\section{The Dynamical Parton Model}\label{Sec.Dynamicalmodel}
The parton distributions of the nucleon are extracted from experimental data by two essentially different approaches\footnote{There are also early attempts \cite{Rojo:2006ce,DelDebbio:2007ee} to extract them using neural networks, which we do not consider any further.} which differ in their choice of the parametrizations of the input distributions at some low input scale $\mu^2_0$. In the common approach, hereafter referred to as ``\emph{standard}'', e.g. \cite{Gluck:1980cp, Duke:1983gd, Eichten:1984eu, Diemoz:1987xu, Martin:1987vw, Harriman:1990hi, Kwiecinski:1990ru, Martin:1994kn, Martin:1994kk, Martin:1995ws, Martin:1996eva, Martin:1998sq, Martin:1999ww, Martin:2001es, Martin:2002dr, Martin:2002aw, Martin:2003sk, Martin:2004ir, Martin:2007bv, Morfin:1990ck, Botts:1992yi, Lai:1994bb, Lai:1996mg, Lai:1999wy, Pumplin:2002vw, Stump:2003yu, Kretzer:2003it, Tung:2006tb, Pumplin:2007wg, Lai:2007dq, Nadolsky:2008zw, Alekhin:1996za, Alekhin:2002fv, Alekhin:2005gq, Alekhin:2006zm, Adloff:2003uh, Chekanov:2002pv, Chekanov:2005nn, Breitweg:1998dz}, the input scale is fixed at some {\em arbitrarily chosen} $\mu^2_0 \geq 1 \text{ GeV}^2$ and the corresponding input distributions are unrestricted, allowing even for negative gluon distributions in the small Bjorken--$x$ region \cite{Breitweg:1998dz, Chekanov:2002pv, Martin:2001es, Martin:2002dr, Martin:2006qv}, i.e., negative cross--sections like $F_{\rm L}(x,Q^2)$.

Alternatively \cite{Altarelli:1973ff, Parisi:1976fz, Novikov:1976dd, Gluck:1977ah,Gluck:1988xx, Gluck:1989ze, Gluck:1991ng, Gluck:1994uf, Gluck:1998xa, Gluck:2007ck, Gluck:2008gs, JimenezDelgado:2008hf} the parton distributions at \mbox{$\mu^2\!>\!1 \text{ GeV}^2$} are QCD radiatively generated from \emph{valence}--like, i.e. \emph{positive definite} ($xf>0$ with $xf\rightarrow0$ for $x\rightarrow0$), \emph{input} distributions for \emph{all} partons at an \emph{optimally determined} input scale \mbox{$\mu^2_0\!<\!1 \text{ GeV}^2$}. This more restrictive ansatz implies, as we will see, more predictive power and less uncertainties concerning the behavior of the parton distributions in the small--$x$ region at $\mu^2\!>\!\mu^2_0$, which is to a large extent due to QCD dynamics.

The dynamical description of the nucleon connects the naive quark model with the parton description of deep inelastic phenomena, i.e., the idea that baryons are bound states of three quarks with scaling violations and the fact that about half of the momentum of the nucleon is carried by the gluons. The original dynamical assumption \cite{Altarelli:1973ff, Parisi:1976fz, Novikov:1976dd, Gluck:1977ah} was that at a certain sufficiently--low resolution scale the nucleon consist of its constituent/valence quarks only, appearing the gluon and sea quarks as a result of bremmstrahlung processes, therefore \emph{radiatively} or \emph{dynamically} generated.

In order to improve the agreement with constraints imposed by data on direct--photon production and some deep inelastic data at medium $x$ (see \cite{Gluck:1989ze, Gluck:1991ng} for more details) the model was subsequently extended to include gluon \cite{Gluck:1989ze} and sea ($\bar{u}\!=\!\bar{d}$) \cite{Gluck:1991ng} \emph{valence}--like input distributions, amounting the underlying physical picture to ``constituent'' gluons and sea quarks comoving with the valence quarks at the low input scale $\mu^2_0$. The outcome of these modifications are \emph{steep} gluon and sea distributions at small $x$ which, as mentioned, due to the valence--like structure of the input distributions and the low value of the input scale (longer evolution distance), are mainly attributable to QCD dynamics. This is in contrast to the fine tuning (or even ad hoc extrapolations into unmeasured regions) required in standard fits, which depend strongly on the assumed ansatz.

\fontsize{11}{17}
\selectfont

These rather unique \emph{predictions} were first confirmed at HERA \cite{Abt:1993cb, Derrick:1993fta} (see also \cite{Gluck:1993im}), which appear to support the dynamical approach to the determination of parton distribution functions, i.e., that the low--$x$ structure of parton distributions can be generated dynamically, almost in a parameter--free way, starting from valence--like distributions at a low input scale. In any case, it does not have any disadvantage\footnote{Except maybe a marginally larger $\chi^2$ for the statistically most significant data sets included in the fits due to the more restrictive ansatz; for the same reason the model is expected to fit comparably (or better) data with lower significance.} respect to ``standard'' approaches, while its predictive power is an important and desirable feature. We will analyze quantitatively the implications of the dynamical approach to parton distribution functions in the next sections.

Besides other criticisms \cite{Gluck:1994uf}, whether or not the standard evolution equations are applicable down to such small scales ($\mu_0^2\sim 0.5 \text{ GeV}^2$) may be a matter of controversy. It should be noted that the predictions of the model are only intended to be compared with physical observables at higher scales, say $\mu^2>2\text{ -- }3 \text{ GeV}^2$; below that, higher--twist effects (although decoupled from the evolution of \emph{covariant} leading--twist distributions themselves) may become relevant and even dominant. Nevertheless the dynamical model provides also a natural link connecting nonperturbative models valid at $\mu^2\!<\!1\text{ GeV}^2$ with the measured distributions at $\mu^2\!>\!2\text{ -- }3\text{ GeV}^2$.

\section{Experimental Input and Analysis Formalism}\label{Sec.DataFormalism}
Following the radiative approach we have updated the previous LO/NLO GRV98 dynamical parton distribution functions of \cite{Gluck:1998xa} and extended these analyses to the NNLO of perturbative QCD. In addition we have made a series of ``standard'' fits in (for the rest) exactly the same conditions as their dynamical counterparts. This allows us to compare the features of both approaches and to test the the dependence in model assumptions. Beyond LO we use the modified minimal substraction ($\overline{\rm MS}$) factorization and renormalization schemes. In order to test the dependence of our results on the specific choice of factorization scheme we have repeated our NLO dynamical analysis in the DIS factorization scheme of Sec.\,\ref{Sec.DISscheme}. The quantitative difference between the $\overline{\rm MS}$ and DIS results turn out to be rather small.

The statistically most significant data that we use are the HERA (H1 and ZEUS) measurements \cite{Adloff:1999ah, Adloff:2000qj, Adloff:2000qk, Adloff:2003uh, Chekanov:2001qu} of the DIS ``reduced'' cross--section of Eq.\,\ref{sigmar}\footnote{Note that the data we use are the radiatively corrected ones as presented by the experimentalists.} for $Q^2\geq 2$ GeV$^2$. Since the experimental extraction of the usual (one--photon exchange) $F^\gamma_2$ from $\frac{d^2\sigma^{\rm NC}}{dx\,dy}$ is (parton) model dependent, we have chosen to work with the full NC framework in order to avoid any further dependence on model assumption. However, it turned out that fitting just to $F^\gamma_2$ gives very similar results. In addition, we have used the fixed target $F_2^p$ data of SLAC \cite{Whitlow:1991uw} , BCDMS \cite{Benvenuti:1989rh} , E665 \cite{Adams:1996gu} and NMC \cite{Arneodo:1996qe}, and the structure function ratios $F_2^n/F_2^p$ of BCDMS \cite{Benvenuti:1989gs}, E665 \cite{Adams:1995sh}  and NMC \cite{Arneodo:1996kd}; both subject to the standard cuts $Q^2 \geq 4$ GeV$^2$ and $W^2\!\geq\!10$ GeV$^2$. The well--known standard target mass corrections (Eq.\,\ref{TMC}) to $F_2$  have been taken into account in the medium to large $x$--region for $Q^2\!<\!100$ GeV$^2$.

\fontsize{11}{18.2}
\selectfont
As indicated in Sec.\,\ref{Sec.Fheavy}, we work in the (three flavors) FFNS, i.e., heavy quarks ($c, b, t$) are not considered as partons and the number of active flavors appearing in the splitting functions and the coefficient functions is fixed to $n_F\!\!=\!\!3$. An extension to the usual (ZM)VFNS will be presented in Sec.\,\ref{Sec.HeavyDIS}. Note that in all our analyses the strong running coupling is governed by a variable number of flavors $n_R$, as is common in the $\overline{\rm MS}$ renormalization scheme. This is perfectly compatible with the chosen value for $n_F$ as detailed in \cite{Gluck:2006ju}.

The heavy quark ($c,\,b$; top quark contributions are negligible) contributions to $F^{\rm NC}_{2,L}$ are theoretically described in the FFNS by the fully predictive fixed--order perturbation theory (cf. Sec.\ref{Sec.Fheavy}). These contributions are quantitatively negligible for $F_{2, \rm L}^{\gamma \rm Z,\,Z}$ in Eq.\,\ref{neutralcurrent}, but they have been included nevertheless; for $F_3^{\gamma\rm Z,\,Z}$ they are negligibly small (cf. Sec.\ref{Sec.Fheavy}) and we have neglected them. The HERA measurements of heavy quark production \big($F_2^{c,\,b}$\big) of \cite{Chekanov:2003rb, Adloff:2001zj, Aktas:2004az, Aktas:2005iw} have also been included in the LO and NLO fits. In the NNLO fits these data have not been included, and the unknown third order coefficient functions in the heavy quark contributions of Eq.\,\ref{Flight_heavy}  have been neglected. For the quark masses we have chosen:
\begin{equation}
m_c = 1.3\,{\rm GeV},\qquad m_b = 4.2\,{\rm GeV},\qquad m_t = 175\,{\rm GeV}\, ,
\end{equation}
which turn out to be the optimal choices for all our subsequent analyses, in particular for heavy quark production. 

Furthermore, the Drell--Yan dimuon pair production data of the E866/NuSea (fixed target) experiment for $\frac{d^2\sigma^{pN}}{dx_F\,dM}$ with $N\!=\!p,\,d$ of \cite{Webb:2003ps} as well as their asymmetry $\frac{\sigma^{pd}}{2\sigma^{pp}}$ measurements \cite{Towell:2001nh} have been used. The description of Drell--Yan cross--sections at LO of QCD is well known \cite{Yndurain, Pink, BargerPhillips} to be systematically out of the ballpark by about 30 to 50\%. This is taken into account in our LO fits by including a phenomenological (constant) K--factor (i.e., $\sigma^{LO}\!\to\! K\sigma^{\text{LO}}$) which is allowed to float, in both cases (dynamical and ``standard'') we obtain $K\!=\!1.46$\,. A further complication in the inclusion of these data is that they are given in terms of $x_F$, whereas the NNLO expressions have been given only in terms of the dilepton rapidity $y$ \cite{Anastasiou:2003yy,Anastasiou:2003ds}. Since experimentally the dilepton $p_T$ is small (below about 1.5 GeV) as compared to the dilepton invariant mass $M \gtrsim 5$ GeV, we have checked that it can be safely neglected and the two distributions can be related using leading order kinematics\footnote{That is, considering $p_T\!=\!0$ so that $\frac{d^2\sigma}{dM\,dx_F}\!=\!\frac{1}{x_1+x_2}\frac{d^2\sigma}{dM\,dy}$
(cf. Sec.\,\ref{Sec.Hadroproduction}).}, as was done in \cite{Alekhin:2006zm}. For our NNLO analysis we used the routine developed in \cite{Anastasiou:2003ds} and improved in \cite{Alekhin:2006zm}.

Finally, the Tevatron high--$p_{\rm T}$ inclusive jet data of D0 \cite{Abbott:2000ew} and CDF \cite{Abulencia:2005yg} have been used together with the \texttt{fastNLO} package of \cite{Kluge:2006xs} for calculating the relevant cross--sections up to NLO. In the NNLO fits the jet data have been omitted due to the lack of its theoretical description at this order.

\begin{figure}
\centering
\includegraphics[width=0.5\textwidth]{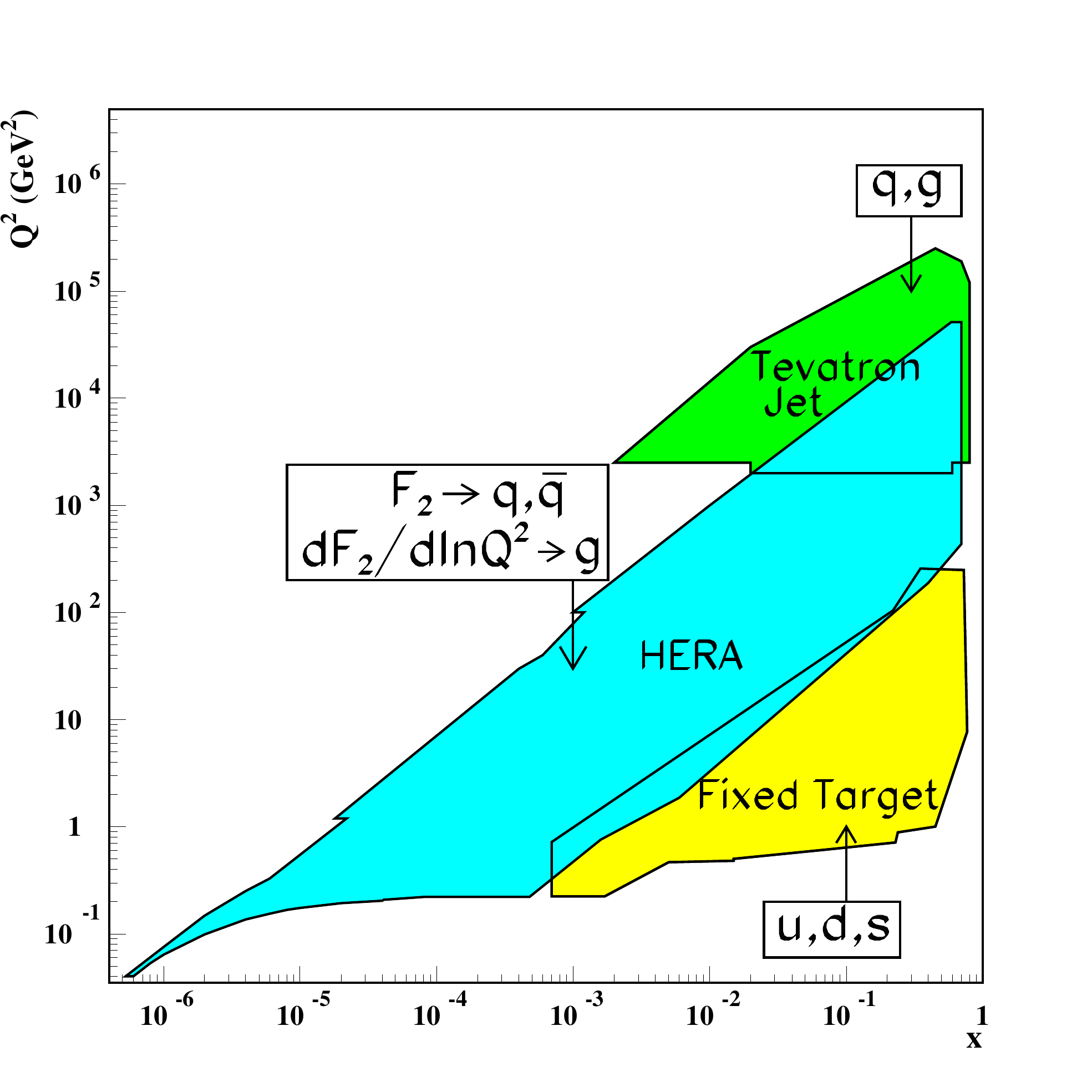}
\caption{Kinematic $x/Q^2$ range of fixed target and collider experiments; also shown are the major constrains that they make on the PDFs. The figure has been taken from \cite{Amsler:2008zz}.\label{Fig201}}
\end{figure}

\fontsize{11}{16.5}
\selectfont

The kinematical range of fixed target and collider experiments are complementary as can be seen in Fig.\,\ref{Fig201}. Note however that in the theoretical description of the data usually convolutions of the parton distributions are involved, i.e., this range is only orientative of the values at which the parton distributions need to be evaluated.

It is worth to mention that in the (NLO) DIS analysis the description of heavy quark contributions, Drell--Yan data and jet data is achieved, using their known theoretical $\overline{\rm MS}$ expression, through the factorization scheme transformations of Sec.\,\ref{Sec.DISscheme}. These transformations allow also for a consistent comparison of our DIS results with the ones obtained in the $\overline{\rm MS}$ scheme.

We include $\alpha_s(M_{\rm Z}^2)$ as a free parameter in our fits, and determine its value together with the parton distributions. It should be noted that there is a certain correlation between the value of the input scale $\mu_0^2$ and the resulting values for $\alpha_s(M_{\rm Z}^2)$, which tends to increase with $\mu_0^2$. In the dynamical approach we performed fits for various values of the input scale keeping $\alpha_s(M_{\rm Z}^2)$ as a free parameter while requiring a valence--like structure $(a_f>0)$  for \emph{all} the input distributions. Then we fixed the best choice for $\mu_0^2\!=\!0.30, \;0.50, \;0.55 \text{ GeV}^2$ for the LO, NLO and NNLO fits respectively and performed the final precision fits and error analyses. In the ``standard'' approach the input scale was fixed to $\mu_0^2\!=\!2 \text{ GeV}^2$.

On the line of \cite{Gluck:1994uf, Gluck:1998xa} we use the following parametrizations for our \cite{Gluck:2007ck} input distributions:
\begin{equation}\label{inputparameters}
 \begin{array}{l}
xu_v(x,\mu_0^2) = N_u\, x^{a_u}(1-x)^{b_u}(1+A_u\, \sqrt{x} + B_u\, x)\\
xd_v(x,\mu_0^2) = N_d\, x^{a_d}(1-x)^{b_d}(1+A_d\, \sqrt{x} + B_d\, x)\\
x\Delta(x,\mu_0^2) = N_\Delta\, x^{a_\Delta}(1-x)^{b_\Delta}(1+A_\Delta\, \sqrt{x} + B_\Delta\, x)\\
x\Sigma(x,\mu_0^2) = N_\Sigma\, x^{a_\Sigma}(1-x)^{b_\Sigma}(1+A_\Sigma\, \sqrt{x} + B_\Sigma\, x)\\
xg(x,\mu_0^2) = N_g\, x^{a_g}(1-x)^{b_g}
 \end{array}
\end{equation}
where $u_v\equiv u-\bar{u}$, $d_v\equiv d - \bar{d}$, $\Sigma\equiv \bar{u} + \bar{d}$ and $\Delta\equiv \bar{d} - \bar{u}$. The distributions are further constrained by quark number and momentum conservation sum rules:
\begin{equation}
\label{sumrules}
\int_0^1 u_v\,dx = 2,\quad \int_0^1 d_v\,dx =1, \quad \int_0^1 x\bigl( u_v+d_v+2(\bar{u}+\bar{d}+\bar{s}) +g \bigr) dx =1.
\end{equation}
which, as usual, we use to determine $N_{u,d,g}$, which therefore are not free parameters in our fits. We have tried different forms for these parametrizations, including ($1 + c\, x^d$) and ($1 + c \sqrt x  + d\,x + e\, x^{1.5}$) for the polynomials, without finding any improvement or  substantial change relative to the classical form Eq.\,\ref{inputparameters}. In particular all of our fits did not require the polynomial for the gluon distribution. The values obtained for the parameters of the input distributions in our different fits are given in Tab.\,\ref{TabA01}.

\fontsize{11}{17}
\selectfont

As in \cite{Gluck:1994uf, Gluck:1998xa}, the ``light'' sea  is no longer symmetric ($\bar{u}\!\neq\!\bar{d}$) as required by the Drell--Yan data, which are instrumental in fixing $\Delta$ (cf. Sec.\,\ref{Sec.Hadroproduction}). Since the data sets we are using are insensitive to the specific choice of the strange quark distributions, we consider a symmetric strange sea. In the dynamical approach the strange densities are entirely radiatively generated  starting from $s(x,\mu_0^2)=\bar{s}(x,\mu_0^2)=0$, while in the ``standard'' approach we choose as usual $s(x,\mu_0^2)\!=\!\bar{s}(x,\mu_0^2)\!=\!\tfrac{1}{4}\big(\bar{u}(x,\mu_0^2)+\bar{d}(x,\mu_0^2)\big)$.

The minimization procedure follows the usual chi--squared method with $\chi^2$ defined as:
\begin{equation}
\label{chi2}
 \chi^2(p) = \sum_{i=1}^N \left( \frac{\text{data}(i) - \text{theory}(i,p)}{\text{error}(i)} \right)^2,
\end{equation}
where $p$ denotes the set of $21$ independent parameters in the fit, including $\alpha_s(M_Z^2)$, and $N$ is the number of data points included; $N\!=\!1739\,(1567)$ for the LO/NLO\,(NNLO) fits. The errors include systematic and statistical uncertainties, being the total experimental error evaluated in quadrature. The (fully correlated) normalization error of each data set is treated separately by allowing a common normalization factor ($N_{\text{set}}$) to float within the experimental error\footnote{These normalization factors multiply the data in our current analyses, i.e. $\text{data}(i) \to N_{\text{set}(i)}\,\times \text{data}(i)$ in Eq.\,\ref{chi2}. In \cite{Gluck:2007ck} a slightly different definition of $\chi^2$, in which these normalization factors entered dividing the theoretical predictions, was used and thus the values quoted for $\chi^2$ in \cite{Gluck:2007ck} are slightly different from those in Tab.\,\ref{TabA01}. These difference are in any case negligible and we have explicitly checked that this does not affect in any case the outcome of the fits.}. The values obtained for these normalization factors in our different fits are given in Tab.\,\ref{TabA02}.

The minimum $\chi^2$ obtained in our fits as well as the contributions stemming only from subsets of data are given in Table \ref{TabA01}. As expected $\chi^2$ is slightly smaller for the ``standard'' fits, in particular for the DIS data which are the statistically most significant sets. In general, the LO fits do not give an appropriate description of the data while at NLO it is already satisfactory ($\chi^2\simeq 1$) and the NNLO fits result in an even better (smaller) $\chi^2$, typically $\chi_{\rm NNLO}^2\simeq 0.9 \chi_{\rm NLO}^2$. We compare extensively our results with the data used in Sec.\,\ref{Sec.Comparisons}.

In view of this improvement in $\chi^2$, one could in principle reduce the number of parameters at NNLO and still obtain a good fit to the data ($\chi^2\simeq 1$), which would result in a reduction of the NNLO uncertainties. As explained in the next section, we have chosen for the uncertainty estimations the (relatively simple) hessian method, keeping the same ansatz in all of our fits in order to make these estimations somehow comparable. Note that the error calculations for parton distributions are rather arbitrary and must be interpreted with care and understood always as a rough estimation, useful only to compare analyses with similar treatments and in particular to disentangle in which regions the distributions are more/less constrained. This would be the case even with a more rigorous error analysis (taking into account correlations, etc.) since, as mentioned, the results obtained depend strongly on the framework used.

Furthermore, these error estimations constitute the propagation of the experimental errors into the input parameters of the distributions, other sources of errors like, for example, the choices of renormalization and factorization scales, which particular solutions of the RGE are employed (cf. Sec.\,\ref{Sec.RGEsolution}) or even which data sets are used, should be treated separately and, in any case, keep always in mind that they are not included in what is usually meant by ``error'' of the parton distributions.

\fontsize{11}{16.5}
\selectfont

\section{Estimation of Uncertainties}\label{Sec.Uncertainties}
The experimental errors of the data induce an uncertainty in the determination of parton distribution functions (and therefore to any quantity calculated using them) which can be estimated now by linear propagation and other methods \cite{Botje:1999dj, Botje:2001fx, Pumplin:2000vx, Stump:2001gu, Pumplin:2001ct}.

Our estimates are based in the so--called Hessian method, which has been discussed in detail in \cite{Pumplin:2000vx, Pumplin:2001ct}. Being the
minimum $\chi^2(p)$ characterized by the set of parameter values $p^0$, this method is based on the approximation of the variations of $\chi^2$ around its minimum by:
\begin{equation}
\Delta \chi^2 = \chi^2(p)-\chi^2(p^0) \simeq \sum_{i=1}^d \frac{\partial \chi^2}{\partial p_i}\Delta p_i + \frac{1}{2}\sum_{i,j=1}^d \frac{\partial \chi^2}{\partial p_i \partial p_j}\Delta p_i \, \Delta p_j\,,
\end{equation}
where $d$ is the number of parameters considered in the error estimation (see below), $\Delta p_i \!\equiv\! p_i - p^0_i$ are displacements from their values at the minimum and the derivatives are also taken at the minimum, where the linear term in the above equation should vanish (meaning that the calculations of physical observables in $\chi^2$ should vary \emph{linearly} around this point in the parameter space). Therefore all the information of $\chi^2$ around its minimum is contained in the matrix of second derivatives or Hessian matrix $H_{ij}$\footnote{We define the Hessian matrix as $H_{ij}\equiv \frac{\partial \chi^2}{\partial p_i \, \partial p_j}$, with no factors of one half.} in this approximation. The idea behind the method is to use this matrix to calculate the variations of the results with the parameters in the ``vicinity'' of the minimum, where the approximation should work. What is meant by ``vicinity'' is characterized by the \emph{tolerance parameter} $T$ via $\Delta \chi^2 \leq T^2$.

If the data used were perfectly uncorrelated and perfectly compatible between them, the tolerance parameter corresponding to 1 $\sigma$ uncertainty in the calculations (stemming, of course, from the 1 $\sigma$ experimental errors in $\chi^2$) should be $T\!=\!1$. As in others QCD fits (cf. \cite{Martin:2002aw, Pumplin:2001ct}), this is not our case because our data come from different observables and experiments and, moreover, the correlated errors are included in the experimental error in Eq.\,\ref{chi2}. As, for example, which deviations from the expected value of $N \pm \sqrt{2N}$ are acceptable for $\chi^2$ in a global QCD fit, which value to choose for the tolerance parameter $T$ is somewhat arbitrary and subject to subjective interpretations. Since it seems reasonable that it somehow scales with the number of data (sets and) points included in the fits, we have chosen $T^2\!=\!\frac{\sqrt{2N}}{(1.65)^2}$; this results in $T\!=\!4.654 \;(4.535)$ for our LO/NLO (NNLO) fits. Here we interpret the (1 $\sigma$) deviation ($\sqrt{2N}$) in $\chi^2$ as more than one standard deviation in the predictions, say 90\%, and we have rescaled conveniently to what we consider as one standard deviation errors; again, this is completely arbitrary.

For the numerical evaluation of the Hessian matrix it is convenient to map the parameter space into a coordinate system where the tolerated ``vicinity'' of the minimum is represented by a hypersphere. In this way the variations of $\chi^2$ are uniform and the instabilities resulting from the numerical difference of the typical changes in $\chi^2$ in different directions of the parameter space are avoided \cite{Pumplin:2000vx}. Since the Hessian matrix is symmetric, it has a complete set of ($d$) orthonormal eigenvectors $V_i^{(k)}\equiv v_{ik}$ with eigenvalues $\epsilon_k$, that is:
\begin{equation}
\sum_{j=1}^d H_{ij} v_{jk} = \epsilon_k v_{ik}, \hspace{2em}
\sum_{i=1}^d v_{ij} v_{ik} = \delta_{jk},\qquad \text{then}\qquad
H_{ij} = \sum_{k=1}^d \epsilon_k v_{ik}\,v_{jk}
\end{equation}
The rescaled eigenvectors of the hessian matrix provide a natural basis for the mentioned transformation, which is achieved via:
\begin{equation}
\Delta p_i = \sum_{k=1}^d M_{ik}z_k, \qquad M_{ik}\equiv \sqrt{\frac{2}{\epsilon_k}} v_{ik},
\end{equation}
where $z_k$ are the new coordinates in term of which the tolerated region is the hypersphere:
\begin{equation}
\Delta \chi^2 = \sum_{k=1}^d z_k^2 \leq T^2.
\end{equation}
In practice the matrix $M_{ij}$ is calculated by an extension of \texttt{MINUIT} called \texttt{ITERATE} \cite{Pumplin:2000vx} which uses an iterative procedure which converge to the eigenvectors and eigenvalues of the Hessian matrix of $\chi^2$ in Eq.\,\ref{chi2}.

\fontsize{11}{16.2}
\selectfont

\begin{figure}[t]
\centering
\includegraphics[width=0.94\textwidth]{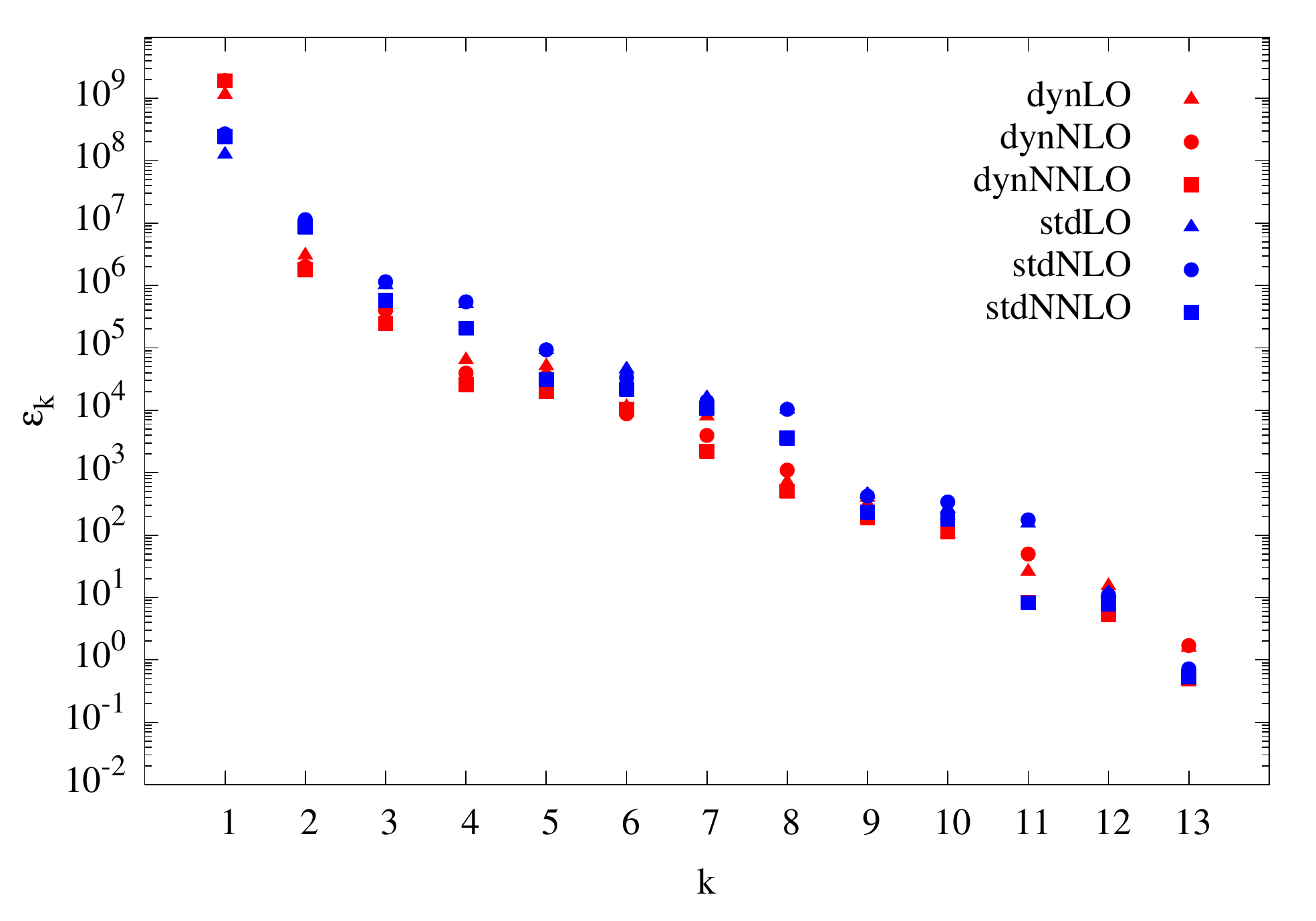}
\caption{Eigenvalues of the Hessian matrix for our LO, NLO\,($\overline{\rm MS}$) and NNLO\,($\overline{\rm MS}$) fits ordered from largest to smallest. In this order their decrease is, as expected \cite{Pumplin:2001ct}, approximately exponential.\label{Fig202}}
\end{figure}

With help of $M_{ij}$ it is possible to calculate the \emph{eigenvector basis sets} consisting in $2d$ sets of parameters $p_i^{\pm j} \!=\! p_i^0 \pm T M_{ij}$ which contain (in conjunction with $T$) the same information than the hessian matrix. Although by this construction the number of parameters entering in the error calculations is (unnecessarily) increased, it turns out to be useful for the propagation of experimental errors into quantities which depend on parton distributions. For example, the derivatives and a unit vector in the direction of the gradient of \emph{any} quantity $X(p)$ are:
\begin{equation}
\frac{\partial X}{\partial z_k}=\frac{X(p^{+k}) - X(p^{-k})}{2T}, \hspace{3em} \hat{D}_k(X) = \frac{X(p^{+k}) - X(p^{-k})}{\sqrt{\sum_j(X(p^{+j}) - X(p^{-j})^2}},
\end{equation}
and we can estimate the uncertainty which follows from the condition $\Delta\chi^2 \!\leq\!T^2$ by a displacement of length $T$ in the direction of the gradient; it is simply given by:
\begin{equation}
\Delta X=\sum_{k=1}^d \frac{\partial X}{\partial z_k}(T\hat{D}_k)=\frac{1}{2}\sqrt{\displaystyle \sum_{k=1}^d \left(X(p^{+k}) - X(p^{-k})\right)^2}.
\end{equation}
Considering that the calculations required for practical applications in collider phenomenology are usually rather involved, the method is indeed quite convenient. Tables with the parameters of these \emph{eigenvector sets} corresponding to our different fits are given in App.~A; the eigenvalues of the hessian matrix (cf. Fig.\,\ref{Fig202}) are also included.

For the parameters entering in the error estimations the expressions above reduce to:
\begin{equation}
\hat{D}_k(p_i)=\frac{M_{ik}}{\sqrt{\sum_j M_{ij}^2}}, \hspace{3em}  \Delta p_i = T \sum_{k=1}^d M_{ik}^2\; ;
\end{equation}
Since $M_{ik}\!\propto\!\sqrt{\frac{2}{\epsilon_k}}$, a large (small) eigenvalue corresponds to a direction in which $\chi^2$ increases rapidly (slowly), making the parameters tightly (poorly) constrained; they are usually referred to as \emph{steep} (\emph{flat}) directions. Note as well that the elements of the unit--length gradient indicate the influence of a particular eigenvector over the error of a quantity; in the case of the fitting parameters they give, by comparison, also information on their correlations. As suggested in \cite{Pumplin:2002vw}, we include in our final error analysis only those parameters that are actually sensitive to the data used, i.e., those parameters which are not close to ``flat'' directions in the overall parameter space. For the used data and our functional form Eq.\,\ref{inputparameters} $d\!=\!13$ such parameters, including $\alpha_s(M_Z^2)$, are identified and are included in our final error analysis; the remaining ill--determined eight polynomial parameters $A_f$ and $B_f$, with \emph{highly--correlated} uncertainties of more than 50\%, were held fixed.

\fontsize{11}{15.7}
\selectfont

At this point it should be mentioned that we have explicitly checked that the linear approximation is appropriate for our results, i.e., that the $\chi^2$ variations around the minimum are approximately quadratic for \emph{all} parameters included in the error analysis. It is also worth to mention that since the value of $T$ is rather arbitrary, in order to compare uncertainty results for distributions with different conventions one has to take into account the different definitions of the tolerance parameter. Fortunately, as is clear from its expression, the results for different conventions are linearly scalable, i.e., $\Delta X_{Ti}\!=\frac{T_i}{T_j}\Delta X_{T_j}$.

\section{The New Generation of Dynamical Parton Distributions}\label{Sec.NewGeneration}

\begin{figure}[b!]
\centering
\includegraphics[width=0.9\textwidth]{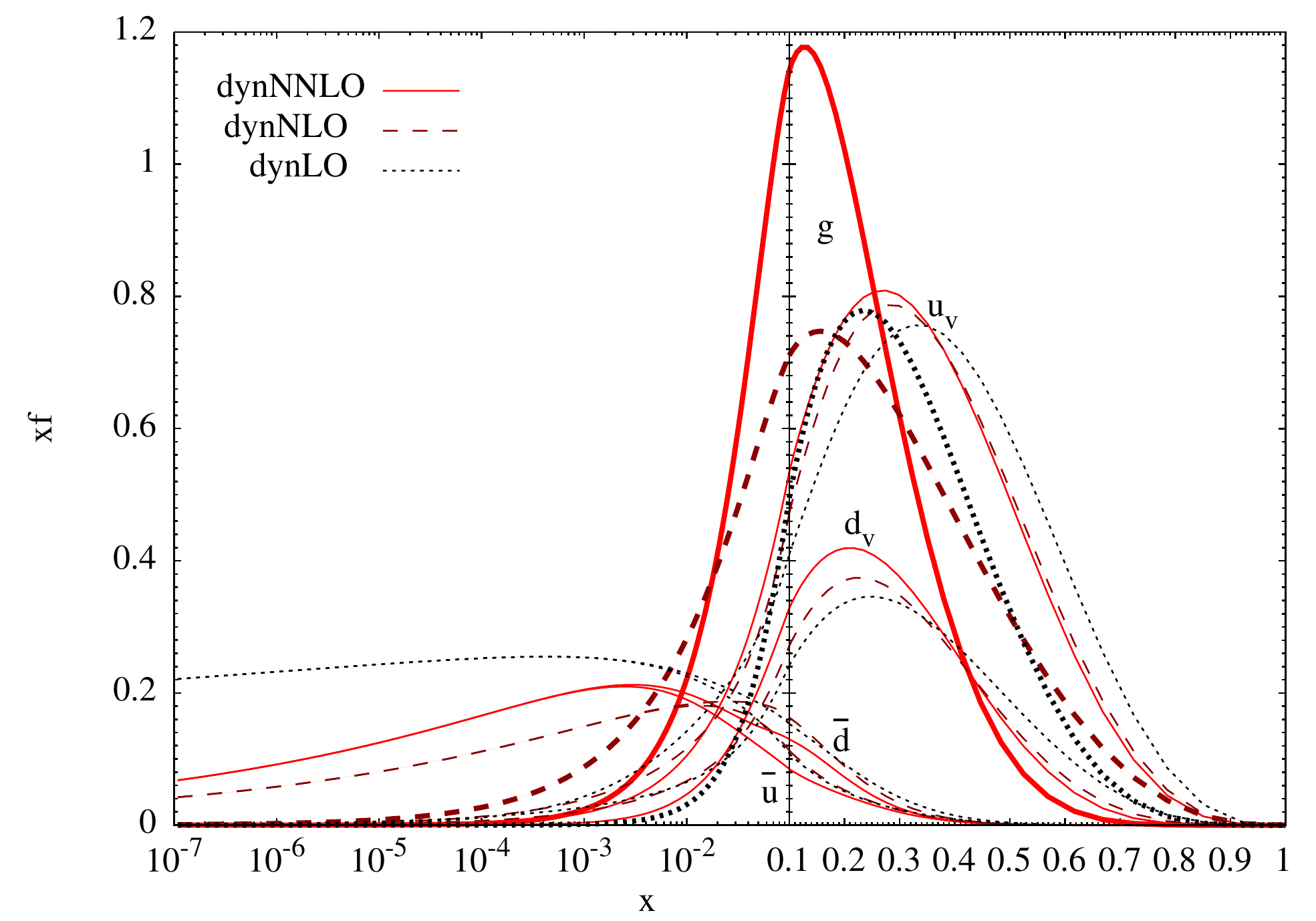}
\caption{The valence--like input densities at $\mu_0^2 = 0.3,\; 0.5,\; 0.55$ GeV$^2$ for our dynamical LO, NLO\,($\overline{\rm MS}$) and NNLO\,($\overline{\rm MS}$) results respectively. The strange sea $s\!=\!\bar{s}$ vanishes at the input scale. Note that the lines of the gluon input distributions have been plotted thicker than the other lines in order to avoid confusion. The strange sea $s\!=\!\bar{s}$ vanish at the input scale.\label{Fig203}}
\end{figure}

In this section we compare our new dynamical parton distributions with the results obtained in recent common ``standard'' analyses, including the results of our ``standard'' fits which, as mentioned, were carried out in exactly the same conditions as their dynamical counterparts. Recall also that in our analyses, as in all presently available distributions, the renormalization and factorization scales were set equal, which simplifies considerably the results but (obviously) prohibits their independent variation; we denote them here and in the following by $Q^2\!\equiv\!\mu^2\!=\!\mu_F^2\!=\!\mu_R^2$.

Our dynamical input distributions are presented in Fig.\,\ref{Fig203} according to the parameters in Table.\,\ref{TabA01}. Being the dominant parameter in the small--x region $a_f$, the dynamical distributions have by construction, i.e., by optimally choosing a low $\mu_0^2$, a valence--like structure ($a_f\!>\!0$) for \emph{all} partons; in other words, not only the valence but also the sea and gluon input distributions vanish at small--$x$. This is in contrast to the gluon and sea distributions of the ``standard'' fits, where the input scale is $\mu_0^2\!=\!2$ GeV$^2$ and thus have $a_f\!\lesssim\!0$ as is usual for common ``standard'' fits. Note that in both cases the valence distributions as well as $\Delta$ have a strong valence--like behavior, being in the former case enforced by the quark number constrains in Eq.\,\ref{sumrules} while in the latter it indicates that the small--x sea turns out to be approximately isospin symmetric. It should also be emphasized that all of our valence--like input distributions as well as the ones for the ``standard'' fits are manifestly \emph{positive} in contrast to the negative gluon distributions in the small--$x$ region believed to be \emph{needed} in other standard fits \cite{Breitweg:1998dz, Chekanov:2002pv, Martin:2001es, Martin:2002dr, Martin:2006qv}.

\fontsize{11}{18}
\selectfont

\begin{figure}[t]
\centering
\includegraphics[width=\textwidth]{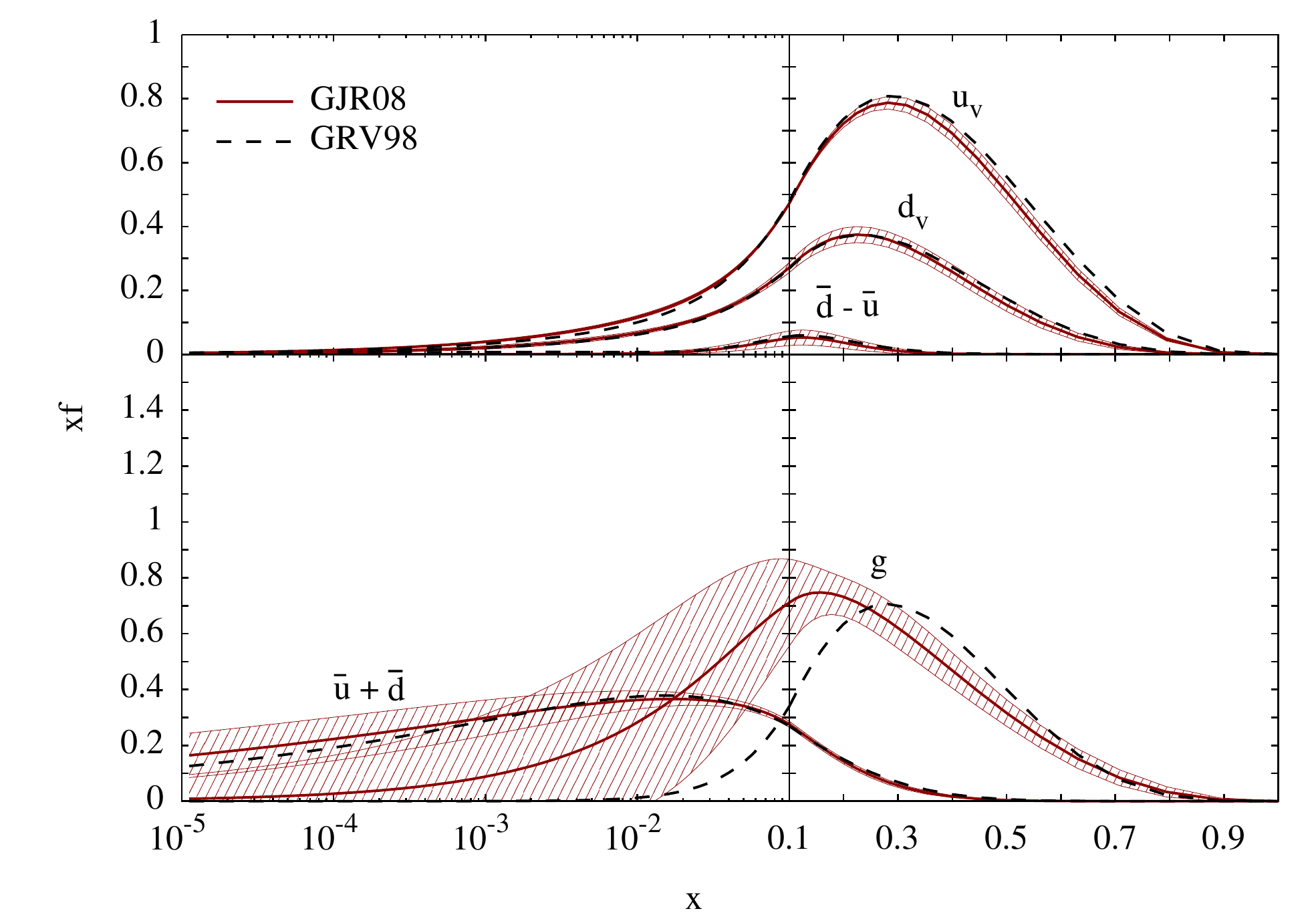}
\caption{Our (GJR08) \cite{Gluck:2007ck} dynamical NLO\,($\overline{\rm MS}$) input densities at the input scale $\mu_0^2\!=\!0.5$ GeV$^2$ together with their $\pm 1\,\sigma$ uncertainty bands. The previous GRV98 \cite{Gluck:1998xa} NLO\,($\overline{\rm MS}$) dynamical distributions are shown at their input scale $\mu_0^2\!=\!0.4$ GeV$^2$ for comparison. The strange sea $s\!=\!\bar{s}$ vanish at the input scale in both analyses. \label{Fig204}}
\end{figure}

Our \cite{Gluck:2007ck} NLO\,($\overline{\rm MS}$) valence--like input distributions together with their $1\,\sigma$ uncertainties are compared with the previous GRV98 \cite{Gluck:1998xa} NLO\,($\overline{\rm MS}$) dynamical distributions at their respective input scale in Fig.\,\ref{Fig204}. As can be seen the results of both analyses turn out to be very similar except for the gluon which peaks at a slightly larger value of $x$. However, such differences are merely within a $2\sigma$ band of the new dynamical results. Both the LO and NLO\,(DIS) results are also similar in both analyses. Note however that the DIS distributions in \cite{Gluck:1998xa} were not determined through a fit in the DIS factorization scheme but rather obtained through the factorization scheme transformations implied by Eq.\,\ref{DIStransformation}. This treatment leads sometimes to a \emph{small} negative gluon distribution in the large $x\!>\!0.5$ region which, however, is compatible within uncertainties with our definite positive gluon distribution resulting from a full analysis using the DIS factorization scheme. Besides this observation, the DIS results are, as mentioned, quite similar in both analysis. Furthermore by transforming our distributions from one scheme to the other we get also very similar results, which confirms the applicability of the transformations of Sec.\,\ref{Sec.DISscheme}.

\begin{figure}[b!]
\centering
\includegraphics[width=1.02\textwidth]{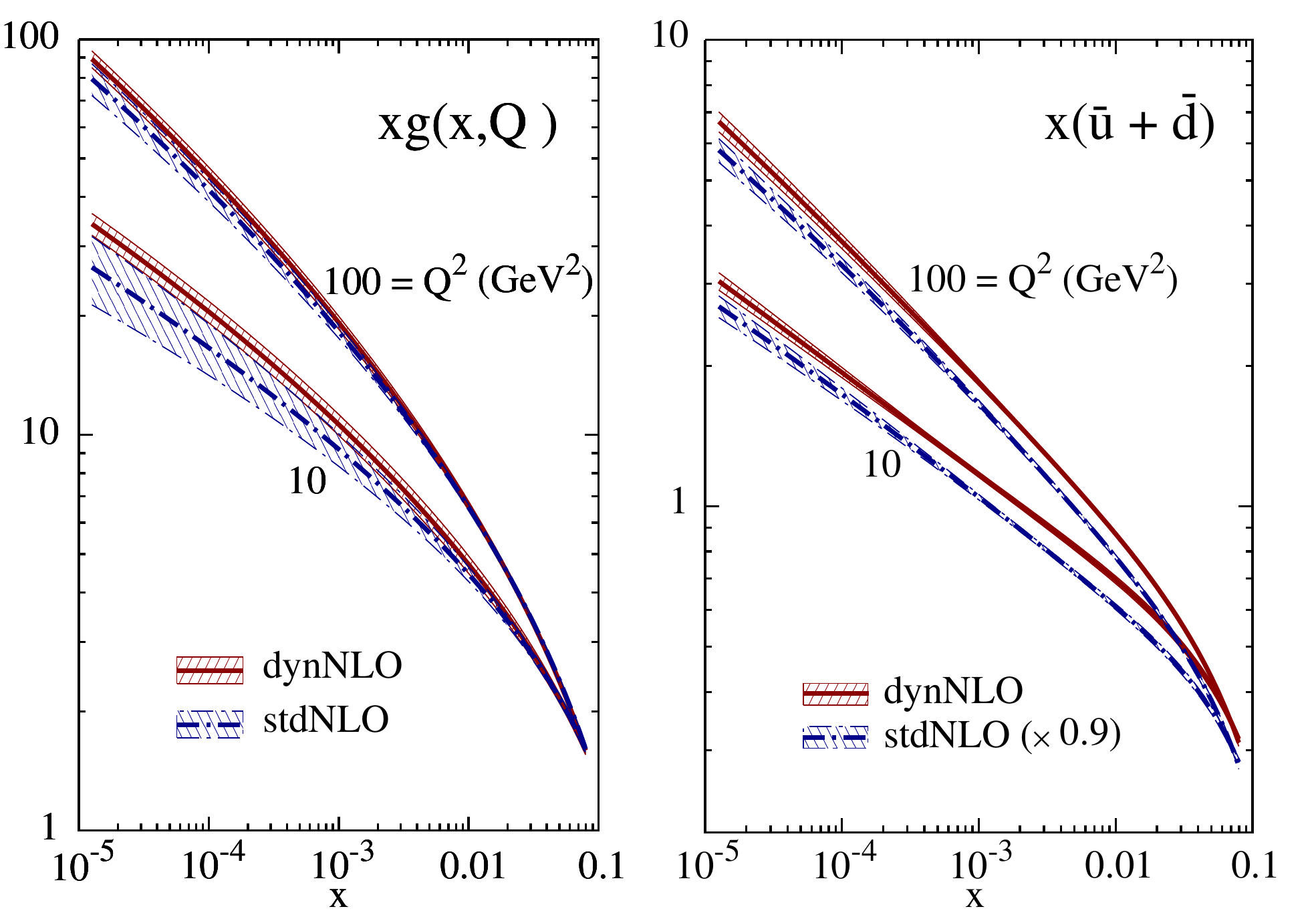}
\caption{The small--$x$ NLO\,($\overline{\rm MS}$) predictions of our dynamically (radiatively) generated gluon and sea--quark distributions, together with their $\pm 1\,\sigma$ uncertainties, as compared to the  results of a ``standard'' analysis carried out under the same conditions. To ease the visibility of the two error bands of $x(\bar{u}+\bar{d})$ we have multiplied the stdNLO results by 0.9 as indicated. The corresponding GRV98 predictions lie within the $1\,\sigma$ band of our new dynNLO results.\label{Fig205}}
\end{figure}

With respect to the uncertainties obtained in our different analyses, it must be noted that the dynamical input implies, as expected, a stronger constrained gluon distribution at larger values of $Q^2$ as compared to the gluon density obtained from ``standard'' fits with a  conventional non--valence--like input at $\mu_0^2\!>\!1$ GeV$^2$, as can be seen in Fig.\,\ref{Fig205} for our NLO\,($\overline{\rm MS}$) results. Since our valence--like sea input has a rather small $a_\Sigma$, i.e., vanishes only slowly as $x\to 0$, the uncertainties of the sea distributions ($\bar{u}+\bar{d}$) turn out to be only marginally smaller than those of the ``standard'' fit where the  sea increases as $x\to 0$ already at the  input scale. Notice that the uncertainties generally decrease as $Q^2$ increases due to the QCD evolution, as observed also in other analyses \cite{Pumplin:2001ct, Martin:2002aw}.

\begin{figure}[p]
\centering
\includegraphics[width=0.88\textwidth]{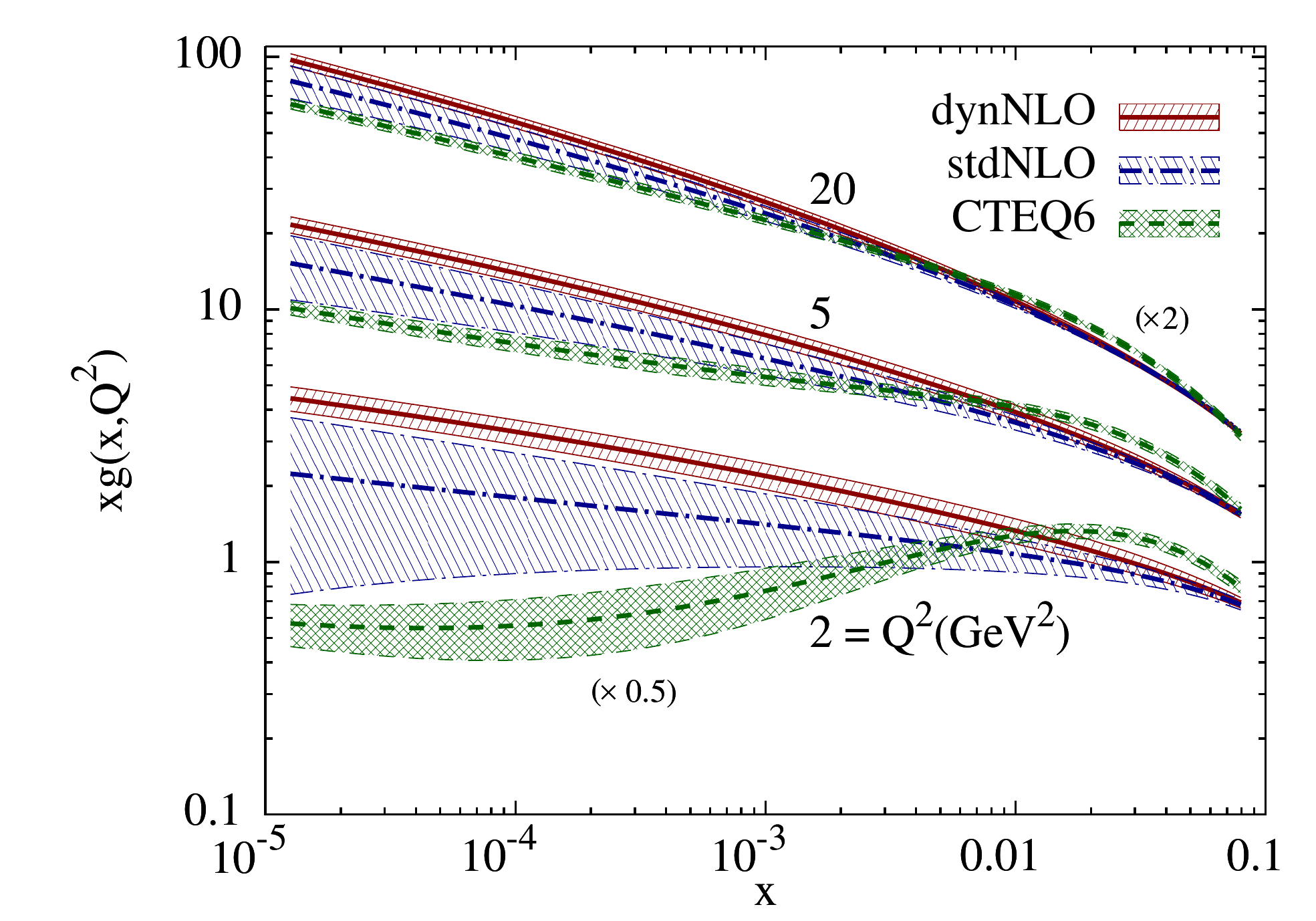}
\caption{Comparing the $\pm 1\,\sigma$ error bands of our dynamical (dyn), standard (std) and CTEQ6 \cite{Pumplin:2002vw} NLO\,($\overline{\rm MS}$) gluon distribution at small $x$. Note that $Q^2\!=\!2$ GeV$^2$ is the input scale of our ``standard'' fit, which is close to the CTEQ6 input scale $\mu_0^2\!=\!m_c^2\simeq 1.7$ GeV$^2$, where the standard CTEQ6 fit employs a valence--like gluon input. The results at $Q^2\!=$ 2 and 20 GeV$^2$ have been multiplied by 0.5 and 2 respectively as indicated in the figure. \label{Fig206}}
\includegraphics[width=0.88\textwidth]{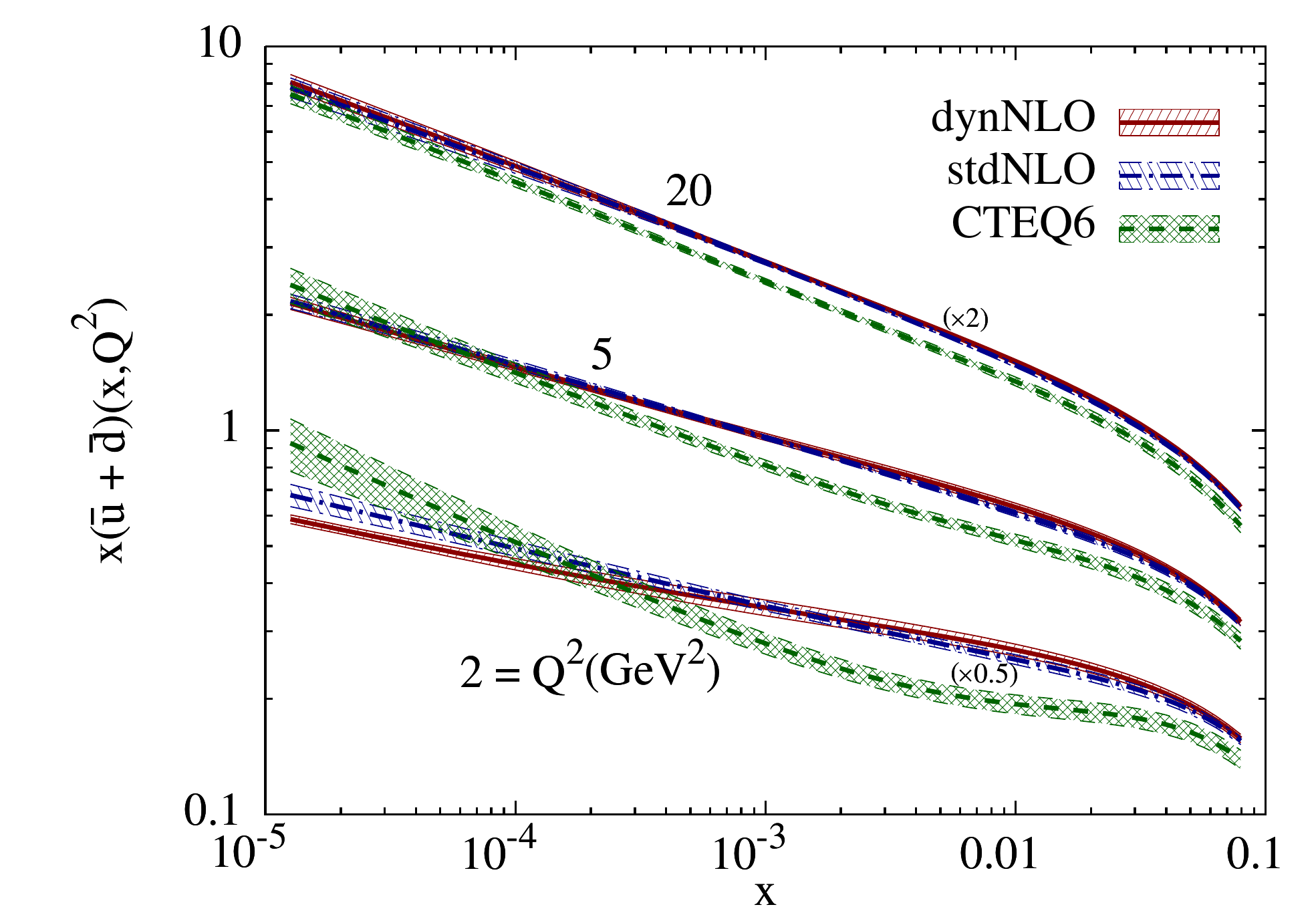}
\caption{As Fig.\,\ref{Fig206} but for the sea quark distributions $\bar{u}+\bar{d}$.\label{Fig207}}
\end{figure}

\fontsize{11}{15}
\selectfont

At this point it should be mentioned that the standard CTEQ6 analysis \cite{Pumplin:2002vw} surprisingly resulted in a valence--like input gluon distribution at a scale as large as $\mu_0^2\!=\!m_c^2\!\simeq\! 1.7$ GeV$^2$. Thus the CTEQ6 gluon distribution is expected to be  similarly tightly constrained at $Q^2\!>\!\mu_0^2$ as our dynamical results. That this is indeed the case is illustrated in Fig.\,\ref{Fig206} where the $1\,\sigma$ uncertainties\footnote{When comparing our uncertainty results with the ones of CTEQ6 \cite{Pumplin:2002vw} where $T=10$ has been assumed, we rescale these CTEQ uncertainties according to $\Delta a_i \!\to\! 0.4654 \Delta a_i$ in order to comply with our $T\!=\!4.654$.} of the CTEQ6 gluon are similar in size to our dynamical NLO\,($\overline{\rm MS}$) result, whereas the ``standard'' fit, with $Q^2\!=\!2$ GeV$^2$ and a non--valence--like structure, results in a  sizeably larger uncertainty. Due to the sizeably different input scales the CTEQ6 gluon falls up to 30 -- 40\% below our dynamical NLO gluon for $x\!<\!10^{-3}$ and $Q^2\!>\!10$ GeV$^2$.

The situation is different for the sea  distribution in the small--$x$ region (cf. Fig.\,\ref{Fig207}); here the CTEQ6 input at $\mu_0^2=m_c^2$ is not valence--like, i.e. increases as $x\!\to\!0$ as expected for a ``standard'' fit, and thus the  $1\,\sigma$ uncertainty is comparable to our ``standard'' NLO fit, both being larger than the $1\,\sigma$ uncertainty obtained from our dynamical NLO fit.

\begin{figure}[t]
\centering
\includegraphics[width=0.7\textwidth]{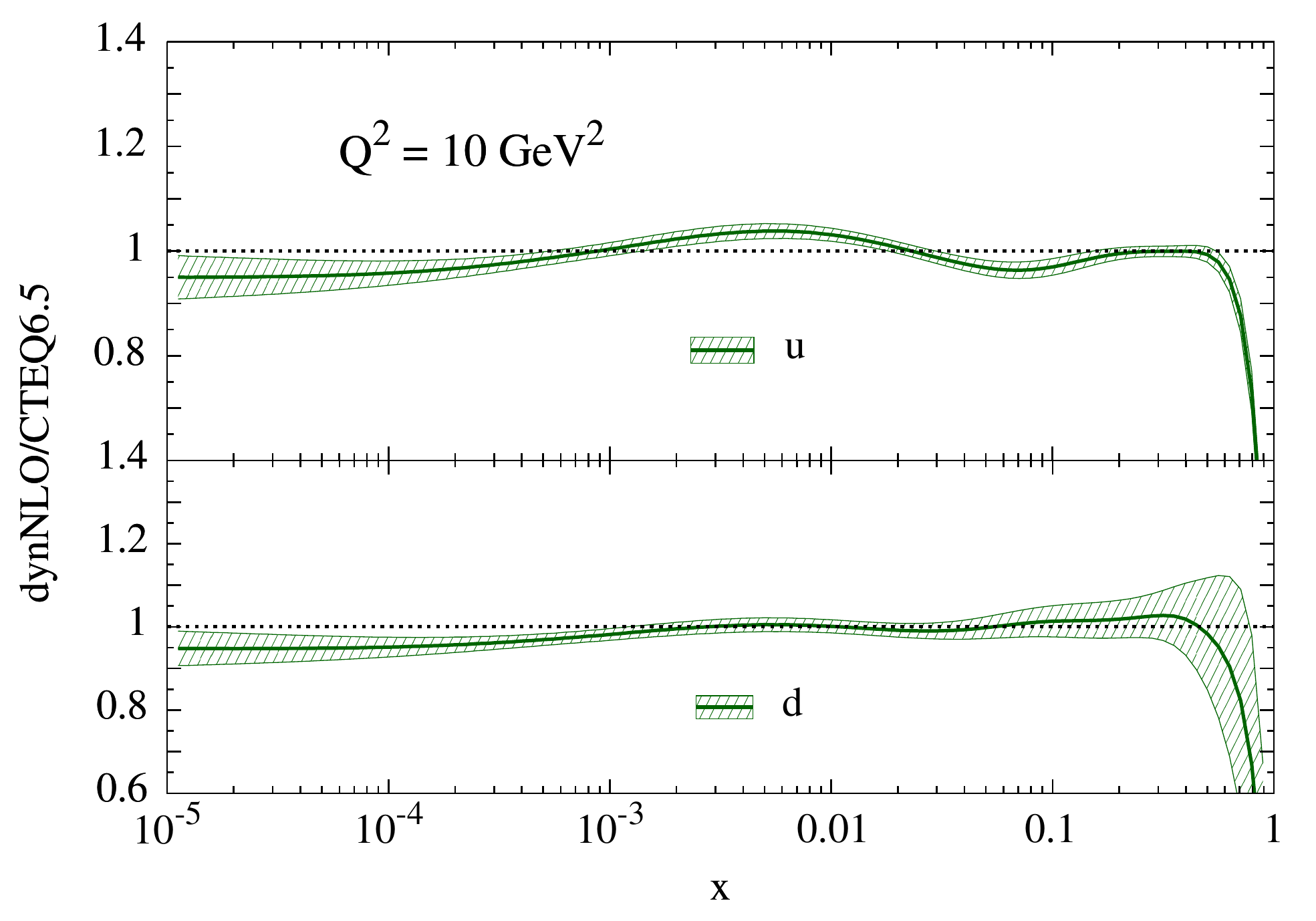}
\caption{Comparing our dynamical NLO\,($\overline{\rm MS}$) $u$ and $d$ parton distributions with the ones of CTEQ6.5 at $Q^2\!=\!10$ GeV$^2$. The shaded areas represent the estimated $\pm\!1\!\sigma$ uncertainty of our analysis.\label{Fig208}}
\end{figure}
\begin{figure}[t]
\centering
\includegraphics[width=0.85\textwidth]{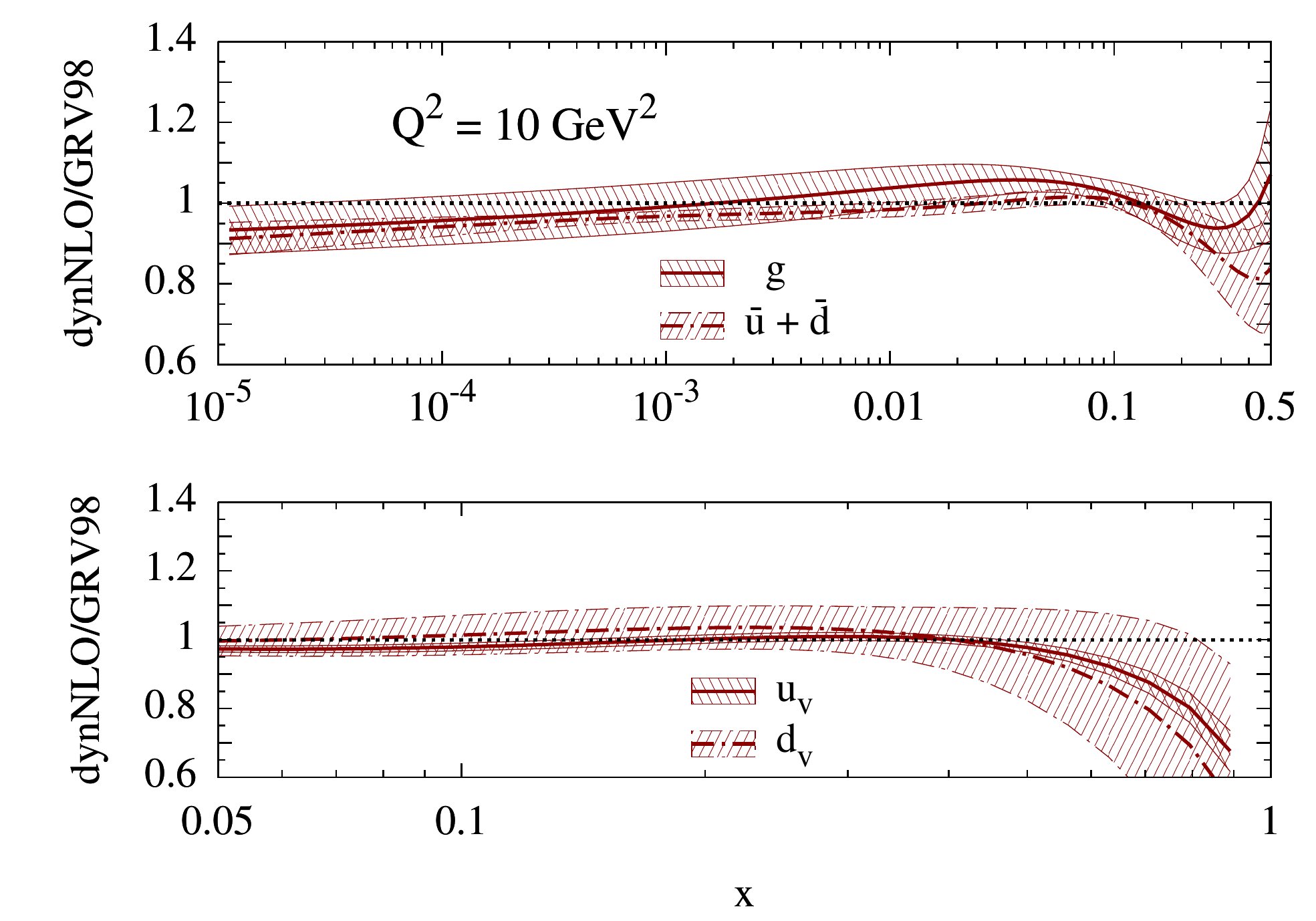}
\caption{Comparing the present dynamical NLO\,($\overline{\rm MS}$) distributions with the previous dynamical ones of GRV98 at \mbox{$Q^2\!=\!10$ GeV$^2$}. The shaded areas represent the estimated $\pm 1\,\sigma$ uncertainty band of our analysis.\label{Fig209}}
\end{figure}

It should be emphasized that the distributions resulting from (GM)VFNS analyses (i.e., the newer sets of CTEQ and MRST) depend strongly on the model assumed for the treatment of heavy quark contributions; which is in contrast to FFNS fits where heavy quark mass effects are fully taken into account in a theoretically unambiguous framework. For example, the inclusion of finite charm mass effects in CTEQ6.5 \cite{Tung:2006tb} reduces the charm contribution to $F_2(x,Q^2)$, which is compensated by larger $u$ and $d$ distributions at small $x$ as compared to the standard CTEQ6 analysis \cite{Pumplin:2002vw}, where heavy quarks have been treated in the zero--mass approximation. Fig.\,\ref{Fig208} illustrates that such ``enhanced'' quark distributions differ very little from our dynamical results. These ratios remain practically unchanged at higher scales, like $Q^2\!=\!M_W^2$ relevant for $W^{\pm}$ production, therefore our predictions for hadronic $W^{\pm}/Z^0$ production cross--sections, for example, at Tevatron and LHC are similar to the ``enhanced'' ones observed in \cite{Tung:2006tb}. We study hadronic weak gauge boson production in more detail in Sec.\,\ref{Sec.WandHiggs}.

Moreover, this holds also true for the GRV98 analysis \cite{Gluck:1998xa}, indeed in the relevant small--$x$ region these ratios would be practically unaltered if the GRV98 distributions were used instead of our current result since both dynamical analyses give rather similar results. Despite the fact that their respective analyses were done with approximately ten years of time difference and meanwhile a substantial increase in the precision of the HERA data has been achieved, the dynamical distributions, i.e. the predictions stemming from the dynamical model, remain stable to within less than about $20\%$. This is illustrated in Fig.\ref{Fig209} where ratios of our dynamical NLO\,($\overline{\rm MS}$) gluon, sea and valence distributions are compared with the previous GRV98 results and at a representative scale $Q^2\!=\!10$ GeV$^2$ relevant for phenomenological applications. As we have seen, this is in contrast to ``standard'' fits, which depend strongly on the particular details of the analyses (e.g. the data included, ansatz used, details on the treatment of quark masses, etc.).

\begin{figure}[t]
\centering
\includegraphics[width=0.92\textwidth]{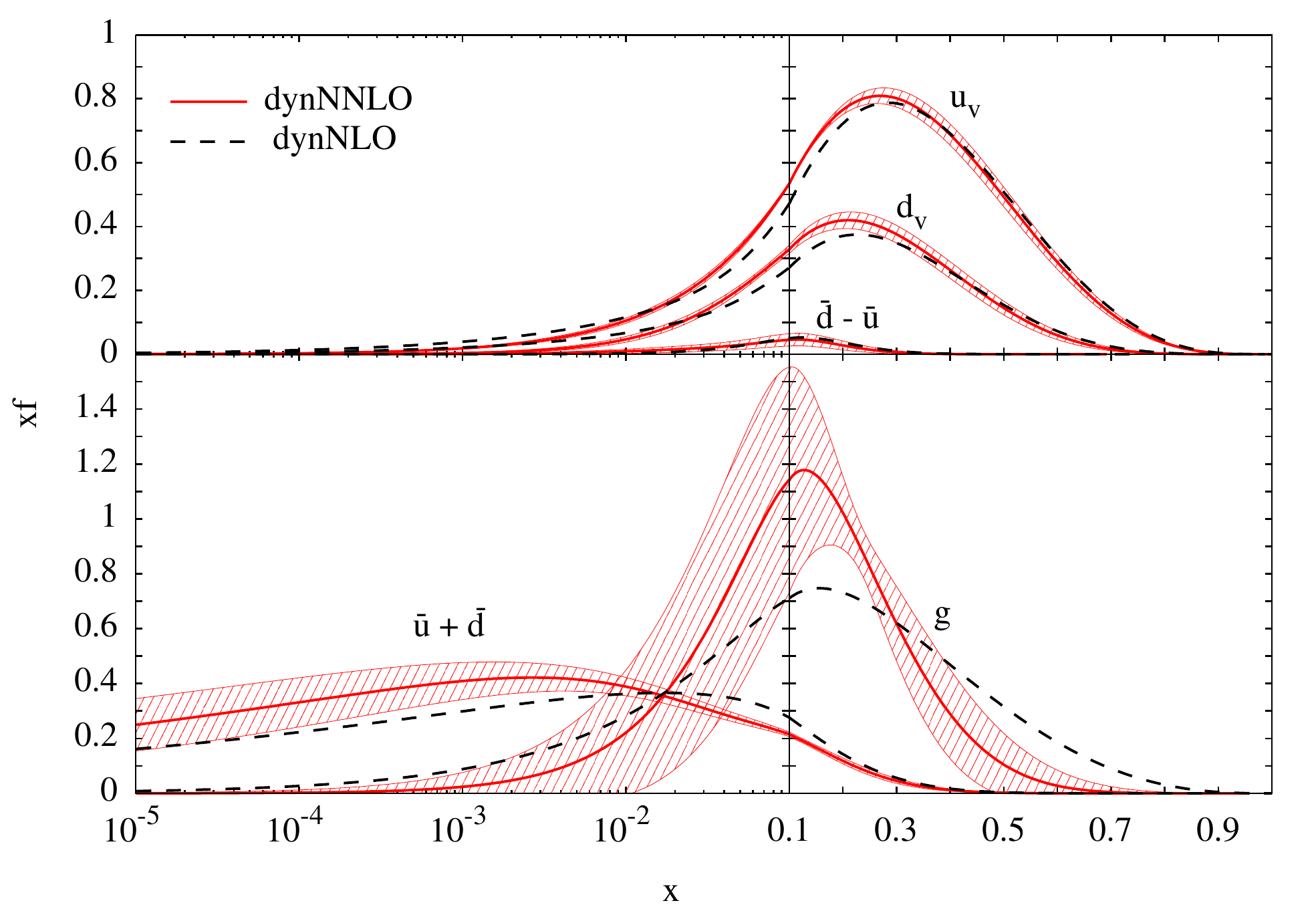}
\caption{The dynamical NNLO\,($\overline{\rm MS}$) input distributions and their $\pm 1\,\sigma$ uncertainty band at $\mu_0^2\!=\!0.55$ GeV$^2$ together with our dynamical NLO\,($\overline{\rm MS}$) input distributions at $\mu_0^2\!=\!0.5$ GeV$^2$. The strange sea $s=\bar{s}$ vanishes at the input scale. \label{Fig210}}
\end{figure}

Turning now our attention to our \cite{JimenezDelgado:2008hf} NNLO results, we present in Fig.\,\ref{Fig210} our dynamical NNLO\,($\overline{\rm MS}$) input distributions at $\mu_0^2\!=\!0.55$ GeV$^2$, according to the parameters in Table \ref{TabA01}, together with their $1\,\sigma$ uncertainties; our NLO\,($\overline{\rm MS}$) input distributions are shown again for comparison. The resulting valence ($u_v$ and $d_v$) distributions are now somewhat enhanced around $x=0.1$ to 0.2 respect to our NLO\,($\overline{\rm MS}$) results, and a strong and clear valence--like small--$x$ behavior of the NNLO gluon input is now required, $a_g=0.994\pm 0.379$, as compared to $a_g\simeq 0.5168 \pm$ 0.4017 at NLO. Furthermore there is also a strong enhancement of the NNLO gluon over the NLO one around $x=0.1$ and a sizeable depletion at larger values of $x$. The $1\,\sigma$ uncertainties at NNLO are comparable to the NLO ones shown in Fig.\,\ref{Fig204}, except for the NLO gluon at $x \gtrsim 0.3$ which is stronger constrained due to the light high--$p_T$ jet data (cf. Sec~\ref{Sec.DataFormalism}) which cannot be consistently included in an NNLO analysis.

\begin{figure}[t]
\centering
\includegraphics[width=0.9\textwidth]{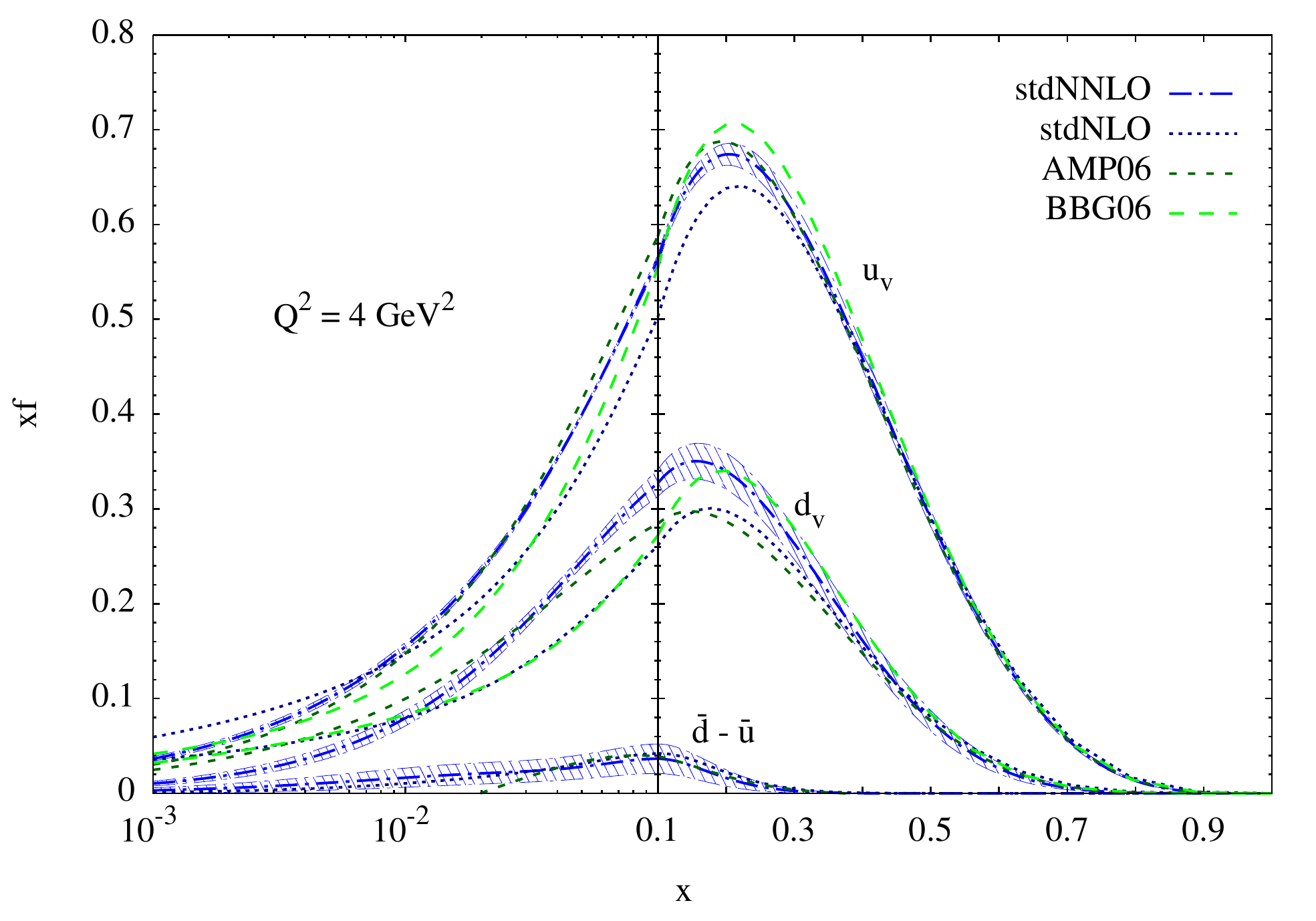}
\caption{Our ``standard'' NNLO\,($\overline{\rm MS}$) valence distributions and their $\pm 1\,\sigma$ uncertainties at $Q^2\!=\!4$ GeV$^2$ together with our ``standard'' NLO\,($\overline{\rm MS}$) results and the ``standard'' NNLO results of AMP06 \cite{Alekhin:2006zm} and BBG06 \cite{Blumlein:2006be}. Our dynamical valence distributions at $Q^2=4$ GeV$^2$ practically coincide with the standard ones shown.\label{Fig211}}
\end{figure}

\fontsize{11}{15.5}
\selectfont

The valence distributions of our NNLO ``standard'' analysis $(\mu_0^2=2$ GeV$^2$) are compared with the ``standard'' NNLO ones of Alekhin, Melnikov and Petriello \cite{Alekhin:2006zm} (AMP06, $\mu_0^2\!=\!9$ GeV$^2$) and the ``standard'' NNLO pure NS analysis of Bl\"umlein, B\"ottcher and Guffanti \cite{Blumlein:2006be} (BBG06, $\mu_0^2=4$ GeV$^2$) in Fig.\,\ref{Fig211} at $Q^2\!=\!4$ GeV$^2$. In the relevant valence $x$--region, $x\!\gtrsim\!0.1$, we confirm the BBG06 results, in particular the enhancement of $xd_v$ with respect to the result of AMP06. In any case we, as well as BBG06, observe a significant enhancement of the NNLO $xu_v$ and $xd_v$ with respect to the NLO results. Our dynamical valence distributions at $Q^2\!=\!4$ GeV$^2$ practically coincide with the standard ones shown in Fig.\,\ref{Fig211}, which indicates that the valence distributions are very robust with respect to the choice of the input scale $\mu_0^2$.

\begin{figure}[p]
\centering
\includegraphics[width=0.85\textwidth]{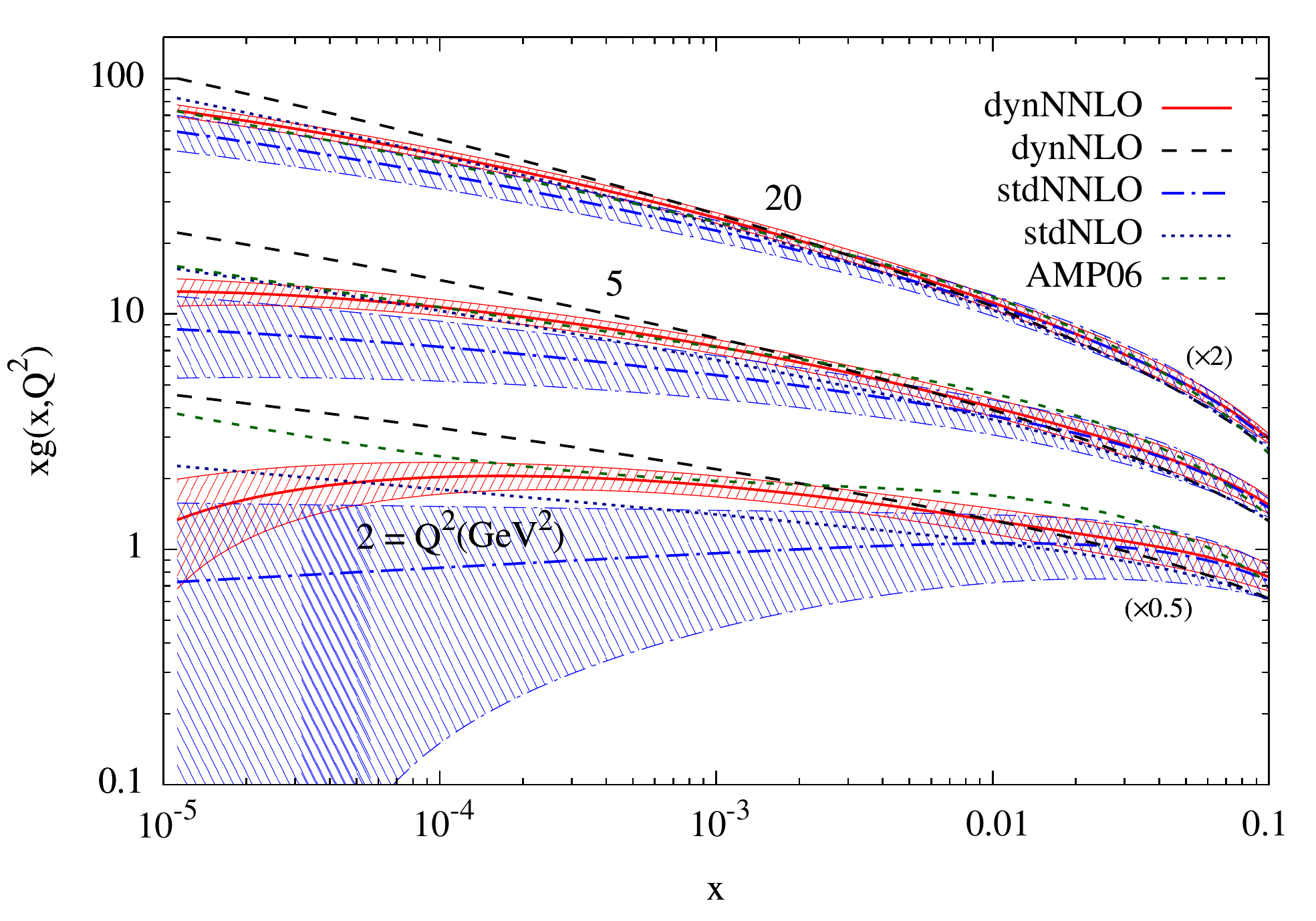}
\caption{Comparing the $\pm 1\,\sigma$ error bands of our dynamical (dyn) and standard (std) NNLO\,($\overline{\rm MS}$) gluon distributions at small $x$ for various fixed values of $Q^2$. Our NLO\,($\overline{\rm MS}$) dynamical and ``standard'' results as well as the ``standard'' NNLO results of AMP06 \cite{Alekhin:2006zm}, based on an input scale $\mu_0^2\!=9\!$ GeV$^2$, are shown as well. Note that $Q^2\!=\!2$ GeV$^2$ is the input scale of our standard fits. The results at $Q^2\!=\!2$ and 20 GeV$^2$ have been multiplied by 0.5 and 2 respectively as indicated in the figure.\label{Fig212}}
\includegraphics[width=0.85\textwidth]{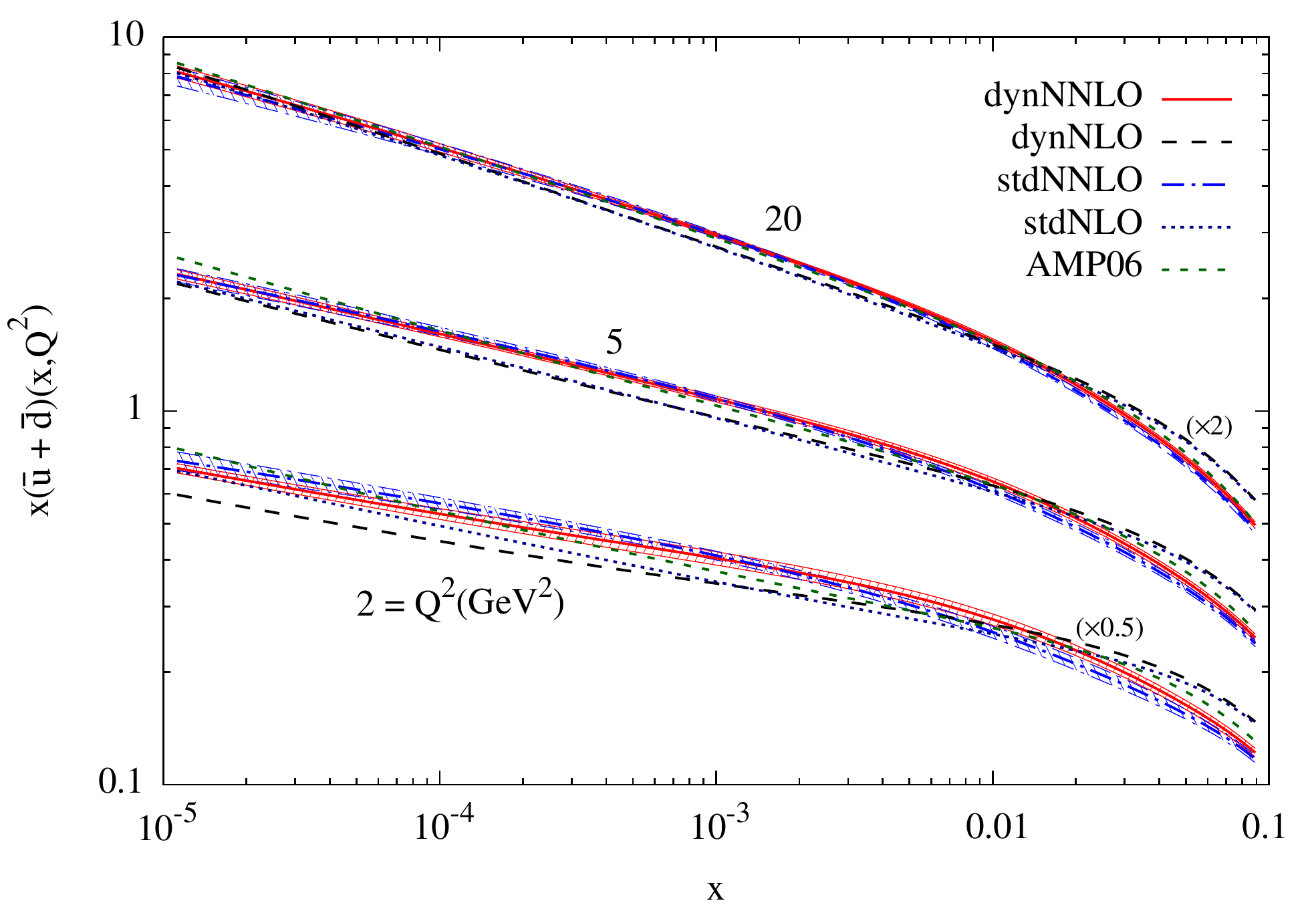}
\caption{As in Fig.~3 but for the sea quark distribution
$x(\bar{u}+\bar{d}$).\label{Fig213}}
\end{figure}

Again, the distinctive valence--like NNLO gluon input at low $\mu_0^2\!<\!1$ GeV$^2$ of the dynamical model (cf. Fig.\,\ref{Fig210}) implies a far stronger constrained gluon distribution at larger values of $Q^2$ as compared to a gluon density obtained from a NNLO ``standard'' fit with a conventional non--valence--like input at $\mu_0^2\!>\!1$ GeV$^2$, namely $\mu_0^2\!=\!2$ GeV$^2$, as can be seen in Fig.\,\ref{Fig212}.

In contrast to our ``standard'' NLO results, the ``standard'' NNLO gluon input at $\mu_0^2\!=\!2$ GeV$^2$ is very weakly constrained at small $x$ ($a_g\!=\!0.0637 \pm 0.1333$, cf.~Table \ref{TabA01}) and therefore momentum conservation (cf. Eq.\,\ref{sumrules}) cannot sufficiently constrain it at larger values of $x$, moreover since high--$p_T$ jet data have not been taken into account as already pointed out. Note as well that this common ``standard'' NNLO input gluon distribution at $\mu_0^2\!=\!2$ GeV$^2$ is also compatible with a valence--like small--$x$ behavior ($a_g\!>\!0$), a tendency already observed in \cite{Martin:2002dr}, and that our dynamical NNLO gluon distribution in Fig.\,\ref{Fig212} remains valence--like, i.e. decreases with decreasing $x$, even at $Q^2\!=\!2$ GeV$^2$. This is mainly caused by the NNLO splitting function $P_{gg}^{(2)}$ (cf. Sec.~1.2 to 1.4) which is \emph{negative} and \emph{more} singular in the small--$x$ region \cite{Vogt:2004mw} than the LO and NLO ones\footnote{For $n_f=3$, $xP_{gg}^{(2)}(x)\sim-3147.66\ln\frac{1}{x} +14737.89$ as $x\to 0$, whereas $xP_{gg}^{(0)}(x)\sim 12$ and $xP_{gg}^{(1)}(x)\sim-81.33$.}.

As in the NLO case, the uncertainties generally decrease as $Q^2$ increases due to the QCD evolution, but the ones of the dynamical predictions in the small--$x$ region remain substantially smaller than the uncertainties of the common ``standard'' results as exemplified in Fig.\,\ref{Fig212}. Furthermore, it is a general feature of \emph{any} NNLO gluon distribution in the small--$x$ region that it falls \emph{below} the NLO one as can be seen in Fig.\,\ref{Fig212} by comparing the solid curves with the long--dashed ones, and the dashed--dotted curves with the dotted ones. In the first dynamical case the NNLO predictions for $x<10^{-3}$ are several $\sigma$ below the NLO ones.

The dynamical sea distribution $x(\bar{u}+\bar{d})$ derives from a less pronounced ($a_{\Sigma}\!<\!a_g$) valence--like input in Fig.\,\ref{Fig210} which vanishes very slowly as $x\!\to\!0$; $a_{\Sigma} \!=\! 0.1374 \pm 0.0501$  (cf.~Table \ref{TabA01}). This implies that the valence--like sea input is similarly increasing with decreasing $x$ down to $x\simeq 0.01$ as the sea input obtained by the common ``standard'' fit where $a_{\Sigma}\!=\!-0.1098\pm 0.0122$ according to Table \ref{TabA01}.  Therefore, the $1\,\sigma$ uncertainty bands of our dynamically predicted sea distributions at larger values of $Q^2$ in Fig.\,\ref{Fig213} are only marginally smaller than the corresponding ones of the ``standard'' fit. In contrast to the evolution of the gluon distribution in Fig.\,\ref{Fig212}, the NNLO sea distributions in Fig.\,\ref{Fig213} lie always \emph{above} the NLO ones in the small--$x$ region, $x\!\lesssim\!10^{-2}$, and at not too large values of $Q^2$.  Here all NNLO sea distributions are rather similar, including the  ```standard'' one of AMP06 \cite{Alekhin:2006zm} (cf. Fig.\,\ref{Fig213}).

\fontsize{11}{15.3}
\selectfont

\section{Determination of the Strong Coupling}\label{Sec.alphas}
As already mentioned, we use the reference value $\alpha_s(M_Z^2)$ as contour condition for the determination of the strong coupling constant; for which we always employ the exact iterative solutions (cf. Sec.\,\ref{Sec.Renormalization}). Instead of using a fixed value for $\alpha_s(M_Z^2)$, typically the LEP measurement or the world average of 0.1176 \cite{Amsler:2008zz}, we include it as a free parameter in our fits and their error estimations. Thus the running coupling is determined in our analyses together with the parton distributions of the nucleon, in particular it is closely related to the gluon distribution which drives the QCD evolution.  We briefly discuss in this section the results obtained and compare them with similar determinations of the strong coupling. For a more detailed and comparative recent discussion of NLO and NNLO results the reader is referred to \cite{Blumlein:2007dk}.

It is worth to mention at this point that our treatment implies that each \emph{eigenvector basis set} (cf. Sec.\,\ref{Sec.Uncertainties}) has a different associated value of $\alpha_s(M_Z^2)$. The use of these values is mandatory for uncertainty studies; note that the difference between values of $\alpha_s(Q^2)$ corresponding to different eigenvector sets (belonging to the same fit) can be as large as 10\% at low scales.

\begin{figure}[b!]
\centering
\includegraphics[width=0.86\textwidth]{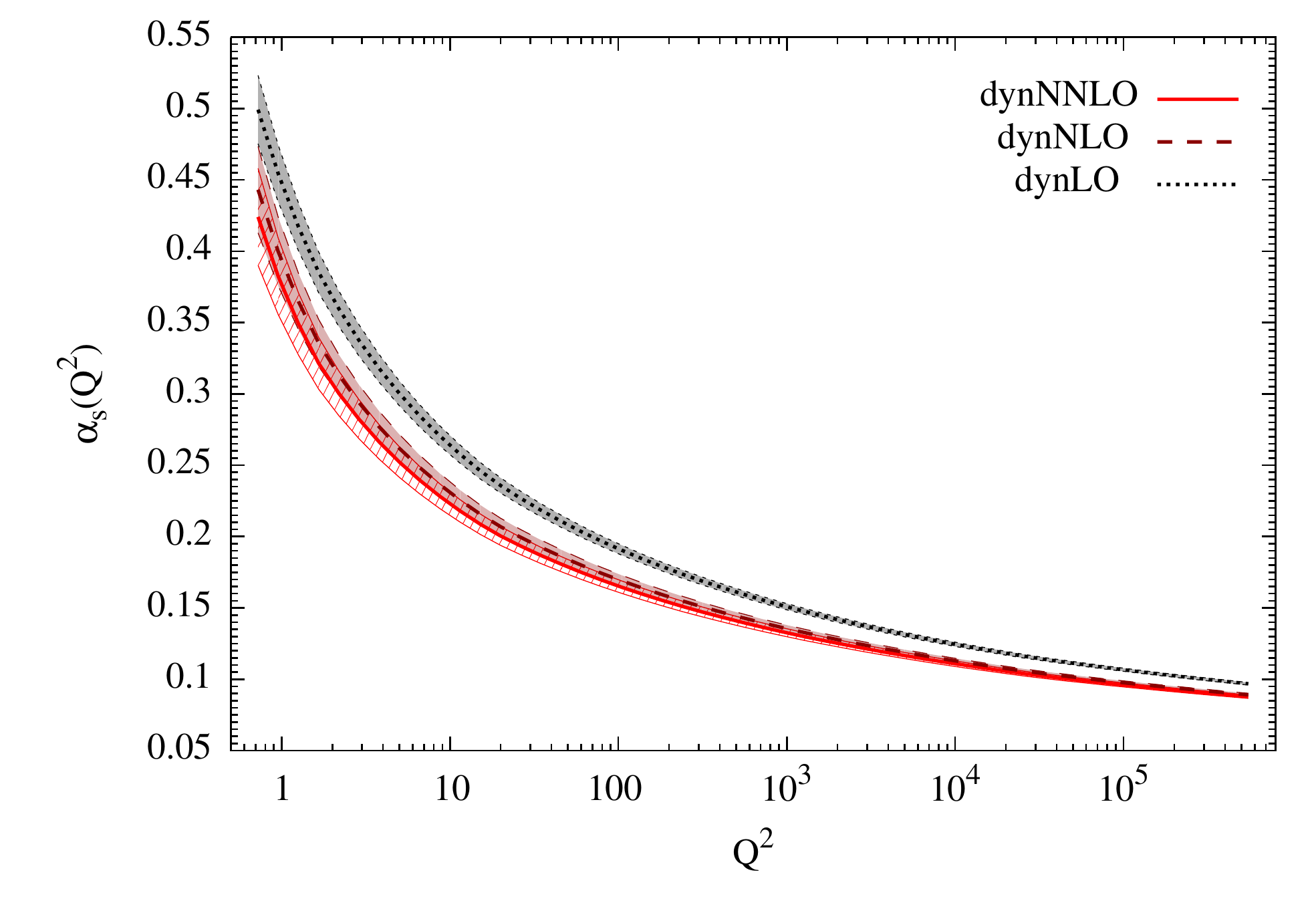}
\caption{Values of $\alpha_s(Q^2)$ resulting from our dynamical fits together with their respective $\pm1\,\sigma$ error bands.\label{Fig214}}
\end{figure}

In Fig.\,\ref{Fig214}, we show the results for the running coupling $\alpha_s(Q^2)$ according to the values obtained for $\alpha_s(M_Z^2)$ in our dynamical fits (cf. Table \ref{TabA01}) together with their respective $\pm1\,\sigma$ error bands. Note, as is well known, that the LO solutions are much \emph{larger} and the results stabilize at NLO. In general, the higher the perturbative order the faster $\alpha_s (Q^2)$ increases as $Q^2$ decreases. In order to compensate for this increase, a NNLO analysis is expected to result in a smaller value for $\alpha_s(M_Z^2)$ than a NLO one as is indeed the case. Both values lie, however, within a $1\,\sigma$ uncertainty. Similar results were obtained in a previous dynamical fit \cite{Gluck:2007sq} which, however, was performed for a restricted set of (mainly small--$x$) DIS data.

\fontsize{11}{15}
\selectfont

According to the $1\,\sigma$ errors for $\alpha_s(M_Z^2)$ quoted in Table~\ref{TabA01}, in general the strong coupling is stronger constrained in the dynamical fits than in their ``standard'' counterparts. This is in principle expected due to the larger evolution distance of the dynamical distributions. It should also be mentioned again that there is a certain correlation between the chosen value of $\mu_0^2$ and the resulting values for $\alpha_s(M_Z^2)$, which increase for increasing $\mu_0^2$.

At NNLO the $\alpha_s$--uncertainty of our dynamical fit is about half as large as the ``standard'' one. Apart from the larger evolution distance, this is due to the strongly constrained valence--like input gluon distribution in the small--$x$ region; consequently the energy--momentum sum rule (cf. Eq.\,\ref{sumrules}) sufficiently constrains $xg(x,\mu_0^2)$ in the medium to large $x$--region in the dynamical case while in the ``standard'' one the gluon distribution, and thus $\alpha_s(M_Z^2)$, is considerably less restricted, moreover since jet data were not included at NNLO.

Keeping in mind that our stated errors always refer to $1\,\sigma$ uncertainties, our ``standard'' NLO fit error for $\alpha_s(M_{\rm Z}^2)$ is compatible with the $2\sigma$ uncertainty stated in the literature (see, e.g.,\cite{Pumplin:2002vw} and the discussion in \cite{Martin:2001es}). Our NLO\,($\overline{\rm MS}$) results are also compatible, within about $1\,\sigma$ uncertainty, with the ones obtained from analyzing only DIS structure functions. The same holds true for our ``standard'' NNLO fit result (cf. Table \ref{TabA01}) which, within errors, is compatible with the one obtained from a ``standard'' fit \cite{Gluck:2006pm} to a restricted set of small--$x$ DIS data.

Previous ``standard'' NNLO fits considering only the flavor non--singlet valence sector of structure functions \cite{Blumlein:2004ip, Blumlein:2006be, Gluck:2006yz} resulted in somewhat smaller values of $\alpha_s(M_Z^2)$ but remain within 1 to $2\sigma$ uncertainty. A similar NS valence analysis \cite{Kataev:1997nc} as well as a full analysis \cite{Santiago:1999pr} being based, however, on incomplete calculations of the moments of 3--loop anomalous dimensions (splitting functions) yielded slightly larger values of $\alpha_s$ at NNLO, however the estimated errors are large enough so as to comply with our result in Table \ref{TabA01}.

As already mentioned several times, data for high--$p_T$ jet production in hadron--hadron scattering should not be included in a consistent NNLO analysis, since such processes are theoretically known only up to NLO (Sec.\,\ref{Sec.Hadroproduction}). Including them nevertheless in a ``standard'' NNLO analysis requires \cite{Martin:2007bv, Martin:2004ir} generally larger values of $\alpha_s(M_Z^2)$.

Our $\alpha_s(M_Z^2)$--uncertainty in Table \ref{TabA01} is also about twice as large as the one obtained in a comparable standard NNLO analysis \cite{Alekhin:2006zm} where the high--$p_T$ jet data have been disregarded for consistency reasons as well. Without these data the gluon distribution is little constrained in the medium-- to large--$x$ region where it plays an important role for the QCD evolution at small values of $x$ due to the convolution with the dominant $P_{gg}^{(k)}$. This $\alpha_s$--uncertainty remains sizeable irrespective of the choice of the input scale $\mu_0^2\!>\!1$ GeV$^2$. Only within a Bayesian treatment of systematic errors, by taking into account point--to--point correlations \cite{Alekhin:2000es,Alekhin:2000ch,Alekhin:2001ih}, the uncertainty of $\alpha_s(M_Z^2)$ turns out to be about two times smaller \cite{Alekhin:2002fv, Alekhin:2005gq, Alekhin:2006zm, Alekhin:2000es, Alekhin:2000ch, Alekhin:2001ih}.

\section{Comparisons with the Data Used}\label{Sec.Comparisons}
\begin{figure}[p]
\centering
\includegraphics[width=\textwidth]{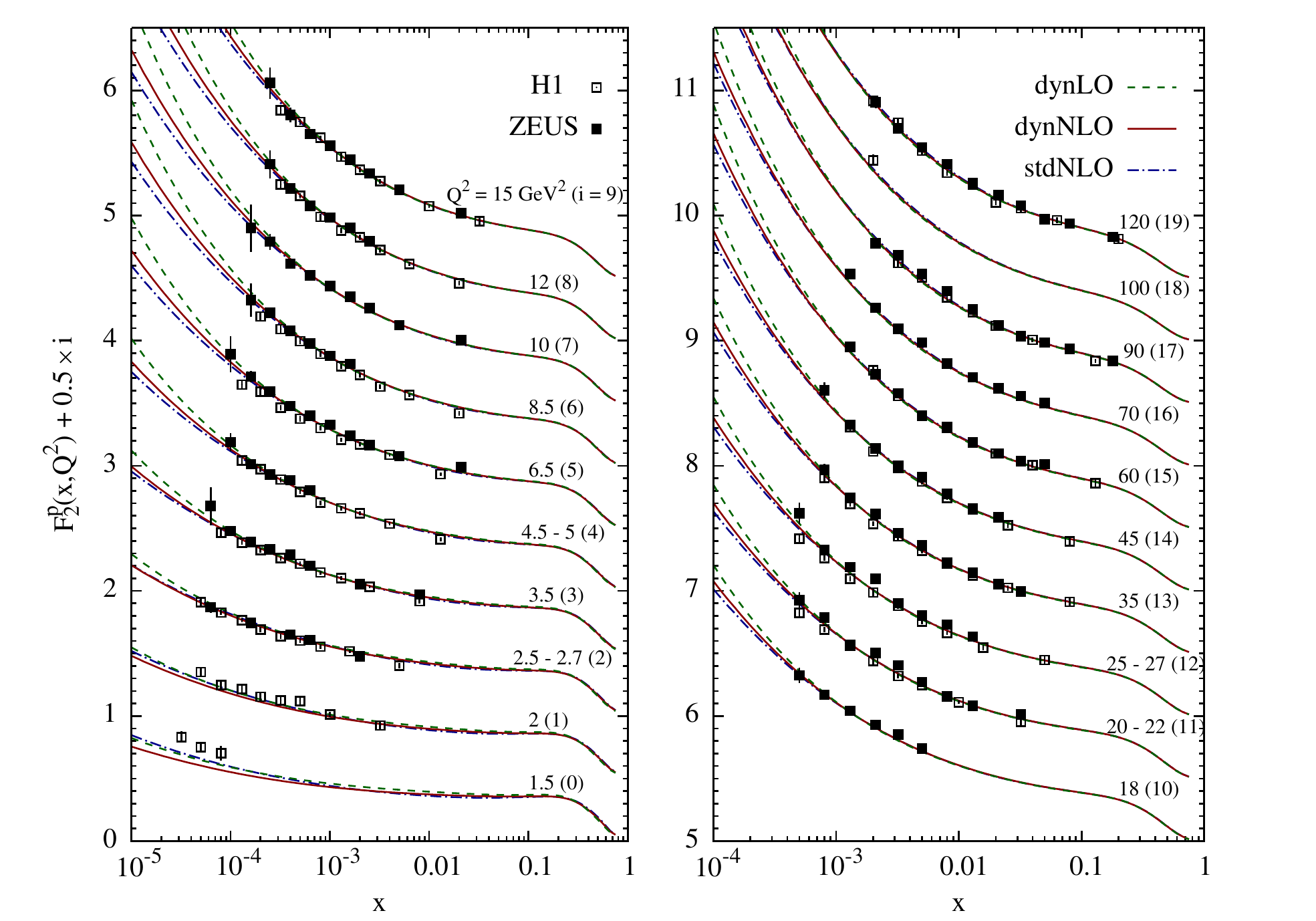}
\includegraphics[width=\textwidth]{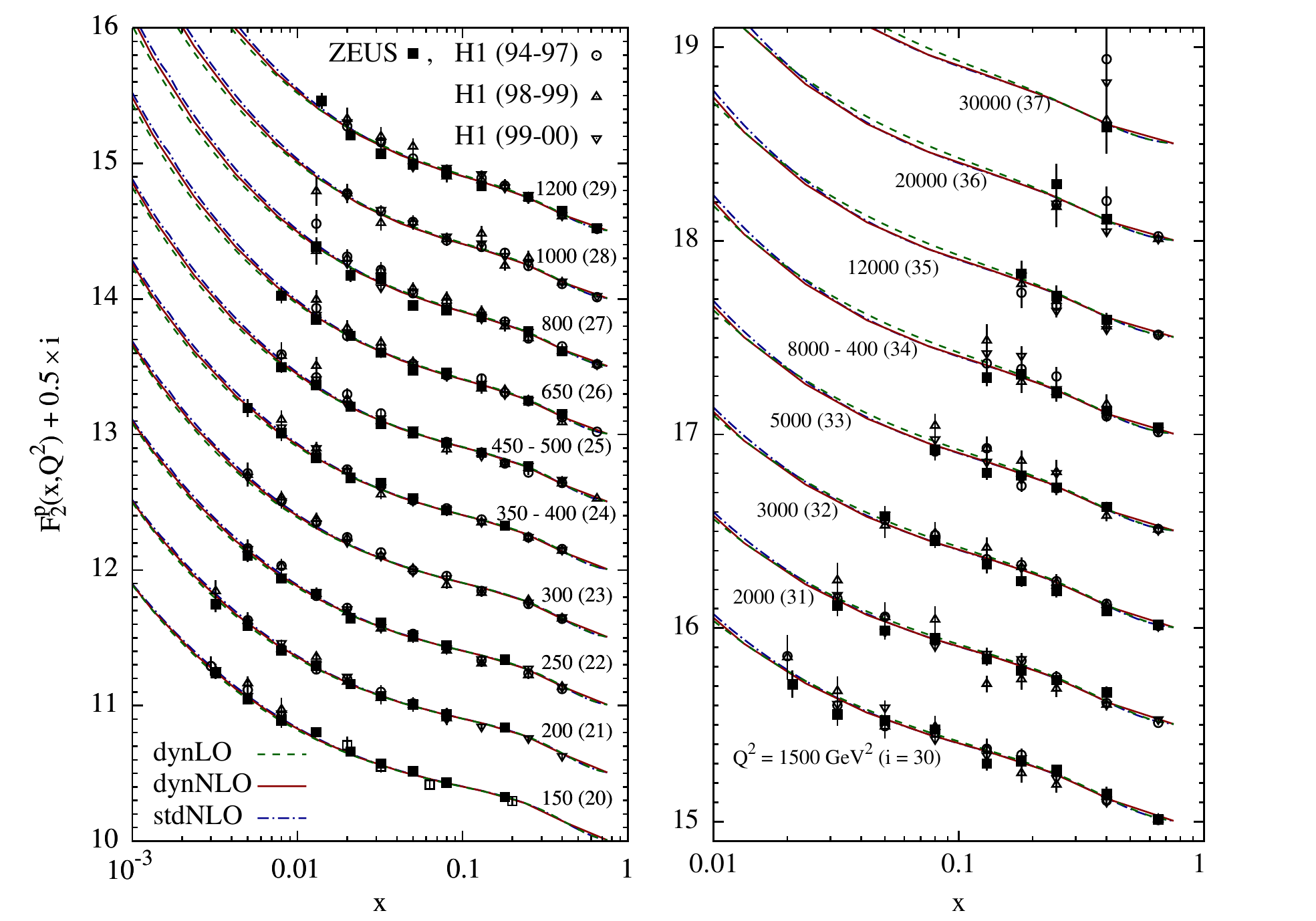}
\caption{Comparison of our dynamical (dyn) and standard (std) NLO\,($\overline{\rm MS}$) results for $F_2(x,Q^2)$ with HERA data for $Q^2\!\geq\! 1.5$ GeV$^2$ \cite{Adloff:1999ah, Adloff:2000qj, Adloff:2000qk, Adloff:2003uh, Chekanov:2001qu}. The dynamical LO results are shown for comparison. To ease the graphical presentation we have plotted $F_2^p(x,Q^2)+0.5\times  i(Q^2)$ with $i(Q^2)$ indicated in parentheses in the figure.\label{Fig215}}
\end{figure}

\begin{figure}[p]
\centering
\includegraphics[width=\textwidth]{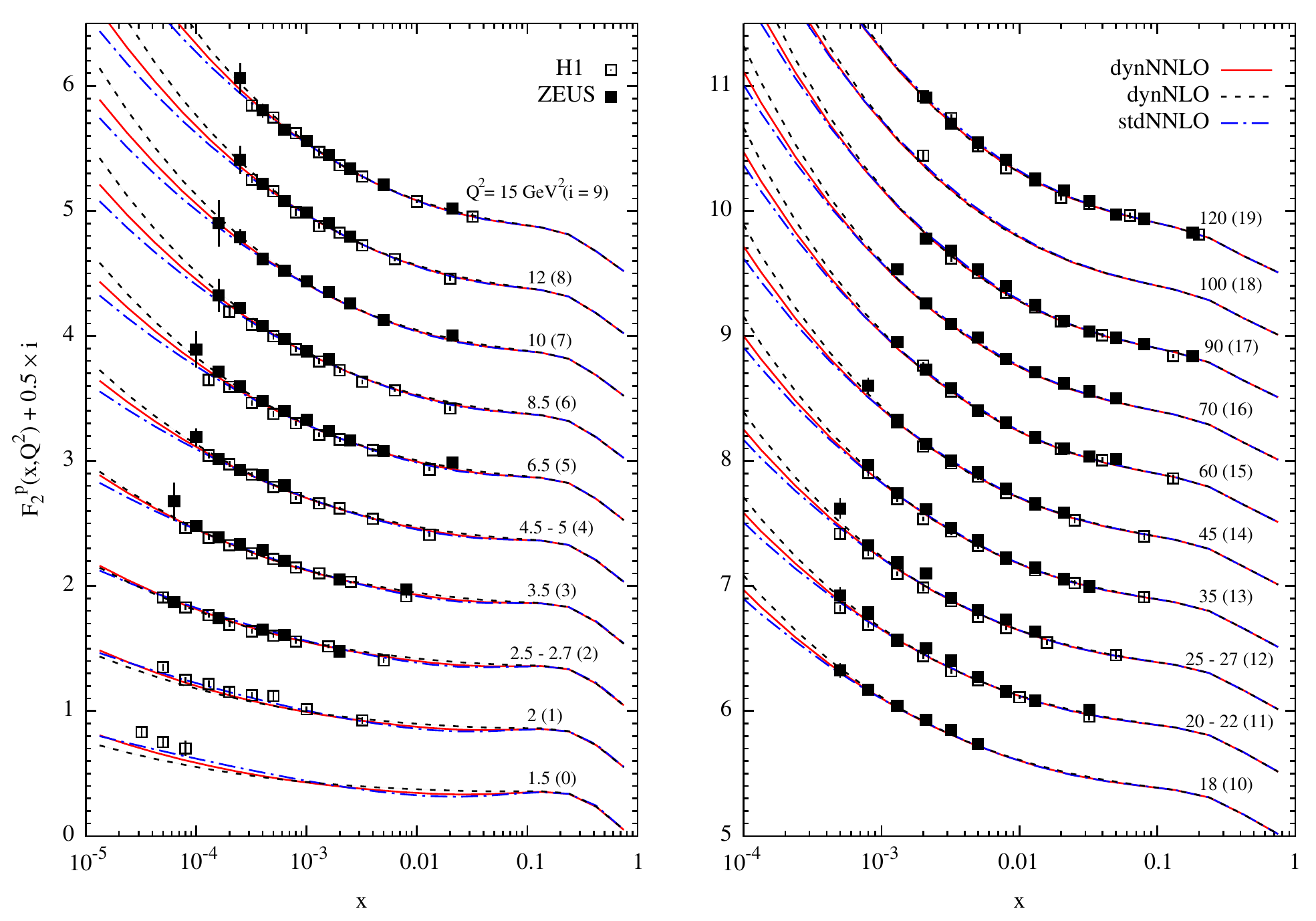}
\includegraphics[width=\textwidth]{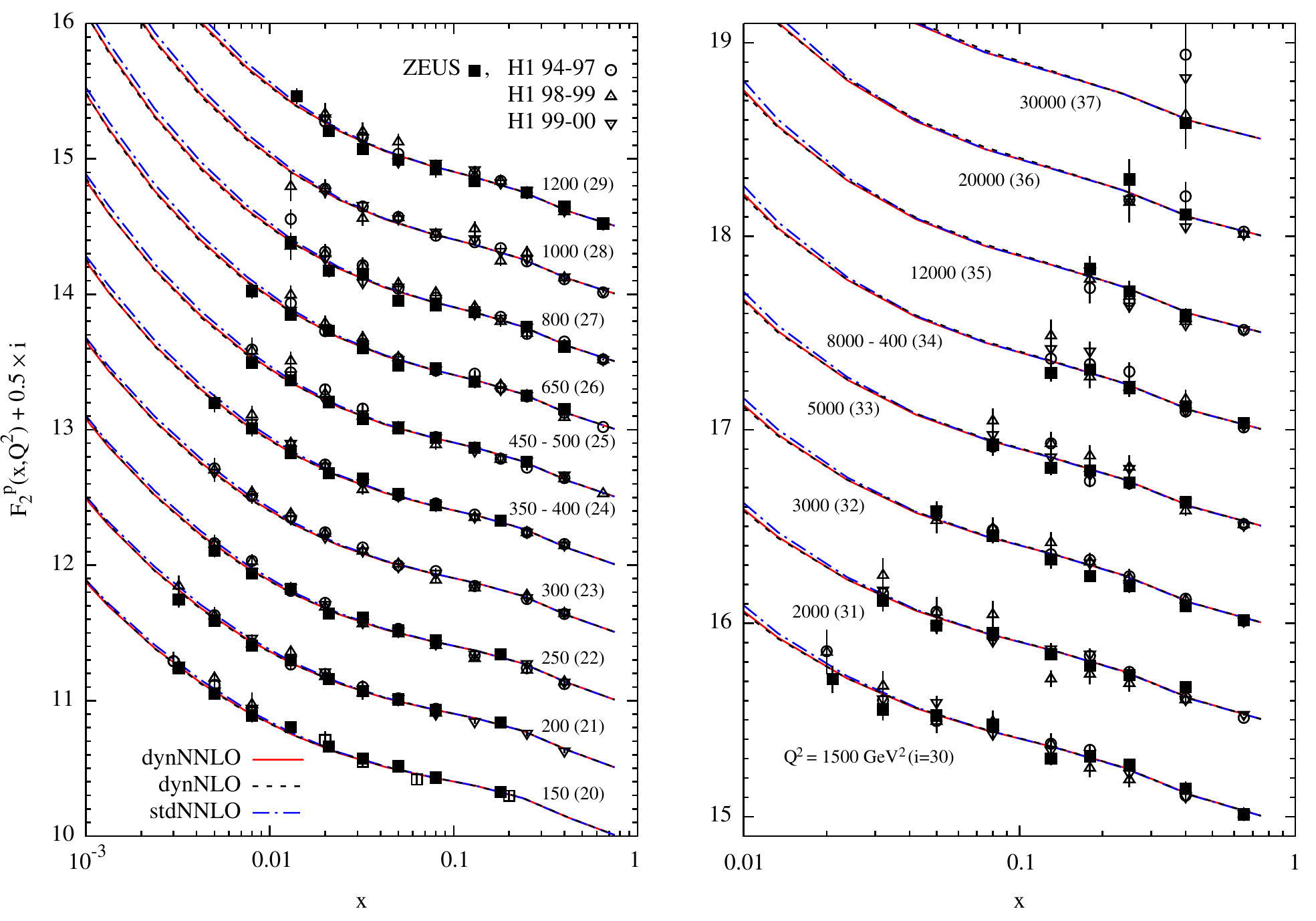}
\caption{As in Fig.\,\ref{Fig215} but for our NNLO\,($\overline{\rm MS}$) results. The dynamical NLO\,($\overline{\rm MS}$) results are shown again for comparison. \label{Fig216}}
\end{figure}

As discussed in Sec.\,\ref{Sec.DataFormalism}, we have used for our analyses the experimentally directly measured ``reduced'' DIS cross--section $\sigma_r^\emph{\rm NC}$ (cf. Eq.\,\ref{sigmar}), although fitting just to the dominant (one--photon exchange only) structure function $F_2^\gamma$ resulted in very similar results. Representative comparisons of our results with the relevant HERA (H1 and ZEUS) data \cite{Adloff:1999ah, Adloff:2000qj, Adloff:2000qk, Adloff:2003uh, Chekanov:2001qu} on this structure function  of the proton, $F_2^p(x,Q^2)$, are presented in Figs.\,\ref{Fig215} (NLO) and \ref{Fig216} (NNLO). As can be seen the data are well described throughout the whole  medium-- to small--$x$ region for $Q^2\!\gtrsim\!$ 2 GeV$^2$ and thus perturbative QCD is here fully operative. It should be reemphasized that, due to our valence--like input, the dynamical small--$x$ results ($x\!\lesssim\!10^{-2}$) are {\em predictions} being mainly generated by the QCD evolution; this is in contrast to a common ``standard'' fit where the gluon and sea \emph{input} distributions do {\em not} vanish as $x\to 0$. At $Q^2\!<\!2$ GeV$^2$ the theoretical results fall below the data in the very small--$x$ region, although the NNLO results are closer to the data than the NLO ones; this is not unexpected for perturbative leading--twist results, since nonperturbative (higher--twist) contributions to  $F_2$ will eventually become relevant, even dominant, for  decreasing values of $Q^2$.

\begin{figure}[b!]
\centering
\includegraphics[width=0.85\textwidth]{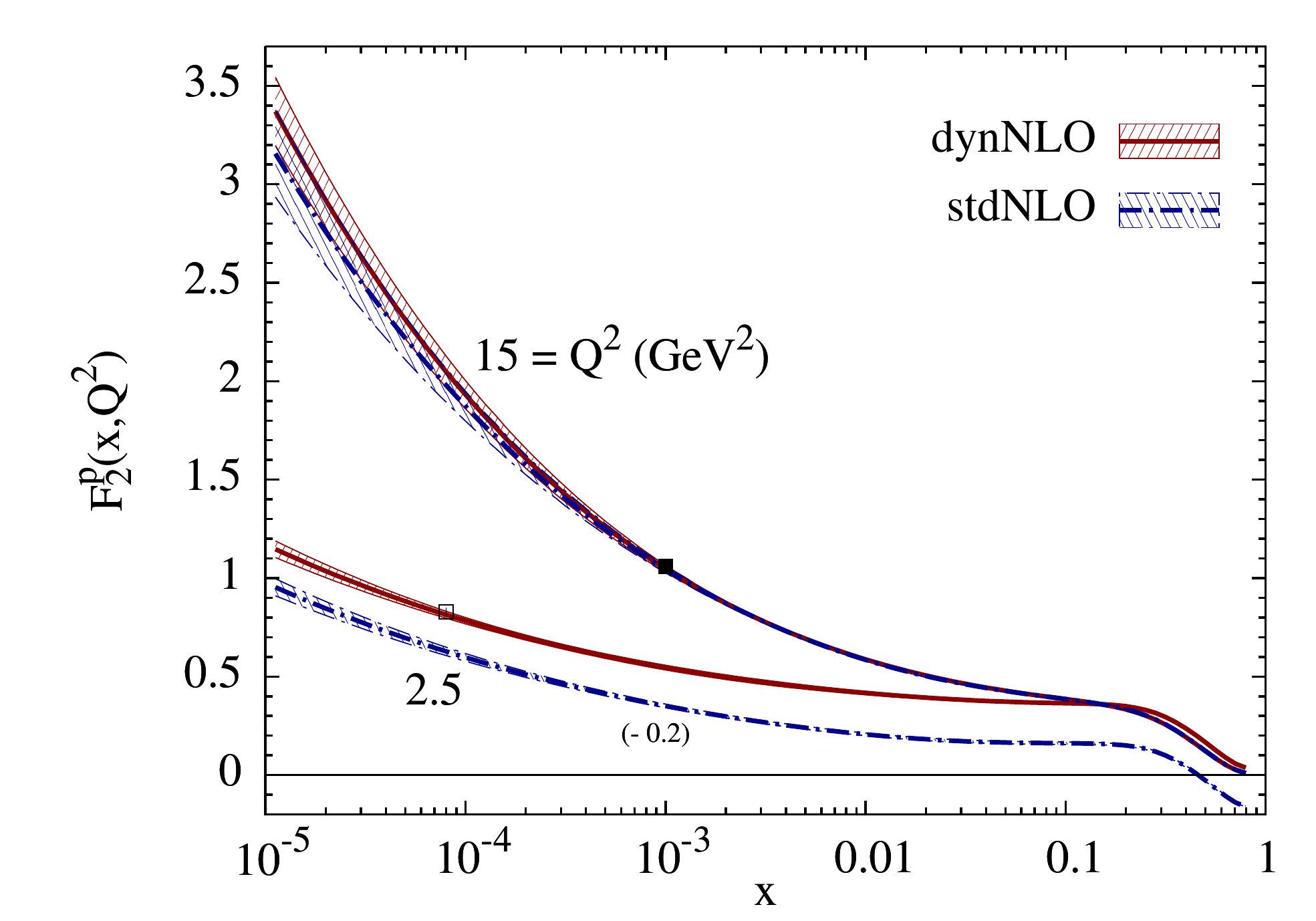}
\caption{Typical $\pm 1\,\sigma$ uncertainty bands of our dynamical and ``standard'' NLO\,($\overline{\rm MS})$ results for two representative values of $Q^2$. To ease the visibility of the two error bands at $Q^2\!=\!2.5$ GeV$^2$ we have subtracted 0.2 from the stdNLO result as indicated. For illustration two H1 and ZEUS data points from Fig.\,\ref{Fig215} with their almost invisible errors are shown as well at $Q^2\!=\!2.5$ and 15 GeV$^2$ respectively. \label{Fig217}}
\vspace{1em}
\end{figure}

\begin{figure}[t]
\centering
\includegraphics[width=0.9\textwidth]{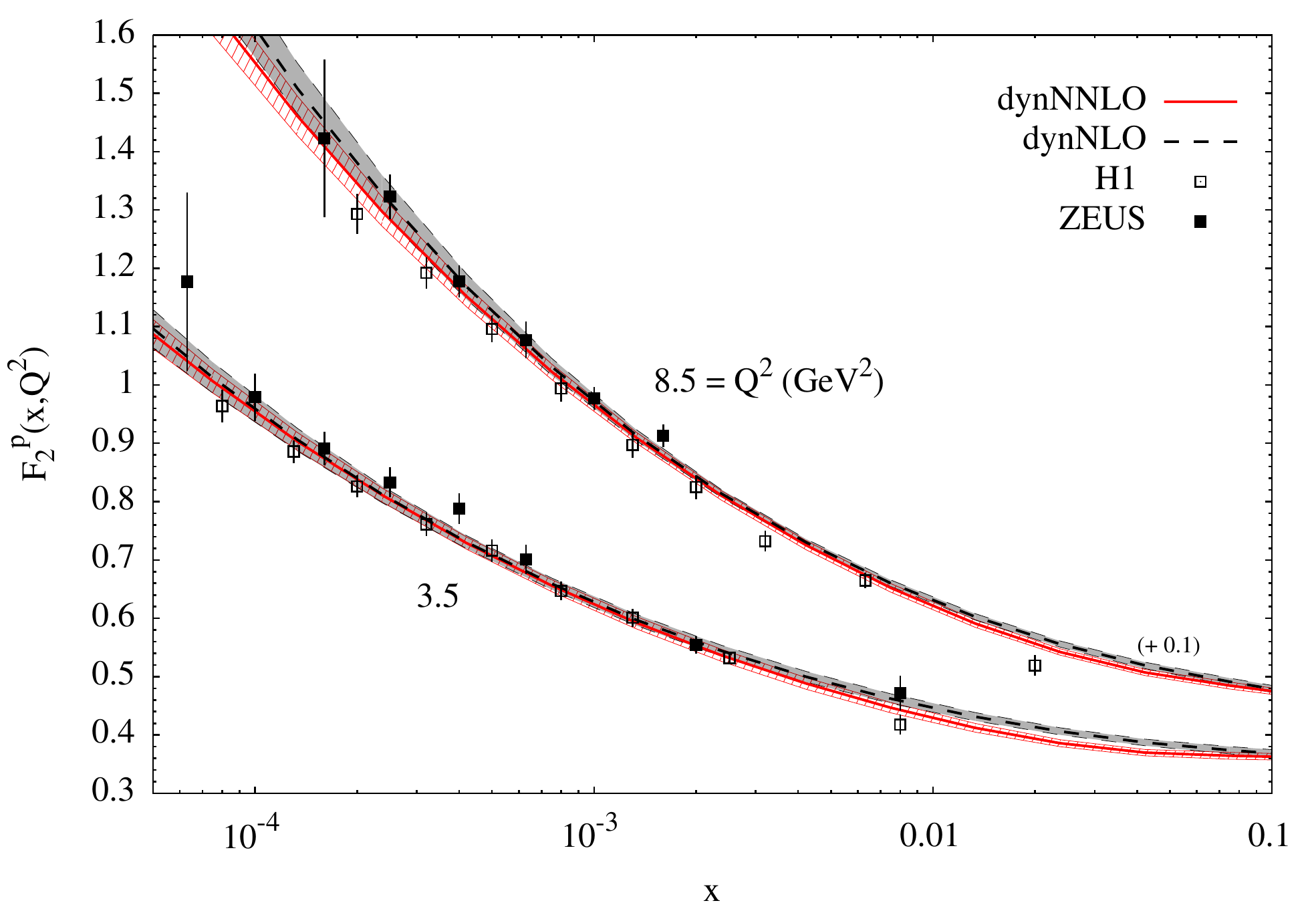}
\caption{Typical $\pm 1\,\sigma$ uncertainty bands of our dynamical NNLO and NLO results for two representative values of $Q^2$. The ``standard'' NNLO and NLO results are very similar. To ease the visibility we have added 0.1 to the results for $Q^2\!=\!8.5$ GeV$^2$ as indicated. \label{Fig218}}
\end{figure}

Furthermore, the more restrictive ansatz of the valence--like input distributions at small--$x$ as well as the sizeably larger evolution distance (starting at $\mu_0^2\!<\!1$ GeV$^2$) generally imply, as for the distributions themselves (cf. Sec.\,\ref{Sec.NewGeneration}), {\em smaller} uncertainties concerning the behavior of structure functions in the small--$x$ region than the corresponding results obtained from the common ``standard'' fits, in particular as $Q^2$ increases. This is illustrated in Fig.\,\ref{Fig217} for our NLO\,($\overline{\rm MS}$) results where the error bands correspond to a $1\,\sigma$ uncertainty. However, since our NLO valence--like sea input has a rather small power of $x$, i.e. vanishes only slowly as $x\!\to \!0$ as already discussed, the uncertainties of the sea dominated $F_2^p(x,Q^2)$ turn out to be not too different from the ``standard'' fit where the sea increases as $x\!\to\!0$, negative power of $x$, already at the input scale $\mu_0^2\!=\!2$ GeV$^2$. The results for our dynamical and standard NNLO fits are respectively rather similar to the NLO ones shown in Fig.\,\ref{Fig217} (cf. Fig.\,\ref{Fig218}).

Although the inclusion of NNLO corrections implies an improved value of $\chi^2$, as mentioned typically $\chi_{\rm NNLO}^2\!\simeq\!0.9 \chi_{\rm NLO}^2$ (cf. Table \ref{TabA01}), present high precision DIS data are not sufficiently accurate to distinguish between the NLO results and the minute NNLO effects of a few percent. This is illustrated in Fig.\,\ref{Fig218} where the experimental (statistical and systematic) errors are far bigger than the differences between the NLO and NNLO results. It should, however, be noticed that the NNLO uncertainty bands are somewhat narrower (reduced) than the ones at NLO; the results are similar for our  ``standard'' fits. By analyzing only the flavor non--singlet valence sector of structure functions, it was already been pointed out \cite{Gluck:2006yz} that NNLO effects cannot be delineated by present data in the medium-- to large--$x$ region, and moreover, uncertainties of NLO and LO analyses (such as higher twists, different factorization schemes and QED contributions to the QCD evolution) turn out to be comparable in size to the NNLO 3--loop contributions \cite{Gluck:2006yz}.

\begin{figure}[t]
\centering
\includegraphics[width=0.495\textwidth]{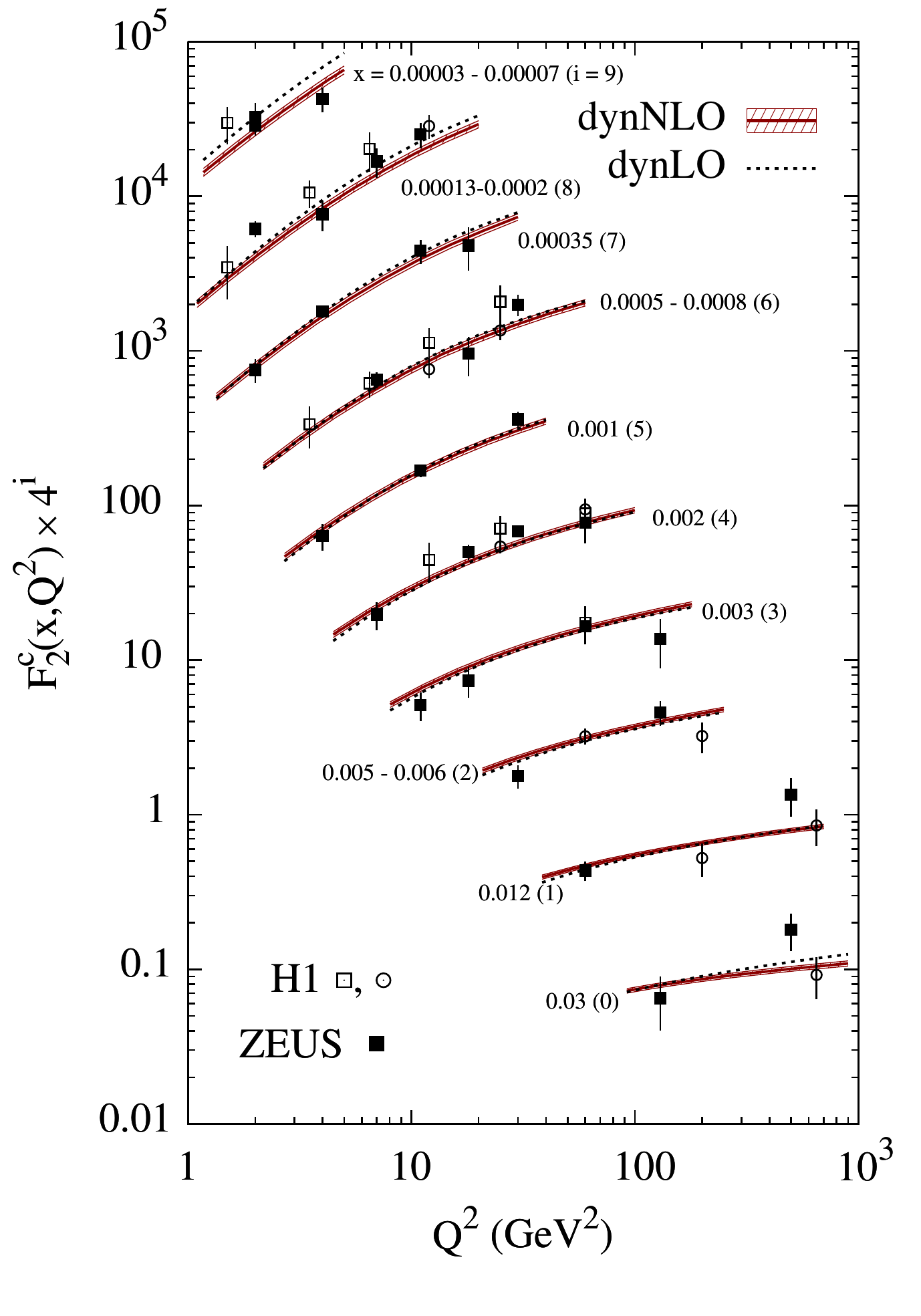}
\includegraphics[width=0.495\textwidth]{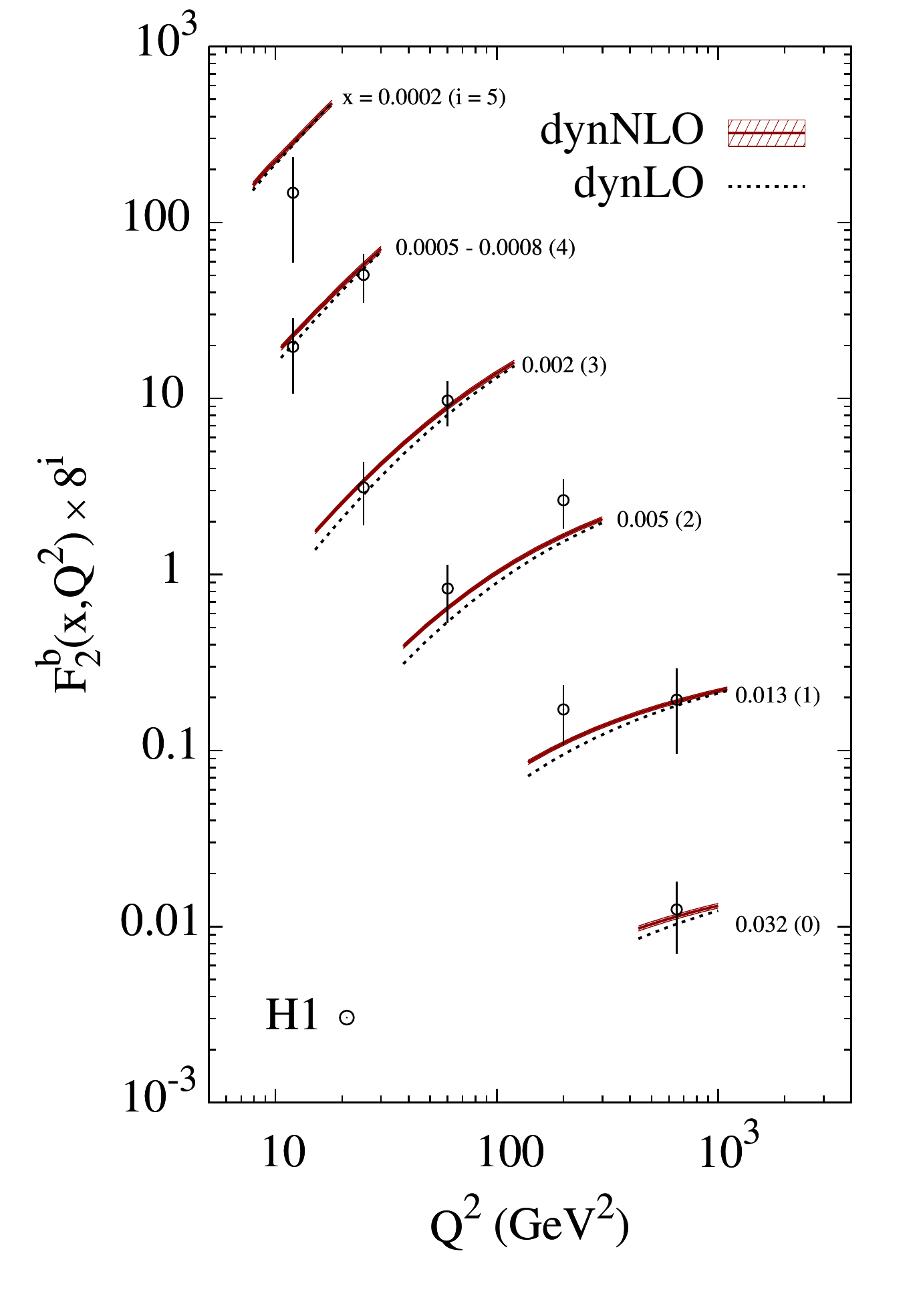}
\caption{The dynamical NLO\,($\overline{\rm MS}$) predictions for $F_2^{c,b}$ in the strict $n_F\!=\!3$ FFNS, choosing $\mu\!=\!4m_h^2$ with  $m_c\!=\!1.3$ GeV and $m_b\!=\!4.2$ GeV, together with the $\pm 1\,\sigma$ uncertainty band. For comparison we also display the central LO predictions which are  entirely due to the $\gamma^*$--gluon fusion subprocess $\gamma^*g\to h\bar{h}$. The charm production data as obtained from $D^*$ measurements are taken from \cite{Adloff:2001zj, Chekanov:2003rb} (solid and open squares) and the H1 direct track measurements from \cite{Aktas:2004az, Aktas:2005iw} (open circles); the bottom production data taken from \cite{Aktas:2004az, Aktas:2005iw}.\label{Fig219}}
\end{figure}
Turning now to heavy quark production in DIS, recent HERA data on charm \cite{Adloff:2001zj, Aktas:2004az, Aktas:2005iw, Chekanov:2003rb} and bottom \cite{Aktas:2004az, Aktas:2005iw} contributions to $F_2(x,Q^2)$ are compared in Fig.\,\ref{Fig219} with our dynamical LO and NLO\,($\overline{\rm MS}$) predictions. Although not shown, it is worth to mention that also in this case the uncertainties of the ``standard'' results turn out to be larger than the ones implied by the dynamical model. The impressive agreement observed in Fig.\,\ref{Fig219} illustrates the reliability of the the $n_F\!=\!3$ FFNS, in which the heavy quarks ($h=c,b,t$) are always produced as {\em final} states in fixed--order perturbation theory via hard production processes initiated by the light partons of the nucleon ($u,d,s$ quarks and the gluon $g$). In other words, the \emph{intrinsic} heavy quark content of the nucleon (if any) is marginal since all the observed heavy quark contributions are well taken into account by \emph{extrinsic} heavy quark production, i.e., originated only from light quark partons in the initial state. Therefore only the $n_F\!=\!3$ light $u,d,s$ quark flavors and gluons constitute the genuine (intrinsic) partons of the proton and (in principle) the heavy $c,b,t$ quark flavors should not be included in the parton structure of the nucleon. This fact together with the perturbative stability of heavy quark production (cf. Sec.\,\ref{Sec.Fheavy}), even at $Q^2\!\gg\!m_h^2$, observed also in Fig.\,\ref{Fig219}, indicate that there is \emph{no need} to resum supposedly ``large logarithms'' $\big(\ln \frac{Q^2}{m_h^2}\big)$; which is of course in contrast to genuine collinear logarithms appearing in light (massless) quark and gluon hard scattering processes. In our opinion the use of the VFNS should therefore be avoided when possible; somewhat dissenting views are summarized in \cite{Vogt:2007vv}. We will have a closer look at the above mentioned perturbative stability of heavy quark production in DIS and determine under which conditions the use of the (ZM)VFNS is appropriate in Secs.\,\ref{Sec.HeavyDIS} and \ref{Sec.WandHiggs}.

\begin{figure}[p]
\centering
\includegraphics[width=0.965\textwidth]{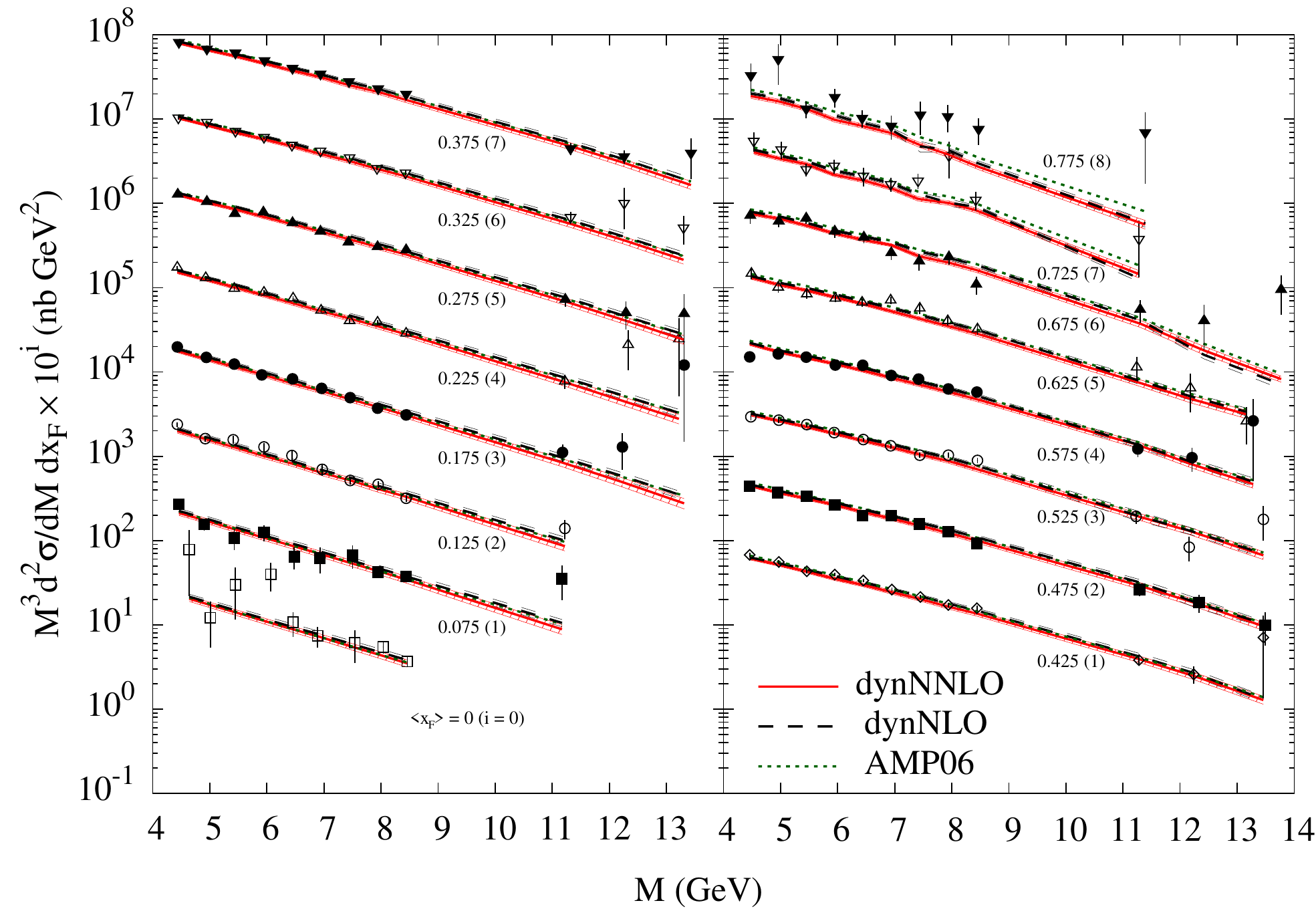}
\includegraphics[width=0.965\textwidth]{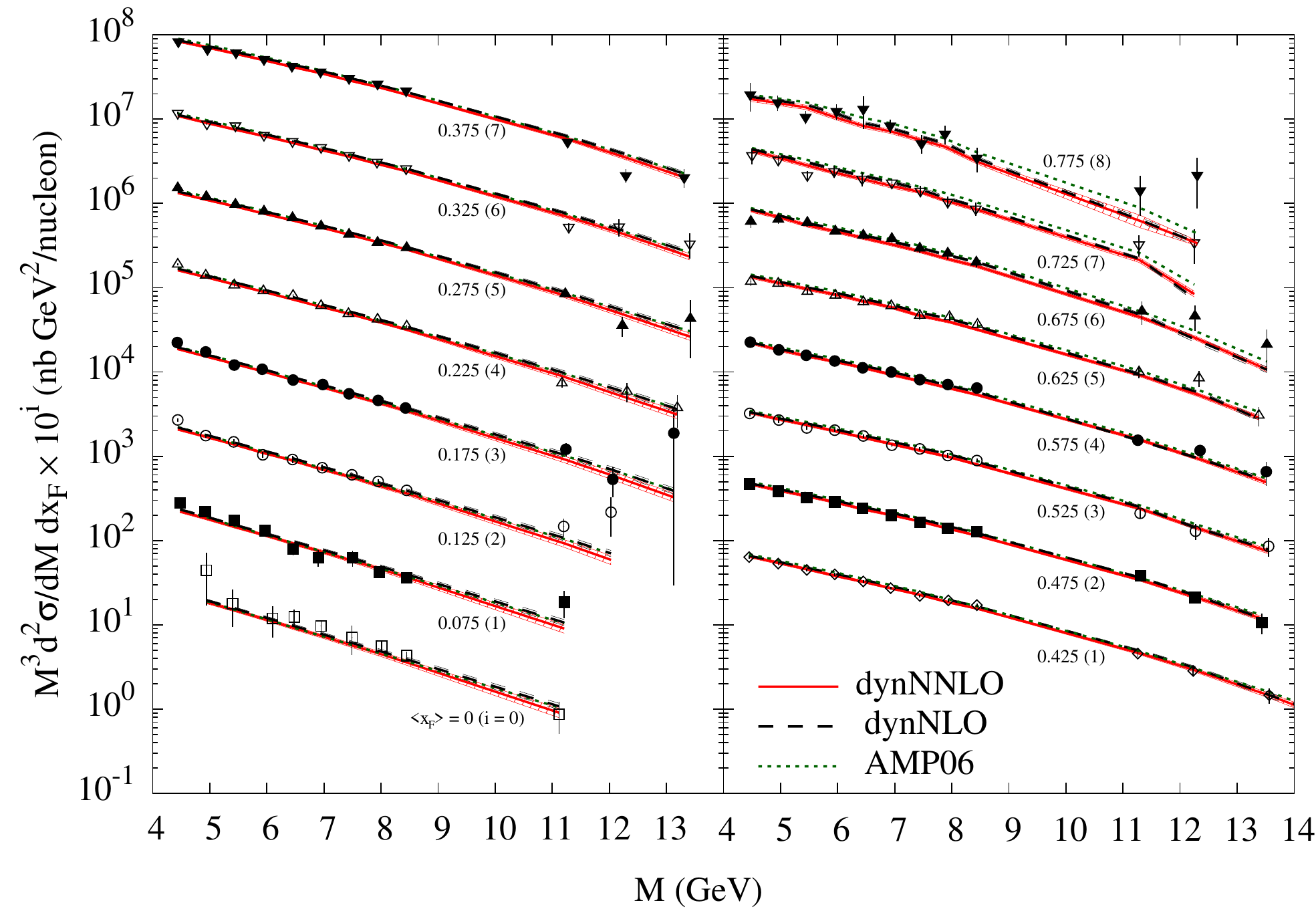}
\caption{Our dynamical NNLO and NLO \cite{Gluck:2007ck} results, together with their $\pm 1\,\sigma$ uncertainties, for Drell--Yan dilepton production in $pp$ and $pd$ collisions respectively for various selected average values of $x_F$ using the data sets of \cite{Webb:2003ps}. For comparison the NNLO AMP06 results \cite{Alekhin:2006zm} are shown as well. To ease the graphical presentation we have multiplied the results for the cross--sections by $10^i$ with $i$ indicated in parentheses in the figure for each fixed average value of $x_F$. \label{Fig220}}
\end{figure}

\begin{figure}[b!]
\centering
\includegraphics[width=0.5\textwidth]{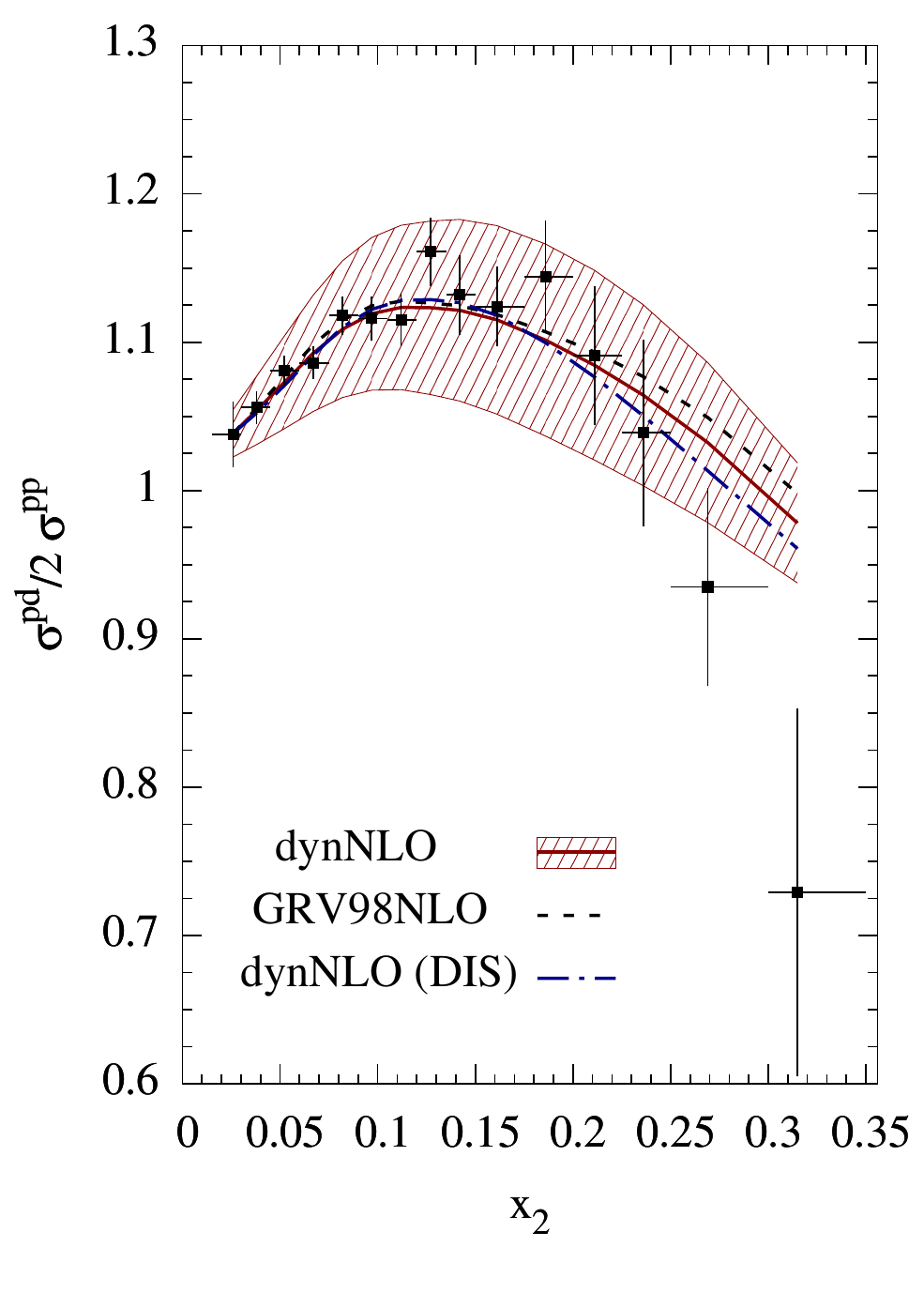}
\caption{Our dynamical NLO result in the $\overline{\rm MS}$ factorization scheme, together with its $\pm 1\,\sigma$ uncertainty, for $\frac{\sigma^{pd}}{2\sigma^{pp}}$, with $\sigma^{pN}\!\equiv\!\frac{d^2 \sigma^{pN}}{dM\,dx_F}$, as a function of the average fractional momentum $x_2$ of the target partons. The GRV98 NLO result \cite{Gluck:1998xa} is shown for comparison. The dynamical NLO\,(DIS) result is shown by the dashed--dotted curve. The data for the dimuon mass range $4.6\leq M_{\mu^+\mu^-}\leq 12.9$ GeV are taken from \cite{Towell:2001nh}.\label{Fig221}}
\end{figure}

As already pointed out (cf. Sec.\,\ref{Sec.Hadroproduction}), the measurements of Drell--Yan dilepton production in $pp$ and $pd$ collisions \cite{Webb:2003ps,Towell:2001nh} are instrumental in fixing $\Delta\equiv\bar{d}-\bar{u}$ (or $\bar{d}/\bar{u}$). In Fig.\,\ref{Fig220} we display our dynamical NLO and NNLO results, together with the $\pm 1\,\sigma$ uncertainties, and the NNLO AMP06 ones from \cite{Alekhin:2006zm} for the differential dimuon mass distributions for $pp$ and $pd$ collisions respectively; the ``standard'' fit results differ only marginally from the dynamical ones. The data are given for various average values of $x_F\!=\!x_1-x_2$, where $x_1$ and $x_2$ refer to the fractional momenta of the quarks in the beam ($p$) and the nucleon target ($N$) respectively; experimentally $x_F\!>\!0$ ($x_1\!>\!x_2$) and consequently the Drell--Yan cross--section is dominated by the annihilation of a beam quark with a target antiquark. In the relevant kinematic region where high--statistics data exist, all three NNLO and NLO results shown agree within 1 $\sigma$.

In order to test the dependence of our results on the specific choice of the factorization scheme, we have repeated our dynamical NLO analysis in the DIS factorization scheme of Sec.\,\ref{Sec.DISscheme}. Since the (light) parton distributions in the DIS scheme are defined via the $F_2$ structure function, it is not very surprising that, in contrast to the DIS parton distributions themselves, the results for physical observables directly related to DIS structure functions are very similar. Indeed the results in the DIS factorization scheme for $F_2^p(x,Q^2)$ in  Figs.\,\ref{Fig215} and \ref{Fig217} are practically indistinguishable from the ones in the $\overline{\text{MS}}$ scheme, as are the results for $F_2^{c,b}(x,Q^2)$ in Fig.\,\ref{Fig219}. Differences become visible only for processes which are not directly related to DIS structure functions, such as  Drell--Yan cross--sections, but there the differences lie well within the $\pm 1\,\sigma$ uncertainty of the NLO\,($\overline{\rm MS}$) results as illustrated in Fig.\,\ref{Fig221} for the DY--asymmetry. There we display our dynamical NLO\,($\overline{\rm MS}$) result for $\sigma^{pd}/2\sigma^{pp}$ together with the $\pm 1\,\sigma$  uncertainty band as well as the previous GRV98 result which agree  in the statistically relevant $x$--region, with $x_2$ referring to the average fractional momentum of the target partons. The dynamical NNLO\,($\overline{\rm MS}$) as well as the results from the ``standard'' NNLO AMP06 fit \cite{Alekhin:2006zm} are only marginally different (cf. \cite{JimenezDelgado:2008hf}); being the differences between \emph{all} the theoretical predictions much smaller than the experimental errors and than the $1\,\sigma$ uncertainty bands obtained in our NLO and NNLO analyses; which differ only marginally as well (cf. \cite{JimenezDelgado:2008hf}).

\begin{figure}[t]
\centering
\includegraphics[width=0.95\textwidth]{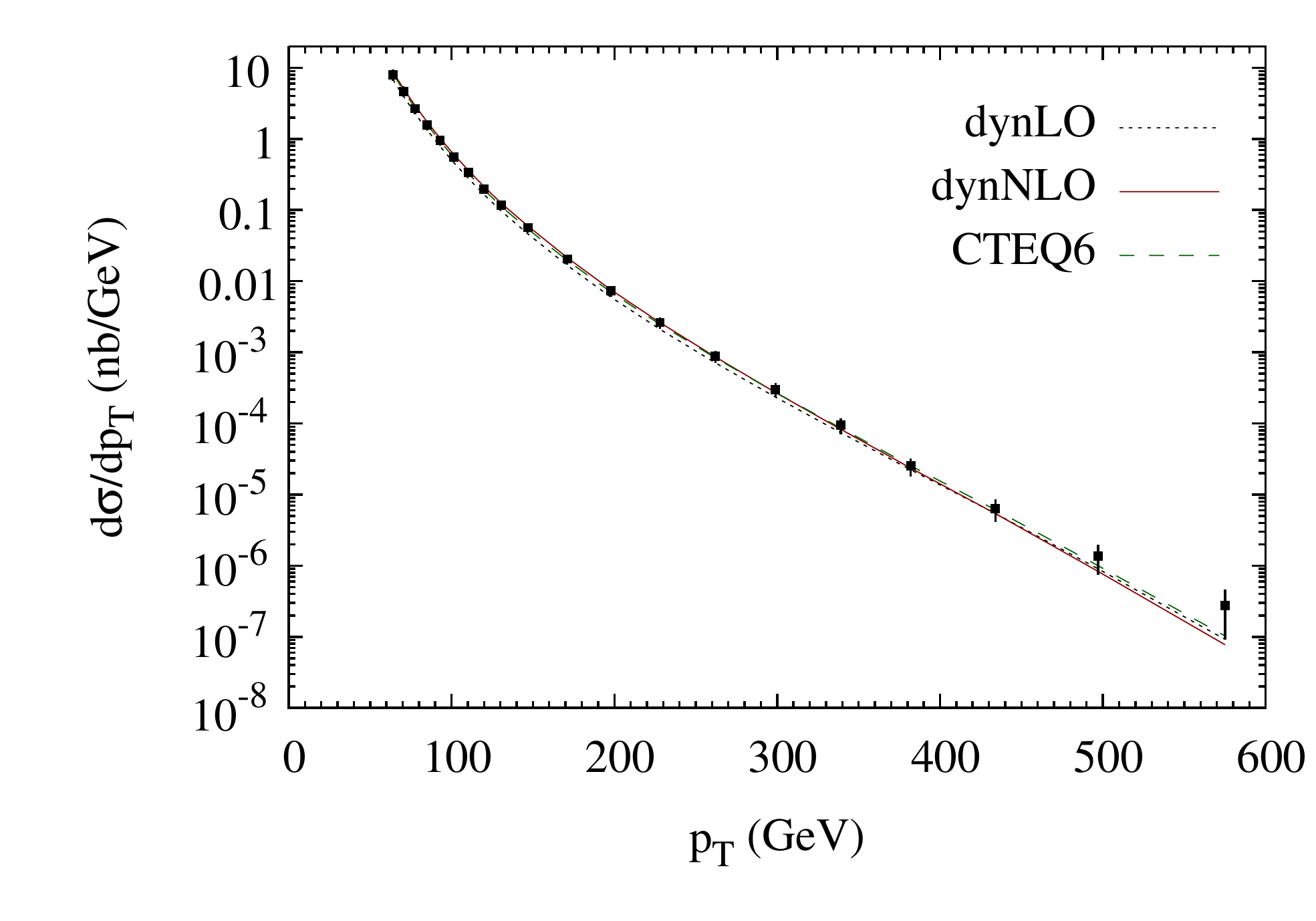}
\caption{The $p\bar{p}$ CDF high--$p_{\rm T}$ inclusive jet data \cite{Abulencia:2005yg} compared with our dynamical LO and NLO\,($\overline{\rm MS}$) results, as well as with the NLO CTEQ6 result. \label{Fig222}}
\end{figure}
Finally the $p\bar{p}$ CDF (Tevatron II) high--$p_{\rm T}$ inclusive jet data \cite{Abulencia:2005yg} are compared in Fig.\,\ref{Fig222} with our dynamical LO and NLO\,($\overline{\rm MS}$) results, as well as  with the ones of CTEQ6 \cite{Pumplin:2002vw}. The small $1\,\sigma$ error bands are almost invisible on the huge logarithmic scale used. Our NLO result almost coincides with the one of CTEQ. There is a  clear improvement at NLO as compared to LO which falls slightly below the data at $p_{\rm T}$ \!$\lesssim$ 300 GeV. Nevertheless the LO high--$p_{\rm T}$ fit corresponds to $\chi^2/\rm dof\simeq 1$, which is only twice as large as at NLO (cf. Table \ref{TabA01}).

\section{Astrophysical Relevance of the Dynamical Predictions}\label{Sec.Astrophysics}

\begin{figure}[t]
\centering
\includegraphics[width=0.9\textwidth]{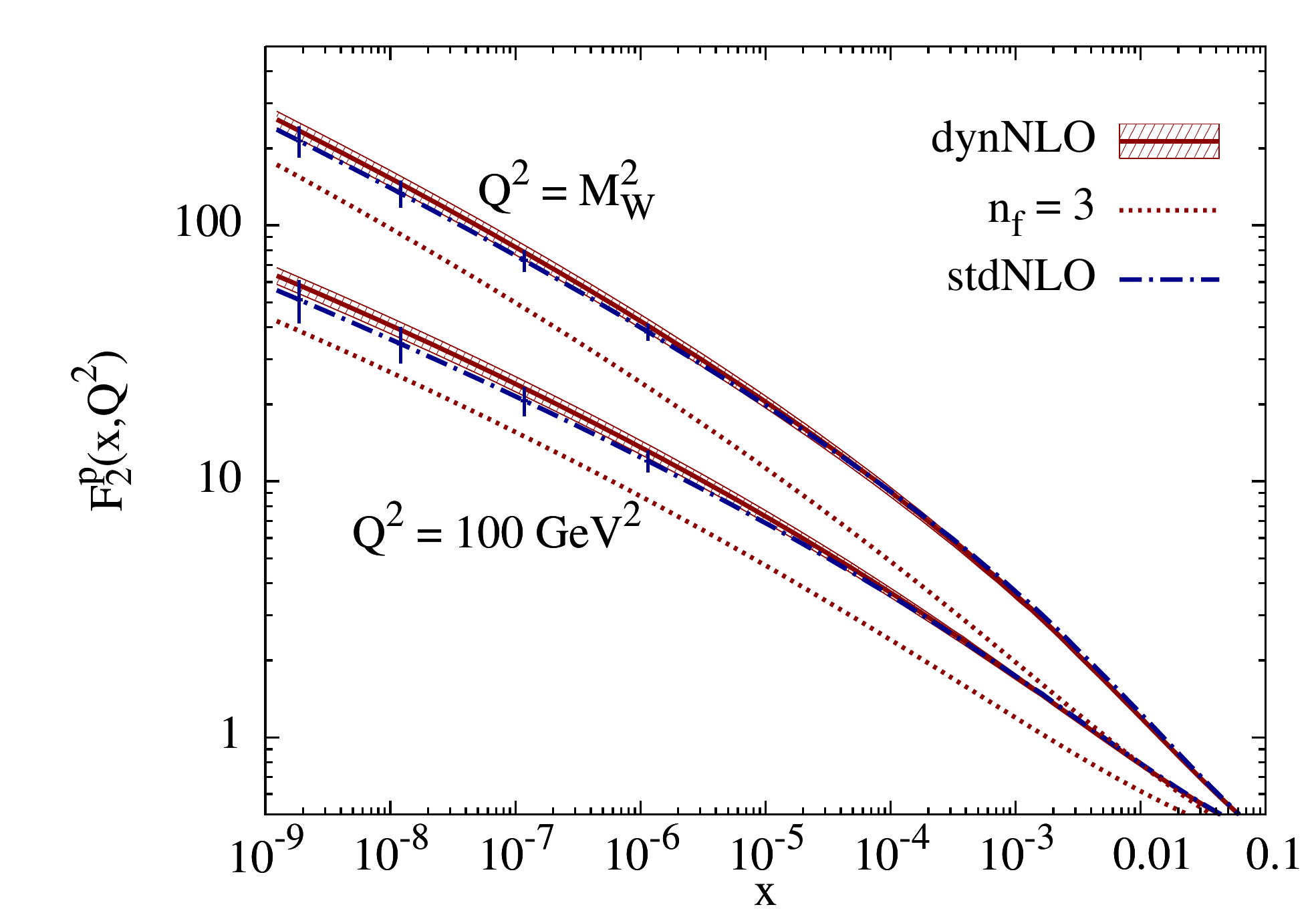}
\caption{Dynamical NLO\,($\overline{\rm MS}$) predictions for $F_2^p(x,Q^2)$, together with their $1\,\sigma$ uncertainties, for  extremely small values of $x$. The $1\,\sigma$ uncertainties of the  standard (std) NLO fit extrapolations are shown by the vertical bars. The dotted curves are the contributions from the light ($n_F\!=\!3$) quarks and gluons to $F_2^p$ for the dynamical (dyn) NLO result. In other words, the difference between the dotted and solid curves is due to NLO heavy quark (charm, bottom) contributions which derive  from photon--gluon (quark) fusion processes. The dynamical GRV98 predictions lie within the $\pm 1\,\sigma$ band of our present dynNLO predictions.
\label{Fig223}}
\end{figure}

The predictive power of the dynamical model is especially important for investigations concerning $\nu N$ cross--sections \cite{Gluck:1998js, Frichter:1994mx, Capelle:1998zz} of ultrahigh energy cosmic neutrinos produced via astrophysical processes, e.g. in active galactic nuclei, black holes or in the decays of very massive particles (see, for example, \cite{Halzen:2000km, Halzen:2006mq} for more details). Here one needs a somewhat reliable knowledge of parton distributions at the weak scale $Q^2\!=\!M_W^2$ down to $x\!\simeq\!10^{-9}$ $(x\!\simeq\!\frac{M_W^2}{2m_NE_\nu}\big)$ at the highest energies of $E_\nu\simeq 10^{12} \text{ GeV}$ which in the ``standard'' approach requires extrapolations into yet unmeasured regions. As our parameter--free small--$x$ predictions for parton distributions at $x<10^{-2}$ are entirely of QCD--dynamical origin and depend rather little on the detailed input parameters at $x\!\gtrsim\!10^{-2}$, it is interesting to study these predictions in kinematic regions not accessible by present DIS experiments. Furthermore, the more  restrictive ansatz of the valence--like input distributions at small--$x$ as well as the sizeably larger evolution distance imply, as we have seen, smaller uncertainties concerning the behavior of structure functions in the small-$x$ region than the corresponding results obtained from the common ``standard'' approach, in particular as $Q^2$ increases.

Since $F_2^p(x,Q^2)$ is, in the very small--$x$ region, dominated by $\bar{u}(x,Q^2)$ and $\bar{d}(x,Q^2)$, as are the CC neutrino--(isoscalar) nucleon cross--sections, the $F_2^p$ structure function can be utilized for estimating the magnitude of uncertainties of the predictions in the extreme small--$x$ region, which are shown in Fig.\,\ref{Fig223}. At $Q^2\!=\!M_W^2$ our dynamical NLO predictions correspond to a $\pm1\,\sigma$ uncertainty of about $\pm$7\% at $x\!=\!10^{-9}$ whereas the uncertainty of the extrapolation of a ``standard'' fit is about twice as large. At smaller scales the uncertainties obviously increase as illustrated in Fig.\,\ref{Fig223} at $Q^2\!=\!100$ GeV$^2$. Taking into account previous extrapolation  ambiguities \cite{Gluck:1998xa}, one can conclude \cite{Gluck:1998js} that the dynamically predicted small--$x$ parton distributions allow neutrino--nucleon cross--sections to be calculable with an accuracy of about 10\% at the highest cosmic neutrino energies. It should be mentioned that an ad hoc fixed power law of $x$ extrapolation of the standard CTEQ6.5 structure functions \cite{Tung:2006tb} to $x\!=\!10^{-8}$ at $Q^2\!=\!M_W^2$ lies, accidentally, only about 10\% below our dynamical NLO prediction in Fig.\,\ref{Fig223}. On the other hand, an alternative parametrization \cite{Berger:2007vf} of present HERA (ZEUS) data which is not QCD oriented but based on analyticity and unitarity gives, when extrapolated to $x\!=\!10^{-8}$, a factor of about 6 smaller a value for $F_2^p(10^{-8},M_W^2)$ than our result in Fig.\,\ref{Fig223}. Since the perturbative dynamical QCD {\em pre}dictions for the small--$x$ behavior of structure functions down to $x\!=\!10^{-5}$ proved to be in agreement with later HERA measurements as discussed in Sec.\,\ref{Sec.Comparisons}, it is hard to  imagine that perturbative QCD dynamics and evolutions should become entirely inappropriate at $x\!=\!10^{-8}$ to $10^{-9}$ at even much larger scales.

\section{The Longitudinal Structure Function}\label{Sec.FL}

\begin{figure}[b!]
\centering
\includegraphics[width=\textwidth]{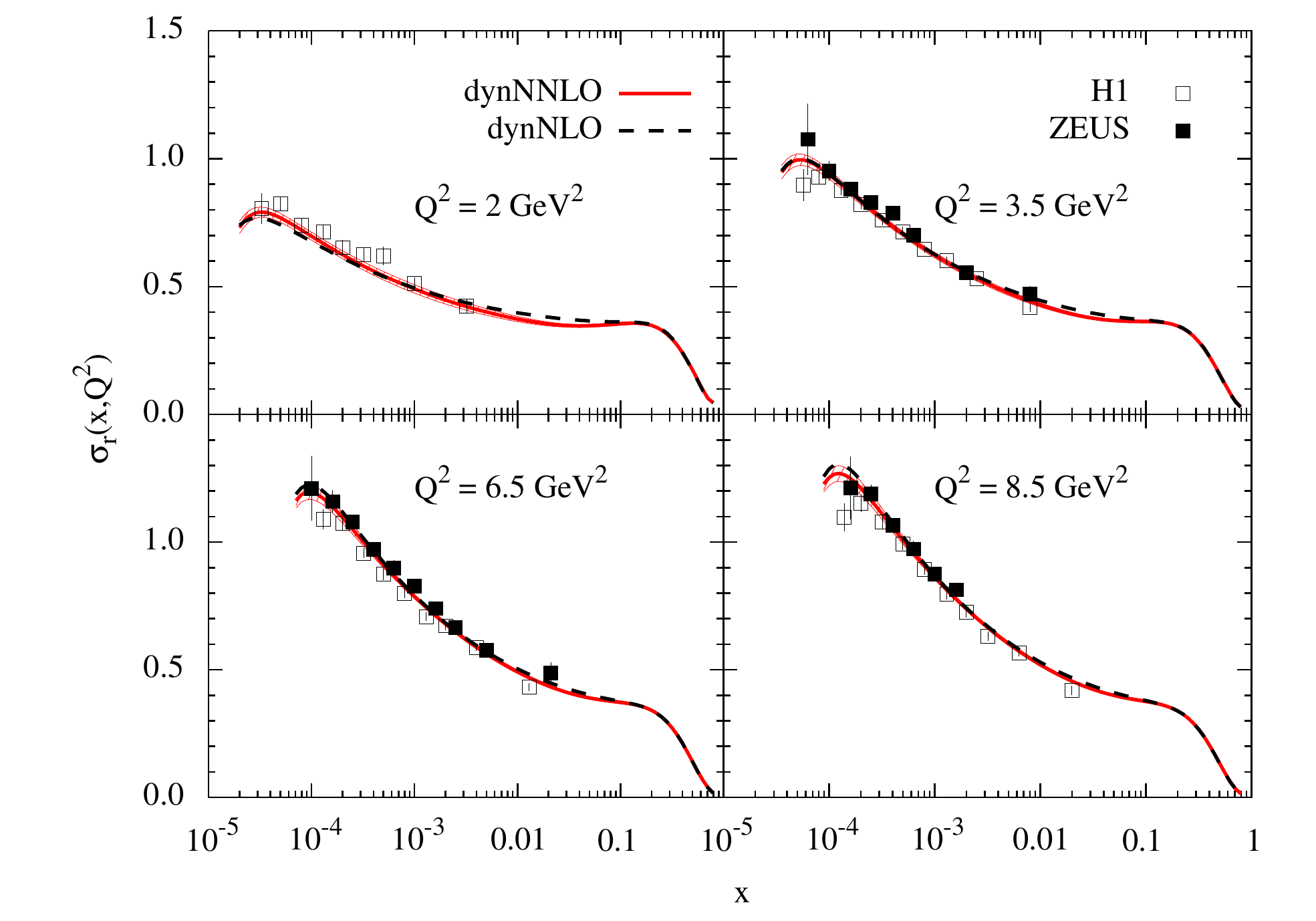}
\caption{The dynamical NNLO predictions, together with their $\pm 1\,\sigma$ uncertainties, for the ``reduced'' DIS cross--section $\sigma_r(x,Q^2) \!= \!F_2 - (y^2/Y_+) F_L$. The uncertainty bands of our previous dynamical NLO results (dashed curves) are very similar in size \cite{Gluck:2007ck} as the ones shown for NNLO. The HERA data for some representative fixed values of $Q^2$ are taken from \cite{Adloff:1999ah, Adloff:2000qj, Adloff:2000qk, Adloff:2003uh, Chekanov:2001qu}.
\label{Fig224}}
\end{figure}

\begin{figure}[t]
\centering
\includegraphics[width=\textwidth]{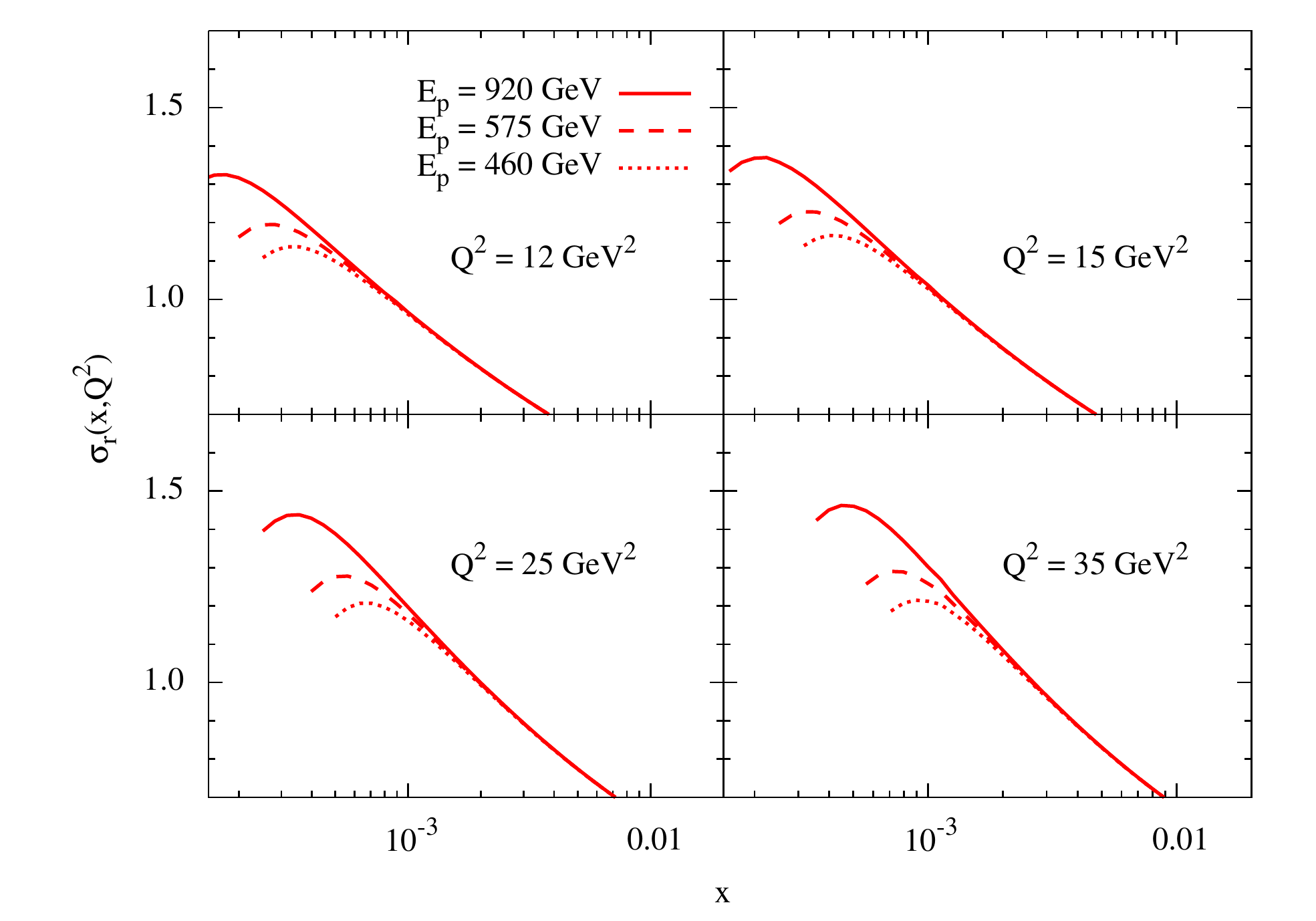}
\caption{Our dynamical NNLO predictions for $\sigma_r(x,Q^2)$ for different proton beam energies $E_p$ relevant for most recent HERA--H1 measurements \cite{Aaron:2008tx}. The $\pm 1\,\sigma$ uncertainty bands are similar to the ones shown in Fig.\,\ref{Fig224}.  Notice that the curves terminate when $y=1$.\label{Fig225}}
\end{figure}

The importance of using the experimentally directly measured ``reduced'' DIS cross--sections $\sigma_r^{\rm NC}$ (cf. Eq.\,\ref{sigmar}) for the determination of parton distributions has been emphasized in \cite{Martin:2006qv}. As mentioned, for not too large values of $Q^2$ this quantity is dominated by the one--photon exchange structure functions, in particular by $F_2$. However, the effect of $F_L$ becomes increasingly relevant as $x$ decreases at a given $Q^2$, where $y$ increases. This is seen in the data as a flattening of the growth of $\sigma_r(x,Q^2)$ as $x$ decreases to very small values, at fixed $Q^2$, leading eventually to a turnover as can be seen in Fig.\,\ref{Fig224}. At the lower values of $Q^2$ in Fig.\,\ref{Fig224} it was not possible in \cite{Martin:2006qv} to reproduce this turnover at NLO. This was mainly due to the negative longitudinal cross--section (negative $F_L(x,Q^2$)) encountered in \cite{Martin:2002dr, Martin:2006qv}. Since all of our cross--sections and subsequently all structure functions are manifestly positive throughout the whole kinematic region considered, our dynamical NLO and NNLO results in Fig.\,\ref{Fig224} are in good agreement with all small--$x$ HERA measurements \cite{Adloff:1999ah, Adloff:2000qj, Adloff:2000qk, Adloff:2003uh, Chekanov:2001qu}. The same holds true for our ``standard'' NLO and NNLO results which, besides having wider uncertainty bands, are almost indistinguishable from the dynamical ones shown in Fig.\,\ref{Fig224}. In Fig.\,\ref{Fig225} we display our NNLO results for the ``reduced'' cross-section at different proton beam energies $E_p$, relevant for most recent H1 measurements \cite{Reimer:2007iy}. The turnover at small $x$ becomes more pronounced at smaller energies because of the larger values of $y$. Our dynamical small--$x$ predictions are fully compatible with the (preliminary) H1 data presented in \cite{Aaron:2008tx}.

\begin{figure}
\centering
\includegraphics[width=0.86\textwidth]{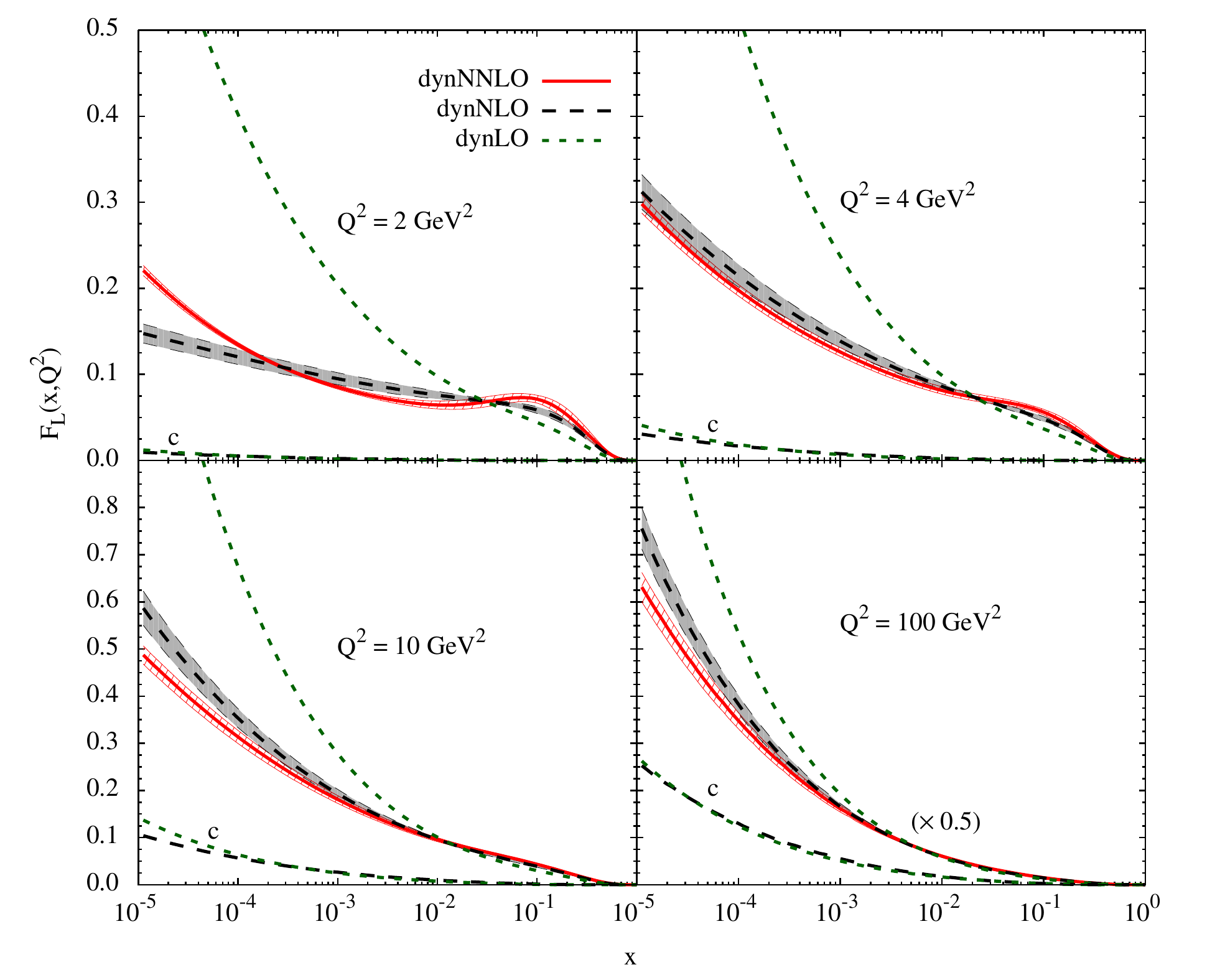}
\includegraphics[width=0.86\textwidth]{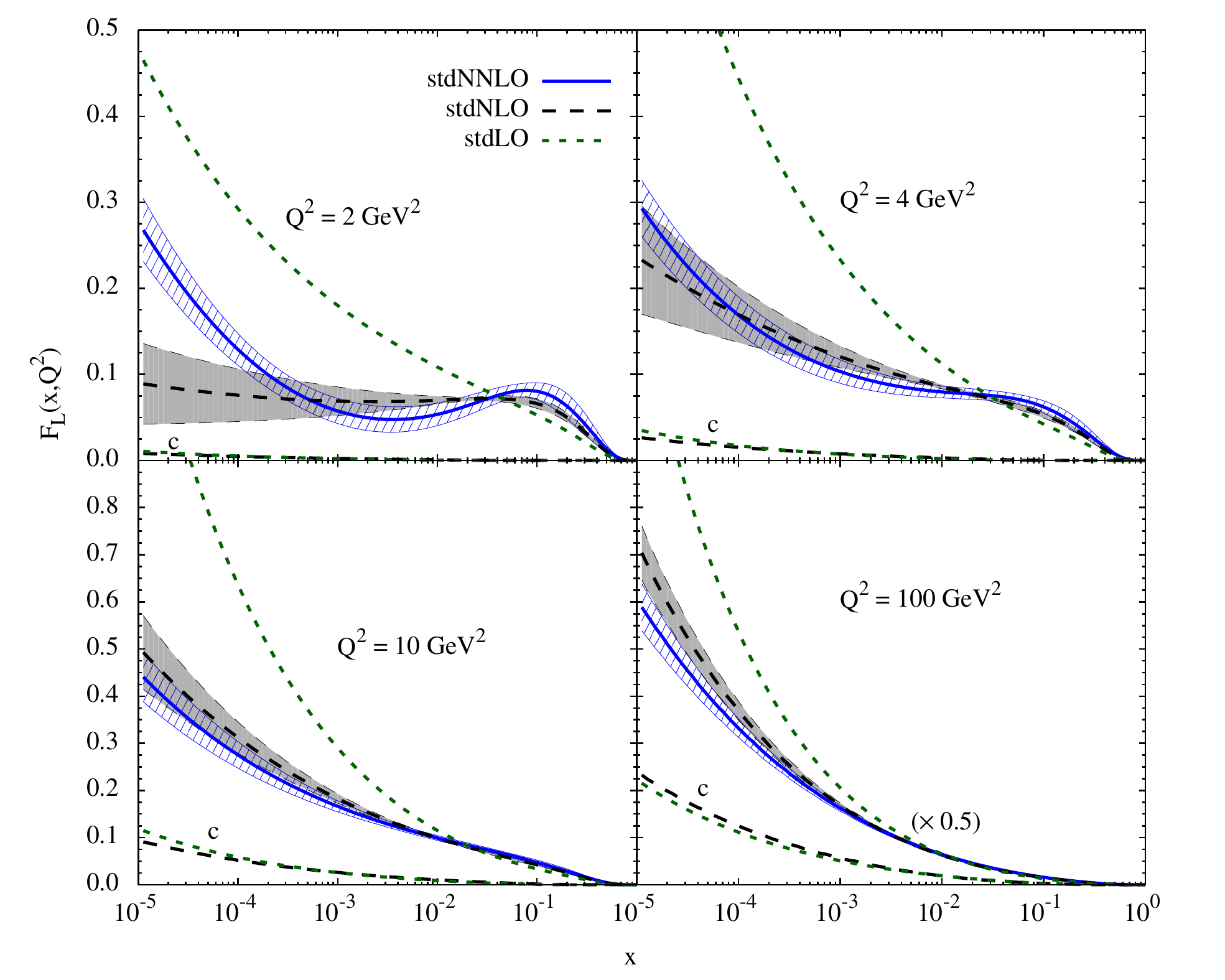}
\caption{Dynamical and ``standard'' parton model LO, NLO and NNLO predictions for $F_L(x,Q^2)$ together with the $\pm 1\,\sigma$ uncertainty bands at NLO and NNLO. The heavy charm ($c$) contributions at LO (short--dashed curves) and NLO (long--dashed curves) are shown as well. The results at $Q^2=100$ GeV$^2$ are multiplied by 0.5 as indicated.
\label{Fig226}}
\end{figure}

Turning now to $F_L$ itself we present our dynamical and ``standard'' LO, NLO and NNLO predictions Fig.\,\ref{Fig226}, together with the small subdominant charm contributions at LO and NLO. These gluon--dominated dynamical predictions become perturbatively stable already at $Q^2\!\gtrsim$ 2--3 GeV$^2$, where precision measurements could even delineate NNLO effects in the very small--$x$ region. This is in contrast to the common ``standard'' results, as has been already observed in \cite{Gluck:2007sq}, but the differences between the NNLO and NLO results are here less distinguishable due to the much larger $1\,\sigma$ uncertainty bands which partly overlap in the very small--$x$ region. It should be noticed that the NLO/NNLO instabilities implied by the ``standard'' fit results obtained in \cite{Martin:2002dr,Martin:2006qv} at $Q^2\!\lesssim$ 5 GeV$^2$ are far more violent than the ones shown in Fig.\,\ref{Fig226} for our ``standard'' results, which is mainly due to the negative $F_L(x,Q^2)$ encountered in \cite{Martin:2002dr,Martin:2006qv}. The perturbative stability in any scenario becomes in general better the larger $Q^2$, typically beyond $4-5$ GeV$^2$ \cite{Martin:2002dr,Georgi:1976ve,Gluck:2007sq,Martin:2006qv}, as evident from Fig.\,\ref{Fig226}. This is due to the fact that the $Q^2$ evolutions eventually force any parton distribution to become sufficiently steep in $x$. It should be mentioned that the sizeable discrepancies between NNLO and NLO predictions at $Q^2\!=\!2$ GeV$^2$ and $x\!\simeq\!10^{-5}$ in Fig.\,\ref{Fig226} are not too surprising since $Q^2\!\simeq\!2$ GeV$^2$ represents somehow a borderline value for the leading twist--2 contribution to become dominant at small--$x$ values. This is further corroborated by the observation that the dynamical NNLO and NLO twist--2 fits slightly undershoot the HERA data for $F_2$ at $Q^2\!\lesssim$ 2 GeV$^2$ in the small--$x$ region (cf. Figs.\,\ref{Fig215} and \ref{Fig216}), which indicates that nonperturbative (higher--twist) contributions to structure functions become relevant there \cite{Gluck:1998xa,Gluck:2007ck}.

\begin{figure}
\centering
\includegraphics[width=0.9\textwidth]{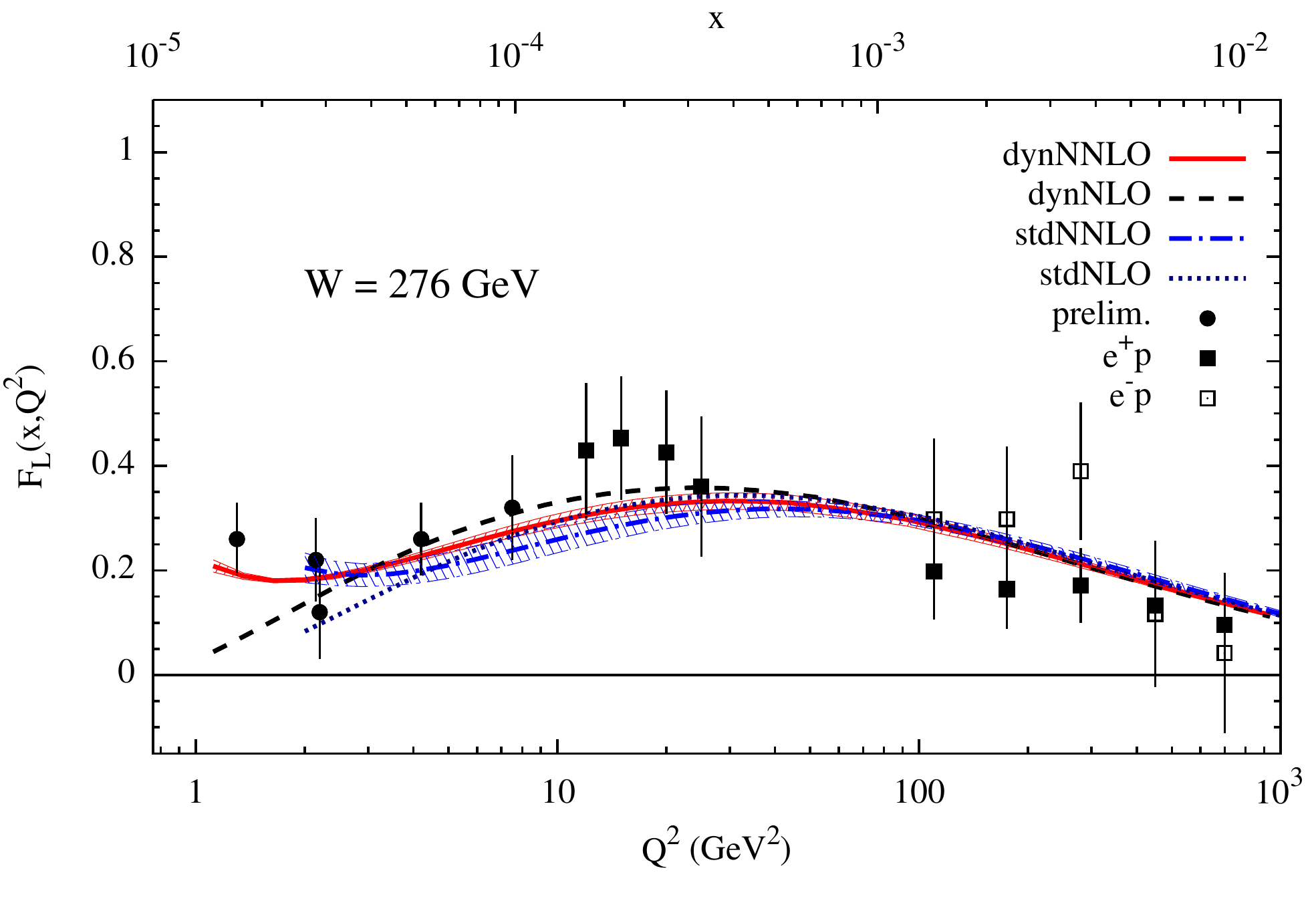}
\caption{Dynamical and ``standard'' NNLO and NLO predictions for $F_L(x,Q^2)$ at a fixed value of $W=276$ GeV. The NLO\,($\overline{\rm MS}$) results are taken from \cite{Gluck:2007ck}. The (partly preliminary) H1 data \cite{Adloff:2000qk, Adloff:2003uh, Latovika:2004ya, Lobodzinska:2003yd} are at fixed $W\simeq 276$ GeV.\label{Fig227}}
\end{figure}
For completeness we finally compare in Fig.\,\ref{Fig227} our (leading--twist) NNLO dynamical and ``standard'' predictions for $F_L(x,Q^2)$, together with their $\pm 1\,\sigma$ error bands, with a representative selection of (partly preliminary) H1 data \cite{Adloff:2000qk, Adloff:2003uh, Adloff:1996yz, Latovika:2004ya, Lobodzinska:2003yd} at fixed $W\simeq 276$ GeV. For comparison we also show in Fig.\,\ref{Fig227} our NLO results which have comparable, although somewhat larger, uncertainty bands (cf. \cite{Gluck:2007ck}); note as well that the uncertainties of the dynamical predictions are again smaller than the corresponding to the ``standard'' fits. All our NNLO and NLO results for $F_L$, being gluon dominated in the small--$x$ region, are in full agreement with present measurements which is in contrast to expectations \cite{Martin:2001es, Martin:2002dr, Martin:2006qv} based on negative parton distributions and structure functions at small values of $x$. To illustrate the manifest positive definiteness of our dynamically generated structure functions at $Q^2\!\geq\! \mu_0^2$, we show $F_L(x,Q^2)$ in Fig.\,\ref{Fig227} down to small values of $Q^2$ although, as mentioned, leading twist--2 predictions need not necessarily be confronted with data below, say 2 GeV$^2$.

\fontsize{11}{16.2}
\selectfont

\section{The Role of Heavy Quark Flavors in DIS}\label{Sec.HeavyDIS}

In this and the following section we compare \emph{fixed flavor number scheme} (FFNS) and (zero-mass) \emph{variable flavor number scheme} (VFNS) parton model predictions at high energy colliders at LO and NLO of QCD. Based on our LO-- and NLO--FFNS dynamical and ``standard'' parton distributions \cite{Gluck:2007ck}, we generate radiatively sets of VFNS parton distributions \cite{Gluck:2008gs}, where also the heavy quark flavors are considered as massless partons within the nucleon. In this section we focus on the role of these distributions in deep inelastic scattering, while in the next one we study their implications for hadron colliders. In particular we show that the VFNS predictions are compatible with the reliable FFNS ones for processes in which the invariant mass of the produced system far exceeds the mass of the participating heavy quark flavor \cite{Gluck:2008gs}.

As explained in Sec.\,\ref{Sec.Fheavy}, the VFNS is characterized by increasing the number of massless partons $n_F$ by one unit at the \emph{unphysical} thresholds $Q^2\!=\!m_h^2$, starting from $n_F\!=\!3$ at $Q^2\!=\!m_c^2$, i.e. $c(x,m_c^2)\!=\!\bar{c}(x,m_c^2)\!=\!0$ up to NLO. Thus the ``heavy'' $n_F\!>\!3$ quark distributions are perturbatively uniquely generated from the $n_F\! -\!1$ ones via the massless renormalization group equations (evolution). The VFNS heavy quark contributions are therefore calculated in analogy to those of the light quarks, e.g. their contributions to DIS structure functions follow Sec.\,\ref{Sec.Flight}. Our choice for the input of the ``heavy'' $h\!=\!c,b$ VFNS distributions are our FFNS distributions at $Q^2\!=\!m_h^2$. The VFNS predictions at scales $Q^2\!\gg\! m_h^2$ should become insensitive to this, somewhat arbitrary, input selection \cite{Alekhin:2002fv}, whose virtues are the fulfillment of the standard sum--rule constraints (cf. Eq.\,\ref{sumrules}) together with reasonable shapes and sizes of the various input distributions.

\begin{figure}[t]
\centering
\includegraphics[width=0.495\textwidth]{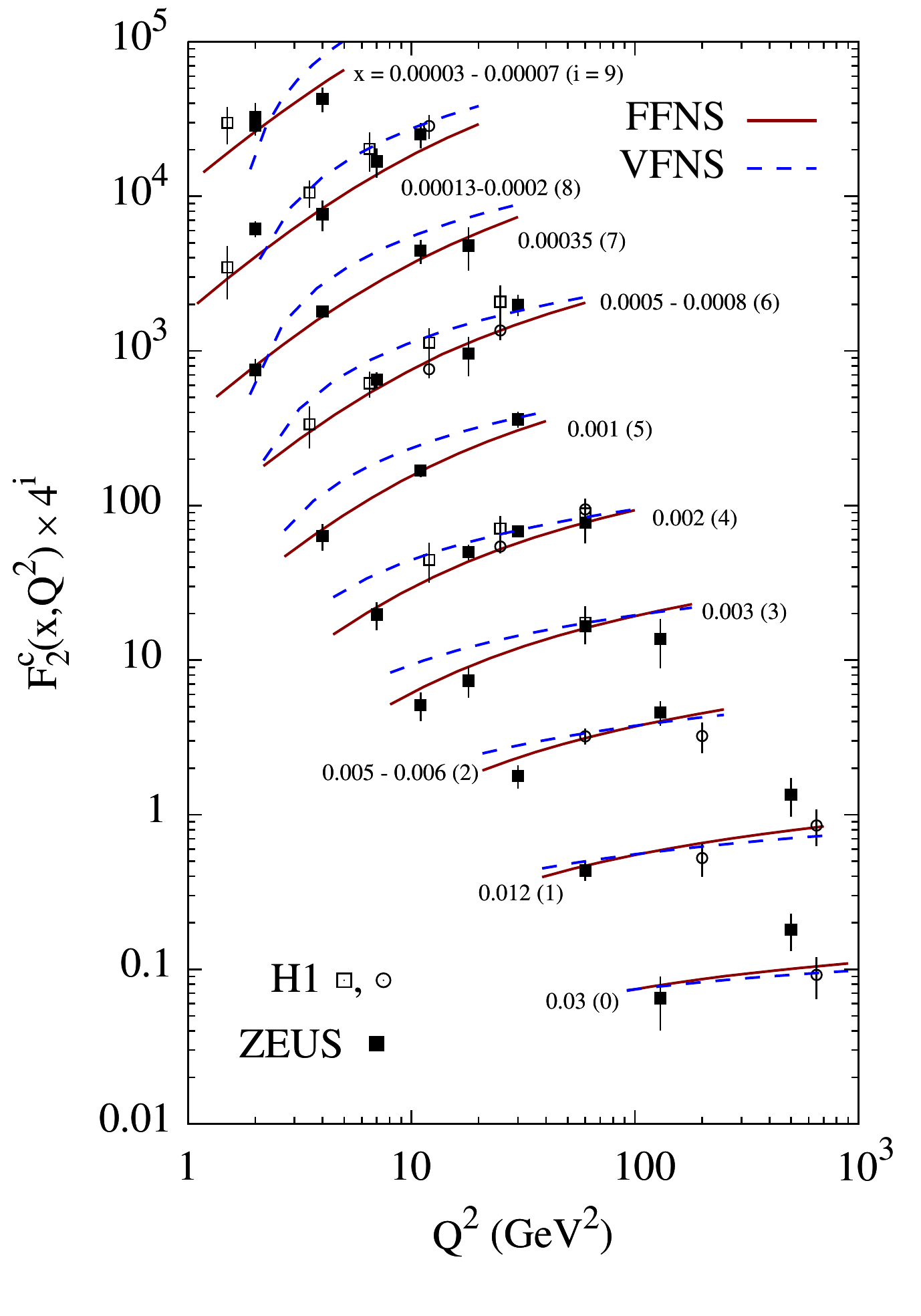}
\includegraphics[width=0.495\textwidth]{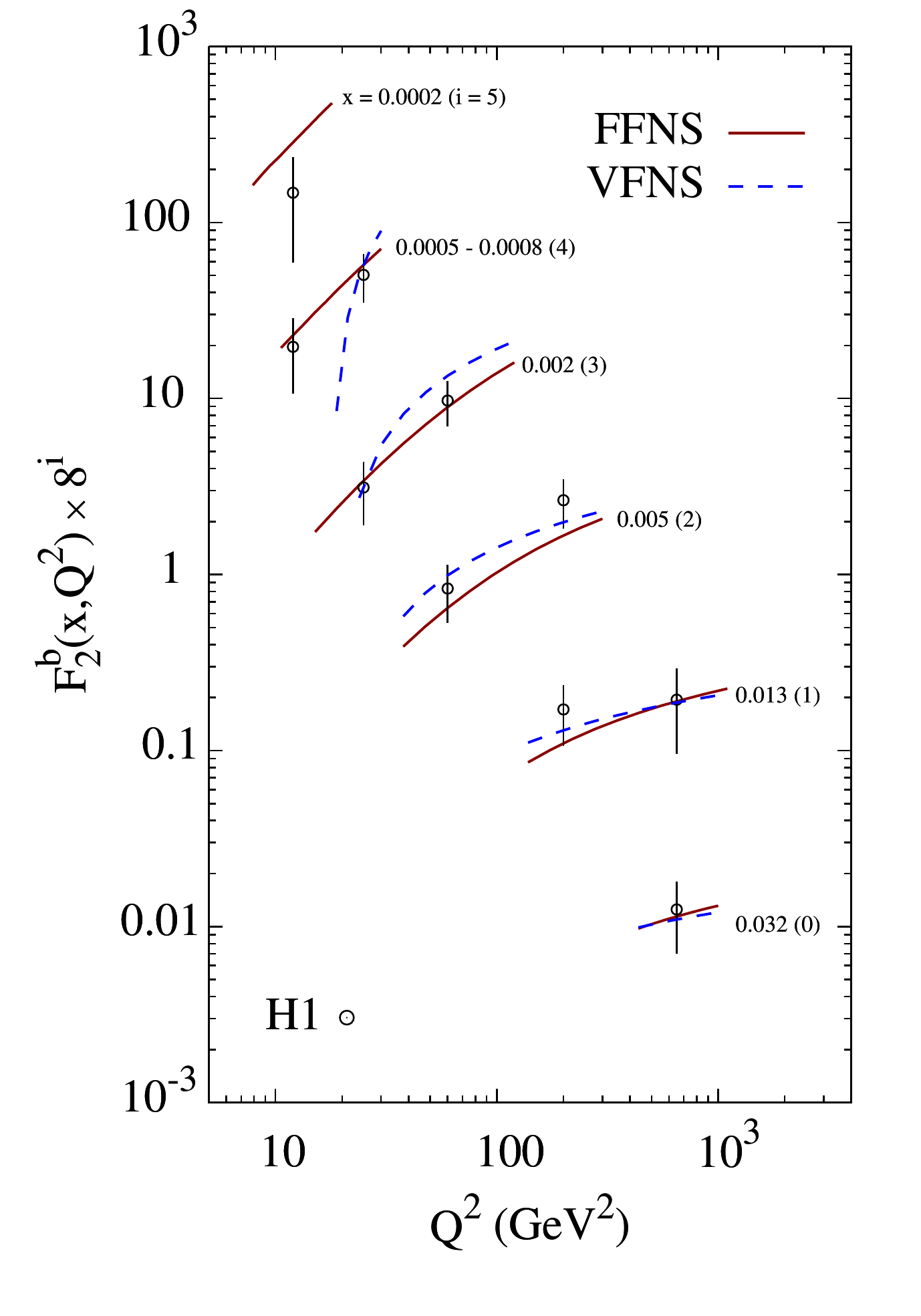}
\caption{Comparing our dynamical NLO\,($\overline{\rm MS}$) calculations for $F_2^{c,b}(x,Q^2)$ in the (zero-mass) VFNS with heavy quark electroproduction data \cite{Adloff:2001zj, Aktas:2004az, Aktas:2005iw, Chekanov:2003rb}. The NLO\,($\overline{\rm MS}$) FFNS predictions are shown for comparison; here $\mu^2\!=\!4m_h^2$, with $m_{c,b}\!=\!1.3,4.2$ GeV, although the results are rather unaffected by this choice \cite{Gluck:1993dpa}.\label{Fig228}}
\end{figure}

Since the deep inelastic structure function $F_2(x,Q^2)$ plays an instrumental role in determining the parton distributions of the nucleon, we consider as a first test of the VFNS ``heavy'' quark distributions the charm and bottom contributions to this quantity. In Fig.\,\ref{Fig228} we compare our VFNS results, with $\mu^2\!=\!Q^2$ as usual\footnote{Notice that here and in the following $\mu\!\equiv\!\mu_F\!=\!\mu_R$ where $\mu_R$ and $\mu_F$ are the renormalization and factorization scales respectively. As mentioned, this choice is dictated by the fact that our (and all other presently available) parton distributions were determined and evolved with $\mu_F\!=\!\mu_R$, i.e. with the commonly adopted standard evolution equations.} (cf. Sec.\,\ref{Sec.Flight}), with recent HERA data on heavy quark electroproduction. It can be seen that the data are, except for higher $Q^2$, very badly described by the VFNS results. This well--known fact is moreover totally expected since the VFNS appear as a resummation of the FFNS contributions, which make sense \emph{only} for $Q^2\!\gg\!m_h^2$. For this same reason it is not common to find comparisons of (ZM)\,VFNS predictions with heavy quark production data, especially near threshold; we show them here nevertheless because while presenting our results we got involved in discussions where we realized that the reader may not be aware of (or even reluctant to accept) this well--established fact. Note as well the excellent agreement of the FFNS predictions (already discussed in Sec.\,\ref{Sec.Comparisons}) and that, despite the common belief that ``non--collinear'' logarithms $\big(\ln\frac{Q^2}{m_h^2}\big)$ \emph{need} to be resummed, for $Q^2\!\gg\!m_h^2$ present data are not sufficiently accurate to discriminate between VFNS and FFNS results in Fig.\,\ref{Fig228}. At this point it is worth to recall that, in contrast to what it is sometimes assumed, the calculations in both schemes do not converge in the limit $Q^2\!\gg\!m_h^2$ since, even in this limit, the FFNS results \emph{always} contain contributions from the (physical) threshold region $\hat{s}\gtrsim\hat{s}_{\rm th}\!=\!4m_h^2$ (cf. Sec.\,\ref{Sec.Fheavy} and \cite{Gluck:1993dpa}).

\begin{figure}[p]
\centering
\includegraphics[width=0.9\textwidth]{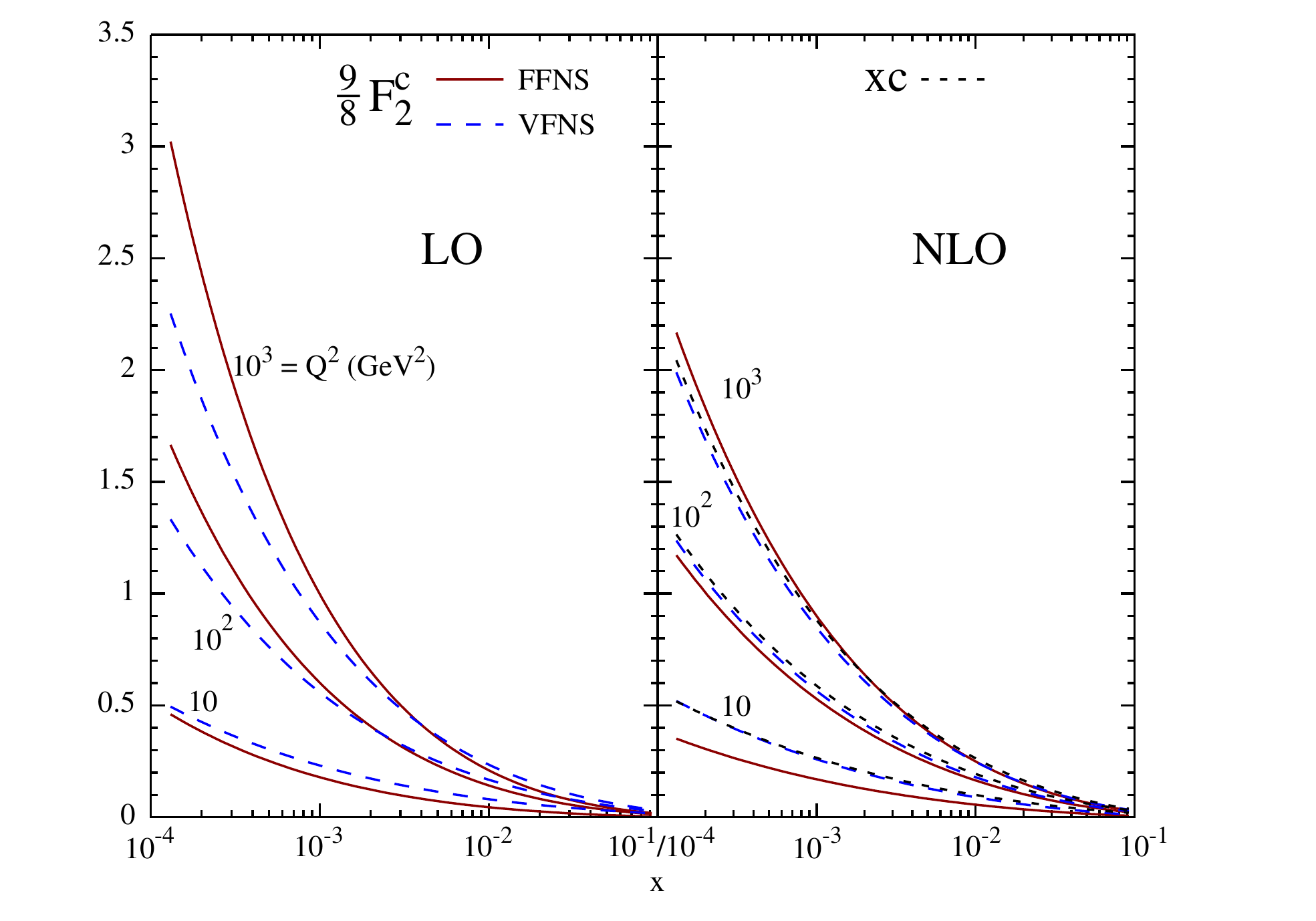}
\caption{The predicted $x$--dependencies of $\frac{9}{8}F_2^c(x,Q^2)$ in the FFNS and VFNS at some typical fixed values of $Q^2$. For the FFNS the renormalization and factorization scales are chosen to be $\mu^2\equiv\mu_F^2=\mu_R^2=Q^2+4m_c^2$ with $m_c\!=\!1.3$ GeV, and, as usual, $\mu^2=Q^2$ for the VFNS. The NLO--VFNS charm distribution is given by $xc(x,Q^2)$ as shown by the short--dashed curves.\label{Fig229}}
\includegraphics[width=0.9\textwidth]{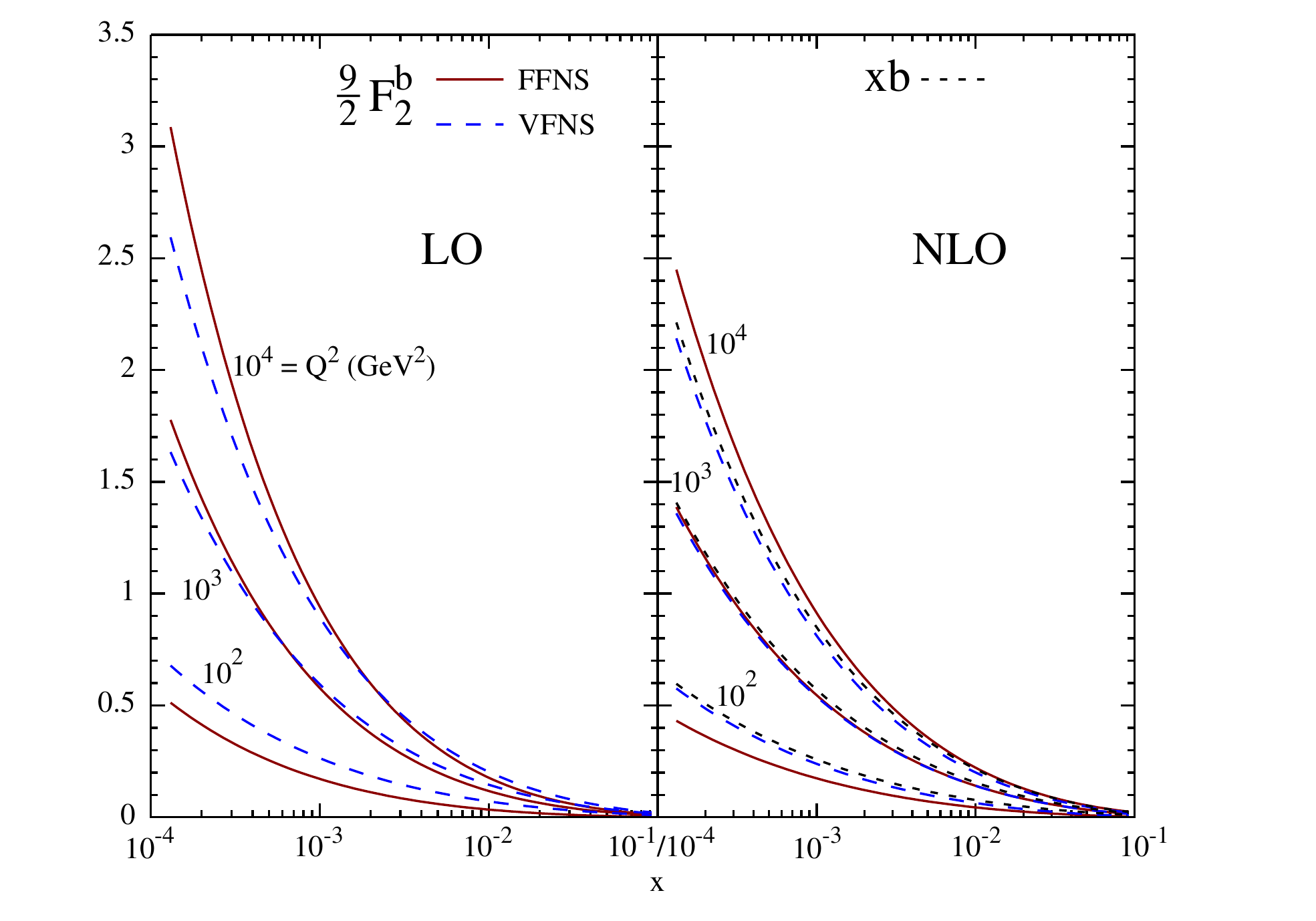}
\caption{As in Fig.\,\ref{Fig229} but for bottom production, i.e.$\frac{9}{2}F_2^b(x,Q^2)$, choosing $\mu^2\equiv\mu_F^2\!=\!\mu_R^2\!=\!Q^2+4m_b^2$ with $m_b\!= \!4.2$ GeV for the FFNS. The short--dashed curves show the NLO--VFNS bottom distribution $xb(x,Q^2)$.\label{Fig230}}
\end{figure}

\fontsize{11}{18.2}
\selectfont

In Figs.\,\ref{Fig229} and \ref{Fig230} we compare the VFNS ($\mu^2\!=\!Q^2$) with the FFNS predictions for $F_2^c(x,Q^2)$ and $F_2^b(x,Q^2)$ respectively, here we choose $\mu^2\!=\!Q^2+4m_h^2$ for the FFNS so that both scales become approximately equal in the limit $Q^2\!\gg\!m_h^2$; as mentioned the FFNS results are not very sensitive to this specific choice of the factorization and renormalization scale \cite{Gluck:1993dpa}. Notice that the NLO--VFNS predictions for $xh$ (short--dashed curves) are very similar to the ones for $(2e_h^2)^{-1}F_2^h$ (dashed curves) despite the fact that $(2e_h^2)^{-1}F_2^h\!=\!(1+\alpha_sC_q) \otimes h+\frac{1}{2}\alpha_s C_g\otimes g$, i.e.\ the ${\cal{O}}(\alpha_s)$ quark and gluon contributions almost cancel. As expected \cite{Gluck:1993dpa} the discrepancies between the predictions for $xh(x,Q^2)$ in the VFNS and for $(2e_h^2)^{-1}F_2^h(x,Q^2)$ in the FFNS in the relevant kinematic region (small $x$, large $Q^2$) never disappear and can amount to as much as about 30\% at very small--$x$, even at $W^2\!=\!Q^2(\frac{1}{x}-1)$ far above threshold, i.e.\ $W^2\gg \hat{s}_{\rm th}$. This is due to the fact that here the ratio of the threshold energy $\sqrt{\hat{s}_{\rm th}}$ of the massive subprocess ($\gamma^*g\to h\bar{h}$, etc.) and the mass of the produced heavy quark $\sqrt{\hat{s}_{\rm th}}/m_h=2$ is not sufficiently high to exclude significant contributions from the threshold region. Even for the lightest heavy quark, $h\!=\!c$, such non--relativistic $(\beta_c=|\vec{p}_c|/E_c \lesssim 0.9$) threshold region contributions to $F_2^h(x,Q^2)$ are sizeable for  $W^2 \lesssim 10^6$ GeV$^2$ due to significant $\beta_c<0.9$ contributions, and the situation becomes worse for $h\!=\!b$ \cite{Gluck:1993dpa}.

\fontsize{11}{17}
\selectfont

\begin{figure}[t]
\centering
\includegraphics[width=0.56\textwidth]{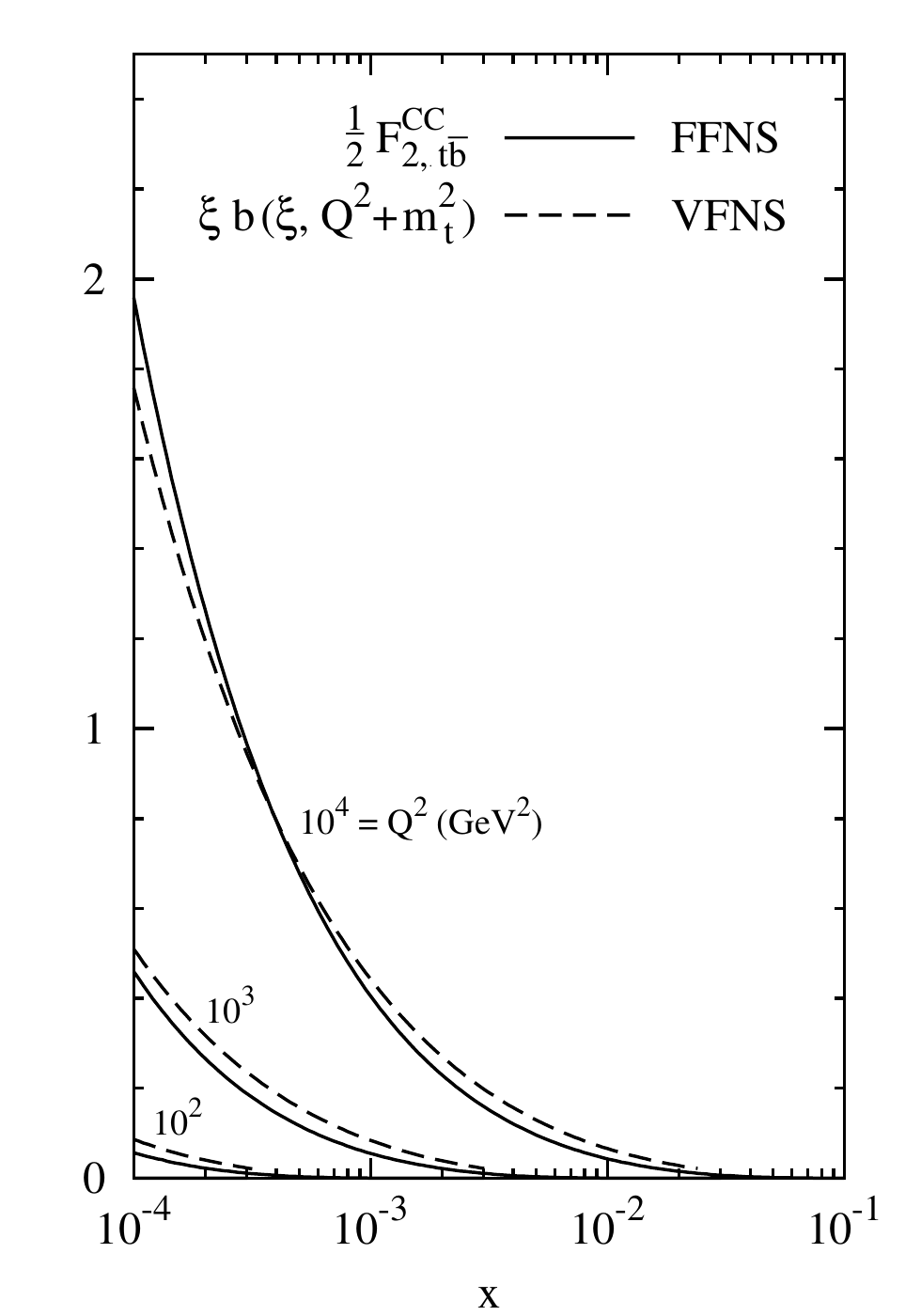}
\caption{LO predictions for the $x$--dependencies of the weak charged--current structure function $\frac{1}{2}F_{2,t\bar{b}}^{\rm CC}(x,Q^2)$ for $t\bar{b}$ production in the FFNS at some typical fixed values of $Q^2$; here the momentum scale is chosen to be\vspace{0.25em} $\mu^2\!\equiv\!\mu_R^2\!=\!\mu_F^2\!=\!Q^2 + (m_t+m_b)^2$ with $m_t\!=\!175$ GeV. These predictions are compared with the bottom distribution $\xi b(\xi,\, Q^2+m_t^2)$ in the VFNS where $\xi = x(1+\frac{m_t^2}{Q^2})$.\label{Fig231}}
\end{figure}

This is in contrast to processes where one of the produced particles is much heavier than the other one, like the weak CC contribution $W^+\,g\to t\,\bar{b}$ to $F_2^{\rm CC}$. Here $\sqrt{\hat{s}_{\rm th}}/m_b\!=\!\frac{m_t+m_b}{m_b}\gg 1$ and the extension of the threshold region where $\beta_{\bar{b}} \lesssim 0.9$, being proportional to $m_b/\sqrt{\hat{s}_{\rm th}}\ll 1$, is strongly reduced as compared to $\frac{m_h}{2m_h}\!=\!0.5$ in the former case of $h\bar{h}$ production. Thus the single top production rate in $W^+\,g\to t\,\bar{b}$ is dominated by the (beyond--threshold) relativistic region where $\beta_{\bar{b}}>0.9$ and therefore is expected to be well approximated by $W^+b\to t$ where $b$ is an effective massless parton within the nucleon. In Fig.\,\ref{Fig231} we compare the LO FFNS\footnote{Notice that the fully massive NLO FFNS QCD corrections to $W^+\,g\to t\,\bar{b}$ are unfortunately not available in the literature.} predictions for $\frac{1}{2}F_{2,t\bar{b}}^{\rm CC}(x,Q^2)$ with the corresponding VFNS ones for $\xi b(\xi,Q^2+m_t^2)$, where the latter refers to the $W^+b\to t$ transition using the slow rescaling variable \cite{Barnett:1976ak, Barbieri:1976rd}  $\xi=x(1+\frac{m_t^2}{Q^2})$ with $m_t\!=\!175$ GeV; for $F_{2,t\bar{b}}^{\rm CC}(x,Q^2)$ we used $\mu^2\equiv\mu_F^2=\mu_R^2\!=\!Q^2+(m_t+m_b)^2$. As expected the differences between the two schemes are here less pronounced than in the case of $c\bar{c}$ and $b\bar{b}$ electroproduction in Figs.\,\ref{Fig229} and \ref{Fig230}.

\fontsize{11}{17}
\selectfont

\section{Weak Gauge Boson and Higgs Production at Tevatron and LHC}\label{Sec.WandHiggs}

\begin{figure}[b!]
\centering
\includegraphics[width=0.9\textwidth]{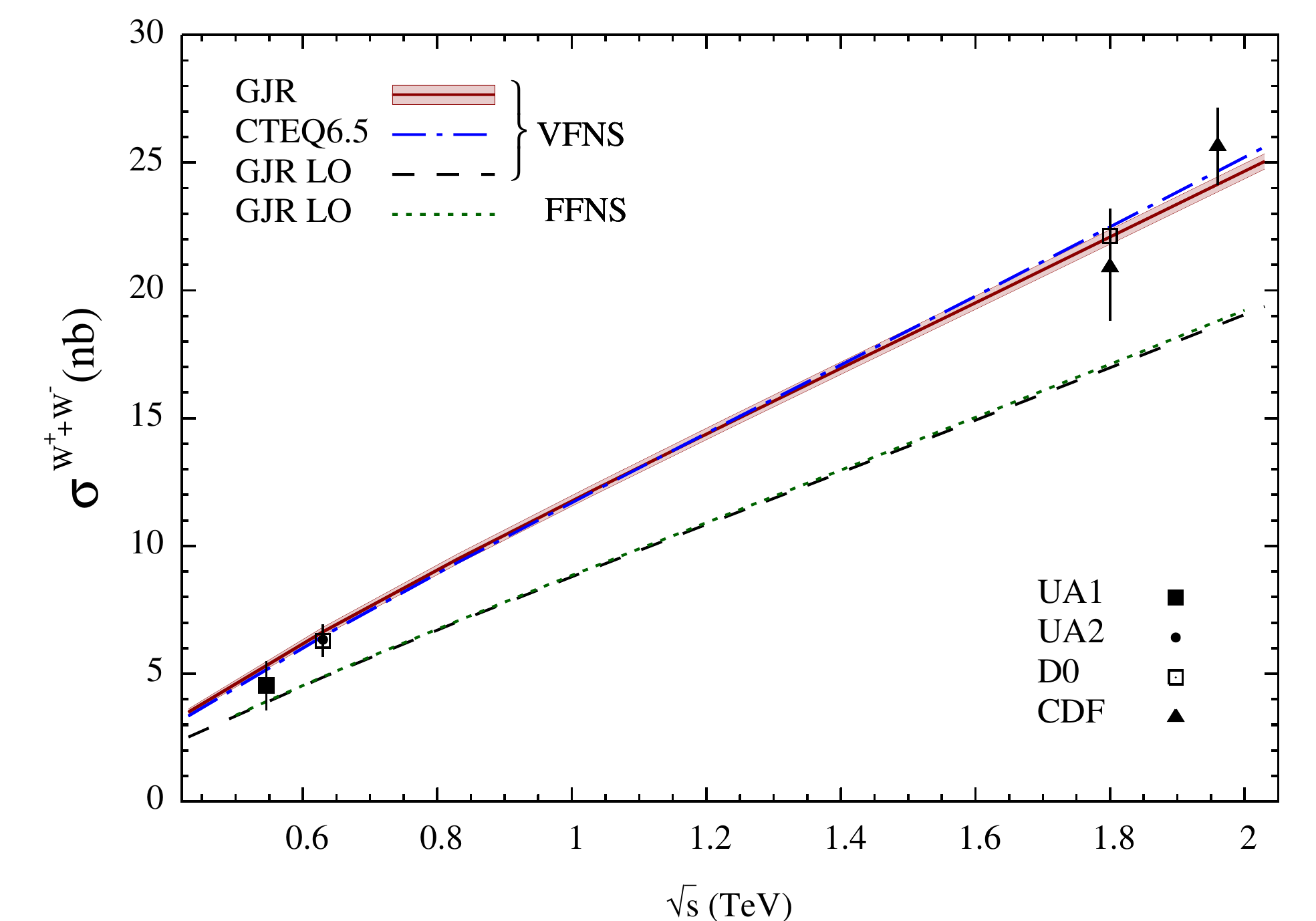}
\caption{Our (GJR) \cite{Gluck:2008gs} predictions for the total $W^+\!+W^-$ production rates at $p\bar{p}$ colliders. The data have been taken from \cite{Albajar:1988ka, Alitti:1991dm, Abbott:1999tt, Abe:1992ys, Abulencia:2005ix} and the NLO--VFNS CTEQ6.5 distributions from \cite{Tung:2006tb}. The adopted momentum scale is $\mu^2\equiv\mu_F^2\!=\!\mu_R^2\!=\!M_W^2$. The scale uncertainty of our NLO GJR predictions, due to varying $\mu$ according to $\frac{1}{2}M_W\leq\mu\leq 2M_W$, amounts to less than 2\% at $\sqrt{s}=1.96$ TeV, for example.
\label{Fig232}}
\end{figure}

The results of Sec.\,\ref{Sec.HeavyDIS} indicate that one may resort to the simpler VFNS with its massless $h(x,Q^2)$ distributions to estimate rather reliably the production rates of heavy quarks, gauge bosons, Higgs scalars, etc. at Tevatron and LHC energies, for processes in which the invariant mass of the produced system far exceeds the mass of the participating heavy quark flavor. As a next test of these VFNS distributions we turn now to hadronic $W^{\pm}$ production and present in Fig.\,\ref{Fig232} our \cite{Gluck:2008gs} LO/NLO FFNS/VFNS predictions for $\sigma(p\,\bar{p}\!\to\!W^{\pm}\,X)$ as compared to the data \cite{Albajar:1988ka, Alitti:1991dm, Abbott:1999tt, Abe:1992ys, Abulencia:2005ix} and to predictions based on the NLO CTEQ6.5 distributions \cite{Tung:2006tb}. Although quantitatively slightly different, the dominant light quark contributions in the FFNS ($u\,\bar{d}\!\to\! W^+$, $u\,\bar{s}\!\to\! W^+$, etc.) are due to the same subprocesses as in the VFNS, but the relevant heavy quark contributions have been calculated via $g\,\bar{s}(\bar{d})\!\to\!\bar{c}$ $W^+$, $g\,u\!\to\! b\,W^+$, etc. as compared to $c\,\bar{s}(\bar{d})\!\to\! W^+$, $\bar{b}\,u\!\to\! W^+$, etc. in the VFNS. Here we again expect the VFNS with its effective massless ``heavy'' quark distributions $h(x,Q^2)$ to be adequate, since non--relativistic contributions from the threshold region in the FFNS are suppressed due to $\sqrt{\hat{s}_{\rm th}}/m_{c,b}\simeq\frac{M_W}{m_{c,b}}\gg 1$. Taking into account that the NLO/LO $K$ factor is expected \cite{Campbell:2005bb} to be in the range of 1.2 -- 1.3, our LO--FFNS predictions in Fig.\,\ref{Fig232} imply equally agreeable NLO predictions as the (massless quark) NLO-VFNS ones shown by the solid and dashed--dotted curves in Fig.\,\ref{Fig232}.

\fontsize{11}{15.2}
\selectfont

\begin{table}[t!]
\centering
\fontsize{7.6}{9}
\selectfont
\renewcommand{\arraystretch}{1.2}
\begin{tabular}{|c|c|c|c||c|c|c|c|}
\multicolumn{8}{|c|}{$\sigma^{\textrm{pp}\rightarrow\textrm{W}X}$ (nb),\textrm{ } $\sqrt{\textrm{s}}$=14 TeV}\\\hline
\multicolumn{4}{|c||}{VFNS: NLO (LO)}&\multicolumn{4}{|c|}{FFNS: NLO (LO)}\\
               & $\textrm{W}^+$ & $\textrm{W}^-$ & $\textrm{W}^+ + \textrm{W}^-$ &
$\textrm{W}^+$ & $\textrm{W}^-$ & $\textrm{W}^+ + \textrm{W}^-$                  & \\\hline
ud  &  87.7 (77.6) & 60.6 (52.6)  & 148.3 (130.3) &  93.3 (81.9) & 64.6 (55.7) & 157.9 (137.6) &     ud            \\
us  &    3.9 (3.3) &  1.3 (1.2)  &   5.3 (4.5)   &  4.2 (3.5) & 1.5 (1.3) & 5.7 (4.8) &     us            \\
ub  &    7.3 (7.0)$\times10^{-4}$ &  2.3 (2.2)$\times10^{-4}$  &   9.6 (9.2)$\times10^{-4}$   &
\mbox{ -~ } (6.5)$\times10^{-4}$ & \mbox{ -~ } (1.8)$\times10^{-4}$ & \mbox{ -~ } (8.3)$\times10^{-4}$ & gu$\rightarrow$bW \\
cd  &    1.3 (1.1) &  2.3 (2.0)  &   3.6 (3.1)   & \mbox{ -~~ } (1.0) & \mbox{ -~~ } (2.0) & \mbox{ -~~ } (3.0) & gd$\rightarrow$cW\\
cs  &  14.7 (12.2) & 14.7 (12.2) &  29.4 (24.3)  & \mbox{ -~~  } (10.6) & \mbox{ -~~ } (10.6) & \mbox{ -~~ } (21.3) & gs$\rightarrow$cW \\
cb  &   1.5 (1.4)$\times10^{-2}$ &  1.5 (1.4)$\times10^{-2}$  &   2.9 (2.7)$\times10^{-2}$   & - & - & - & - \\\hline
total & 107.5 (94.2) & 79.1 (67.9) & 186.5 (162.1) & $\simeq$ 111.4 (97.0) & $\simeq$ 81.2 (69.6) & $\simeq$ 192.7 (166.7) & total    \\\hline
\end{tabular}
\caption{NLO\,(LO) VFNS and FFNS predictions for $W^{\pm}$ production at LHC. The uncertainties implied by different scale choices are summarized in
Eqs.\,\ref{LHC_NLO} and \ref{LHC_LO}. The total NLO--FFNS rates are obtained by adopting an expected \cite{Campbell:2005bb} $K$--factor of 1.2 for the subleading gluon initiated LO rates involving the heavy $c$ and $b$ quarks.\label{Tab201}}
\end{table}

In Table \ref{Tab201} we present our VFNS and FFNS predictions for $W^{\pm}$ production at LHC and compare the relevant subprocess contributions to $\sigma(pp\!\to\! W^{\pm} X)$ at $\sqrt{s}\!=\!14$ TeV. The light quark fusion contributions in the $ud$ and $us$ sector turn out to be somewhat larger in the FFNS than in the VFNS which is due to the fact that the $u,d,s$ (and gluon) distributions are evolved for fixed $n_F\!=\!3$ in the FFNS. More interesting, however, are the subprocesses involving heavy quarks. Here the LO--VFNS is capable to reproduce, to within less than 15\%, the LO--FFNS predictions based on the gluon--induced fixed--order subprocesses $g\,u\!\to\!b\,W$, $g\,d\!\to\!c\,W$ and in particular on the sizeable CKM non--suppressed $g\,s\!\to\!c\,W$ contribution. As mentioned above, the NLO corrections to these latter heavy quark contributions cannot be calculated for the time being.  However, since these contributions amount to about only 15\% of the total FFNS results for $W^{\pm}$ production (being dominated by the light $ud$ and $us$ fusions in Table \ref{Tab201}), we can {\em safely} employ the expected \cite{Campbell:2005bb} $K$ factor of $K\!\simeq\! 1.2$ for the relevant $g\,d\!\to\!c\,W$ and $g\,s\!\to\!c\,W$  LO contributions in Table \ref{Tab201} for obtaining the total NLO--FFNS predictions without committing any significant error. The resulting total rate for $W^++W^-$ production at LHC of $192.7\!\pm\!4.7$ nb is comparable to our NLO--VFNS prediction in Table \ref{Tab201} of $186.5\pm 4.9$ nb where we have added the $\pm 1\,\sigma$ uncertainties implied by our dynamical parton distributions. Using our ``standard'' FFNS parton distributions instead of the dynamical ones for generating the VFNS distributions, the dynamical NLO--VFNS prediction slightly increases to 190.7 nb. This latter prediction reduces to 181.0 nb when using the smaller scale $\mu^2\!=\!\frac{1}{4}M_W^2$. The scale uncertainties of our predictions are defined by taking $\frac{1}{2}M_W\leq \mu\leq 2\,M_W$, using $M_W=80.4$ GeV, which gives rise to the upper limits ($\mu=2M_W$) and lower limits $(\mu=\frac{1}{2}M_W)$ of our predicted cross--sections.  In this way we obtain the following total uncertainty estimates of our NLO predictions at LHC:

\begin{equation}\label{LHC_NLO}
\sigma(pp\to W^+ +W^-+X) = \Bigg\{
\renewcommand{\arraystretch}{1.5}
 \begin{array}{c} 
  186.5 \pm 4.9_{\rm pdf}\,\,_{-5.5}^{+4.8}
            \mid_{\rm scale} {\rm nb}\quad ({\rm VFNS}) \\
  192.7 \pm 4.7_{\rm pdf}\,\,_{-4.8}^{+3.8}
            \mid_{\rm scale} {\rm nb}\quad ({\rm FFNS}) \end{array}
\end{equation}
and, for completeness, at LO
\begin{equation}\label{LHC_LO}
\sigma(pp\to W^+ +W^-+X) = \Bigg\{
\renewcommand{\arraystretch}{1.5}
 \begin{array}{c} 
  162.1 \pm 3.9_{\rm pdf}\,\,_{-21.8}^{+20.3}
            \mid_{\rm scale} {\rm nb}\quad ({\rm VFNS}) \\
  166.7 \pm 4.0_{\rm pdf}\,\,_{-19.0}^{+17.3}
            \mid_{\rm scale} {\rm nb}\quad ({\rm FFNS}) \end{array}
\end{equation}
where the subscript ``pdf'' refers to the $1\,\sigma$ uncertainties of our parton distribution functions. For comparison, the NLO--VFNS prediction of CTEQ6.5 \cite{Tung:2006tb} is 202 nb with an uncertainty of 8\%, taking into account a pdf uncertainty of slightly more than $2\sigma$. Similarly, MRST \cite{Martin:2001es, Martin:2002dr} predict about 194 nb. From these results we conclude that, for the time being, the total $W^{\pm}$ production rate at LHC can be safely predicted within an uncertainty of about 10\% irrespective of the factorization scheme.

\fontsize{11}{18}
\selectfont

\begin{figure}[p]
\centering
\includegraphics[width=0.83\textwidth]{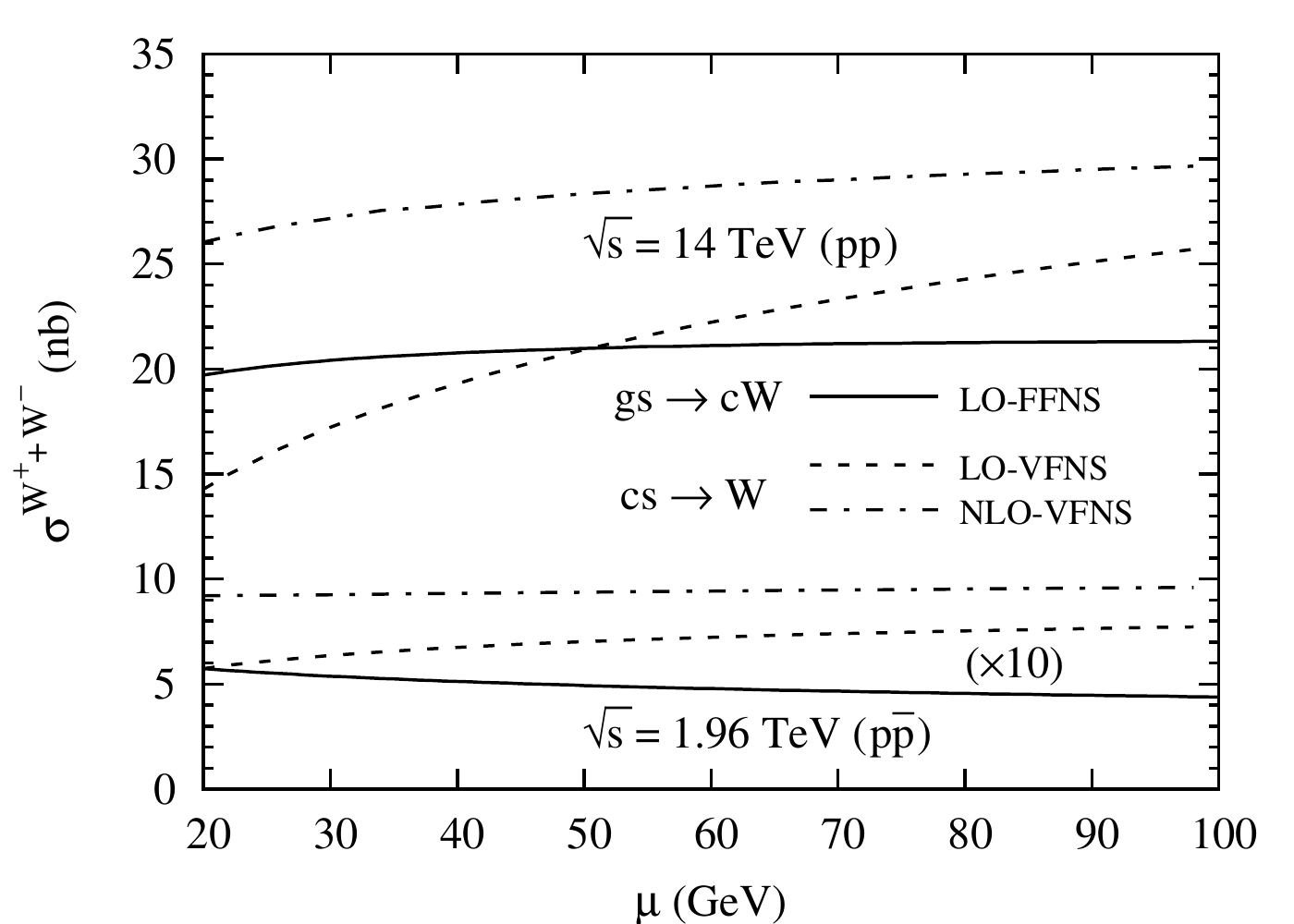}
\caption{The scale ($\mu^2\equiv\mu_F^2\!=\!\mu_R^2$) dependence of the LO--FFNS contribution to the total $W^+ +W^-$ production rates due to the subprocesses $g\,s\!\to\! c\,W$ and $g\,d\!\to\! c\,W$ compared to the LO and NLO ones in the VFNS due to $c\,s\!\to\! W$ and $g\,d\!\to\! c\,W$ fusion respectively. The results refer to the $pp$--LHC ($\sqrt{s}\!=\!14$ TeV) and to the $p\bar{p}$--Tevatron ($\sqrt{s}\!=\!1.96$ TeV) with the latter ones being multiplied by the indicated factor of 10 as indicated.\label{Fig233}}
\includegraphics[width=0.83\textwidth]{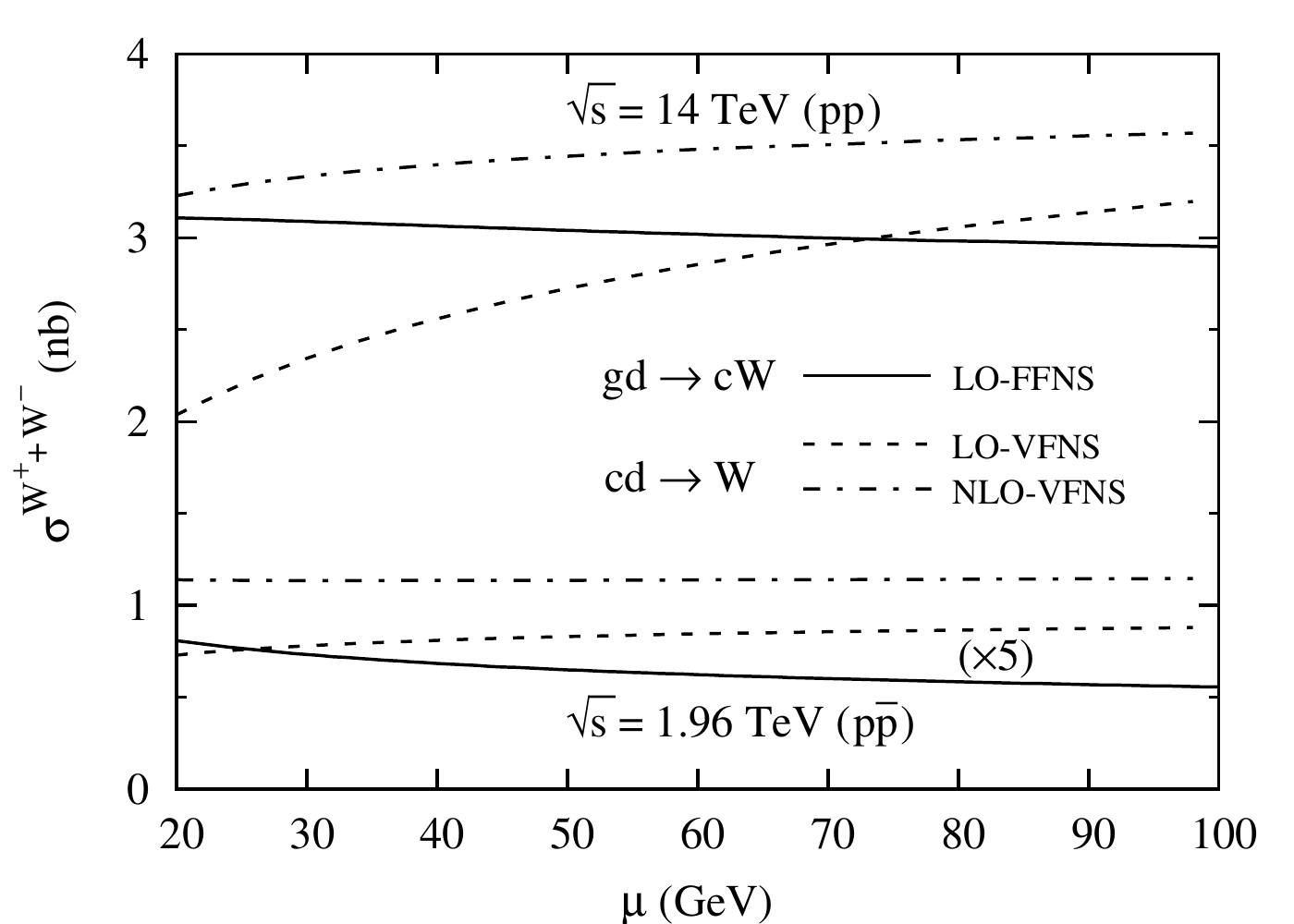}
\caption{As in Fig.\,\ref{Fig233} but for the FFNS subprocess $g\,d\!\to\! c\,W$ to be compared with $c\,d\!\to\! W$ in the VFNS. The results for the Tevatron ($\sqrt{s}\!=\!1.96$ TeV) are multiplied by a factor of 5 as indicated.\label{Fig234}}
\end{figure}

It is also interesting to study the dependence of the FFNS predictions  for the contributions to $W^{\pm}$ production involving heavy quarks on the chosen scale $\mu$ as shown in Figs.\,\ref{Fig233} and \ref{Fig234}. In these figures we compare the $g\,s\!\to\!c\,W$ initiated production rates in the FFNS with the quark fusion $c\,s\!\to\!W$ ones in the VFNS and similarly the $g\,d\!\to\!c\,W$ ones with the $c\,d\!\to\!W$ fusion respectively. These factorization scheme dependencies are rather mild for the LO--FFNS predictions, in contrast to the situation for the LO--VFNS predictions which stabilize, as expected, at NLO. The mild $\mu$ dependence is similar to the situation encountered in $tW$ production \cite{Campbell:2005bb} via the subprocess $g\,b\!\to\!t\,W^-$.

\begin{figure}[t]
\centering
\includegraphics[width=0.8\textwidth]{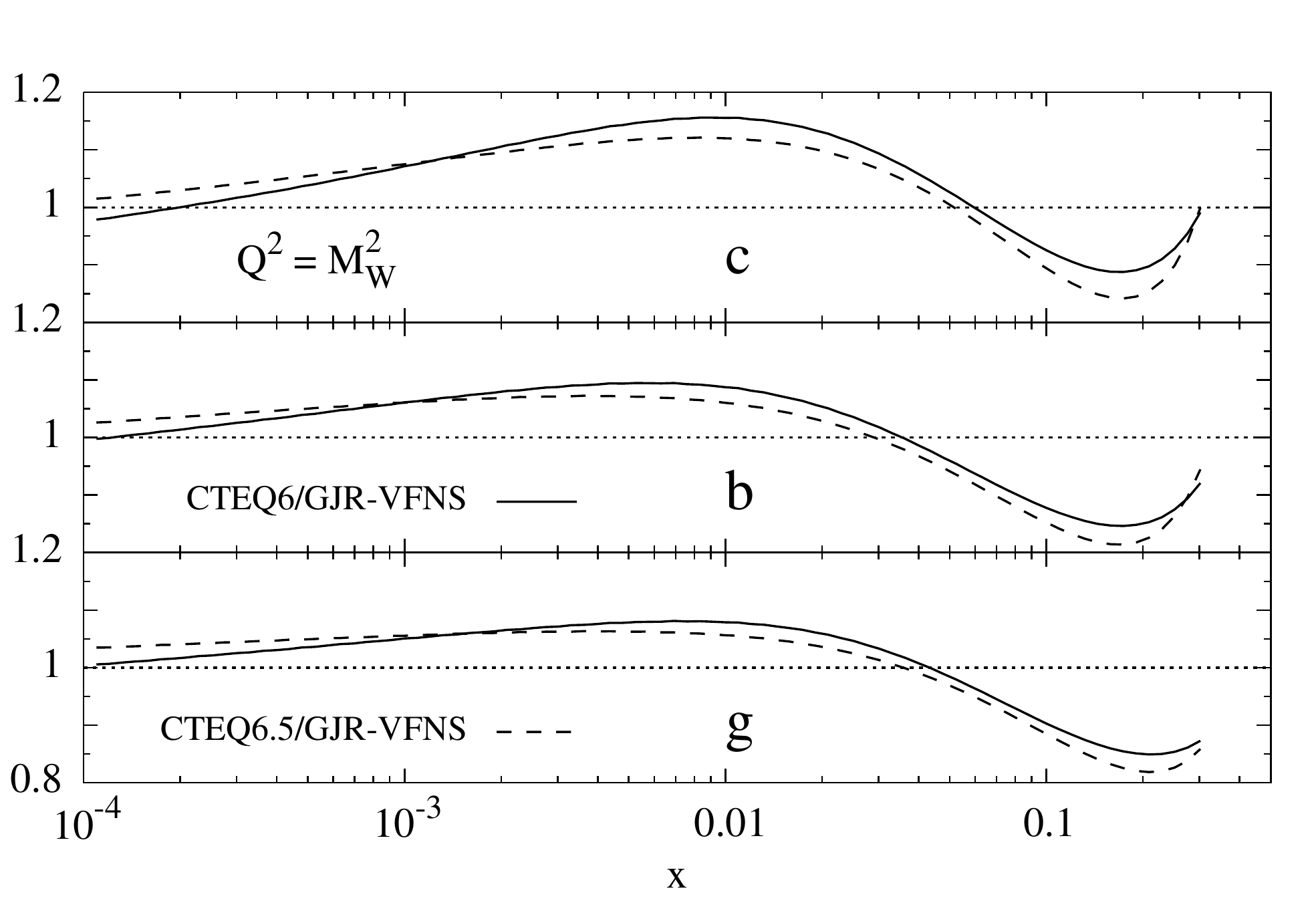}
\caption{Comparing our present (GJR-VFNS) dynamical parton distributions generated in the VFNS at NLO\,($\overline{\rm MS}$) with the ones of CTEQ6 \cite{Pumplin:2002vw} and CTEQ6.5 \cite{Tung:2006tb} at $Q^2=M_W^2$.\label{Fig235}}
\end{figure}

A similar situation, where the invariant mass of the produced system sizeably exceeds the mass of the participating heavy quarks, is encountered in (heavy) Higgs $H$ production accompanied by two heavy $b$--quarks, for example. Here $H\!=\!H_{\rm SM}^0;\,\, h^0,\, H^0,\, A^0$ denotes the  SM Higgs--boson or a light scalar $h^0$, a heavy scalar $H^0$ and a  pseudoscalar $A^0$ of supersymmetric theories with $M_H\!\gtrsim\!100$ GeV. In the FFNS the dominant production mechanism starts with the LO subprocess  $g\,g\to b\,\bar{b}\,H$ $(q\,\bar{q}\to b\,\bar{b}\,H$ is marginal), to be compared with the $b\,\bar{b}$ fusion subprocess in the VFNS starting with $b\,\bar{b}\to H$ at LO. Again, $\sqrt{\hat{s}_{\rm th}}/m_b=\frac{2m_b+M_H}{m_b}\simeq \frac{M_H}{m_b}\gg 1$ in the FFNS which indicates that the simpler LO and NLO (NNLO) VFNS  $b\bar{b}$ fusion subprocesses may serve as a reasonable effective approximation.  Within the scale uncertainties it turns out that the  FFNS and VFNS predictions at NLO are compatible \cite{Dittmaier:2003ej, Dawson:2004sh, Campbell:2004pu, Dawson:2005vi}. This result holds for scale choices\footnote{Note that the {\em independent} variation of $\mu_F$ and $\mu_R$ considered in \cite{Dittmaier:2003ej,Dawson:2004sh,Campbell:2004pu,Dawson:2005vi} is, as mentioned before, not strictly compatible with the utilized parton distributions determined and evolved according to $\mu_R=\mu_F$.} $\mu_{R,F}=(\frac{1}{8}\,{\rm to}\,\frac{1}{2})\sqrt{\hat{s}_{\rm th}}$ with $\frac{1}{4}\sqrt{\hat{s}_{\rm th}}$ being considered as a suitable ``central'' choice in $b(x,\mu_F^2)$ for calculations based on the $b\bar{b}$ fusion  process in the VFNS. It should, however, be mentioned that the VFNS tends to somewhat overestimate \cite{Dittmaier:2003ej,Dawson:2004sh, Campbell:2004pu, Dawson:2005vi} the exact FFNS Higgs--boson production rates by about$^{10}$ 10 to 20\%.

Finally let us note that all our results and comparisons  concerning the VFNS hold irrespective of the specific parametrizations used for the ``heavy'' $h(x,Q^2)$ distributions. When comparing our VFNS distributions \cite{Gluck:2008gs} with the ones of CTEQ6.5 \cite{Tung:2006tb}, the relevant ratios vary for $10^{-4}$ $\lesssim x \lesssim 0.1$ at most between  0.9 -- 1.1 at LO and NLO; similar results hold for other VFNS distributions, e.g., those of \cite{Alekhin:2002fv}. This is illustrated more quantitatively in Fig.\,\ref{Fig235}, where we compare our $c$, $b$ and gluon distributions with the ones of CTEQ6 \cite{Pumplin:2002vw} and CTEQ6.5 \cite{Tung:2006tb} in the relevant region of $x\lesssim 0.3$ at $Q^2\!=\!M_W^2$. The ratios for the light $u$-- and $d$--distributions are even closer to 1 than the ones shown in Fig.\,\ref{Fig235}, typically between 0.95 and 1.05.

\fontsize{11}{17}
\selectfont

\section{Isospin Violations and the NuTeV Anomaly}\label{Sec.NuTeV}

\begin{figure}[t]
\centering
\includegraphics[width=0.65\textwidth]{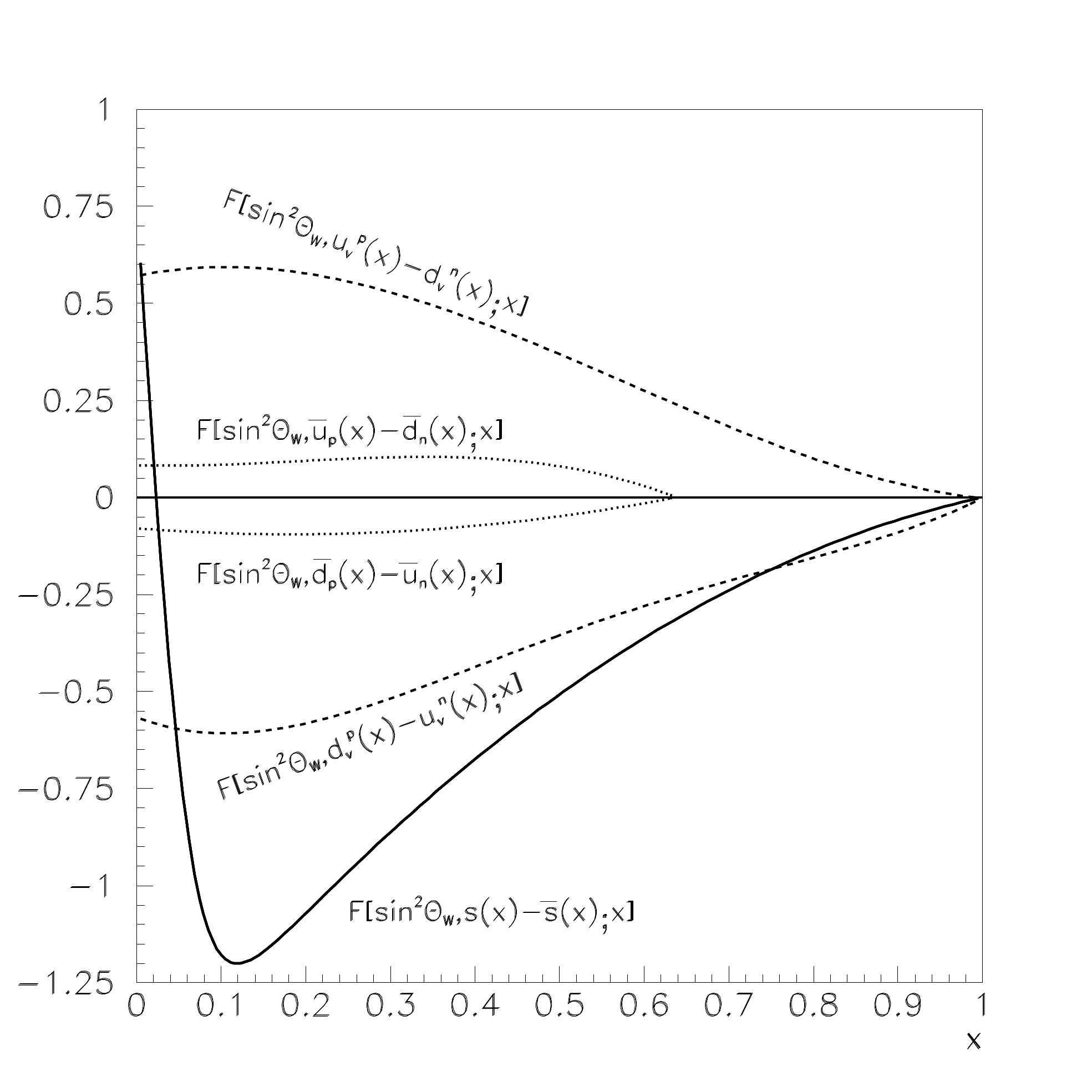}
\caption{The functionals describing the shift in the NuTeV $s_W^2$ measurement caused by isospin and strange sea asymmetries in the nucleon as presented in \cite{Zeller:2002du}.\label{Fig236}}
\end{figure}

The NuTeV collaboration has reported \cite{Zeller:2001hh} a measurement of the Weinberg angle $s_W^2\!\equiv\! \sin^2\!\theta_W$ which is approximately three standard deviations above the world average of other electroweak measurements (``anomaly''), i.e. $\sin^2\theta_W\!=\!0.2277\pm 0.0013\pm 0.0009$ as compared to the world average of $\sin^2\theta_W \!=\! 0.2228 \pm 0.0004$. Although the NuTeV group has included several uncertainties in their original analysis, e.g. effects due to a non--isoscalar target, higher--twists, charm production, etc.; several sources, spanning from \emph{new} physics to nuclear effects, have been proposed to explain this discrepancy (see, for instance, \cite{Davidson:2001ji, Londergan:2003pq, Zeller:2002du, McFarland:2002sk}). Among them
isospin-symmetry violating contributions of the parton distributions in the nucleon and effects caused by the strange sea asymmetry $s\neq \bar{s}$ turn out to be relevant. According to the experimental methods used for the extraction of the Weinberg angle from measurements of
\begin{equation}
R^{\nu(\bar{\nu})}(x,Q^2)\equiv \frac{d^2\sigma_{\rm NC}^{\nu(\bar{\nu})N}(x,Q^2)}{d^2\sigma_{\rm CC}^{\nu(\bar{\nu})N}(x,Q^2)}\;,
\end{equation}
the NuTeV collaboration have presented \cite{Zeller:2002du} functionals $F[s_W^2,\delta q;x]$ (cf. Fig.\,\ref{Fig236}) which permit the evaluation of the shifts in the NuTeV result for $s_W^2$ caused by a \emph{momentum} distribution asymmetry $\delta q(x,Q^2)$ as:
\begin{equation}
\label{DsW2}
\Delta s_W^2 =  \int_0^1 F[s_W^2,\delta q;x]\,\delta q(x,Q^2)\, dx
\end{equation}
where the asymmetries are evaluated at $Q^2\!\simeq \!10$ GeV$^2$ appropriate for the NuTeV experiment.

\begin{figure}[p]
\centering
\includegraphics[width=0.65\textwidth]{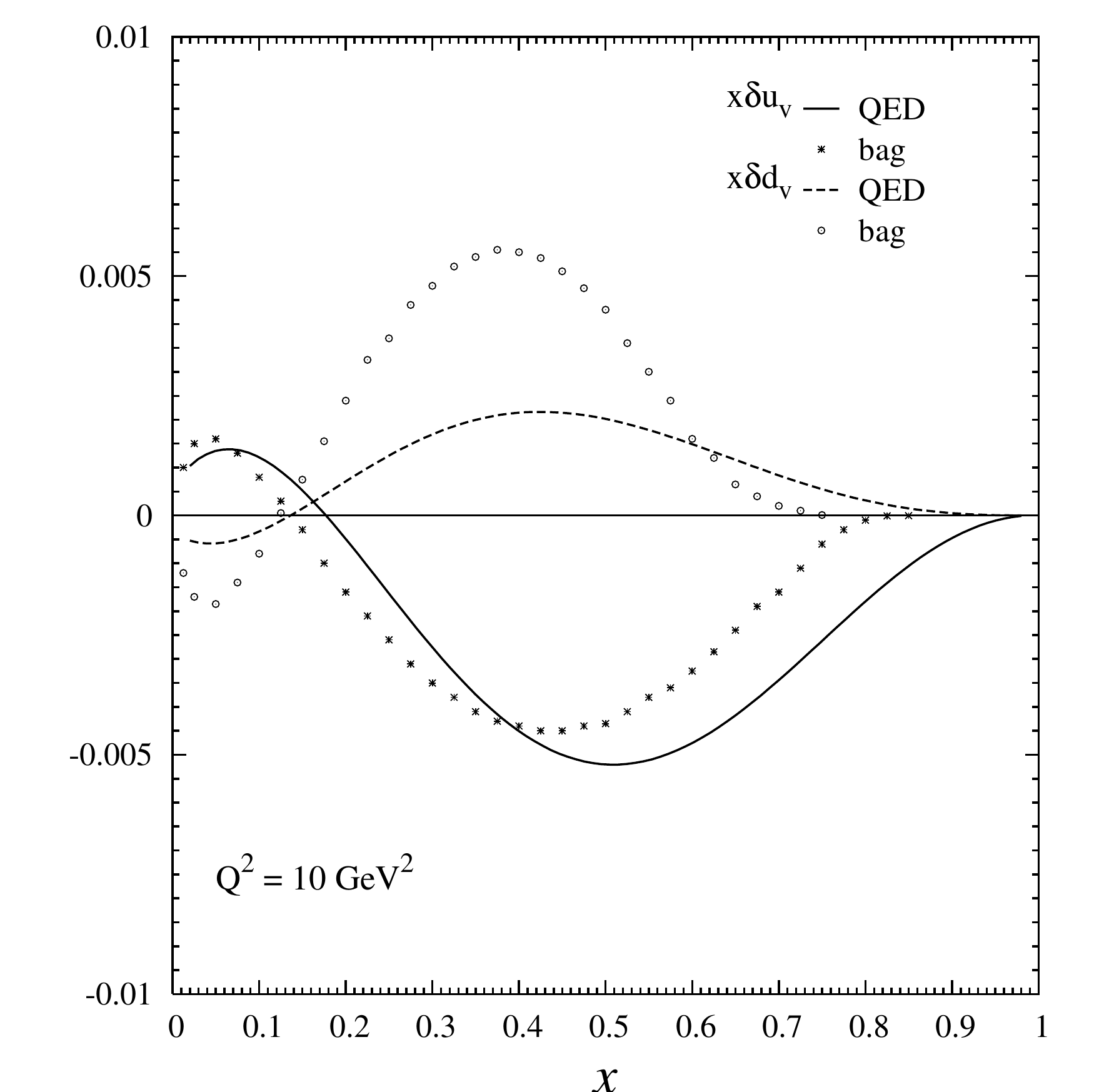}
\caption{The isospin violating ``majority'' $\delta u_v$ and ``minority'' $\delta d_v$ valence quark distributions evaluated at $Q^2\!=\!10$ GeV$^2$. Our radiative QED predictions are calculated according to Eq.\,\ref{isoviol}; the bag model estimates have been taken from \cite{Londergan:2003pq}.\label{Fig237}}
\includegraphics[width=0.65\textwidth]{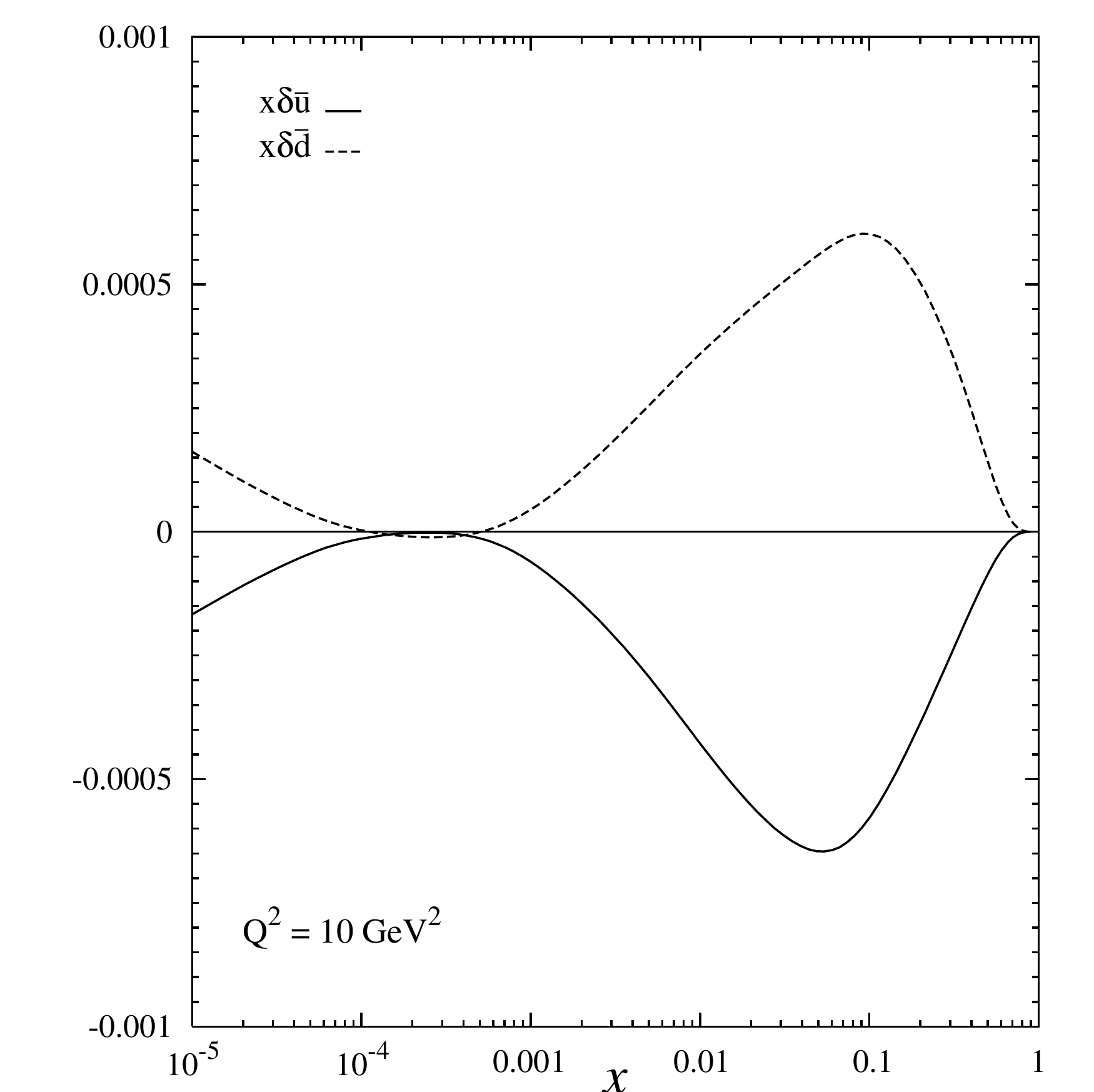}
\caption{The isospin violating sea distributions $\delta\bar{u}$ and $\delta\bar{d}$ at $Q^2\!=\!10$ GeV$^2$ calculated according to Eq.\,\ref{isoviol} with $u_v$, $d_v$ replaced by $\bar{u},\, \bar{d}$ respectively.\label{Fig238}}
\end{figure}

We \cite{Gluck:2005aq, Gluck:2005xh} shall approach the issue of isospin violations following Sec.\,\ref{Sec.Isospinviolations} within the framework of the dynamical (radiative) parton model and obtain predictions which depend on a single free parameter, required by the nonperturbative contributions and chosen to be $m_q\!=\!10$ MeV, i.e. of the order of the current quark masses, which are the usual kinematical lower bound for a photon emitted by a quark (similar to the electron mass $m_e$ for a photon radiated off an electron \cite{Frixione:1993yw}). We integrate Eq.\,\ref{QEDcorr} as:
\begin{eqnarray}\label{isoviol}
\delta u_v(x,Q^2) & = & \frac{\alpha}{4\pi}(e_u^2-e_d^2)
      \int_{m_q^2}^{Q^2}d\ln q^2 
       \int_x^1\frac{dy}{y}\,\, 
         P^{\gamma (0)}_{qq}\big(\tfrac{x}{y}\big) u_v(y,\, q^2)\nonumber\\
\delta d_v(x,Q^2) & = & -\frac{\alpha}{4\pi}(e_u^2-e_d^2)
      \int_{m_q^2}^{Q^2}d\ln q^2 
       \int_x^1\frac{dy}{y}\,\, 
         P^{\gamma (0)}_{qq}\big(\tfrac{x}{y}\big) d_v(y,\, q^2),
\end{eqnarray}
and similarly for $\delta\bar{u}$ and $\delta\bar{d}$. To evaluate this equation we use the LO parton distributions $q_v(x,q^2)$ and $\bar{q}(x,q^2)$ of the dynamical (radiative) parton model from \cite{Gluck:1998xa}\footnote{Note that our new dynamical distributions where not available at the time of making our original analysis \cite{Gluck:2005xh} of the NuTeV anomaly. Our results are however very similar to the GRV98 ones (cf. Sec.\,\ref{Sec.NewGeneration}).}. The parton distributions at  $q^2\!<\!\mu_{\rm LO}^2$, where $\mu_{\rm LO}^2\!=\!0.26$ GeV$^2$ is the input scale in \cite{Gluck:1998xa}, are taken to equal their values at the input scale $\mu_{\rm LO}^2$. The resulting valence and sea asymmetries at $Q^2\!=\!10$ GeV$^2$ are presented in Figs.\,\ref{Fig237} and \ref{Fig238} respectively. The valence asymmetries are compared with nonperturbative bag model results from \cite{Londergan:2003pq} which arise from an entirely different origin, namely through the mass differences $\delta m\!=\!m_d-m_u$ and $\delta M\!=\!M_n - M_p$. As can be seen, our radiative QED predictions and the bag model estimates are comparable for $\delta u_v$ but differ considerably for $\delta d_v$. Although our method differs somewhat from a similar evaluation \cite{Martin:2004dh} of the QED isospin violations, our resulting $\delta q_v(x,Q^2)$ turn out to be quite similar as already anticipated there \cite{Martin:2004dh}.

Turning now to the impact of our results on the NuTeV anomaly we present in Table \ref{Tab202} the corrections to $s_W^2$ evaluated according to Eq.\,\ref{DsW2}. For comparison, the QED valence isospin asymmetries of \cite{Martin:2004dh} imply, via Eq.\,\ref{DsW2}, contributions to $\Delta s_W^2$ similar to ours, namely - 0.00090 and - 0.00043 due to $\delta u_v$ and $\delta d_v$, respectively. Also shown in Table \ref{Tab202} are the additional contributions to $\Delta s_W^2$ stemming from the nonperturbative hadronic bag model calculations. In addition we use a nonperturbative estimate \cite{Alwall:2004rd} for the effects caused by the strange sea asymmetry. Both additional contributions are comparable in size to our radiative QED results; the total correction becomes:
\begin{eqnarray}
\Delta s_W^2|_{\rm total} & = & \Delta s_W^2|_{\rm QED} + 
     \Delta s_W^2|_{\rm bag} + \Delta s_W^2|_{\rm strange}\nonumber\\
& = & -0.0011\,\,\, - \,\,\,0.0015 \,\,\,-\,\,\,0.0017\;\; = \; -0.0043
\end{eqnarray}
Thus the NuTeV measurement  of $\sin^2\theta_W = 0.2277(16)$ would be shifted to $\sin^2\theta_W = 0.2234(16)$, which is in agreement with the standard value 0.2228(4).

\begin{table}[t!]
\centering
\renewcommand{\arraystretch}{1.5}
\begin{tabular}{l||ccccc}
$\Delta s_W^2$ & $\delta u_v$ & $\delta d_v$ & $\delta\bar{u}$ 
    & $\delta\bar{d}$ & total\\
\hline
QED & -0.00071 & -0.00033 & -0.000019 & -0.000023 & -0.0011\\
\hline
bag & -0.00065 & -0.00081 & --- & --- & -0.0015\\
\hline
\end{tabular}
\caption{The QED corrections to $\Delta s_W^2$ evaluated according to Eq.\,\ref{DsW2} using Eq.\,\ref{isoviol}. The nonperturbative bag model estimates \cite{Rodionov:1994cg} are taken from \cite{Londergan:2003pq}; different nonperturbative approaches give similar results \cite{Londergan:2003pq}.\label{Tab202}}
\end{table}

\fancyhead[CO]{}
\fancyhead[CE]{}

\fontsize{11}{15}
\selectfont

\chapter*{Summary and Conclusions}
\setcounter{chapter}{-1}
\setcounter{section}{0}
\addcontentsline{toc}{chapter}{Summary and Conclusions}
\chaptermark{Summary and Conclusions}
Utilizing recent NC DIS structure function measurements and data on hadronic Drell--Yan dilepton production and high--$p_{\rm T}$ inclusive jet production, we have updated the previous GRV98 \cite{Gluck:1998xa} global fits for the dynamical parton distributions of the nucleon at LO and NLO of perturbative QCD and extended our analyses with an appropriate uncertainty analysis for each of our fits. The former dynamical distributions are compatible with the new ones at the 1 to $2\sigma$ level of the newly estimated uncertainties. Furthermore, we have extended our analyses to the NNLO, where the high--$p_T$ jet data as well as the heavy quark electroproduction data have been disregarded for consistency reasons, i.e. since NNLO corrections have not yet been calculated.

The small--$x$ structure of the dynamical distributions is generated entirely radiatively from {\em valence--like}, {\em definite positive}, input distributions at an optimally chosen input scale $\mu_0^2\!<\!1$ GeV$^2$. Our NLO results are stable with respect to the choice of factorization scheme ($\overline{\rm MS}$ versus DIS) and the NNLO predictions are perturbatively stable with respect to the NLO ones and are in agreement with all present measurements for $Q^2 \!\gtrsim\! 2$ GeV$^2$. In general, the NNLO corrections imply an improved value of $\chi^2$, typically $\chi_{\rm NNLO}^2\simeq 0.9\,\chi_{\rm NLO}^2$. However, even though the dynamical NNLO uncertainties are somewhat smaller than the NLO ones, it turned out that present DIS precision data are still not sufficiently accurate to distinguish between NLO results and the minute NNLO effects of a few percent. The stability of these \emph{dynamical} results guarantees reliable calculations of cross--sections for, e.g., heavy quark, $W^{\pm}$, $Z^0$, and high--$p_{\rm T}$ jet production at hadron colliders like Tevatron and in particular LHC.

Our dynamical distributions have been compared with conventional ``standard'' ones obtained from non--valence--like positive gluon and sea input distributions at some arbitrarily chosen higher input scale $\mu_0^2\!>\!1$ GeV$^2$. For this purpose we have also performed common ``standard'' fits in exactly the same conditions as the dynamical ones but assuming $\mu_0^2\!=\!2$ GeV$^2$. Notice that, contrary to the dynamical approach, the finite small--$x$ behavior of the input gluon and sea distributions is here fitted, and not radiatively generated by QCD evolutions. For this reason as well as because of the sizably different evolution distances, i.e. sizably different $\mu_0^2$, the uncertainties of these less constrained ``standard'' distributions are larger than those of their dynamical counterparts, in particular in the small--$x$ region.

The more constrained ansatz of the dynamical model implies a certain predictive power, in particular in the present experimentally unexplored extremely small--$x$ region relevant for evaluating ultrahigh energy neutrino--nucleon cross--sections in astrophysical applications. Here we provide predictions down to $x\!\simeq\!10^{-9}$ at the weak scale $Q^2\!=\!M_W^2$ as required \cite{Gluck:1998js, Frichter:1994mx, Capelle:1998zz} for the highest cosmic neutrino energies of $10^{12}$ GeV. These predictions are strongly constrained within the dynamical parton model and are entirely of QCD-dynamical origin in the very small--$x$ region. Since previous predictions \cite{Gluck:1991ng,Gluck:1993im,Gluck:1989ze} for the small--$x$ region based on the dynamical parton model and the data available at the time were subsequently confirmed \cite{Abt:1993cb, Derrick:1993fta} at HERA, the presently available very precise small--$x$ data \cite{Adloff:1999ah, Adloff:2000qj, Adloff:2000qk, Adloff:2003uh, Chekanov:2001qu} utilized here allows us to be quite confident about the reliability of our improved small--$x$ predictions within the framework of the successful dynamical parton model.

Our predictions for the longitudinal structure $F_L$ and results for the ``reduced'' DIS cross--section $\sigma_r$ are in agreement with all HERA data, including the most recent H1 measurements, in particular in the small--$x$ region and down to $Q^2\!=\!2$ GeV$^2$. The dynamical NLO/NNLO predictions for $F_L$ are positive throughout the whole kinematic region considered and become perturbatively stable already at $Q^2\!=\!2-3$ GeV$^2$, where future precision measurements could even delineate NNLO effects in the very small--$x$ region, where the NNLO uncertainty bands are smaller than the NLO ones. This is in contrast to the common ``standard'' approach, however instabilities and differences are here less distinguishable due to the much larger uncertainties.

In order to analyze the role of heavy quark flavors in high--energy colliders, we have generated radiatively sets of (zero-mass) VFNS parton distributions based on our FFNS LO and NLO dynamical parton distributions. We have confronted these VFNS and FFNS predictions in situations where the invariant mass of the produced system ($h\bar{h},\, t\bar{b},\, cW,\, tW$, Higgs--bosons, etc.) does not exceed or exceeds by far the mass of the participating heavy flavor. In the former case the VFNS predictions deviate from the FFNS ones by up to about 30\% even at $Q^2\gg m_c^2$ while in the latter case these deviations are appreciably smaller, within about 10\%, which is within the margins of renormalization and factorization scale uncertainties. Our results indicate that the simpler VFNS with its effective treatment of heavy quarks ($c$, $b$) as massless partons can be employed for calculating processes where the invariant mass of the produced system is sizeably larger than the mass of the participating heavy quark flavor. Taking into account the uncertainties of parton distributions and those due to scale choices, the total $W^{\pm}$ production rate at LHC can be predicted within an uncertainty of about 10\%, which becomes significantly smaller at the Tevatron. Similarly the Higgs production rates at LHC are predicted with an uncertainty of 10 -- 20\% where the VFNS production rates exceed the FFNS ones by about 20\% at LHC.

Finally, we have evaluated the modifications to the standard isospin symmetric parton distributions due to QED ${\cal{O}}(\alpha)$ photon bremsstrahlung corrections within the dynamical model and compared them with nonperturbative bag model calculations, where the violation of isospin symmetry arises from entirely {\em{different}} (hadronic) sources, predominantly through quark and target mass differences. Taken together these two isospin violating effects and effects caused by a strangeness asymmetry ($s\!\neq\!\bar{s}$) in the nucleon the discrepancy between the large result for
$\sin^2\theta_W$ as derived from deep inelastic $\nu(\bar{\nu})\,N$ data (``NuTeV anomaly'') and the world average of other measurements is entirely removed.

\texttt{FORTRAN} codes (grids) containing our dynamical parton distribution functions along with the uncertainties of the most relevant sets can be obtained on request\footnote{\href{mailto:pjimenez@het.physik.uni-dortmund.de}{pjimenez@het.physik.uni-dortmund.de}} or directly downloaded from \url{http://doom.physik.uni-dortmund.de/pdfserver}; they are also included in the \emph{Les Houches Accord PDF Interface} \cite{Whalley:2005nh}. We hope that they will be of utility for the analysis and understanding of the (new) results which will be coming soon in High-Energy/Particle Physics.
\cleardoublepage
\fancyhead[CO]{}
\fancyhead[CE]{}
\fancyfoot[CE,CO]{\thepage}
\chapter*{A. Tables of Numerical Values}\label{numbers}
\setcounter{chapter}{4}
\setcounter{section}{0}
\setcounter{figure}{0}
\setcounter{table}{0}
\setcounter{equation}{0}
\setcounter{footnote}{0}
\chaptermark{Tables of Numerical Values}
\renewcommand{\thechapter}{A}
\addcontentsline{toc}{chapter}{A. Tables of Numerical Values}
We present here the numerical values obtained in our different fits and error analyses. Tab.\,\ref{TabA01} represent a summary of our results; there we include the central values obtained for the parameters in our fits, $\alpha_s(M_Z^2)$ and the parameters of the input distributions, and their errors. It also contains the minimum values of $\chi^2$ obtained for the complete fits as well as for different combinations of data which we find interesting to quote separately. We did not find that useful to include the $\chi^2$ obtained for all individual data sets; nevertheless the interested reader can obtain them on request\footnote{\href{mailto:pjimenez@het.physik.uni-dortmund.de}{pjimenez@het.physik.uni-dortmund.de}}. In Tab.\,\ref{TabA02} we quote the values obtained for the overall normalization factors for each individual data set in our different fits, as already explained (cf. Sec.\,\ref{Sec.DataFormalism}), this factors were allowed to float within the experimental normalization uncertainty of each data set.

Finally, in Tables \ref{TabA03} to \ref{TabA09} we give all the information necessary for uncertainty studies (cf. Sec.\,\ref{Sec.Uncertainties}). Each of these tables contains the eigenvalues of the hessian matrix and the eigenvector basis sets for one of our fits; the central values are included again for comparison. Recall that with these values it is possible to calculate the uncertainty estimated for any quantity as well as its derivatives at the central values of the parameters in the linear approximation. The reader interested may also reconstruct the coefficients $\hat{D}_k(p_i)$, which give information on the correlations of the different parameters, or even the hessian matrix and therefore the (approximated) $\chi^2$ profiles of our fits.
\clearpage
\fancyfoot[CE,CO]{}
\input{TabA01}

\input{TabA02}
\input{TabA03}
\input{TabA04}
\input{TabA05}
\input{TabA06}
\input{TabA07}
\input{TabA08}
\input{TabA09}
\cleardoublepage
\fancyfoot[CE,CO]{\thepage}

\fontsize{11}{15.8}
\selectfont

\chapter*{B. Splitting Functions}
\setcounter{chapter}{5}
\setcounter{section}{0}
\setcounter{figure}{0}
\setcounter{table}{0}
\setcounter{equation}{0}
\setcounter{footnote}{0}
\chaptermark{Splitting Functions}
\renewcommand{\thechapter}{B}
\addcontentsline{toc}{chapter}{B. Splitting Functions}
We present here the expressions for the splitting functions which have been implemented in our code for the solutions of the RGE (cf. Sec.\,\ref{Sec.RGEsolution}). Together with the well--known one-loop splitting functions \cite{Gross:1973ju, Georgi:1974sr}, and the two-loop ones \cite{Floratos:1977au, Floratos:1978ny, GonzalezArroyo:1979df, Floratos:1981hs}, the expressions for the 3--loop splitting functions \cite{Moch:2004pa, Vogt:2004mw} are necessary for the evolution of parton densities up to NNLO. For these latter quantities we use the $x$--space parametrizations presented also in \cite{Moch:2004pa, Vogt:2004mw}, and transform them to $n$--space using, e.g., \cite{Blumlein:1998if}. For the $n_F^2$ contribution in the non--singlet sector we use, however, the exact $n$--space expression as presented in \cite{Gracey:1993nn}.

These quantities can be expressed in terms of simple functions and harmonic sums \cite{GonzalezArroyo:1979df, Blumlein:1998if}; the functions that we use are defined in terms of integer $n$ values as \cite{GonzalezArroyo:1979df}:
\begin{equation}
\begin{array}{l}
\displaystyle S_m(n)=\sum_{i=1}^n\frac{1}{i^m},\qquad \tilde{S}(n)=\sum_{i=1}^n\frac{(-1)^i}{i^2}S_1(i)\\[1em]
\displaystyle S^\prime_m(\tfrac{n}{2})=S_m\Big(\!{\rm int}\big(\tfrac{n}{2}\!\big)\Big)=2^{m-1}\sum_{i=1}^n\frac{1+(-1)^i}{i^m}.
\end{array}
\end{equation}
Their analytic continuations to complex--$n$ space can be expressed in terms of logarithmic derivatives of Euler's $\Gamma$-function. For the first harmonic sums they are given by:
\begin{equation}
\begin{array}{l}
\displaystyle S_1(n)=\gamma_E + \psi^{(0)}(n+1), \quad \gamma_E=0.577216\ldots\;,\\[1em]
\displaystyle S_2(n)=\zeta(2) -\psi^{(1)}(n+1), \quad \zeta(2)=\tfrac{\pi^2}{6}, \\[1em]
\displaystyle S_3(n)=\zeta(3) -\tfrac{1}{2}\psi^{(2)}(n+1), \quad \zeta(3)=1.202057\ldots\;,
\end{array}
\end{equation}
where $\psi^{(l)}(n)\!=\!\frac{d^{(l+1)}\,\Gamma(n)}{d\,n^{(l+1)}}$. Since in principle $S^\prime_m(\frac{n}{2})$ and $\tilde{S}(n)$ are defined for even (odd) $n$, there are two analytic continuations, which we denote + (-); they are given by:
\begin{equation}
\begin{array}{l}
\displaystyle S^\prime_{m,+}(\tfrac{n}{2})=S_m(\tfrac{n}{2}), \qquad S^\prime_{m,-}(\tfrac{n}{2})=S_m(\tfrac{n-1}{2})\\
\displaystyle \tilde{S}_\pm(n)=-\tfrac{5}{8}\zeta(3) \pm \left( \tfrac{S_1(n)}{n^2} -\tfrac{\zeta(2)}{2}\big(\psi(\tfrac{n+1}{2})-\psi(\tfrac{n}{2})\big)
+ \int_0^1dx\,x^{n-1}\frac{{\rm Li}_2(x)}{1+x}\right).
\end{array}
\end{equation}
For the calculation of the above integral we use in practice the following parametrization \cite{Gluck:1989ze}:
\begin{equation}
\int_0^1dx\,x^{n-1}\frac{{\rm Li}_2(x)}{1+x} \simeq + \tfrac{1.01}{n+1} - \tfrac{0.846}{n+2} + \tfrac{1.155}{n+3} - \tfrac{1.074}{n+4} +\tfrac{0.55}{n+5×}
\end{equation}
Finally, we list the expressions used for the splitting functions in the remaining of the appendix\footnote{Note that the numerical values of the QCD color factors are explicitly given, in particular for the 3--loop functions they are included in the parametrizations.}:
\input{splittings}

\cleardoublepage
\phantomsection
\addcontentsline{toc}{chapter}{Bibliography}
\fontsize{11}{13.9}
\selectfont
\input{bibliography}
\fontsize{11}{16}
\selectfont
\chapter*{Acknowledgements}
First and foremost I would like to thank Prof.~Reya for choosing me as his student and giving me the opportunity of getting my Ph.D. at the  University of Dortmund. I am deeply indebted to both Prof.~Reya and Prof.~Gl\"uck for teaching me all the things that I have learned in these four years, both about physics and about the ``world'' of High--Energy/Particle Physics. I would like to extend these thanks to the people from the ``Vrije Universiteit Amsterdam'', P.~Mulders, D.~Boer and C.~Bomhof, who keep offering me their advice and support.

I am particularly grateful to some of my colleagues at the theoretical High--Energy Physics departments of the ``Technische Universit\"at Dortmund'', for helping me at some point and/or for advice and friendship in general, in approximately chronological order: C.~Pisano, A.~Mukherjee, A.~Kartavtsev, G.~Piranishvili, C.~Schuck, S.~Rakshit, D.~van Dyk, H.~Sedello, J.~Laamanen, S.~Schacht, O.~Micu \ldots\,. I would also like to extend these thanks to the rest of my colleagues, and in particular to our secretary, Mrs. Laurent, for her diligence.

Finally, I also want to thank the following people for letting me used their codes and/or for valuable discussions, again in approximately chronological order: C.~Pisano, J.~Smith, A.~Vogt, M.~Whalley and S.~Alekhin. In some cases it may have been part of their job, but still, I am thankful to them for making my life easier.

This work has been supported in part by the  ``Bundesministerium f\"ur Bildung und Forschung''.
\clearpage
\thispagestyle{empty}
\cleardoublepage
\end{document}

%% file: TabA01.tex
 \begin{sidewaystable}[p]
\footnotesize
\centering
\begin{tabular}{|c|c|c|c|c||c|c|c|}
\cline{2-8}
\multicolumn{1}{c|}{} &                                  \multicolumn{4}{c||}{dynamical}                      &                         \multicolumn{3}{|c|}{standard}          \rule{0em}{1.3em}\\
\cline{2-8}
\multicolumn{1}{c|}{} &         LO         &         NLO ({$\overline{\text{MS}}$})
                                                                  &         NLO (DIS)    &         NNLO ({$\overline{\text{MS}}$})
                                                                                                              &         LO          &
                                                                                                                    NLO ($\overline{\text{MS}}$)
                                                                                                                                                          &         NNLO ($\overline{\text{MS}}$)\\
\hline
$\chi^2_\text{total}$ &        1.276        &        1.038        &        1.070        &        0.986        &        1.134        &        0.991        &        0.947         \\
$\chi^2_\text{DIS}$   &        1.183        &        1.002        &        0.985        &        0.904        &        1.022        &        0.914        &        0.873         \\
$\chi^2_\text{DY}$    &        1.634        &        1.329        &        1.533        &        1.180        &        1.499        &        1.320        &        1.119         \\
$\chi^2_\text{HQ}$    &        0.811        &        0.930        &        0.959        &         ---         &        0.987        &        1.115        &         ---          \\
$\chi^2_\text{Jet}$   &        0.996        &        0.255        &        0.191        &         ---         &        0.857        &        0.372        &         ---          \\
\hline
$\alpha_s$            & 0.1263 $\pm$ 0.0015 & 0.1145 $\pm$ 0.0018 & 0.1135 $\pm$ 0.0019 & 0.1124 $\pm$ 0.0020 & 0.1339 $\pm$ 0.0030 & 0.1178 $\pm$ 0.0021 & 0.1158 $\pm$ 0.0035 \rule{0em}{1.0em}\\
\hline
$N_u$                 & 3.3434              & 1.2757              & 0.4341              & 4.4049              & 1.8815              & 0.5889              & 3.2350              \\
$a_u$                 & 0.6135 $\pm$ 0.0140 & 0.4960 $\pm$ 0.0184 & 0.3069 $\pm$ 0.0174 & 0.7875 $\pm$ 0.0307 & 0.4852 $\pm$ 0.0105 & 0.3444 $\pm$ 0.0099 & 0.6710 $\pm$ 0.0216 \\
$b_u$                 & 3.1866 $\pm$ 0.0760 & 3.4525 $\pm$ 0.0773 & 2.3124 $\pm$ 0.1602 & 3.6857 $\pm$ 0.0690 & 3.5621 $\pm$ 0.0675 & 3.7312 $\pm$ 0.0566 & 3.9293 $\pm$ 0.0575 \\
$A_u$                 &-3.3631              &-2.0704              & 0.8040              &-1.1483              &-2.2792              &-0.1740              &-0.5302              \\
$B_u$                 & 7.6775              & 13.225              & 12.163              & 4.5921              & 7.8837              & 17.997              & 3.9029              \\
\hline
$N_d$                 & 0.3016              & 0.7893              & 0.1766              & 13.824              & 0.3422              & 0.2585              & 13.058              \\
$a_d$                 & 0.3473 $\pm$ 0.0522 & 0.5165 $\pm$ 0.0699 & 0.2514 $\pm$ 1.4779 & 1.1778 $\pm$ 0.1068 & 0.3251 $\pm$ 0.0413 & 0.2951 $\pm$ 0.0410 & 1.0701 $\pm$ 0.0860 \\
$b_d$                 & 3.6160 $\pm$ 0.3205 & 4.6006 $\pm$ 0.4019 & 3.3688 $\pm$ 0.2456 & 5.6754 $\pm$ 0.4645 & 4.2769 $\pm$ 0.3223 & 4.8682 $\pm$ 0.3520 & 6.2177 $\pm$ 0.4459 \\
$A_d$                 &-0.7803              &-1.8488              &-0.4417              &-2.2415              &-0.6076              &-1.0552              &-2.5830              \\
$B_d$                 & 18.572              & 14.179              & 23.866              & 3.5917              & 15.429              & 26.536              & 3.8965              \\
\hline
$N_\Delta$            & 4.6430 $\pm$ 4.7776 & 4.0918 $\pm$ 4.3574 & 8.5986 $\pm$ 0.2744 & 8.6620 $\pm$ 8.8217 & 9.4877 $\pm$ 9.3066 & 7.2847 $\pm$ 7.3505 & 8.1558 $\pm$ 8.3289 \\
$a_\Delta$            & 1.4409 $\pm$ 0.3692 & 1.5483 $\pm$ 0.3778 & 1.4181 $\pm$ 0.2829 & 1.2963 $\pm$ 0.3490 & 1.1918 $\pm$ 0.3204 & 1.2773 $\pm$ 0.3329 & 1.1328 $\pm$ 0.3350 \\
$b_\Delta$            & 12.870 $\pm$ 2.7713 & 16.854 $\pm$ 3.2734 & 15.224 $\pm$ 0.0653 & 19.057 $\pm$ 3.2768 & 17.948 $\pm$ 3.1069 & 18.756 $\pm$ 3.2474 & 21.043 $\pm$ 3.5084 \\
$A_\Delta$            &-2.8689              &-2.7767              &-6.2906              &-6.8745              &-6.6444              &-6.3187              &-7.6334              \\
$B_\Delta$            & 9.3879              & 24.257              & 16.243              & 19.402              & 15.486              & 18.306              & 20.054              \\
\hline
$N_\Sigma$            & 0.6333 $\pm$ 0.0953 & 0.8627 $\pm$ 0.1922 & 0.9348 $\pm$ 0.3710 & 1.2316 $\pm$ 0.2749 & 0.2914 $\pm$ 0.0236 & 0.2295 $\pm$ 0.0182 & 0.4250 $\pm$ 0.0396 \\
$a_\Sigma$            & 0.0224 $\pm$ 0.0293 & 0.1450 $\pm$ 0.0607 & 0.1516 $\pm$ 0.0331 & 0.1374 $\pm$ 0.0501 &-0.1444 $\pm$ 0.0123 &-0.1573 $\pm$ 0.0120 &-0.1098 $\pm$ 0.0122 \\
$b_\Sigma$            & 8.0034 $\pm$ 0.6713 & 9.6252 $\pm$ 0.7050 & 6.5321 $\pm$ 0.3293 & 10.843 $\pm$ 0.9727 & 8.8290 $\pm$ 0.4900 & 8.8819 $\pm$ 0.5009 & 10.341 $\pm$ 0.6556 \\
$A_\Sigma$            &-2.0277              &-1.7699              &-1.2910              &-4.5634              &-1.6165              & 0.8704              &-3.0946              \\
$B_\Sigma$            & 6.5419              & 7.9169              & 1.6333              & 11.940              & 11.010              & 8.2179              & 11.613              \\
\hline
$N_g$                 & 19.592              & 3.1367              & 19.447              & 23.034              & 1.1043              & 1.3667              & 3.0076              \\
$a_g$                 & 1.3902 $\pm$ 0.6770 & 0.5168 $\pm$ 0.4017 & 0.9146 $\pm$ 0.0377 & 0.9940 $\pm$ 0.3794 &-0.2250 $\pm$ 0.0474 &-0.1050 $\pm$ 0.0767 & 0.0637 $\pm$ 0.1333 \\
$b_g$                 & 4.5219 $\pm$ 1.1414 & 2.7961 $\pm$ 0.7487 & 6.6235 $\pm$ 0.3218 & 6.7892 $\pm$ 2.6667 & 3.7574 $\pm$ 0.4961 & 3.3358 $\pm$ 0.3759 & 5.4473 $\pm$ 2.2254 \\
\hline
\end{tabular}
\caption{Minimum values of $\chi^2$ and the values of $\alpha_s(M_Z^2)$ together with the parameters of the input distributions obtained in our fits. $\chi^2_\text{total}$ is the total $\chi^2$ divided by the number of degrees of freedom ($N-21$) while $\chi^2_\text{DIS,\,DY,\,HQ,\,Jet}$ refer only to the the indicated measurements and are divided by the respective number of data points used ($N_\text{DIS,\,DY,\,HQ,\,Jet}\!=\!1178,390,61,110$). In particular ''DIS`` refers only to the HERA measurements of $\sigma_r$ and the measurements of $F_2$ in fixed target experiments; the measurements of heavy quark production in DIS are grouped under the subscript HQ. The input scale is $Q_0^2\!=\!0.3 ,\; 0.5, \; 0.55 \text{ GeV}^2$ for our LO, NLO, NNLO \emph{dynamical} fits, while it is  $Q_0^2\!=\!2 \text{ GeV}^2$ for their ''\emph{standard}`` counterparts. The $\pm 1 \sigma$ uncertainties of the parameters included in the error analysis are given as well.\label{TabA01}}
\end{sidewaystable}

%% file: TabA02.tex
\begin{table}[p]
\footnotesize
\centering
\begin{tabular}{|c|c|c||c|c|c|c||c|c|c|}
\cline{4-10}
\multicolumn{3}{c|}{} & \multicolumn{4}{c||}{dynamical} & \multicolumn{3}{c|}{standard} \rule{0em}{1.3em}\\
\cline{3-10}
\multicolumn{2}{c|}{} & $\Delta_{\rm exp}$
& \makebox[5em][c]{LO} & NLO ({$\overline{\text{MS}}$}) & NLO (DIS) & NNLO ({$\overline{\text{MS}}$})
& \makebox[5em][c]{LO} & NLO ($\overline{\text{MS}}$)               & NNLO ($\overline{\text{MS}}$)\\
\hline
\cite{Adloff:1999ah} &$\sigma_{r}^{\rm NC}$          &1.5 &  -1.50 &  -1.50 &  -1.50 &  -1.50 &   0.11 &  -1.48 &  -1.50  \\
\hline
\cite{Adloff:2000qj} &$\sigma_{r}^{\rm NC}$          &1.8 &  -1.75 &  -1.80 &  -1.80 &  -1.80 &   0.23 &  -1.31 &  -1.11  \\
\hline
\cite{Adloff:2000qk} &$\sigma_{r}^{\rm NC}$          &1.7 &   1.50 &   1.70 &   1.70 &   1.08 &   1.70 &   1.70 &   1.32  \\
\hline
\cite{Adloff:2003uh} &$\sigma_{r}^{\rm NC}$          &1.5 &  -1.50 &  -1.50 &  -1.50 &  -1.50 &   0.77 &  -0.80 &  -0.63  \\
\hline
\cite{Chekanov:2001qu} &$\sigma_{r}^{\rm NC}$          &2   &  -1.95 &  -1.73 &  -1.66 &  -2.00 &  -0.58 &  -1.36 &  -1.43  \\
\hline
\cite{Whitlow:1991uw} &$F_2^p$                        &2   &  -0.95 &  -0.75 &  -1.11 &  -0.79 &   0.89 &   0.23 &   0.13  \\
\hline
\end{tabular}
\caption{Overall normalization factors (in percent) multiplying the data as obtained in our different fits. Obviously, a quoted value of $\Delta_{\rm set}$ means that each data point of the set is to be multiplied by a factor $N_{\rm set}\!=\!1+0.01\Delta_{\rm set}$ in order to obtain the $\chi^2$ values quoted in Tab.~\ref{TabA01}.\label{TabA02}}
\end{table}

%% file: TabA03.tex
\begin{sidewaystable}[p]
\footnotesize
\centering
\begin{tabular}{|c||c|c|c|c|c|c|c|c|c|c|c|c|c|c|}
\hline
\,    k &
\raisebox{-0.5em}{ --- } \vspace{-0.3em} &
\,    1 &
\,    2 &
\,    3 &
\,    4 &
\,    5 &
\,    6 &
\,    7 &
\,    8 &
\,    9 &
\,   10 &
\,   11 &
\,   12 &
   13 \\
   -k &
      &
   -1 &
   -2 &
   -3 &
   -4 &
   -5 &
   -6 &
   -7 &
   -8 &
   -9 &
  -10 &
  -11 &
  -12 &
  -13 \\
\hline\hline
\raisebox{-0.5em}{ $a_u$} \vspace{-0.3em} & 
\raisebox{-0.5em}{  0.613509} \vspace{-0.3em} & 
  0.613502 &
  0.611564 &
  0.621897 &
  0.611956 &
  0.617888 &
  0.615121 &
  0.609651 &
  0.608492 &
  0.612221 &
  0.618900 &
  0.616491 &
  0.609380 &
  0.612140 \\
 & 
 & 
  0.613516 &
  0.615454 &
  0.605121 &
  0.615062 &
  0.609130 &
  0.611897 &
  0.617367 &
  0.618526 &
  0.614797 &
  0.608118 &
  0.610526 &
  0.617637 &
  0.614877 \\
\hline
\raisebox{-0.5em}{ $b_u$} \vspace{-0.3em} & 
\raisebox{-0.5em}{  3.186557} \vspace{-0.3em} & 
  3.186559 &
  3.187021 &
  3.185745 &
  3.183642 &
  3.203232 &
  3.229955 &
  3.162033 &
  3.138295 &
  3.193351 &
  3.170311 &
  3.168302 &
  3.184420 &
  3.181860 \\
 & 
 & 
  3.186555 &
  3.186093 &
  3.187368 &
  3.189472 &
  3.169882 &
  3.143159 &
  3.211080 &
  3.234819 &
  3.179763 &
  3.202802 &
  3.204812 &
  3.188693 &
  3.191254 \\
\hline
\raisebox{-0.5em}{ $a_d$} \vspace{-0.3em} & 
\raisebox{-0.5em}{  0.347256} \vspace{-0.3em} & 
  0.347255 &
  0.347258 &
  0.347272 &
  0.370050 &
  0.359667 &
  0.339854 &
  0.359382 &
  0.357875 &
  0.385852 &
  0.356530 &
  0.345044 &
  0.359405 &
  0.349127 \\
 & 
 & 
  0.347257 &
  0.347255 &
  0.347241 &
  0.324463 &
  0.334846 &
  0.354659 &
  0.335131 &
  0.336638 &
  0.308661 &
  0.337983 &
  0.349469 &
  0.335108 &
  0.345386 \\
\hline
\end{tabular}
\caption{Eigenvalues of the hessian matrix and parameters of the \emph{eigenvector sets} obtained in our dynamical LO fit; the central values are included as well for comparison.\label{TabA03}}
\end{sidewaystable}

%% file: TabA04.tex
\begin{sidewaystable}[p]
\footnotesize
\centering
\begin{tabular}{|c||c|c|c|c|c|c|c|c|c|c|c|c|c|c|}
\hline
\,    k &
\raisebox{-0.5em}{ --- } \vspace{-0.3em} &
\,    1 &
\,    2 &
\,    3 &
\,    4 &
\,    5 &
\,    6 &
\,    7 &
\,    8 &
\,    9 &
\,   10 &
\,   11 &
\,   12 &
   13 \\
   -k &
      &
   -1 &
   -2 &
   -3 &
   -4 &
   -5 &
   -6 &
   -7 &
   -8 &
   -9 &
  -10 &
  -11 &
  -12 &
  -13 \\
\hline\hline
\raisebox{-0.5em}{ $a_u$} \vspace{-0.3em} & 
\raisebox{-0.5em}{  0.496026} \vspace{-0.3em} & 
  0.496022 &
  0.499214 &
  0.503278 &
  0.498062 &
  0.502876 &
  0.492343 &
  0.496398 &
  0.502445 &
  0.492529 &
  0.507449 &
  0.497092 &
  0.491085 &
  0.494250 \\
 & 
 & 
  0.496030 &
  0.492838 &
  0.488775 &
  0.493990 &
  0.489177 &
  0.499710 &
  0.495654 &
  0.489608 &
  0.499523 &
  0.484603 &
  0.494961 &
  0.500968 &
  0.497803 \\
\hline
\raisebox{-0.5em}{ $b_u$} \vspace{-0.3em} & 
\raisebox{-0.5em}{  3.452497} \vspace{-0.3em} & 
  3.452498 &
  3.451817 &
  3.451394 &
  3.464875 &
  3.479555 &
  3.433107 &
  3.417764 &
  3.402781 &
  3.444825 &
  3.444864 &
  3.422363 &
  3.452773 &
  3.448937 \\
 & 
 & 
  3.452496 &
  3.453177 &
  3.453600 &
  3.440119 &
  3.425439 &
  3.471887 &
  3.487230 &
  3.502213 &
  3.460169 &
  3.460130 &
  3.482631 &
  3.452221 &
  3.456057 \\
\hline
\raisebox{-0.5em}{ $a_d$} \vspace{-0.3em} & 
\raisebox{-0.5em}{  0.516506} \vspace{-0.3em} & 
  0.516505 &
  0.516522 &
  0.516671 &
  0.546668 &
  0.504925 &
  0.529469 &
  0.521365 &
  0.534719 &
  0.570075 &
  0.527560 &
  0.510827 &
  0.533069 &
  0.521482 \\
 & 
 & 
  0.516506 &
  0.516490 &
  0.516341 &
  0.486344 &
  0.528086 &
  0.503543 &
  0.511647 &
  0.498293 &
  0.462937 &
  0.505451 &
  0.522185 &
  0.499943 &
  0.511530 \\
\hline
\end{tabular}
\caption{As in Tab.~\ref{TabA03} but for our dynamical NLO ($\overline{\rm MS}$) fit.\label{TabA04}}
\end{sidewaystable}

%% file: TabA05.tex
\begin{sidewaystable}[p]
\footnotesize
\centering
\begin{tabular}{|c||c|c|c|c|c|c|c|c|c|c|c|c|c|c|}
\hline
\,    k &
\raisebox{-0.5em}{ --- } \vspace{-0.3em} &
\,    1 &
\,    2 &
\,    3 &
\,    4 &
\,    5 &
\,    6 &
\,    7 &
\,    8 &
\,    9 &
\,   10 &
\,   11 &
\,   12 &
   13 \\
   -k &
      &
   -1 &
   -2 &
   -3 &
   -4 &
   -5 &
   -6 &
   -7 &
   -8 &
   -9 &
  -10 &
  -11 &
  -12 &
  -13 \\
\hline\hline
\raisebox{-0.5em}{ $a_u$} \vspace{-0.3em} & 
\raisebox{-0.5em}{  0.306867} \vspace{-0.3em} & 
  0.306860 &
  0.310103 &
  0.311130 &
  0.306209 &
  0.308928 &
  0.304746 &
  0.301103 &
  0.301468 &
  0.308682 &
  0.305389 &
  0.305187 &
  0.309550 &
  0.306309 \\
 & 
 & 
  0.306873 &
  0.303630 &
  0.302603 &
  0.307524 &
  0.304805 &
  0.308987 &
  0.312631 &
  0.312265 &
  0.305051 &
  0.308345 &
  0.308546 &
  0.304183 &
  0.307424 \\
\hline
\raisebox{-0.5em}{ $b_u$} \vspace{-0.3em} & 
\raisebox{-0.5em}{  2.312378} \vspace{-0.3em} & 
  2.312379 &
  2.311831 &
  2.312747 &
  2.315337 &
  2.332138 &
  2.305158 &
  2.241694 &
  2.259061 &
  2.331879 &
  2.224455 &
  2.323001 &
  2.347497 &
  2.313101 \\
 & 
 & 
  2.312376 &
  2.312924 &
  2.312008 &
  2.309418 &
  2.292617 &
  2.319597 &
  2.383061 &
  2.365694 &
  2.292877 &
  2.400301 &
  2.301754 &
  2.277258 &
  2.311654 \\
\hline
\raisebox{-0.5em}{ $a_d$} \vspace{-0.3em} & 
\raisebox{-0.5em}{  0.251431} \vspace{-0.3em} & 
  0.251430 &
  0.251405 &
  0.252146 &
  0.271020 &
  0.247062 &
  0.259514 &
  0.258067 &
  0.255641 &
  0.281805 &
  0.249529 &
  0.257780 &
  0.252656 &
  0.252166 \\
 & 
 & 
  0.251432 &
  0.251457 &
  0.250717 &
  0.231843 &
  0.255800 &
  0.243349 &
  0.244795 &
  0.247221 &
  0.221057 &
  0.253334 &
  0.245083 &
  0.250206 &
  0.250696 \\
\hline
\end{tabular}
\caption{As in Tab.~\ref{TabA03} but for our dynamical NLO (DIS) fit.\label{TabA05}}
\end{sidewaystable}

%% file: TabA06.tex
\begin{sidewaystable}[p]
\footnotesize
\centering
\begin{tabular}{|c||c|c|c|c|c|c|c|c|c|c|c|c|c|c|}
\hline
\,    k &
\raisebox{-0.5em}{ --- } \vspace{-0.3em} &
\,    1 &
\,    2 &
\,    3 &
\,    4 &
\,    5 &
\,    6 &
\,    7 &
\,    8 &
\,    9 &
\,   10 &
\,   11 &
\,   12 &
   13 \\
   -k &
      &
   -1 &
   -2 &
   -3 &
   -4 &
   -5 &
   -6 &
   -7 &
   -8 &
   -9 &
  -10 &
  -11 &
  -12 &
  -13 \\
\hline\hline
\raisebox{-0.5em}{ $a_u$} \vspace{-0.3em} & 
\raisebox{-0.5em}{  0.787468} \vspace{-0.3em} & 
  0.787465 &
  0.784777 &
  0.797611 &
  0.795914 &
  0.781447 &
  0.784423 &
  0.793819 &
  0.797631 &
  0.781535 &
  0.804210 &
  0.784518 &
  0.771879 &
  0.789575 \\
 & 
 & 
  0.787472 &
  0.790160 &
  0.777326 &
  0.779023 &
  0.793490 &
  0.790514 &
  0.781118 &
  0.777306 &
  0.793402 &
  0.770727 &
  0.790419 &
  0.803058 &
  0.785362 \\
\hline
\raisebox{-0.5em}{ $b_u$} \vspace{-0.3em} & 
\raisebox{-0.5em}{  3.685688} \vspace{-0.3em} & 
  3.685689 &
  3.686390 &
  3.683478 &
  3.718515 &
  3.670539 &
  3.666422 &
  3.652960 &
  3.649911 &
  3.673404 &
  3.686581 &
  3.676319 &
  3.706735 &
  3.679074 \\
 & 
 & 
  3.685687 &
  3.684986 &
  3.687898 &
  3.652860 &
  3.700837 &
  3.704953 &
  3.718416 &
  3.721465 &
  3.697972 &
  3.684795 &
  3.695057 &
  3.664641 &
  3.692302 \\
\hline
\raisebox{-0.5em}{ $a_d$} \vspace{-0.3em} & 
\raisebox{-0.5em}{  1.177823} \vspace{-0.3em} & 
  1.177822 &
  1.177790 &
  1.177904 &
  1.196838 &
  1.215055 &
  1.191256 &
  1.198524 &
  1.200643 &
  1.254898 &
  1.215521 &
  1.209642 &
  1.165585 &
  1.181551 \\
 & 
 & 
  1.177823 &
  1.177855 &
  1.177741 &
  1.158807 &
  1.140590 &
  1.164389 &
  1.157121 &
  1.155003 &
  1.100748 &
  1.140124 &
  1.146003 &
  1.190060 &
  1.174094 \\
\hline
\end{tabular}
\caption{As in Tab.~\ref{TabA03} but for our dynamical NNLO ($\overline{\rm MS}$) fit.\label{TabA06}}
\end{sidewaystable}

%% file: TabA07.tex
\begin{sidewaystable}[p]
\footnotesize
\centering
\begin{tabular}{|c||c|c|c|c|c|c|c|c|c|c|c|c|c|c|}
\hline
\,    k &
\raisebox{-0.5em}{ --- } \vspace{-0.3em} &
\,    1 &
\,    2 &
\,    3 &
\,    4 &
\,    5 &
\,    6 &
\,    7 &
\,    8 &
\,    9 &
\,   10 &
\,   11 &
\,   12 &
   13 \\
   -k &
      &
   -1 &
   -2 &
   -3 &
   -4 &
   -5 &
   -6 &
   -7 &
   -8 &
   -9 &
  -10 &
  -11 &
  -12 &
  -13 \\
\hline\hline
\raisebox{-0.5em}{ $a_u$} \vspace{-0.3em} & 
\raisebox{-0.5em}{  0.485240} \vspace{-0.3em} & 
  0.485150 &
  0.484787 &
  0.491357 &
  0.483906 &
  0.485505 &
  0.490707 &
  0.485338 &
  0.482954 &
  0.482081 &
  0.488477 &
  0.482270 &
  0.482949 &
  0.484030 \\
 & 
 & 
  0.485331 &
  0.485693 &
  0.479124 &
  0.486575 &
  0.484976 &
  0.479774 &
  0.485143 &
  0.487527 &
  0.488400 &
  0.482004 &
  0.488211 &
  0.487532 &
  0.486450 \\
\hline
\raisebox{-0.5em}{ $b_u$} \vspace{-0.3em} & 
\raisebox{-0.5em}{  3.562131} \vspace{-0.3em} & 
  3.562148 &
  3.562213 &
  3.561557 &
  3.563007 &
  3.565184 &
  3.577978 &
  3.600729 &
  3.538180 &
  3.535688 &
  3.599779 &
  3.551151 &
  3.561512 &
  3.562353 \\
 & 
 & 
  3.562114 &
  3.562049 &
  3.562706 &
  3.561255 &
  3.559079 &
  3.546284 &
  3.523533 &
  3.586082 &
  3.588574 &
  3.524484 &
  3.573111 &
  3.562751 &
  3.561909 \\
\hline
\raisebox{-0.5em}{ $a_d$} \vspace{-0.3em} & 
\raisebox{-0.5em}{  0.325093} \vspace{-0.3em} & 
  0.325081 &
  0.325088 &
  0.325096 &
  0.325602 &
  0.346250 &
  0.324753 &
  0.319837 &
  0.335290 &
  0.351953 &
  0.340123 &
  0.318459 &
  0.336759 &
  0.325878 \\
 & 
 & 
  0.325106 &
  0.325099 &
  0.325091 &
  0.324585 &
  0.303937 &
  0.325433 &
  0.330350 &
  0.314897 &
  0.298233 &
  0.310063 &
  0.331728 &
  0.313427 &
  0.324309 \\
\hline
\end{tabular}
\caption{As in Tab.~\ref{TabA03} but for our "standard" LO fit.\label{TabA07}}
\end{sidewaystable}

%% file: TabA08.tex
\begin{sidewaystable}[p]
\footnotesize
\centering
\begin{tabular}{|c||c|c|c|c|c|c|c|c|c|c|c|c|c|c|}
\hline
\,    k &
\raisebox{-0.5em}{ --- } \vspace{-0.3em} &
\,    1 &
\,    2 &
\,    3 &
\,    4 &
\,    5 &
\,    6 &
\,    7 &
\,    8 &
\,    9 &
\,   10 &
\,   11 &
\,   12 &
   13 \\
   -k &
      &
   -1 &
   -2 &
   -3 &
   -4 &
   -5 &
   -6 &
   -7 &
   -8 &
   -9 &
  -10 &
  -11 &
  -12 &
  -13 \\
\hline\hline
\raisebox{-0.5em}{ $a_u$} \vspace{-0.3em} & 
\raisebox{-0.5em}{  0.344398} \vspace{-0.3em} & 
  0.344357 &
  0.343869 &
  0.350081 &
  0.345654 &
  0.344269 &
  0.349966 &
  0.348086 &
  0.342753 &
  0.346743 &
  0.343755 &
  0.346281 &
  0.341964 &
  0.343070 \\
 & 
 & 
  0.344440 &
  0.344928 &
  0.338715 &
  0.343143 &
  0.344528 &
  0.338831 &
  0.340711 &
  0.346044 &
  0.342054 &
  0.345042 &
  0.342516 &
  0.346833 &
  0.345727 \\
\hline
\raisebox{-0.5em}{ $b_u$} \vspace{-0.3em} & 
\raisebox{-0.5em}{  3.731177} \vspace{-0.3em} & 
  3.731183 &
  3.731258 &
  3.730421 &
  3.731742 &
  3.733646 &
  3.762185 &
  3.711322 &
  3.713921 &
  3.755515 &
  3.718474 &
  3.759098 &
  3.728188 &
  3.730805 \\
 & 
 & 
  3.731170 &
  3.731095 &
  3.731933 &
  3.730612 &
  3.728707 &
  3.700168 &
  3.751032 &
  3.748432 &
  3.706838 &
  3.743879 &
  3.703255 &
  3.734165 &
  3.731548 \\
\hline
\raisebox{-0.5em}{ $a_d$} \vspace{-0.3em} & 
\raisebox{-0.5em}{  0.295074} \vspace{-0.3em} & 
  0.295068 &
  0.295051 &
  0.295079 &
  0.295296 &
  0.316132 &
  0.293811 &
  0.300743 &
  0.305102 &
  0.286040 &
  0.324254 &
  0.304417 &
  0.303864 &
  0.296393 \\
 & 
 & 
  0.295079 &
  0.295096 &
  0.295068 &
  0.294851 &
  0.274015 &
  0.296336 &
  0.289404 &
  0.285045 &
  0.304108 &
  0.265893 &
  0.285730 &
  0.286283 &
  0.293754 \\
\hline
\end{tabular}
\caption{As in Tab.~\ref{TabA03} but for our "standard" NLO ($\overline{\rm MS}$) fit.\label{TabA08}}
\end{sidewaystable}

%% file: TabA09.tex
\begin{sidewaystable}[p]
\footnotesize
\centering
\begin{tabular}{|c||c|c|c|c|c|c|c|c|c|c|c|c|c|c|}
\hline
\,    k &
\raisebox{-0.5em}{ --- } \vspace{-0.3em} &
\,    1 &
\,    2 &
\,    3 &
\,    4 &
\,    5 &
\,    6 &
\,    7 &
\,    8 &
\,    9 &
\,   10 &
\,   11 &
\,   12 &
   13 \\
   -k &
      &
   -1 &
   -2 &
   -3 &
   -4 &
   -5 &
   -6 &
   -7 &
   -8 &
   -9 &
  -10 &
  -11 &
  -12 &
  -13 \\
\hline\hline
\raisebox{-0.5em}{ $a_u$} \vspace{-0.3em} & 
\raisebox{-0.5em}{  0.671037} \vspace{-0.3em} & 
  0.671003 &
  0.670631 &
  0.679071 &
  0.671452 &
  0.673766 &
  0.679071 &
  0.668731 &
  0.683373 &
  0.670638 &
  0.670962 &
  0.661626 &
  0.662003 &
  0.671028 \\
 & 
 & 
  0.671070 &
  0.671443 &
  0.663002 &
  0.670622 &
  0.668307 &
  0.663002 &
  0.673342 &
  0.658700 &
  0.671435 &
  0.671112 &
  0.680448 &
  0.680070 &
  0.671046 \\
\hline
\raisebox{-0.5em}{ $b_u$} \vspace{-0.3em} & 
\raisebox{-0.5em}{  3.929296} \vspace{-0.3em} & 
  3.929303 &
  3.929383 &
  3.927740 &
  3.930061 &
  3.947894 &
  3.964171 &
  3.917265 &
  3.918270 &
  3.894149 &
  3.937717 &
  3.941855 &
  3.927778 &
  3.925812 \\
 & 
 & 
  3.929289 &
  3.929209 &
  3.930853 &
  3.928531 &
  3.910698 &
  3.894421 &
  3.941327 &
  3.940322 &
  3.964443 &
  3.920875 &
  3.916737 &
  3.930814 &
  3.932780 \\
\hline
\raisebox{-0.5em}{ $a_d$} \vspace{-0.3em} & 
\raisebox{-0.5em}{  1.070050} \vspace{-0.3em} & 
  1.070047 &
  1.070038 &
  1.070095 &
  1.069775 &
  1.100577 &
  1.049610 &
  1.077231 &
  1.090571 &
  1.114848 &
  1.121840 &
  1.042448 &
  1.081022 &
  1.071756 \\
 & 
 & 
  1.070054 &
  1.070062 &
  1.070005 &
  1.070326 &
  1.039524 &
  1.090491 &
  1.062870 &
  1.049529 &
  1.025252 &
  1.018260 &
  1.097653 &
  1.059079 &
  1.068345 \\
\hline
\end{tabular}
\caption{As in Tab.~\ref{TabA03} but for our "standard" NNLO ($\overline{\rm MS}$) fit.\label{TabA09}}
\end{sidewaystable}

%% file: splittings.tex
\newcommand{\n}{n}
\newcommand{\ns}{n^2}
\newcommand{\nc}{n^3}
\newcommand{\nfo}{n^4}
\newcommand{\nfi}{n^5}
\newcommand{\nsi}{n^6}
\newcommand{\nse}{n^7}
\newcommand{\nei}{n^8}
\newcommand{\nni}{n^9}
\newcommand{\npOne}{\left(n + 1\right)}
\newcommand{\npOnes}{\left(n + 1\right)^2}
\newcommand{\npOnec}{\left(n + 1\right)^3}
\newcommand{\nmOne}{\left(n - 1\right)}
\newcommand{\nmOnes}{\left(n - 1\right)^2}
\newcommand{\npTwo}{\left(n + 2\right)}
\newcommand{\npTwos}{\left(n + 2\right)^2}
\newcommand{\npTwoc}{\left(n + 2\right)^3}
\newcommand{\npThree}{\left(n + 3\right)}
\newcommand{\npThrees}{\left(n + 3\right)^2}
\newcommand{\npFour}{\left(n + 4\right)}
\newcommand{\SOne}{S_1(n)}
\newcommand{\SOnes}{S_1^2(n)}
\newcommand{\STwo}{S_2(n)}
\newcommand{\SThree}{S_3(n)}
\newcommand{\SFour}{S_4(n)}
\newcommand{\SPpTwo}{S^\prime_{2,+}({n\over2})}
\newcommand{\SPmTwo}{S^\prime_{2,-}({n\over2})}
\newcommand{\SPpThree}{S^\prime_{3,+}({n\over2})}
\newcommand{\SPmThree}{S^\prime_{3,-}({n\over2})}
\newcommand{\STp}{\tilde{S}_+(n)}
\newcommand{\STm}{\tilde{S}_-(n)}
\newcommand{\SPpmTwo}{S^\prime_{2,\pm}({n\over2})}
\newcommand{\SPpmThree}{S^\prime_{3,\pm}({n\over2})}
\newcommand{\STpm}{\tilde{S}_\pm(n)}
\newcommand{\AZero}{\left(-S_1(n-1)\right)}
\newcommand{\BOne}{\left(-\frac{S_1(n)}{n}\right)}
\newcommand{\BOneM}{\left(-\frac{S_1(n-1)}{n-1}\right)}
\newcommand{\BOneOne}{\left(-\frac{S_1(n+1)}{n+1}\right)}
\newcommand{\BOneTwo}{\left(-\frac{S_1(n+2)}{n+2}\right)}
\newcommand{\BTwo}{\frac{S_1^2(n)+S_2(n)}{n}}
\newcommand{\BTwoM}{\frac{S_1^2(n-1)+S_2(n-1)}{n-1}}
\newcommand{\BTwoOne}{\frac{S_1^2(n+1)+S_2(n+1)}{n+1}}
\newcommand{\BThree}{\left(-\frac{S_1^3(n)+3S_1(n)S_2(n)+2S_3(n)}{n}\right)}
\newcommand{\BThreeOne}{\left(-\frac{S_1^3(n+1)+3S_1(n+1)S_2(n+1)+2S_3(n+1)}{n+1}\right)}
\newcommand{\BFour}{\left(\frac{S_1^4(n)+6S_1^2S_2(n)+8S_1(n)S_3(n)+3S_2^2(n)+4S_4(n)}{n}\right)}
\newcommand{\CZero}{\frac{1}{n}}
\newcommand{\COne}{\frac{1}{n+1}}
\newcommand{\CTwo}{\frac{1}{n+2}}
\newcommand{\CThree}{\frac{1}{n+3}}
\newcommand{\CFour}{\frac{1}{n+4}}
\newcommand{\CM}{\frac{1}{n-1}}
\newcommand{\DOne}{\left(-\frac{1}{n^2}\right)}
\newcommand{\DOneM}{\left(-\frac{1}{(n-1)^2}\right)}
\newcommand{\DOneOne}{\left(-\frac{1}{(n+1)^2}\right)}
\newcommand{\DTwo}{\frac{2}{n^3}}
\newcommand{\DTwoOne}{\frac{2}{(n+1)^3}}
\newcommand{\DThree}{\left(-\frac{6}{n^4}\right)}
\newcommand{\DThreeOne}{\left(-\frac{6}{(n+1)^4}\right)}
\newcommand{\DFour}{\frac{24}{n^5}}
\newcommand{\DFourOne}{\frac{24}{(n+1)^5}}
\newcommand{\EOne}{\left(\frac{S_1(n)}{n^2} + \frac{S_2(n)-\zeta(2)}{n}\right)}
\newcommand{\EOneOne}{\left(\frac{S_1(n+1)}{(n+1)^2} + \frac{S_2(n+1)-\zeta(2)}{n+1}\right)}
\newcommand{\ETwo}{\left(-2\frac{S_1(n)}{n^3} + 2\frac{\zeta(2)-S_2(n)}{n^2} - 2\frac{S_3(n)-\zeta(3)}{n}\right)}
\renewcommand{\arraystretch}{1.3}
\begin{equation}
\begin{array}{l}
P_{\rm qq}^{(0)} = {4\over3}\bigg\{3 + \frac{2}{\n\npOne} - 4\SOne\bigg\}, \qquad \quad
P_{\rm qg}^{(0)} = 2n_F\bigg\{\frac{\ns + \n + 2}{\n\npOne\npTwo}\bigg\} \hspace{7em}
\end{array}
\end{equation}

\begin{equation}
\begin{array}{l}
P_{\rm gq}^{(0)} = {8\over3}\bigg\{\frac{\ns + \n + 2}{\nmOne\n\npOne}\bigg\},\qquad
P_{\rm gg}^{(0)} = 3\bigg\{{11\over3} + \frac{4}{\n\nmOne} + \frac{4}{\npOne\npTwo} - 4\SOne\bigg\} - {2\over3}n_F \hspace{1em}
\end{array}
\end{equation}

\begin{equation}
\begin{array}{l}
P_{\rm NS}^{\pm(1)} =
-{16\over9}\bigg\{
  8\SOne\frac{2\n + 1}{\ns\npOnes}
+ 8\left[2\SOne -\frac{1}{\n\npOne}\right]\left[\STwo - \SPpmTwo\right]
+ 12\STwo \\[0.1em] \hspace{3em}
+ 32\STpm
- 4\SPpmThree
- {3\over2}
- 4\left[\frac{3\nc + \ns - 1}{\nc\npOnec}\right]
\mp 8\left[\frac{2\ns + 2\n + 1}{\nc\npOnec}\right]
\bigg\} \\[0.1em] \hspace{3em}
- 4\bigg\{
+ {268\over9}\SOne
- 4\left[2\SOne - \frac{1}{\n\npOne}\right]\left[2\STwo - \SPpmTwo\right]
- {44\over3}\STwo \\[0.1em] \hspace{3em}
- 16\STpm
+ 2\SPpmThree
- {17\over6}
- {2\over9}\left[\frac{151\nfo + 236\nc + 88\ns + 3\n + 18}{\nc\npOnec}\right]
+ 4\left[\frac{2\ns + 2\n + 1}{\nc\npOnec}\right]
\bigg\} \\[0.1em] \hspace{3em}
- {2\over3}n_F\bigg\{
- {80\over9}\SOne
+ {16\over3}\STwo
+ {2\over3}
+ {8\over9}\left[\frac{11\ns + 5\n - 3}{\ns\npOnes}\right]
\bigg\}
\end{array}
\end{equation}

\begin{equation}
\begin{array}{l}
P_{\rm PS}^{(1)} = + {16\over3}\n_F\bigg\{\frac{5\nfi + 32\nfo + 49\nc + 38\ns + 28\n + 8}{\nmOne\nc\npOnec\npTwos}\bigg\} \hspace{15.8em}
\end{array}
\end{equation}

\begin{equation}
\begin{array}{l}
P_{\rm qg}^{(1)} =
+ 6n_F\bigg\{
  2\left[\STwo - \SPpTwo - \SOnes\right]\left[\frac{\ns + n + 2}{\n\npOne\npTwo}\right]
+ 8\SOne\left[\frac{2\n + 3}{\npOnes\npTwos}\right] \\[0.1em] \hspace{3em}
+ 2\left[\frac{\nni + 6\nei + 15\nse + 25\nsi + 36\nfi + 85\nfo + 128\nc + 104\ns + 64\n + 16}{\nmOne\nc\npOnec\npTwoc}\right]
\bigg\} \\[0.1em] \hspace{3em}
+ {8\over3}n_F\bigg\{
  \left[2\SOnes - 2\STwo + 5\right]\left[\frac{\ns + \n + 2}{\n\npOne\npTwo}\right]
- 4\frac{\SOne}{\ns} \\[0.1em] \hspace{3em}
+ \frac{11\nfo + 26\nc + 15\ns + 8\n + 4}{\nc\npOnec\npTwo}
\bigg\}
\end{array}
\end{equation}

\fontsize{11}{14.8}
\selectfont

\begin{equation}
\begin{array}{l}
P_{\rm gq}^{(1)} =
+{32\over9}\bigg\{
  \left[-2\SOnes + 10\SOne - 2\STwo\right]\left[\frac{\ns + n + 2}{\nmOne\n\npOne}\right]
- 4\frac{\SOne}{\npOnes} \\[0.1em] \hspace{3em}
- \frac{12\nsi + 30\nfi + 43\nfo + 28\nc - \ns - 12\n -4}{\nmOne\nc\npOnec}
\bigg\} \\[0.1em] \hspace{3em}
+ 16\bigg\{
 \left[\SOnes + \STwo - \SPpTwo\right]\left[\frac{\ns + n + 2}{\nmOne\n\npOne}\right]
- \SOne\left[\frac{17\nfo + 41\ns - 22\n - 12}{3\nmOnes\ns\npOne}\right] \\[0.1em] \hspace{3em}
+ \frac{109\nni + 621\nei + 1400\nse + 1678\nsi + 695\nfi - 1031\nfo - 1304\nc - 152\ns + 432\n + 144}{9\nmOnes\nc\npOnec\npTwos}
\bigg\} \\[0.1em] \hspace{3em}
+ {32\over9}n_F\bigg\{
  \left[\SOne - {8\over3}\right]\left[\frac{\ns + n + 2}{\nmOne\n\npOne}\right]
+ \frac{1}{\npOnes}
\bigg\}
\end{array}
\end{equation}

\begin{equation}
\begin{array}{l}
P_{\rm gg}^{(1)} =
- {3\over2}n_F\bigg\{
- {80\over9}\SOne
+ {16\over3}
+ {8\over9}\left[\frac{38\nfo + 76\nc + 94\ns + 56\n + 12}{\nmOne\ns\npOnes\npTwo}\right]
\bigg\} \\[0.1em] \hspace{3em}
- {2\over3}n_F\bigg\{
  4
+ 8\left[\frac{2\nsi + 4\nfi + \nfo - 10\nc - 5\ns -4\n - 4}{\nmOne\nc\npOnec\npTwo}\right]
\bigg\} \\[0.1em] \hspace{3em}
- 9\bigg\{
  {268\over9}\SOne
+ 32\SOne\left[\frac{2\nfi + 5\nfo + 8\nc + 7\ns - 2\n - 2}{\nmOnes\ns\npOnes\npTwos}\right]
- {32\over3} \\[0.1em] \hspace{3em}
+ 16\SPpTwo\left[\frac{\ns + n + 1}{\nmOne\n\npOne\npTwo}\right]
- 8\SOne\SPpTwo
+ 16\STp
- 2\SPpThree \\[0.1em] \hspace{3em}
- {2\over9}\left[\frac{457\nni + 2742\nei + 6040\nse + 6098\nsi + 1567\nfi - 2344\nfo - 1632\nc + 560\ns + 1488\n + 576}{\nmOnes\nc\npOnec\npTwoc}\right]
\bigg\}
\end{array}
\end{equation}

\begin{equation}
\begin{array}{l}
P_{\rm NS}^{+(2)} =
+ 1174.898  \AZero
+ 1295.384
+ 714.1  \BOne
- 522.1  \CThree \\[0.1em] \hspace{3em}
+ 243.6  \CTwo 
- 3135  \COne
+ 1641.1  \CZero
+ 1258  \DOne
+ 294.9  \DTwo \\[0.1em] \hspace{3em}
+ {800\over27}  \DThree
+ {128\over81}  \DFour
+ 563.9  \EOne \\[0.1em] \hspace{3em}
+ 256.8  \ETwo \\[0.1em] \hspace{3em}
+ n_F\bigg\{
- 183.187  \AZero  
- 173.924
- {5120\over81}  \BOne 
+ 44.79  \CThree \\[0.1em] \hspace{3em}
+ 72.94  \CTwo 
+ 381.1  \COne 
- 197.0  \CZero
- 152.6  \DOne
- {2608\over81}  \DTwo \\[0.1em] \hspace{3em}
- {192\over81}  \DThree
- 56.66  \EOne
- 1.497  \DThreeOne
\bigg\} \\[0.1em] \hspace{3em}
- n_F^2\frac{32}{3}\bigg\{
+ {17\over72}
- {2\over27}  S_1(n)
- {10\over27}  S_2(n)
+ {2\over9}  S_3(n)
- \left[\frac{12n^4 + 2n^3 - 12n^2 - 2n + 3}{27n^3(n+1)^3}\right]
\bigg\}
\end{array}
\end{equation}

\begin{equation}
\begin{array}{l}
P_{\rm NS}^{-(2)} =
+ 1174.898  \AZero
+ 1295.470
+ 714.1  \BOne
- 433.2  \CThree \\[0.1em] \hspace{3em}
+ 297.0  \CTwo
- 3505.\COne
+ 1860.2  \CZero
+ 1465.2  \DOne
+ 399.2  \DTwo \\[0.1em] \hspace{3em}
+ {320\over9}  \DThree
+ {116\over81}  \DFour
+ 684.0  \EOne \\[0.1em] \hspace{3em}
+ 251.2  \ETwo \\[0.1em] \hspace{3em}
+ n_F\bigg\{
- 183.187  \AZero
- 173.933
- {5120\over81}  \BOne
+ 34.76  \CThree \\[0.1em] \hspace{3em}
+ 77.89  \CTwo
+ 406.5  \COne
- 216.62  \CZero
- 172.69  \DOne
- {3216\over81}  \DTwo \\[0.1em] \hspace{3em}
- {256\over81}  \DThree
- 65.43  \EOne
- 1.136  \DThreeOne
\bigg\} \\[0.1em] \hspace{3em}
- n_F^2\frac{32}{3}\bigg\{
+ {17\over72}
- {2\over27}  S_1(n)
- {10\over27}  S_2(n)
+ {2\over9}  S_3(n)
- \left[\frac{12n^4 + 2n^3 - 12n^2 - 2n + 3}{27n^3(n+1)^3}\right]
\bigg\}
\end{array}
\end{equation}

\begin{equation}
\begin{array}{l}
P_{\rm NS}^{S(2)} = n_F\bigg\{
- 163.9  \left[\BOneM-\BOne\right]
- 7.208  \left[\BOneOne-\BOneTwo\right] \\[0.1em] \hspace{3em}
+ 4.82  \left[\CThree-\CFour\right]
- 43.12  \left[\CTwo-\CThree\right]
+ 44.51  \left[\COne-\CTwo\right] \\[0.1em] \hspace{2.89em}
+ 151.49  \left[\CZero-\COne\right]
+ 178.04  \DOne
+ 6.892  \DTwo
- {40\over27}  \left[2\DThree - \DFour\right] \\[0.1em] \hspace{3em}
- 173.1  \EOne
+ 46.18  \ETwo
\bigg\}
\end{array}
\end{equation}

\fontsize{11}{16}
\selectfont

\renewcommand{\arraystretch}{1.7}

\begin{equation}
\begin{array}{l}
P_{\rm PS}^{(2)} =
+ n_F\bigg\{
- {3584\over27}  \left[\DOneM-\DOne\right]
- 506  \left[\CM-\CZero\right]  \\[0.1em] \hspace{3em}
+ {160\over27}  \left[\DFour-\DFourOne\right] 
- {400\over9}  \left[\DThree-\DThreeOne\right]  \\[0.1em] \hspace{3em}
+ 131.4  \left[\DTwo-\DTwoOne\right]
- 661.6  \left[\DOne-\DOneOne\right] \\[0.1em] \hspace{3em}
- 5.926  \left[\BThree-\BThreeOne\right] \\[0.1em] \hspace{3em}
- 9.751  \left[\BTwo-\BTwoOne\right]
- 72.11  \left[\BOne-\BOneOne\right] \\[0.1em] \hspace{3em}
+ 177.4  \left[\CZero-\COne\right]
+ 392.9  \left[\COne-\CTwo\right]
- 101.4  \left[\CTwo-\CThree\right] \\[0.1em] \hspace{3em}
- 57.04  \left[\EOne-\EOneOne\right]
\bigg\} \\[0.1em] \hspace{3em}
+ n_F^2\bigg\{
+ {256\over81}  \left[\CM-\CZero\right]
+ {32\over27}  \left[\DThree-\DThreeOne\right]  \\[0.1em] \hspace{3em}
+ 17.89  \left[\DTwo-\DTwoOne\right]
+ 61.75  \left[\DOne-\DOneOne\right]  \\[0.1em] \hspace{3em}
+ 1.778  \left[\BTwo-\BTwoOne\right]
+ 5.944  \left[\BOne-\BOneOne\right]  \\[0.1em] \hspace{3em}
+ 100.1  \left[\CZero-\COne\right]
- 125.2  \left[\COne-\CTwo\right]
+ 49.26  \left[\CTwo-\CThree\right]  \\[0.1em] \hspace{3em}
- 12.59  \left[\CThree-\CFour\right]
- 1.889  \left[\EOne-\EOneOne\right]
\bigg\}
\end{array}
\end{equation}

\fontsize{11}{15}
\selectfont

\begin{equation}
\begin{array}{l}
P_{qg}^{(2)} =
+ n_F\bigg\{
- {896\over3}  \DOneM
- 1268.3  \CM
+ {536\over27}  \DFour
- {44\over3}  \DThree \\[0.1em] \hspace{3em}
+ 881.5  \DTwo
+ 424.9  \DOne
+ {100\over27}  \BFour \\[0.1em] \hspace{3em}
- {70\over9}  \BThree
- 120.5  \BTwo
+ 104.42  \BOne \\[0.1em] \hspace{3em}
+ 2522  \CZero
- 3316   \COne
+ 2126   \CTwo
+ 1823   \EOne \\[0.1em] \hspace{3em}
- 25.22  \ETwo
- 252.5  \DThreeOne
\bigg\} \\[0.1em] \hspace{3em}
+ n_F^2\bigg\{
+ {1112\over243}  \CM
- {16\over9}  \DFour
- {376\over27}  \DThree
- 90.8  \DTwo
- 254.0  \DOne \\[0.1em] \hspace{3em}
+ {20\over27}  \BThree
+ {200\over27}  \BTwo
- 5.496  \BOne
- 252.0  \CZero \\[0.1em] \hspace{3em}
+ 158.0  \COne
+ 145.4  \CTwo
- 139.28  \CThree
- 53.09  \EOne \\[0.1em] \hspace{3em}
- 80.616  \ETwo
- 98.07  \DTwoOne
+ 11.70  \DThreeOne
\bigg\}
\end{array}
\end{equation}

\begin{equation}
\begin{array}{l}
P_{gq}^{(2)} =
+ 1189.3  \DOneM
+ 6163.1  \CM
- {4288\over81}   \DFour
+ {1568\over9}  \DThree
- 1794  \DTwo \\[0.1em] \hspace{3em}
+ 4033  \DOne
+ {400\over81}  \BFour \\[0.1em] \hspace{3em}
+ {2200\over27}  \BThree
+ 606.3  \BTwo
+ 2193  \BOne \\[0.1em] \hspace{3em}
- 4307  \CZero
+ 489.3  \COne
+ 1452  \CTwo
+ 146  \CThree \\[0.1em] \hspace{3em}
- 447.3  \ETwo
- 972.9  \DTwoOne \\[0.1em] \hspace{3em}
+ n_F\bigg\{
+ 71.082  \DOneM
- 46.41  \CM
+ {128\over27}  \DFour
+ {704\over81}  \DThree \\[0.1em] \hspace{3em}
+ 20.39  \DTwo
+ 174.8  \DOne
- {400\over81}  \BThree \\[0.1em] \hspace{3em}
- 68.069  \BTwo
- 296.7  \BOne
- 183.8  \CZero
+ 33.35  \COne \\[0.1em] \hspace{3em}
- 277.9  \CTwo
+ 108.6  \DTwoOne
- 49.68  \EOne
\bigg\} \\[0.1em] \hspace{3em}
+ n_F^2\frac{1}{27}\bigg\{
+ 64\left[- \CM + \CZero + 2  \COne\right] \\[0.1em] \hspace{3em}
+ 320\left[\BOneM - \BOne + 0.8  \BOneOne\right] \\[0.1em] \hspace{3em}
+ 96\left[\BTwoM - \BTwo + 0.5  \BTwoOne\right]
\bigg\}
\end{array}
\end{equation}

\begin{equation}
 \begin{array}{l}
P_{gg}^{(2)} =
+ 2675.8  \DOneM
+ 14214  \CM
- 144  \DFour
+ 72  \DThree
- 7471  \DTwo 
+ 274.4  \DOne \\[0.1em] \hspace{3em}
- 20852  \CZero
+ 3968  \COne
- 3363  \CTwo
+ 4848  \CThree
+ 7305  \EOne \\[0.1em] \hspace{3em}
+ 8757  \ETwo 
+ 3589  \BOne 
+ 4425.894 \AZero \\[0.1em] \hspace{3em}
+ 2643.521  
+ n_F\bigg\{
+ 157.27  \DOneM 
+ 182.96  \CM
+ {512\over27}  \DFour
+ {832\over9}  \DThree \\[0.1em] \hspace{3em}
+ 491.3  \DTwo
+ 1541  \DOne 
- 350.2  \CZero 
+ 755.7  \COne
- 713.8  \CTwo
+ 559.3  \CThree \\[0.1em] \hspace{3em}
+ 26.15  \EOne
- 808.7  \ETwo \\[0.1em] \hspace{3em}
- 320  \BOne
- 528.723
- 412.172  \AZero
\bigg\} 
+ n_F^2\bigg\{
- {680\over243}  \CM
- {32\over27}  \DThree \\[0.1em] \hspace{3em}
+ 9.680  \DTwo
- 3.422  \DOne
- 13.878  \CZero 
+ 153.4  \COne
- 187.7  \CTwo 
+ 52.75  \CThree \\[0.1em] \hspace{3em}
- 115.6  \EOne
+ 85.25  \EOneOne \\[0.1em] \hspace{3em}
- 63.23  \ETwo
+ 6.4630
- {16\over9}  \AZero
\bigg\}
\end{array}
\end{equation}